%% file: paper.tex
\documentclass[prd]{revtexmy}

\usepackage{bm}
\usepackage{epsfig}
\usepackage{amssymb}
\usepackage{amsmath}
\usepackage{feynarts}
\usepackage{slashbox}
\usepackage{calrsfs}
\usepackage{multirow}

\begin{document}

\def\be{\begin{equation}}
\def\ee{\end{equation}}
\def\bea{\begin{eqnarray}}
\def\eea{\end{eqnarray}}
\def\cpi{$\pi^\pm$}
\def\npi{$\pi^0$}
\def\cka{$K^\pm$}
\def\nka{$K_S^0$}
\def\pr{$p/\overline{p}$}
\def\lam{$\Lambda/\overline{\Lambda}$}

\newcommand\ds{\displaystyle}
\def\nSGL{\line(3,1){15.5} \hspace{-0.53cm} {\rm SGL}}
\def\nDL{\line(2,1){12.5} \hspace{-0.38cm} {\rm DL}}

\title{The hadronization of partons
}
\author{S.\ Albino}
\affiliation{{II.} Institut f\"ur Theoretische Physik, Universit\"at Hamburg,\\
             Luruper Chaussee 149, 22761 Hamburg, Germany}
\date{\today}
\begin{abstract}
We review the description of inclusive single unpolarized light hadron production using fragmentation functions
in the framework of the factorization theorem.
We summarize the factorization of quantities into perturbatively calculable quantities
and these universal fragmentation functions,
and then discuss some improvements beyond the standard fixed order approach.
We discuss the extraction of fragmentation functions for light charged (\cpi, \cka\ and \pr) and neutral
(\nka\ and \lam) hadrons using these theoretical tools through
global fits to experimental data from reactions at various colliders, in particular
from accurate $e^+ e^-$ reactions at LEP, and the subsequent successful
predictions of other experimental data, such as data gathered at HERA, the Tevatron and RHIC 
from these fitted fragmentation functions as allowed by factorization universality.
These global fits also impose competitive constraints on $\alpha_s(M_Z)$.
Emphasis is placed on the need for accurate data from $ep$ and $pp(\overline{p})$ reactions
in which the hadron species is identified
in order to constrain the separate fragmentation functions of each quark flavour and hadron species.
\end{abstract}


\maketitle

\pagebreak
\tableofcontents
\pagebreak

\section{Introduction \label{intro}}

Fragmentation functions (FFs) constitute one of the most important free inputs required for a comprehensive description
of most collider processes to which perturbative QCD is applicable, being 
a necessary ingredient in any sufficiently complete calculation of processes involving detected hadrons in the final state. 
They quantify the hadronization of quarks and gluons which must eventually occur in every process
in which hadrons are produced.
While parton distribution functions (PDFs), another of the important free inputs, are relevant for collisions involving
at least one hadron, most importantly at present for $pp$ collisions at the LHC, 
FFs are relevant in principle for all collisions, even those without any initial state hadrons such as
electron-positron collisions at the future ILC. 
Furthermore, while knowledge and application of PDFs is limited to the types of hadrons that can be practically used in the initial
state, being almost exclusively nucleons,
FFs can be constrained by, and/or used for the predictions of, measurements of the inclusive productions of
neutral and charged hadron species ranging from the almost massless to the very heavy. 
Such a large range of processes provides a large range of information on hadronization,
and hence provides an important contribution to our understanding of non perturbative physics in general,
and allows for a particularly incontrovertible phenomenological determination of the applicability
and limitations of the various approximations used in the context of QCD factorization.
Such data are also sufficiently accurate
to allow for competitive extractions of $\alpha_s(M_Z)$, the strong coupling constant of QCD evolved for convenience to the $Z$-pole
mass $M_Z$ and the remaining of the most important free inputs, 
which improves the accuracy of perturbative QCD calculations in general and imposes constraints on new physics.
We note here that while other inputs such as higher twist, multi-hadron FFs, 
fracture functions, and so on become important in certain kinematic limits, FFs
are always necessary for a complete description of inclusive hadron production.

In this article, we review the progress in understanding the hadronization of
partons, embodied in FFs, and their application to inclusive hadron production. 
The concept of fragmentation was first introduced by Feynman and Field in 1977 \cite{Field:1976ve} to explain 
the limited transverse momenta and energy fraction scaling of hadrons in jets produced in $e^+ e^-$ collisions,
as well as the presence of a few high transverse momentum ($p_T \gtrsim 2$) GeV hadrons in hadron-hadron collisions.
In the latter case, the collision of a pointlike constituent particle of one hadron with that of another produces a pair of
particles whose directions of motion are opposite to one another but at any angle to the parent particles, including large angles.
These elementary particles must eventually hadronize to produce the high $p_T$ hadrons.
Intuitively, any inclusive single hadron production processes may be predicted by first calculating the equivalent
partonic process (i.e.\ replacing the produced hadron by a parton and the inclusive sum over hadron final states
by partonic ones), then allowing the produced parton to hadronize.
This {\it parton model} is made more precise by the factorization theorem.

Our discussion will be limited to inclusive single unpolarized hadron production, 
being better understood than its polarized counterpart.
Final state partons will on average hadronize to hadrons of all species 
that are not kinematically forbidden by the particular reaction,
but, of all the charged hadrons, partons will mostly hadronize to the 3 lightest ones, being
\cpi, \cka\ and \pr, so their FFs are the most phenomenologically well constrained.
(For a concise summary of the properties of some known mesons and baryons, see Ref.\ \cite{wiki}.)
The species of neutral meson and neutral baryon most likely to be produced in the hadronization process are $\pi^0$,
and $n/\overline{n}$ respectively, but their FFs are accurately approximated by the FFs for \cpi\  and \pr
due to SU(2) (nuclear) isospin symmetry between the $u$ and $d$ quark flavours.
The next lightest species of meson and baryon is \nka\ and \lam\ respectively,
whose FFs have therefore also been greatly studied.
($K^0_L$ is not usually observed in experiment because it takes a long time to decay into charged particles.)
Together with $\eta$, $f_0$, $\rho^\pm$, $\rho$, $\omega$, $K^{*\pm}$, $\eta'$, $a_0$ and $\phi$ mesons, 
whose FFs are unfortunately either unknown or rather poorly known at present, 
the particles mentioned in this paragraph complete the list of known hadrons which have a mass less than or equal to the 
\lam\ mass and which are the most copiously produced in hadron production, 
and thus measurements of their production lead to a rather comprehensive picture of the hadronic final state.
In this review we shall focus on the productions of these particles only, 
with the exceptions of the productions of those mesons just listed.

Much of the techniques for global fitting of PDFs can be carried over to global fitting of FFs,
for example the treatment of systematic errors on the experimental data and the propagation
of experimental errors to the FFs, as well as the techniques used for the minimization of chi-squared.
On the theoretical side, much of the formalism of both the {\it fixed order} (FO) calculations and its improvement
via e.g.\ large $x$ resummation and incorporation of heavy quarks and their masses, is also very similar.

To date, the subject of fragmentation has been important in various areas of phenomenology \cite{:2008afa}, 
and we give an incomplete list of examples:
Phenomenological constraints on the nucleon PDFs for polarized quarks and antiquarks separately, 
which are important for the determination of the source of nucleon spin, can be obtained from 
measurements of semi-inclusive DIS with polarized initial state particles for the inclusive production of single unpolarized light
charged hadrons using knowledge of the FFs for these particles \cite{deFlorian:1997ie}.
Valence quark PDFs can be similarly extracted, but from data 
for which the initial state particles are unpolarized \cite{Gronau:1973gc}.
In both cases, the differences between FFs for positively and negatively charged particles,
are required, which are relatively poorly constrained by data from e.g.\ $pp$ reactions,
and not by accurate $e^+ e^-$ reaction data.
Polarized $pp(\overline{p})$ reaction data from e.g.\ RHIC can also impose constraints on the nucleon PDF 
for the polarized gluon \cite{deFlorian:2002az},
and measurements of photoproduction in polarized $ep$ reaction at e.g.\ the proposed eRHIC \cite{Deshpande:2005wd} 
or LHeC \cite{Dainton:2006wd} colliders can impose
constraints on the photon PDF for polarized partons \cite{Jager:2003vy}. 
FFs provide a consistency check on transverse momentum dependent FFs, 
which replace FFs when the transverse momentum of the produced hadron
is measured in addition to the longitudinal,
since formally the FFs are reproduced by integrating them over the transverse momentum.
The suppression of $\pi^0$ production in heavy ion collisions (Au Au) relative to $pp$ collisions \cite{Adler:2003qi} 
measured by the PHENIX collaboration at RHIC, 
and to be further investigated by the ALICE, ATLAS and CMS collaborations at the LHC, provides 
information on the much anticipated scenario of a quark-gluon plasma (QGP) 
that filled the universe the first ten millionths of a second
after the big bang and created the primordial matter.
Such studies may therefore contribute to our understanding of both cosmology and the physics of non perturbative QCD.
Photons produced by partonic fragmentation \cite{Koller:1978kq,Gluck:1992zx} 
contribute significantly to the photonic background of various direct photon signals in DIS,
such as that of Higgs boson production from hadrons.
Fragmentation of a quark to a lepton pair \cite{Braaten:2001sz} or, equivalently, to a virtual photon \cite{Qiu:2001nr},
is perturbatively calculable if the invariant mass is much greater than $\Lambda_{\rm QCD}$, 
in contrast to FFs for hadrons and the photon,
so measurement of the polarization of the virtual photon that decays to the lepton pair of the Drell-Yan cross section
can be used to test models of the formation of $J/\psi$ particles,
whose production from fragmentation is otherwise similar to that of the virtual photon.
In Ref.\ \cite{Hirai:2007ww} it has been proposed to identify exotic hadrons such as tetraquarks
by identifying properties of FFs which are typical for quarks whose flavour is favoured, i.e.\ 
that are of the same flavour as any of the produced hadron's valence quarks.
Some studies of the fragmentation of squarks and gluinos into hadrons have been performed in Ref.\ \cite{Vlassopulos:1985ph} 
and much more recently into supersymmetric hadrons in Ref.\ \cite{Chang:2006ns}, which may be relevant at the LHC.
The supersymmetric extension of the DGLAP evolution of FFs for light charged hadrons has been studied 
in Ref.\ \cite{Fodor:2000za} in the context of ultra high energy cosmic rays.
FFs for light charged hadrons are required for the calculation of hadronic signatures 
of black hole production at the LHC \cite{Mocioiu:2003gi}.

The rest of this review is structured as follows. 
The basic results of the QCD factorization theorem, which forms the starting point of all calculations of inclusive
hadron production,
are given in section \ref{resoffacttheorem},
in particular the separation from the overall cross section of the process dependent parts, 
which can be calculated using the FO approach to perturbation theory.
(The derivation of these results is outlined in appendix \ref{outlinederivfactTheo}.)
Then in section \ref{epemXS} we discuss the extraction of FFs from accurate $e^+ e^-$ data and from
various well motivated non perturbative assumptions.
The ability of calculations using these fitted FFs to reproduce measurements is studied in section \ref{predfromFFs}.
Then we turn to more recent progress:
Improvements to the standard FO approach that have not been incorporated into nearly all global fits are given 
in section \ref{impstand}, namely hadron mass effects and large $x$ resummation.
The treatment of experimental errors and their propagation to fitted quantities, 
which has been rather comprehensively implemented in PDF fits but to a somewhat lesser degree in FF, 
is discussed in section \ref{treaterrs}.
The 3 most recent global fits, in which the implementation of these improvements can be found,
are discussed and compared in section \ref{currglobalfits}.
Finally, in section \ref{sglres} we examine the improvement of the standard FO approach at small $x$ by resummation
of {\it soft gluon logarithms}, which has so far been successfully tested at LO, 
and we consider what is needed for a full treatment of soft gluon logarithms to NLO.
In section \ref{outlook} we predict the future experimental and theoretical progress of FF extraction,
and give a summary of the current progress in section \ref{summary}.
The LO splitting functions are given in appendix \ref{LOsplitfunc} for reference,
the relevant Mellin space formulae in appendix \ref{mellinspace},
and a summary of all data that can be reasonably reliably calculated and which can be used to provide constraints on FFs 
in appendix \ref{sumofexp}.
Some of these issues have also been discussed very recently in Ref.\ \cite{Arleo:2008dn},
with an aim towards the modification of fragmentation in QCD media.

\section{Results of the QCD factorization theorem \label{resoffacttheorem}}

Intuitively, the detected hadron $h$ of inclusive single hadron production events 
is produced in the jet formed by a parton $a$ produced at a short distance $1/M_f$, and carries away a fraction $z$ of 
the momentum of $a$.
This ``probing scale'' $M_f$ should be of order of the {\it energy scale} $E_s$ of the process,
whose precise choice is somewhat conventional.
The probability density in $z$ for this to take place is given by the FF $D_a^h(z,M_f^2)$.
The hadronic cross section is then equal to the cross section with $h$ replaced by $a$, hereafter referred to as
the partonic cross section, weighted with
the FF of $a$ and summed and integrated over all degrees of freedom such as $a$ and $z$, 
i.e.\ eq.\ (\ref{genformofinchadprodfromffs}) below.
According to the factorization theorem, which follows from QCD in a model independent way, 
the leading twist component of any inclusive single hadron production cross section takes this intuitive form,
where the FFs are process independent or {\it universal} among different initial states.
The factorization theorem also asserts that all processes in the partonic cross section that have energy scale 
below the {\it factorization scale} $M_f$
factor out of the partonic cross section and are accounted for by the FFs, and that
the resulting factorization scale dependence of these FFs may be calculated perturbatively.
In this section, we highlight the results of the factorization theorem.
An outline of the formal derivation of the factorization theorem is given 
in appendix \ref{outlinederivfactTheo} for the interested reader.

\subsection{Factorized cross sections \label{factXS}}

\begin{figure}[h!]
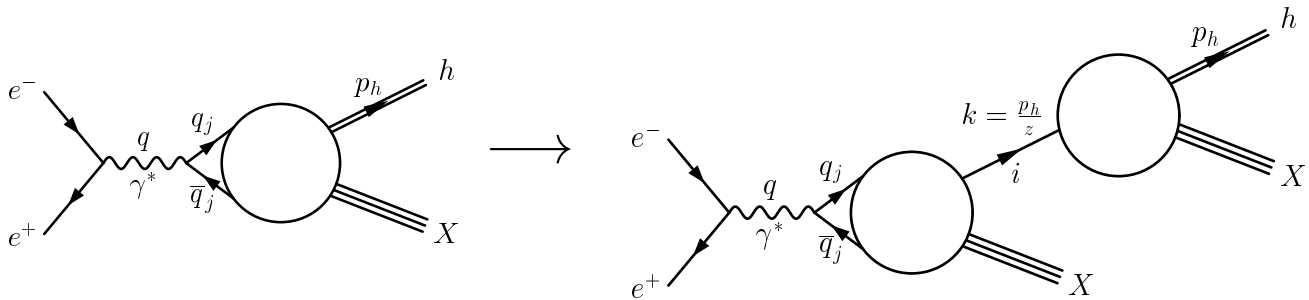

\parbox{.35\linewidth}{
\begin{center}
\includegraphics[width=6cm]{epem.epsi}
\end{center}
}\parbox{.09\linewidth}{\huge $\longrightarrow$}
\parbox{.55\linewidth}{
\begin{center}
\includegraphics[width=9cm]{epemfact.epsi}
\end{center}}
\caption{Amplitude for the process $e^+ e^- \rightarrow \gamma^* \rightarrow q_j +\overline{q}_j \rightarrow h+X$ (left).
Time flows from left to right.
The virtual photon momentum $q$ obeys $q^2=s$, where $\sqrt{s}$ is the center-of-mass energy.
The blob contains all QCD processes consistent with its external propagators or legs.
Also shown is a schematic illustration of the leading twist component of this process after factorization (right),
involving an intermediate real parton $i$ that fragments to the detected hadron $h$. 
The top right blob and its external lines represents the FF $D_i^h$, while the rest of the diagram
is the equivalent partonic cross section, i.e.\ with $h$ replaced by $i$.\label{epem}}
\end{figure}

In general, as well as $E_s$, an inclusive single hadron production process depends on 
the fraction $x$ of the available momentum or energy carried away by the detected hadron $h$.
For example, in $e^+ e^- \rightarrow \gamma^* \rightarrow h+X$, whose kinematics are specified in Fig.\ \ref{epem} (left),
$x=2p_h /\sqrt{s}$ and $E_s=\sqrt{s}$.
We will assume for now that it is reasonable to neglect the effect of the
hadron mass $m_h$, which is of $O(m_h^2/p_h^2)$ when $m_h \ll p_h$, where $p_h$ is the momentum of the detected hadron.
When the effect of hadron mass, to be discussed in section \ref{hadmass}, is taken into account in the calculations, 
the scaling variable $x$ of the factorization theorem and 
the fractions $x_p$ and $x_E$ of the available momentum and energy respectively of the process 
taken away by the detected hadron must be distinguished from one another.
Note that the cross section may depend on other variables in addition to $x$ and $E_s$, but we will not
indicate this explicitly unless necessary.
The factorization theorem asserts that the cross section takes the form of a {\it convolution}
\be
d\sigma^h(x,E_s^2)=\sum_{i=-n_f}^{n_f} \int_x^1 dz 
d\sigma^i\left(\frac{x}{z},\frac{E_s^2}{M_f^2},\frac{m_k^2}{E_s^2},a_s(M_f^2)\right)D^h_i(z,M_f^2),
\label{genformofinchadprodfromffs}
\ee
up to higher twist terms, which are suppressed relative to the overall cross section by a factor $O(\Lambda_{\rm QCD}/E_s)$ or more.
The parton label $i=0$ for the gluon, 
while $i=(-)I$ for (anti)quarks, where $I=1,\ldots,6$ corresponds respectively to the flavours $d$, $u$, $s$, $c$, $b$ and $t$.
The fact that $n_f \neq 6$ necessarily will be explained below.
$m_i$ is a renormalization scheme-dependent mass associated with parton $i$, 
which will be taken to be its pole mass, and $a_s=\alpha_s/(2\pi)$ is the expansion parameter in perturbative series.
The $d\sigma^i$ are the equivalent partonic cross sections obtained by replacing 
the detected hadron $h$ with a real on-shell parton $i$ moving in the same direction but with momentum $p_h/z$, 
and the sum over unobserved hadrons replaced with a sum over unobserved partons.
In e.g.\ $e^+ e^- \rightarrow h+X$, $d\sigma^i$ is completely calculable in perturbation theory, 
while for processes involving initial state hadrons, such as $ep \rightarrow e+h+X$ at HERA and
$pp(\overline{p}) \rightarrow h+X$ at the LHC (RHIC), $d\sigma^i$ will be convolutions of perturbatively calculable quantities with
PDFs for each initial state hadron, a result which also follows from the factorization theorem.
The $d\sigma^i$ are otherwise perturbatively calculable if
all subprocesses with energy scale below some arbitrary factorization scale $M_f \gg \Lambda_{\rm QCD}$ 
are factored out of them and into the FFs $D^h_i$ (and PDFs if applicable) according to the factorization theorem.
While the $d\sigma^i$ differ from process to process, the FFs are {\it universal}
and therefore, through them, measurements of one process impose constraints on others in which the same hadron is produced.

In more detail,
the perturbative expansion of the unfactorized $d\sigma^i$ in the limit $E_s \gg m_i$ contains {\it potential mass singularities},
being logarithms of the form $\ln (m_i/E_s)$ raised to powers of integers and, 
because the largest power in a given order in 
$a_s$ grows with the order, they may spoil the convergence of the series even when $m_i \neq 0$.
These potential mass singularities, which arise from energy processes much below $E_s$, may be factored out of the $d\sigma^i$.
There is clearly some freedom in choosing whether to place each process of energy scale around $M_f$ in
$d\sigma^i$ or $D^h_i$, and this choice defines the {\it factorization scheme}.
(This is a physical definition --- in practice the scheme is fixed by the choice of subtraction terms in the factorization.)
Although the leading twist component of $d\sigma^h$ 
is formally independent of the choices of factorization scale and scheme,
its perturbative approximation discussed above will depend on these choices, 
which must therefore reflect the physics of the overall process.
The theoretical error in this approximation can also be estimated by varying these choices.
The potential mass singularities from (anti)quarks with mass $m_J \gg E_s$ 
are dampened by factors of $E_s/m_J$, i.e.\ they approximately decouple \cite{Appelquist:1974tg}
and so must not be factorized to avoid introducing large uncanceled counterterms in $d\sigma^i$.
Therefore, the {\it active partons}, being those partons
whose potential mass singularities are factored out of the $d\sigma_i$ and into the FFs,
and to which the summation over $i$ in eq.\ (\ref{genformofinchadprodfromffs}) is restricted so that
$n_f$ labels the number of species of active quarks,
should be limited to include only those partons whose masses $m_i \lesssim E_s$.
This will always include the gluon and, 
since perturbative QCD is only valid in the region $E_s \gg \Lambda_{\rm QCD}$,
also the {\it light quarks}, defined to be those quarks whose masses are of $O(\Lambda_{\rm QCD})$ or less, i.e.\ 
the 3 lightest quarks $d$, $u$ and $s$ or $I=1,2,3$.
The intrinsic light hadron PDF of a quark is expected to be of $O(m_h/m_J)$ or less \cite{Witten:1975bh}, 
but the same property may not necessarily hold for FFs.
Therefore, it may also be necessary to always include some of the {\it heavy quarks}, 
defined to be those quark whose mass $m_J \gg \Lambda_{\rm QCD}$, i.e.\ 
$c$, $b$ and $t$ or $I=4,5,6$, in the list of active partons, in which 
case the cross section will not be perturbatively calculable if $m_J \ll E_s$.
For example, it could happen that intrinsic charm quark fragmentation is deemed important in a cross section,
but it cannot be incorporated into the cross section's calculation if the only appropriate scheme is the 3 flavour one.
This problem would be avoided if a method was known for correctly incorporating the intrinsic FF of quark $J$ 
in a cross section when this quark is not active.
This issue will be discussed further at the end of subsection \ref{nfchange} in the light of 
matching conditions between quantities defined with different numbers of active partons.

Note that there is another type of potential mass singularity appearing in the cross section, 
which behaves like a power of $\ln (E_s/m_J)$ for any heavy quark mass $m_J \gg E_s$. These can be absorbed into
the strong coupling constant $a_s$ by using a renormalization scheme for which quark $J$ is not active.

The $n_f${\it th scheme} is a renormalization and factorization scheme for which
the number of active quark flavours is $n_f$.
Results are usually presented in the $n_f$th Collins-Wilczek-Zee (CWZ) scheme \cite{Collins:1978wz},
also known as the decoupling scheme,
which reduces to the $\overline{\rm MS}$ scheme for only $n_f$ massless quark flavours 
in the limit that the $n_f$ active quark masses vanish and the remaining quarks' masses become infinite so that
they completely decouple from the theory.
The unsubtracted partonic cross sections with heavy quarks and their masses included
and their subtraction terms in the CWZ schemes
have been calculated for $e^+ e^-$ and $ep$ reactions in Refs.\ \cite{Nason:1993xx,Kneesch:2007ey} 
and \cite{Kretzer:1998nt} respectively.
The inclusion of heavy quarks and their masses in the partonic cross sections for $pp(\overline{p})$ reactions 
has been calculated in Ref.\ \cite{Nason:1989zy}, and their subtraction terms in Ref.\ \cite{Kniehl:2004fy}.

\subsection{DGLAP evolution \label{DGLAPevolution}}

Roughly speaking, factorization replaces each logarithm $\ln (m_i/E_s)$ 
in the partonic cross section with $\ln (M_f/E_s)$.
These artefacts of factorization may spoil the accuracy of the perturbation series 
in the same way that the potential mass singularities did unless we ensure $M_f = O(E_s)$, 
in which case the $M_f$ dependence of the $D^h_i(z,M_f^2)$ must also be known.
Fortunately, unlike for the $z$ dependence, this $M_f$ dependence is perturbatively calculable, provided
$M_f \gg \Lambda_{\rm QCD}$: Writing the $M_f^2$ dependence of the FFs
in the form of the Dokshitzer-Gribov-Lipatov-Altarelli-Parisi (DGLAP) equation \cite{Gribov:1972ri,Lipatov:1974qm},
\be
\frac{d}{d\ln M_f^2}D_i^h(z,M_f^2)=\sum_{j=-n_f}^{n_f} 
\int_z^1 \frac{dz'}{z'}P_{ij}\left(\frac{z}{z'},a_s(M_f^2)\right)D_j^h(z',M_f^2),
\label{DGLAP}
\ee
with $i=-n_f,\ldots,n_f$,
then the {\it splitting functions} $P_{ij}(z,a_s)$ are each perturbatively calculable as a series in $a_s$. 
Specifically, denoting the square matrix with components $P_{ij}$ by $P$, their expansion in $a_s$ takes the form
\be
P(z,a_s)=\sum_{n=1}^\infty P^{(n-1)}(z) a_s^n,
\label{FOexpanofP}
\ee
where the $P^{(n-1)}(z)$ are non-singular even in the limits for which any active parton mass vanishes.
In the $n_f$th CWZ scheme, the $P_{ij}$ are independent of all parton masses, and thus can be obtained
by taking the limit discussed at the end of subsection \ref{factXS} and performing factorization in the $\overline{\rm MS}$ scheme.
Equations (\ref{DGLAP}) and (\ref{FOexpanofP}) serve similar purposes to the evolution of $a_s$,
\be
\frac{d}{d\ln \mu^2}a_s(\mu^2)=\beta(a_s(\mu^2)),
\label{betafuncdef}
\ee
and the perturbative expansion of the $\beta$ function,
\be
\beta(a_s)=-\sum_{n=2}^{\infty}\beta_{n-2} a_s^n,
\label{pseriesforbeta}
\ee
respectively, which are used to resum powers of the logarithm $\ln \mu$ in perturbatively calculated quantities, 
where $\mu$ is the renormalization scale.
In other words, the ultimate purpose of the DGLAP equation is to resum powers of the logarithm $\ln M_f$ 
for all $M_f \gg \Lambda_{\rm QCD}$ in the partonic cross section.
The physical interpretation of eq.\ (\ref{DGLAP}) is most easily made from its solution, which takes the form
\be
D_i^h(z,M_f^2)=\sum_{j=-n_f}^{n_f} \int_z^1 \frac{dz'}{z'} E_{ij}\left(\frac{z}{z'},a_s(M_f^2),a_s(M_0^2)\right)D_j^h(z',M_0^2).
\label{explevolofD}
\ee
Each quantity $E_{ij}(z,a_s(M_f^2),a_s(M_0^2))$, which also obeys eq.\ (\ref{DGLAP}) on taking $D_i^h \rightarrow E_{ik}$,
may be interpreted as the FF of parton $i$ at resolution scale $M_f$ into parton $j$ at resolution scale $M_0$ 
carrying a fraction $z$ of parton $i$'s momentum.
It is subject to the boundary condition $E_{ij}(z,a_s,a_s)=\delta_{ij}\delta(1-z)$
and depends only on the $P^{(n-1)}(z)$ of eq.\ (\ref{FOexpanofP}), although
its analytic dependence on $z$ may not be calculated.
We will see in subsection \ref{anmelDGLAP} that an analytic form 
for the perturbative calculation of $E$ may be obtained in Mellin space.

\subsection{Changing the number of active partons \label{nfchange}}

So far, everything we have discussed applies for a given assignment of partons as active.
Using always the same FFs at some initial input value $M_f =M_0$ of the factorization scale,
we would like to be able to calculate cross sections at any energy $E_s$ (in this review, $M_f$, $M_0$ and $E_s$
are much greater than $\Lambda_{\rm QCD}$ unless otherwise stated).
However, to ensure that the perturbative calculation of $d\sigma^i$ gives a reliable approximation,
all partons of mass $m_i \ll M_f =O(E_s)$ must be assigned as active, even if they are not active at $M_f=M_0$.
Consequently, this assignment must be allowed to vary and the relationship between quantities
defined in a scheme in which the lightest $n_f$ quarks are active (and of course the gluon, which must always be active)
must be related to similar quantities for which $n_f+1$ quarks are active. 
This is trivial for the factorized partonic cross sections $d\sigma^i$,
and the necessary matching conditions for $a_s$ are known \cite{Marciano:1983pj}.
The matching conditions between the FFs of the $n_f$th scheme, $D_i^h$ with $i=-n_f,\ldots,n_f$, and the 
FFs of the $n_f+1$th scheme, $D_i^{h\prime}$ with $i=-(n_f+1),\ldots,n_f+1$, take the form
\be
D_i^{h\prime}(z,M_f^2)=\sum_{j=-n_f}^{n_f} \int_z^1 \frac{dz'}{z'} 
A_{ij}\left(\frac{z}{z'},\frac{m_k^2}{M_f^2},a_s(M_f^2)\right)D_j^h(z',M_f^2)
\label{mathcondffs}
\ee
where the matrix $A$ is perturbatively calculable for $M_f =O(m_{n_f+1})$, where this matching should therefore be done.
Otherwise, the precise choice of this {\it matching threshold} is arbitrary because it is non physical, and 
should be distinguished from non arbitrary {\it physical thresholds} such as that for the production of a charm quark.
In the CWZ scheme, $A$ depends on $m_{n_f+1}$ but not the other $m_k$ and is now known to NLO \cite{Cacciari:2005ry}.
The only non zero components at NLO are $A_{gg}$, which vanishes if the matching is done at $M_f=m_{n_f+1}$, 
and $A_{n_f+1,g}$. The latter quantity is not needed if the whole FF of the $n_f+1$th flavour quark, both intrinsic and extrinsic,
is treated as one single function to be fitted to data, which is usually the case.
This is in contrast to the spacelike case: because intrinsic heavy quark PDFs for the proton can normally be neglected, 
the quantity $A_{n_f+1,g}$ is necessary for obtaining the PDF of the $n_f+1$th flavour quark,
being almost completely extrinsic.

According to eq.\ (\ref{mathcondffs}), a heavy quark FF with a negligible intrinsic component 
(i.e.\ its FF vanishes when it is not active) is fully determined from the other FFs.
Therefore, if the intrinsic components of the heavy quarks' FFs are negligible, their FFs are perturbatively completely determined
by only the gluon and the light quarks in the $n_f=3$ scheme:
Using eq.\ (\ref{mathcondffs}) to convert to $n_f=4$ scheme gives the charm quark FF in terms of the gluon and light quark FFs,
i.e.\ charm quark fragmentation proceeds via perturbative fragmentation to a gluon or light quark, which
then fragments non perturbatively to the detected hadron.
Similarly the bottom quark FF for $n_f=5$ active quark flavours is fully determined from the gluon, light and charm quark FFs.

On the other hand, it may not be a good approximation to neglect the intrinsic heavy quark FFs even for light hadrons,
and the intrinsic charm FF is certainly not negligible for charmed hadrons such as $D$ mesons.
In many calculations for light hadron production, the charm quark FF $D_4^{h\prime}(z,M_f^2)$
is treated as an unknown function at the matching scale where $M_f =O(m_4)$,
just as the gluon and light quark flavour FFs at the initial scale $M_f =M_0$ are,
and the gluon and all 4 quark FFs are evolved from there according to the DGLAP equation of the $n_f=4$ scheme.
A similar procedure is performed for the bottom quark FF.
While this incorporation of intrinsic charm in the $n_f=4$ scheme is consistent with the factorization scheme,
the conventional approach of setting the complete charm quark FF to zero in the cross section calculated in the $n_f=3$ scheme
is not, because intrinsic charm quark fragmentation effects should not depend on the choice of scheme
(although the precise definition of the intrinsic charm quark FF itself is scheme dependent).
Consequently, this approach contains an inconsistency: When the energy, and therefore the factorization, scale is close 
to the charm quark mass so that both the 3 and 4 flavour number schemes should be valid, 
they will lead to very different cross sections in the case that the intrinsic charm quark FF is large.
In practice, this inconsistency of the 3 flavour scheme, and 4 flavour scheme when one also considers a possibly important
bottom quark FF, does not matter since most cross sections of interest are of a sufficiently high energy scale
that the 5 flavour scheme should suffice for all calculations.

\subsection{Various treatments of quarks with non negligible mass}

In the above discussion, we have assumed that all quarks have non negligible but finite masses.
In practice, the energy scale is usually not close to any quark mass, so all perturbative results are usually approximated
by the zero mass variable flavour number ``scheme'' (ZM-VFNS), where the masses $m_j$ of all quarks above $M_f = O(E_s)$
are set to infinity so that these quarks decouple from the theory, 
which introduces a relative error of $O(E_s^2/m_j^2)$ to the cross section,
while all quarks with masses below are treated as active and their masses $m_i$ are set to zero,
which introduces a relative error of $O(m_i^2/E_s^2)$.
Such an approach therefore fails when the energy scale is of the order of a quark mass, 
but is otherwise a reasonable approximation.
However, the procedure we have been discussing in the previous subsections is more general and is called the 
general mass variable flavour number scheme (GM-VFNS), where both quarks and their masses
are treated in a similar fashion to their treatment 
in the Aivazis-Collins-Olness-Tung (ACOT) scheme \cite{Aivazis:1993pi,Collins:1998rz} of spacelike factorization.
Such a procedure has for example been explicitly implemented for heavy flavour hadrons 
in photo- \cite{Kramer:2001gd} and hadroproduction \cite{Kniehl:2004fy} in $p\overline{p}$ reactions,
and more recently for $e^+ e^-$ reactions \cite{Kneesch:2007ey}.
The usual simplifications and improvements that have been made to the ACOT scheme since its inception
may also be applied to fragmentation, at least in principle.
For example, assuming that the intrinsic fragmentation of heavy quarks is negligible,
the power suppressed terms in the NLO factorized cross section for the inclusive production of this heavy quark 
appearing in the overall hadronic cross section
are in fact arbitrary, since the only purpose of the former cross section in this case is to complete the cancellation 
of all potential mass singularities 
in the overall hadronic cross section such that there are no singularities in the massless limit \cite{Collins:1998rz}.
In the S-ACOT scheme \cite{Kramer:2000hn}, these power suppressed terms are set to zero for simplicity,
while in the ACOT($\chi$) scheme \cite{Tung:2001mv}, the accuracy of the perturbative calculation
relative to the original ACOT scheme is improved: The heavy quark production cross section
is again chosen to be that for the production of a massless quark as in the S-ACOT scheme, 
but its LO part is also multiplied by a $z$ dependent step function which
is equal to that multiplying the gluon production cross section's potential mass singularity,
to ensure it and its counterterm only ever appear at the same time.

\subsection{Analytic Mellin space solution of the DGLAP equation \label{anmelDGLAP}}

The integrations over $z$ on the right hand sides of eqs.\ (\ref{genformofinchadprodfromffs}), (\ref{DGLAP}) 
and (\ref{explevolofD}) are known as convolutions, and are a typical feature of results of the factorization theorem.
The Mellin transform, discussed in appendix \ref{mellinspace}, which does not destroy any information
because this transform is invertible using the inverse Mellin transform of eq.\ (\ref{invmelltrans}),
converts these convolutions into simple products, 
and therefore analytic work is usually performed in Mellin space.
For example, eq.\ (\ref{DGLAP}) becomes
\be
\frac{d}{d\ln M_f^2}D(N,M_f^2)=P(N,a_s(M_f^2))D(N,M_f^2),
\label{DGLAPmellin}
\ee
where for further simplification we are now also omitting parton labels $i$, $j$ and the more 
trivial hadron label $h$, it being understood in eq.\ (\ref{DGLAPmellin}) 
that the matrix $P$ acts on the column vector $D$ according to the usual matrix product definition.
The elements of $P(N,a_s(M_f^2))$ for integer $N$ are equal to the 
anomalous dimensions of the twist two, spin $N$ gauge invariant operators in $D$, 
which are needed in the operator product expansion.
In the solution to eq.\ (\ref{DGLAPmellin}), 
\be
D(N,M_f^2)= E(N,a_s(M_f^2),a_s(M_0^2))D(N,M_0^2),
\label{explevolofDmellin}
\ee
which is the Mellin transform of eq.\ (\ref{explevolofD}),
the elements of $E$ may be expressed analytically in terms of $N$, $a_s(M_f^2)$ and $a_s(M_0^2)$
up to the same accuracy as $P$ \cite{Furmanski:1981cw,Ellis:1993rb}:
Define the series
\be
U(N,a_s)={\bm 1}+\sum_{n=1}^\infty U^{(n)}(N)a_s^n
\label{expanU}
\ee
such that
\be
E(N,a_s,a_0)=U(N,a_s)E_{\rm LO}(N,a_s,a_0)U^{-1}(N,a_0),
\label{waytosolvedglap}
\ee
where
\be
E_{\rm LO}(N,a_s,a_0)=\exp\left[-\frac{P^{(0)}(N)}{\beta_0}\ln \frac{a_s}{a_0}\right]
\label{formalresforELO}
\ee
is the LO result for $E$, being an exact solution to eq.\ (\ref{DGLAPmellin}) with $P=a_s P^{(0)}$.
The purpose of $U^{-1}(N,a_0)$ in eq.\ (\ref{waytosolvedglap}) is to ensure the boundary condition
$E(N,a_s,a_s)={\bm 1}$.
To all orders, eq.\ (\ref{DGLAPmellin}) can be converted via eq.\ (\ref{waytosolvedglap}) to an evolution equation for $U$:
\be
\frac{dU}{da_s}=-\frac{R}{\beta_0}+\frac{1}{\beta_0 a_s}\left[U,P^{(0)}\right],
\label{UfromP2}
\ee
where $R=\sum_{n=1}^\infty a_s^{n-1} R^{(n)}=-\beta_0(P/\beta+P^{(0)}/(\beta_0 a_s))U$.
The exponentiation in eq.\ (\ref{formalresforELO}) and the
commutator in eq.\ (\ref{UfromP2}) are handled by choosing a specific basis for the FFs 
in a factorization scheme in which the symmetries of QCD are obeyed.
Such a basis will be given in subsection \ref{symmetries}.

As a final remark, the Mellin space formalism makes clear the importance 
of the DGLAP equation in the application of perturbation theory to cross section calculations:
Any set of functions $D_i$ of the variables $0< z<1$ and $0<M_f^2 < \infty$ obeys the {\it form} of the 
DGLAP equation, eq.\ (\ref{DGLAP}), by choosing
$P(N,a_s(M_f^2))=(dD(N,M_f^2)/d\ln M_f^2) D^{-1}(N,M_f^2)$, which follows from its Mellin transform, eq.\ (\ref{DGLAPmellin}).
(Here, $D$ can be regarded as a matrix whose columns consist of the $D_i$ for a given hadron species which varies
from column to column, i.e.\ there is some freedom in the definition of $P$.) 
The importance of the formalism behind the DGLAP equation is that $P$ is constrained in QCD to 
depend on purely partonic graphs, and furthermore to be perturbatively calculable
(these results follow from eq.\ (\ref{defofsplitfuncfromdiags})),
and therefore the $M_f$ dependence of the FFs is completely and calculably constrained.
In particular, when expanded in $a_s$ it takes the form in eq.\ (\ref{FOexpanofP}) with non singular coefficients. 
However, eq.\ (\ref{DGLAP}) is not the only way to evolve the FFs in $M_f$.
Alternatives to eq.\ (\ref{DGLAP}) may be preferable for certain physical reasons, such as the double logarithmic approximation (DLA),
to be discussed in subsection \ref{DLA}, which is more apt for the small $x$ region, 
or the evolution proposed in Ref.\ \cite{Dokshitzer:2005bf}, whose equivalent splitting functions may exhibit certain physical 
properties not seen in the usual DGLAP splitting functions beyond LO, 
such as the Gribov-Lipatov relations and the expected large $x$ behaviour beyond leading order
due to purely multi-parton quantum fluctuations when the physical
strong coupling constant is used as the expansion parameter.
Such equations are similar to the DGLAP equation but with both occurrences of $M_f$ in eq.\ (\ref{DGLAP})
multiplied by (different) powers of $z$ to ensure that the inverse 
of this modified scale truly represents the fluctuation lifetimes of 
successive virtual parton states pertinent to the kinematic region being studied.
However, if required, it is always possible to recast such alternative evolution equations back into the form of eq.\ (\ref{DGLAP}),
and thereby use them to obtain an alternative expansion for $P$ in the kinematic region of interest to that in eq.\ (\ref{FOexpanofP}).
To put this in an alternative, but equivalent, way, eq.\ (\ref{FOexpanofP}) is not the only possibility for approximating $P$:
In general, $P$ in certain limits of $N$ and $a_s$ may be better approximated
in the form $P(N,a_s)=\sum_n X^n R^{(n)}(Y)$, where $X$ and $Y$ are each in general a suitably chosen function of both $N$ and $a_s$.
We will see an example of such an alternative expansion in subsection \ref{unifform}, namely eq.\ (\ref{expanpsglinmel})
(where $\omega=N-1$).

\subsection{Symmetries \label{symmetries}}

Although FFs are not physical, 
the factorization scheme should be chosen such that they respect the symmetries of QCD
in order to keep results as simple as possible, and we will assume that this has been done. 
Indeed, the commonly used $\overline{\rm MS}$ scheme is such a scheme.
Consider first the charge conjugation symmetry of QCD.
Writing the cross section or FF for the production of $h^+$ (or $h^-$) as $O^{h^+\ ({\rm or\ }h^-)}$,
where $O=D_i$ or $d\sigma$ respectively,
and then defining charge-sign unidentified and charge-sign asymmetry quantities as $O^{h^\pm}=O^{h^+}+O^{h^-}$
and $O^{\Delta_c h^\pm}=O^{h^+}-O^{h^-}$ respectively (or more generally as 
$O^{h/\bar{h}}=O^h+O^{\bar{h}}$ and $O^{\Delta_c h/\bar{h}}=O^h+O^{\bar{h}}$ which includes also neutral hadrons), 
this symmetry implies that each of these two combinations of cross sections only depends on FFs that have been 
combined in the same way (or they vanish).
The calculation of such cross sections is therefore simpler than that of the 
production of hadrons of a given charge-sign, which depends on both charge-sign unidentified 
{\it and} charge-sign asymmetry FFs.
Lorentz invariance implies a similar feature for the polarization of the hadron and of the partons,
assuming that there are only 2 possibilities for the hadron's polarization.

We now study the simplifications to the evolution of FFs when the charge conjugation symmetry is taken into account.
Letting $D_{q_I}^h=D_I^h$ ($D_{\bar{q}_I}^h=D_{-I}^h$) denote the FF for the hadron $h$ 
of the (anti)quark of flavour $I=1,\ldots,n_f$, this symmetry is accounted for by the results that follow from 
charge conjugation symmetry,
\be
D_{q_I}^{h^+({\rm or}\ h^-)}=D_{\bar{q}_I}^{h^-({\rm or}\ h^+)}.
\label{chargeconjsymm}
\ee
Equation (\ref{chargeconjsymm}) implies that charge-sign unidentified FFs 
$D_{q_I[{\rm or}\ \bar{q}_I]}^{h^\pm}=D_{q_I/\bar{q}_I}^{h^+({\rm or}\ h^-)}$ (the separate brackets ``$[]$''
and ``$()$'' imply that the changes $q_I\rightarrow \bar{q}_I$ and $h^+ \rightarrow h^-$ respectively may be made,
and may be made independently of one another)
where, omitting the superscript ``$h^+({\rm or}\ h^-)$'' for brevity from now on,
the sums
\be
D_{q_I/\bar{q}_I}=D_{q_I}+D_{\bar{q}_I}.
\label{defofchuffs}
\ee
Due to charge conjugation symmetry, these FFs and the gluon FF mix only with each other on evolution, but not with the valence
quark FFs defined in eq.\ (\ref{defofvalquarks}) below.
Furthermore, in a scheme for which $P$ is explicitly independent of quark masses, such as the CWZ scheme,
the SU$(n_f)$ symmetry for $n_f$ active quark flavours of equal mass that follows from QCD
implies that the gluon FF will mix via eq.\ (\ref{DGLAP}) 
with the quark FFs combined into one {\it singlet} quark FF
\be
D_\Sigma=\frac{1}{n_f}\sum_{I=1}^{n_f} D_{q_I/\bar{q}_I},
\label{defofsing}
\ee
In other words, eq.\ (\ref{DGLAP}) is obeyed with $D=(D_{\Sigma},D_g)^T$ and 
\begin{eqnarray}
P=\left(\begin{array}{cc}P_{\Sigma \Sigma} & P_{\Sigma g}\\ P_{g\Sigma} & P_{gg} \end{array}\right).
\end{eqnarray}
This SU$(n_f)$ symmetry also implies that every {\it non singlet} quark FF, being
any linear combination of the $D_{q_I/\bar{q}_I}$ that vanishes when they are all equal,
will only mix with itself on evolution, i.e.\ 
it obeys eq.\ (\ref{DGLAP}) but with $P$ reduced to the single quantity $P_{\rm NS}$ which is the same for
all non singlets.
The non singlets can be chosen such that they and the singlet form a linearly independent set of $n_f$ FFs,
so that after evolution the FFs of quarks of each flavour, or any other alternative basis of FFs, 
can be extracted by taking appropriate linear sums.
A common choice of the set of non singlet FFs is
\be
D_{q_I,{\rm NS}}=D_{q_I/\bar{q}_I}-D_\Sigma.
\label{defofnonsing}
\ee

Equation (\ref{chargeconjsymm}) also implies that charge-sign asymmetry FFs 
$D_{q_I[{\rm or}\ \bar{q}_I]}^{\Delta_c h^\pm}=[-](-)D_{\Delta_c q_I/\bar{q}_I}^{h^+({\rm or}\ h^-)}$
where, again omitting the superscript ``$h^+({\rm or}\ h^-)$'', the differences
\be
D_{\Delta_c q_I/\bar{q}_I}=D_{q_I} -D_{\bar{q}_I},
\label{defofvalquarks}
\ee
which we refer to as the {\it valence quark} FFs.
Although valence quark FFs are the same as charge-sign asymmetry FFs, we distinguish between them depending on the context ---
when working with the symmetries of quarks as we do in this subsection, we shall refer to them as valence quark FFs.
Due to charge conjugation symmetry, valence quark FFs mix with each other on evolution but not with the
summed FFs defined in eq.\ (\ref{defofchuffs}) above.
As for charge-sign unidentified FFs, this mixing is further simplified by the SU$(n_f)$ symmetry introduced above:
The {\it singlet} valence quark FF
\be
D_{\Delta_c \Sigma}=\frac{1}{n_f}\sum_{I=1}^{n_f} D_{\Delta_c q_I/\bar{q}_I}
\ee
and the {\it non singlet} valence quark FFs, which, similarly to the definition of non singlet quark FFs in 
eq.\ (\ref{defofnonsing}), we take as
\be
D_{I,\Delta_c {\rm NS}}=D_{\Delta_c q_I/\bar{q}_I}-D_{\Delta_c \Sigma},
\label{achoiceofvalenceNS}
\ee
will obey eq.\ (\ref{DGLAP}) but with $P$ reduced to the single quantity 
$P_{\Delta_c \Sigma}$ and $P_{\Delta_c {\rm NS}}$ respectively.
As for $P_{\rm NS}$, $P_{\Delta_c {\rm NS}}$ is the same quantity for all non singlet valence quark FFs.
At NLO, $P_{\Delta_c \Sigma}=P_{\Delta_c {\rm NS}}$, which leads to the simpler evolution 
in which each valence quark FF only mixes with itself:
$D_{\Delta_c q_I/\bar{q}_I}$ obeys eq.\ (\ref{DGLAP}) with $P=P_{\Delta_c \Sigma}$ (or $P_{\Delta_c {\rm NS}}$).

In the non singlet and valence quark sector, $E_{\rm LO}$
can be calculated in the form given in eq.\ (\ref{formalresforELO}) and the commutator in eq.\ (\ref{UfromP2}) vanishes.
In the singlet quark and gluon sector, eq.\ (\ref{formalresforELO}) and this commutator 
are both handled by the ``diagonalization'' \cite{Furmanski:1981cw,Ellis:1993rb} of the $2\times 2$ matrix
$P^{(0)}=\lambda_+ M^+ +\lambda_- M^-$,
where $\lambda_\pm=(P_{qq}^{(0)}+P_{gg}^{(0)}\pm \sqrt{(P_{qq}^{(0)}-P_{gg}^{(0)})^2-4 P_{qg}^{(0)} P_{gq}^{(0)}})/2$
are the eigenvalues of $P^{(0)}$ and $M^\pm =(P^{(0)}-\lambda_\mp {\bm 1})/(\lambda_\pm -\lambda_\mp)$
are projection operators, i.e.\ they obey $\sum_i M^i ={\bm 1}$ where $i=\pm$, $M^\pm M^\mp=0$ and $M^\pm M^\pm= M^\pm$.
Then eq.\ (\ref{formalresforELO}) reduces to the calculable form
$E_{\rm LO}(N,a_s,a_0)=\sum_i M^i(N)(a_s/a_0)^{-(\lambda_i(N)/\beta_0)}$, and eq.\ (\ref{UfromP2}) is equivalent
to the result
\be
U^{(n)}=\sum_{ij} \frac{1}{\lambda_j -\lambda_i -\beta_0 n} M^i R^{(n)} M^j,
\label{coeffsinU}
\ee
where the $U^{(n)}$ are defined in eq.\ (\ref{expanU}), and the $R^{(n)}$ immediately after eq.\ (\ref{UfromP2}).
Note that the right hand side of eq.\ (\ref{coeffsinU}) only depends on the $U^{(m)}$ for $m=1,\ldots,n-1$,
so $U$ is constructed order by order.
In general, the zeroes in the numerator of eq.\ (\ref{coeffsinU}) lead to singularities in $E$ for complex values of $N$.
In the calculation of the cross section via the inverse Mellin transform 
defined by eq.\ (\ref{invmelltrans}), it is not necessary that the contour $C$ should lie to the right of these singularities.
Although this arbitrariness due to the dependence on the choice of $C$ is of higher order than the order of the calculation,
for numerical purposes it is preferable to eliminate the singularities, by choice of the spurious higher order terms.
For example, at NLO, $E$ is free of such singularities when eq.\ (\ref{waytosolvedglap}) 
is calculated in the form $E=E_{\rm LO}+U^{(1)}E_{\rm LO}a_s-E_{\rm LO}U^{(1)}a_0$,
which we note obeys the desired boundary condition $E(N,a,a)={\bm 1}$.
Alternatively, the singularities can be made to cancel simply by choosing $U^{-1}(N,a_0)$ in eq.\ (\ref{waytosolvedglap})
to be exactly equal to the inverse of $U(N,a_0)$ \cite{Blumlein:1997em}, instead of expanding it in $a_0$.

\subsection{Sum rules \label{sumrules}}

Intuitively, FFs as probability densities will be constrained by conservation laws.
For example, from momentum conservation, 
the momentum of a parton $i$ must equal the total momentum of all hadrons to which it fragments,
giving the {\it momentum sum rule}
\be
\sum_h \int_0^1 dz \ zD_i^h(z,M_f^2)=1
\label{momsumrule}
\ee
for every parton $i$. 
Similarly, charge conservation implies the {\it charge sum rule}
\be
\sum_h \int_0^1 dz \ e_h D_i^h(z,M_f^2) =e_i,
\label{chargesumrule}
\ee
where $e_{h(i)}$ is the electric charge of hadron $h$ (parton $i$).
In fact, whether this probability density interpretation is correct,
eqs.\ (\ref{momsumrule}) and (\ref{chargesumrule}) hold in the $\overline{\rm MS}$ scheme \cite{Collins:1981uw}: 
They are true for the bare FFs $D_{Bi}^h$ of appendix \ref{outlinederivfactTheo},
for which the probability density interpretation is valid,
which is seen by applying the operation $\sum_h \int_0^1 dz (z) \times$ to the matrix element definition 
of the bare quark FFs in eq.\ (\ref{matelemforquarkFFs}) (and the similar definition for the bare gluon FF),
and identifying the number operator for hadrons of species $h$ in an infinitesimal region of momentum space,
$dN_h=(2\pi)^{d-1} (dP^+/2P^+) d^{d-2} {\bm P}_T a_h^\dagger (P) a_h (P)$.
Then eqs.\ (\ref{momsumrule}) and (\ref{chargesumrule}) follow for factorized FFs at all values of $M_f$ in schemes for 
which no subtraction is made on bare quantities that are free of divergences, such as the $\overline{\rm MS}$ scheme.
Such intuitive but theoretically solid results may be regarded as ``physical''.

In practice, eqs.\ (\ref{momsumrule}) and (\ref{chargesumrule}) are not directly used to impose a precise constraint between FFs.
Firstly, the summation here is over all hadron species $h$, so that they impose no constraint 
in global fits of FFs in the case of a specific hadron.
Secondly, FFs at low $z$ are poorly constrained because soft gluon logarithms render the FO approximation
of the splitting functions inaccurate at small $z$.
This is in contrast to the momentum sum rule for PDFs, which imposes a constraint between them for any given hadron
because the summation is over parton species $i$ instead of over hadron species, 
and PDFs are better understood at small momentum fraction. 
Instead, eq.\ (\ref{momsumrule}) may serve as an upper bound on FFs and therefore as a check on phenomenological extractions
of them because,
when the sum over $h$ is limited to just a few hadrons and the lower bound for the integral over $z$ is 
replaced by a value sufficiently greater than zero,
it will be less than one if all FFs are positive.
However, this positivity condition only follows from the probabilistic interpretation for FFs, which is not quite correct.

Equations (\ref{momsumrule}) and (\ref{chargesumrule}) do impose constraints on partonic cross sections, 
examples of which will be seen in subsection \ref{epemtheory},
which give a useful check on their perturbative calculations.

\subsection{Properties of splitting functions \label{propsplitfunc}}

The LO coefficients of the splitting functions (the $n=1$ coefficients $P^{(0)}$ in eq.\ (\ref{FOexpanofP})) 
of the quark singlet and gluon sector are given in appendix \ref{LOsplitfunc}.
Because of the Gribov-Lipatov relations \cite{Gribov:1972ri}, they are equal to 
the spacelike ones after the interchange $P^{(0)}_{\Sigma g}\leftrightarrow P^{(0)}_{g\Sigma}$ is made.
The LO coefficients of the splitting functions 
$P_{\Delta_c {\rm NS}}$, $P_{\Delta_c \Sigma}$, $P_{\Sigma \Sigma}$ and $P_{\rm NS}$ are equal to one another.
The NLO coefficients (the $n=2$ coefficients $P^{(1)}$ in eq.\ (\ref{FOexpanofP})) have been calculated in Ref.\ \cite{Curci:1980uw} 
(see also Refs.\ \cite{Gluck:1992zx,Ellis:1991qj} for the correction to a misprint therein).
Although the simple Gribov-Lipatov relation does not hold beyond LO, 
the timelike and spacelike splitting functions are related by analytic continuation 
of the form factor from the spacelike to the timelike case \cite{Curci:1980uw,Stratmann:1996hn,Mitov:2006ic}.
Of the NNLO coefficients $P^{(2)}$, the non singlet and singlet valence, the non singlet and the 
two diagonal singlet splitting functions have been calculated in Ref.\ \cite{Mitov:2006ic} using this continuation.
Only the off-diagonal splitting functions $P^{(2)}_{\Sigma g}$ and $P^{(2)}_{g \Sigma}$ need to be
calculated before a full NNLO timelike evolution of all FFs will be possible.
The resulting reduction in the theoretical error on all calculations when upgraded from NLO to NNLO
should lead to a reduction on the total errors on FFs obtained in global fits, although this
reduction will be generally somewhat smaller than the current experimental errors on FFs.

Using the momentum sum rule of eq.\ (\ref{momsumrule}) in eq.\ (\ref{DGLAPmellin}) with $N=2$ and all hadron 
species $h$ summed over gives a constraint on the splitting functions in the singlet and gluon sector:
\be
\int_0^1 dx x (2P_{\Sigma \Sigma}^{(n)}(x)+P_{\Sigma g}^{(n)}(x))=\int_0^1 dx x (2P_{g \Sigma}^{(n)}(x)+P_{g g}^{(n)}(x))=0.
\ee
The factors of 2 account for the identical contributions of quarks and antiquarks.
Similar momentum sum rule constraints exist for the spacelike splitting functions after the interchange
$P_{\Sigma g}\leftrightarrow P_{g\Sigma}$ is made.
Similarly, the charge sum rule of eq.\ (\ref{chargesumrule}) and eq.\ (\ref{DGLAPmellin}) with $N=1$ implies
that the valence quark splitting functions obey
\be
\int_0^1 dx P_{\Delta_c {\rm NS}}^{(n)}(x)=\int_0^1 dx P_{\Delta_c \Sigma}^{(n)}(x)=0,
\ee
which are similar to the valence sum rule constraints on the spacelike splitting functions.

\subsection{The simplest case: $e^+ e^- \rightarrow h+X$ \label{epemtheory}}

We are now in a position to highlight the main features of the perturbative calculation of the above process,
which serves as a simple illustration of the calculation of factorized cross sections in general.
In this subsection we assume for simplicity that there are only $n_f$ flavours of quarks, which are all massless.
This process proceeds via $e^+ e^- \rightarrow \gamma^*,Z \rightarrow q_J +\overline{q}_J\rightarrow h+X$.
For clarity, we will neglect the $Z$ boson for now and discuss the modifications due to its effects later.
Furthermore, the (anti)quark $q_J$ ($\overline{q}_J$) at the electroweak vertex, called the {\it primary quark}, 
of specific flavour $J$ is often tagged in experiment, and we will assume for generality that this is the case. 
The process is calculated by factorization of the modulus squared of the diagram in Fig.\ (\ref{epem}) (left),
in which the resulting partonic kinematics are also shown (right).

We work to LO in electroweak theory. 
Taking $d\sigma^h(x,E_s^2)\rightarrow d\sigma^{h}_{q_J}(x,s)$ and
$d\sigma^I(z,\ldots)\rightarrow d\sigma^I_{q_J}(z,\ldots)$ in eq.\ (\ref{genformofinchadprodfromffs}),
where $q_J$ is the tagged quark,
and then making the replacement $z \rightarrow x/z$ to ensure only differentials in $z$ and not $x/z$ appear,
the cross section can be written
\be
\begin{split}
\frac{d\sigma^h_{q_J}}{dx}(x,s)=&
\int_x^1 \frac{dz}{z}\Bigg[
\frac{d\sigma^{\rm NS}_{q_J}}{dz}\left(z,s,M_f^2\right)D_{q_J/\bar{q}_J}^h\left(\frac{x}{z},M_f^2\right)
+\frac{1}{n_f}\sum_{I=1}^{n_f}\frac{d\sigma^{\rm PS}_{q_J}}{dz}\left(z,s,M_f^2\right)
D_{q_I/\bar{q}_I}^h\left(\frac{x}{z},M_f^2\right)\\
&+\frac{d\sigma^g_{q_J}}{dz}\left(z,s,M_f^2\right)D_g^h\left(\frac{x}{z},M_f^2\right)\Bigg]
\end{split}
\label{XSfromFFs}
\ee
up to higher twist terms of $O(\Lambda_{\rm QCD}/\sqrt{s})$ or less.
Each partonic cross section may be written 
\be
\frac{d\sigma^X_{q_J}}{dz}\left(z,s,M_f^2\right)
=\sigma_0 (s) N_c Q_{q_J}(s)C_X\left(z,a_s(s),\ln \frac{M_f^2}{s}\right)\ {\rm for}\ X={\rm NS},\ {\rm PS}\ {\rm and}\ g.
\label{partXSfromcoefffuncs}
\ee
The quantity $\sigma_0=4\pi\alpha^2/(3s)$ is the leading order (LO) cross section for the 
process $e^+ e^- \rightarrow \mu^+ +\mu^-$, 
and the coupling of the tagged quark $q_J$ at the photon vertex is accounted for by $Q_{q_J}(s)=e_e^2 e_{q_J}^2$,
where $e_{q_J}^2$ is the electric charge of the quark $q_J$ and $e_e^2$ that of the electron/positron.
Note that $Q_{q_J}$ becomes dependent on $s$ when the effects of the $Z$ boson, which will be discussed later, are included.
The $C_X$ are the coefficient functions whose NLO \cite{Altarelli:1979kv} and NNLO \cite{Rijken:1996vr} terms are known.
For the choice $M_f^2=s$, the $C_X(z,a_s,0)=C_X(z,a_s)$ are given to NLO by
\be
\begin{split}
C_{\rm NS}(z,a_s)= &\delta(1-z)+a_s C_F
\Bigg[\left(\frac{2\pi^2}{3}-\frac{9}{2}\right)\delta(1-z)-\frac{3}{2}\left[\frac{1}{1-z}\right]_+ \nonumber \\
& +(1+z^2)\left[\frac{\ln (1-z)}{1-z}\right]_++1+2\frac{1+z^2}{1-z}\ln z +\frac{3}{2}(1-z)\Bigg],\\
C_{\rm PS}(z,a_s)=&O(a_s^2)\ \ \ {\rm and}\\
C_g(z,a_s)=&a_s  C_F \left[2\frac{1+(1-z)^2}{z}\left(\ln (1-z)+2\ln z\right)\right].
\label{coeffNLO}
\end{split}
\ee
Note that the pure singlet contribution only enters at NNLO.
The coefficient functions in the case where the masses of partonic heavy quark are not neglected 
and the remaining heavy quarks are not decoupled (i.e.\ their masses are not set to infinity)
are presented in Ref.\ \cite{Nason:1993xx}.

The non singlet partonic cross section $d\sigma^{\rm NS}_{q_J}/dz$ contains 
only and all those contributions in which the ``detected'' (fragmenting) quark $q_J/\bar{q}_J$ is part of the same quark line
as that for the tagged $q_J/\bar{q}_J$ connected to the electroweak vertex. 
The pure singlet partonic cross section $d\sigma^{\rm PS}_{q_J}/dz$ contains all other contributions,
i.e.\ those for which the tagged $q_J/\bar{q}_J$ that goes through the electroweak vertex is 
not part of the same quark line as the quark which fragments.
It is independent of the flavour of the ``detected'' 
quark and gives the same result when this quark is replaced by an antiquark.
Finally, $d\sigma^g_{q_J}/dz$ contains all contributions in which the ``detected'' parton is a gluon.

In terms of singlets and non singlets,
\be
\begin{split}
\frac{d\sigma^h_{q_J}}{dx}(x,s)=&
\int_x^1 \frac{dz}{z}\Bigg[\frac{d\sigma^{\rm NS}_{q_J}}{dz}\left(z,s,M_f^2\right)
D^h_{q_J,{\rm NS}}\left(\frac{x}{z},M_f^2\right)
+\frac{d\sigma^{\rm S}_{q_J}}{dz}\left(z,s,M_f^2\right)D^h_\Sigma\left(\frac{x}{z},M_f^2\right)\\
&+\frac{d\sigma^g_{q_J}}{dz}\left(z,s,M_f^2\right)D_g^h\left(\frac{x}{z},M_f^2\right)\Bigg],
\label{XSfromFFsnoqtag}
\end{split}
\ee
where the singlet $D^h_\Sigma$ and non singlets $D^h_{q_J,{\rm NS}}$ are defined
in eqs.\ (\ref{defofsing}) and (\ref{defofnonsing}) respectively, and the singlet partonic cross sections
\be
\frac{d\sigma^{\rm S}_{q_J}}{dz}=\frac{d\sigma^{\rm NS}_{q_J}}{dz}+\frac{d\sigma^{\rm PS}_{q_J}}{dz}.
\ee

The full untagged cross section can always be obtained by summing the tagged quarks in eq.\ (\ref{XSfromFFs}) over all flavours,
\be
\frac{d\sigma^h}{dx}(x,s)=\sum_{J=1}^{n_f} \frac{d\sigma^h_{q_J}}{dx}(x,s).
\label{untagXSfromtagXS}
\ee
Conversely, eq.\ (\ref{XSfromFFs}) may be obtained from eq.\ (\ref{untagXSfromtagXS}) simply by 
setting all $Q_{q_I}=0$ except $Q_{q_J}$. 
Thus the tagged cross sections $d\sigma^h_{q_J}/dx$ are physical at least in the sense that
they are factorization scheme and scale independent.

Now including the effects of the $Z$ boson, i.e.\ 
for all processes $e^+ e^- \rightarrow \gamma^*,Z \rightarrow q_J +\overline{q}_J\rightarrow h+X$,
\be
\frac{d\sigma^h_{q_J}}{dx}(x,s)= \frac{d\sigma^h_{\gamma^*,Z;q_J}}{dx}(x,s)+\frac{d\sigma^h_{q_J,F}}{dx}(x,s).
\ee
where $d\sigma^h_{\gamma^*,Z;q_J}/dx$ contains the contributions from all processes which couple
to the virtual photon and which couple to the $Z$ boson in the same way as to the virtual photon.
It is therefore equal to the $d\sigma^h_{q_J}/dx$ with virtual photon effects only but with 
the electroweak coupling $Q_{q_J}(s)$ modified to \cite{Amsler:2008zz}
\be
Q_{q_J}=e_e^2 e_{q_J}^2 +2e_e v_e e_{q_J} v_{q_J} \frac{s(s-M_Z^2)}{(s-M_Z^2)^2+M_Z^2 \Gamma_Z^2}
+(v_e^2+a_e^2)(v_{q_J}^2+a_{q_J}^2)\frac{s^2}{(s-M_Z^2)^2+M_Z^2 \Gamma_Z^2},
\label{elecweakfacttot}
\ee
where the decay width of the $Z$ boson $\Gamma_Z=2.4952$ GeV \cite{Amsler:2008zz} and
the vector and axial-vector couplings of a fermion $f$ to the $Z$ boson are given respectively by
\be
\begin{split}
v_f=&\frac{T_{3,f}-2e_f \sin^2 \theta_W}{2\sin \theta_W \cos \theta_W}\ {\rm and}\\
a_f=&\frac{T_{3,f}}{2\sin \theta_W \cos \theta_W},
\end{split}
\ee
with $T_{3,f}$ the third component of weak isospin of the fermion's left-handed component and $\theta_W$ the electroweak mixing angle.
The cross sections $d\sigma^h_{q_J,F}/dx$ account for all other processes, 
and contain contributions from $Z$ boson exchange with purely axial coupling to the quarks at the
$Z$ boson vertex \cite{Binnewiesthesis}.
Its form after factorization is the same as the form of $d\sigma^h_{q_J}/dx$ in eq.\ (\ref{XSfromFFs}), i.e.\ 
\be
\begin{split}
\frac{d\sigma^h_{q_J,F}}{dx}(x,s)=&
\int_x^1 \frac{dz}{z}\Bigg[
\frac{d\sigma^{\rm NS}_{q_J,F}}{dz}\left(z,s,M_f^2\right)D_{q_J/\bar{q}_J}^h\left(\frac{x}{z},M_f^2\right)
+\frac{1}{n_f}\sum_{I=1}^{n_f}\frac{d\sigma^{\rm PS}_{q_J,F}}{dz}\left(z,s,M_f^2\right)
D_{q_I/\bar{q}_I}^h\left(\frac{x}{z},M_f^2\right)\\
&+\frac{d\sigma^g_{q_J,F}}{dz}\left(z,s,M_f^2\right)D_g^h\left(\frac{x}{z},M_f^2\right)\Bigg]
\end{split}
\label{XSFfromFFs}
\ee
where
\be
\frac{d\sigma^X_{q_J,F}}{dz}\left(z,s,M_f^2\right)=\sigma_0 (s) N_c Q_{q_J}^F(s)C_{X,F}\left(z,a_s(s),\ln \frac{M_f^2}{s}\right)
\ee
with the electroweak coupling
\be
Q_{q_J}^F(s)=(v_e^2+a_e^2)a_{q_J}^2\frac{s(s-M_Z^2)}{(s-M_Z^2)^2+M_Z^2 \Gamma_Z^2}.
\label{QFexp}
\ee
In the untagged cross section, the replacement $a_{q_J}^2 \rightarrow a_{q_J}\sum_{I=1}^{n_f}a_{q_I}$ must be made in 
eq.\ (\ref{QFexp}) in order to account for interference effects between diagrams 
with different quark flavours at the $Z$ boson vertex.
Since $a_{{\mathcal Q}_u}=-a_{{\mathcal Q}_d}$, 
where ${\mathcal Q}_u$ and ${\mathcal Q}_d$ is any up and down type quark respectively,
it is the non zero mass differences within each generation that are responsible for these types of contributions.
The coefficient functions $C_{F,{\rm NS}}$, $C_{F,{\rm PS}}$ and $C_{F,g}$
are all proportional to $C_F T_R$, and so must vanish for photons due to Furry's theorem \cite{furry}.
The series for $C_{F,{\rm NS}}$ begins at $O(a_s^2)$, while the series for $C_{F,{\rm PS}}$ and $C_{F,g}$
begin at higher order, i.e.\ $d\sigma^h_{q_J,F}/dx$ contributes at NNLO and so is neglected in NLO calculations.

In a single event in $e^+ e^- \rightarrow \gamma^*,Z\rightarrow q_J +\overline{q}_J\rightarrow h+X$, 
the number of hadrons of species $h$ produced 
with energy or momentum fraction between $x$ and $x+dx$ is 
\be
N^h_{q_J}(x,s)dx=\frac{dx}{\sigma(s)}\frac{d\sigma^h_{q_J}}{dx} (x,s),
\label{NhfromXS}
\ee
where, choosing $\mu=\sqrt{s}$, the total cross section 
\be
\sigma(s)=\sum_I \sigma_0(s) N_c Q_{q_I}(s) \left(1+\frac{3}{2}C_F a_s(s)\right).
\label{totqtagXS}
\ee
From this result we can obtain two important sum rules.
Firstly, by energy or momentum conservation, the total energy of the final state,
$\sum_{h,J} \int_0^1 dx N^h_{q_J}(x,s) E_h$, must equal the energy of the initial state, $\sqrt{s}$.
Using $x=2E_h/\sqrt{s}$, this equality of energies is equivalent to
\be
\sigma_{q_J}(s)=\frac{1}{2}\sum_h \int_0^1 dx x \frac{d\sigma^h_{q_J}}{dx}(x,s),
\label{physmomsumrule}
\ee
where $\sigma_{q_J}(s)$ is the total cross section when quark $q_J$ is tagged, 
given by eq.\ (\ref{totqtagXS}) with all $Q_{q_I}$ set to zero except that for $I=J$.
In the $\overline{\rm MS}$ scheme, the momentum sum rule of eq.\ (\ref{momsumrule}) and the non singlet, 
singlet and gluon coefficient functions for $N=2$ also imply eq.\ (\ref{physmomsumrule}).
The total electric charge of the final state when quark $q_J$ is tagged is
\be
\sum_h \int_0^1 dx\ e_h\ \frac{1}{\sigma(s)}\frac{d\sigma^h_{q_J}}{dx}(x,s)=0,
\label{physchargesumrule}
\ee
which follows from eq.\ (\ref{chargesumrule}).

The cross section that we have been considering so far in this subsection can be decomposed into cross sections 
in which the vector boson is transversely and longitudinally
polarized with respect to the direction of the detected hadron, denoted by the subscripts $T$ and $L$ respectively:
\be
\frac{d\sigma^h_{q_J}}{dx}=\frac{d\sigma^h_{q_J,T}}{dx}+\frac{d\sigma^h_{q_J,L}}{dx}.
\label{XSdeomptotransandlong}
\ee
As usual, $d\sigma^h_{q_J,T}/dx$ and $d\sigma^h_{q_J,L}/dx$ are factorized in the same way as $d\sigma^h_{q_J}/dx$ 
in eq.\ (\ref{XSfromFFs}).
The corresponding coefficient functions are defined as before via
\be
\frac{d\sigma^X_{q_J,\eta}}{dz}\left(z,s,M_f^2\right)
=\sigma_0 (s) Q_{q_J}(s)C_{\eta,X}\left(z,a_s(s),\ln \frac{M_f^2}{s}\right)\ {\rm for}\ X={\rm NS},\ {\rm PS}\ {\rm and}\ g,
\ {\rm and}\ {\rm for}\ \eta=T\ {\rm and}\ L.
\ee
To $O(a_s)$,
\be
\begin{split}
C_{L,{\rm NS}}(z,a_s)=&a_s C_F,\\
C_{L,{\rm PS}}(z,a_s)=&O(a_s^2)\ \ \ {\rm and}\\
C_{L,g}(z,a_s)=&a_s C_F\left[4\frac{1-z}{z}\right].
\end{split}
\ee
Therefore, these coefficient functions up to and including the $O(a_s^2)$ terms
are required for a NLO calculation of $d\sigma^h_{q_J,L}/dx$.
This means that the LO part of the longitudinal component of $d\sigma^h_{q_J,F}/dx$, 
being of $O(a_s^2)$, is required for consistency in a NLO calculation of $d\sigma^h_{q_J,L}/dx$.
Note that the transverse coefficient functions $C_{T,X}=C_X-C_{L,X}$.

The asymmetric cross section $d\sigma^h_{q_j,A}/dx$, which is due to parity-violating effects of the $Z$ boson,
appears in the differential cross section \cite{Mele:1990cw,Nason:1993xx,Rijken:1996vr}
\be
\frac{d^2 \sigma^h_{q_J}}{dx d \cos \theta}=\frac{3}{8}(1+\cos^2 \theta)\frac{d\sigma^h_{q_J,T}}{dx}
+\frac{3}{4}\sin^2 \theta \frac{d\sigma^h_{q_J,L}}{dx}+\frac{3}{4}\cos \theta \frac{d\sigma^h_{q_J,A}}{dx}
\label{genformofdsigbydxdcostheta}
\ee
where $\theta$ is the scattering angle of the produced hadron. 
From a standard tensor analysis, eq.\ (\ref{genformofdsigbydxdcostheta}) is the most general form 
for inclusive single hadron production from a spin 1 vector boson.
Note that eq.\ (\ref{XSdeomptotransandlong}) is reproduced on integrating over $\cos \theta$.
Unlike $d\sigma^h_{q_J,T}/dx$, $d\sigma^h_{q_J,L}/dx$ and $d\sigma^h_{q_J,F}/dx$,
the factorized $d\sigma^h_{q_J,A}/dx$ depends only on the valence quark FFs:
\be
\frac{d\sigma^h_{q_J,A}}{dx}(x,s)=\int_x^1 \frac{dz}{z}
\frac{d\sigma_{q_J,A}}{dz}\left(z,s,M_f^2\right)D_{\Delta_c q_J/\bar{q}_J}^h\left(\frac{x}{z},M_f^2\right).
\ee
Here, the partonic cross sections are given by
\be
\frac{d\sigma_{q_J,A}}{dz}\left(z,s,M_f^2\right)=\sigma_0 (s) Q_{q_J}^A(s)C_A\left(z,a_s(s),\ln \frac{M_f^2}{s}\right),
\ee
where the electroweak coupling
\be
Q_{q_J}^A(s)=2a_e a_{q_J}\left(e_e \frac{s(s-M_Z^2)}{(s-M_Z^2)^2+M_Z^2 \Gamma_Z^2}
+2v_e v_{q_J}\frac{s^2}{(s-M_Z^2)^2+M_Z^2 \Gamma_Z^2}\right)
\ee
and, to NLO,
\be
C_A(z,a_s)=C_{T,{\rm NS}}(z,a_s)-a_s C_F (1-z).
\ee

\section{Global fitting of fragmentation functions from $e^+ e^-$ reaction data \label{epemXS}}

A comprehensive review of measurements of inclusive single hadron production in $e^+ e^-$ reactions, 
$e^+ e^-\rightarrow \gamma^*,Z \rightarrow h+X$, up to the year 1995 is given in Ref.\ \cite{Lafferty:1995jt},
and, to the best of the author's knowledge, all data to the present day 
can be obtained in numerical form from Ref.\ \cite{durhamHEPDATA}.
Usually, the normalized cross section
\be
F^h_{S_A}(x,s)=\frac{\sum_{q_J\in S_A}\frac{d\sigma_{q_J}^h}{dx}(x,s)}
{\sum_{q_J\in S_A}\sigma_{q_J}(s)}
\label{genformofdata}
\ee
is measured, where $S_A$ is the set of all tagged quarks. 
Equation (\ref{physmomsumrule}) requires the normalization of this cross section to be such that 
\be
\int_0^1 dx \frac{x}{2} F^h_{S_A}(x,s)=1.
\label{normofepemXS}
\ee
At LO, the results of subsection \ref{epemtheory} give
\be
F^h_{S_A}(x,s)=\frac{\sum_{q_J\in S_A}Q_{q_J}D_{q_J/\bar{q}_J}^h(x,s)}{\sum_{q_J\in S_A}Q_{q_J}},
\label{epemmeasuredatLO}
\ee
i.e.\ the measured cross section is essentially an FF or a charge-weighted sum of FFs.
In practice, cross sections are measured in an $x$ bin of finite width.
Writing the range of the bin as $x_l<x<x_h$, such cross sections must be calculated as
\be
\langle F^h_{S_A}\rangle (x_l,x_h,s)=\frac{1}{x_h-x_l}\int_{x_l}^{x_h}dx F^h_{S_A}(x,s).
\ee
Fortunately, by working in Mellin space this bin averaging can be calculated analytically:
From eq.\ (\ref{invmelltrans})
\be
\langle F^h_{S_A}\rangle (x_l,x_h,s)=\frac{1}{x_h-x_l}
\frac{1}{2\pi i}\int_C dN \frac{x_h^{1-N} -x_l^{1-N}}{1-N} F^h_{S_A}(N,s).
\ee
The large amount of accurate data for inclusive single light charged and neutral hadron production from 
these reactions, in particular from LEP,
has been used to accurately constrain many of the degrees of freedom for unpolarized 
charge-sign unidentified FFs for light hadrons through global fits.
The experimental data sets from $e^+ e^-$ reactions relevant for present global fits of FFs 
for \cpi, \cka, \pr, \nka\ and \lam\ particles
are summarized in Tables \ref{PionResults} --- \ref{LambdaResults}.
The normalization uncertainty common to all data points is also given and therefore, in order to 
be correctly implemented, should be treated separately
from the statistical uncertainty which varies from data point to data point.
The method to do this will be discussed in subsection \ref{syserrinchi2}.
Measurements in which the quark at the electroweak boson vertex is identified, or ``tagged'',
as either a light, $b$ or $c$ flavour quark
have been performed by the TPC \cite{Aihara:1986mv}, DELPHI \cite{Abreu:1998vq} and SLD \cite{Abe:1998zs} collaborations, 
and as either a $u$, $d$, $s$, $b$ or $c$ flavour quark by the OPAL collaboration \cite{Abbiendi:1999ry},
which allows FFs of quarks with the same electroweak couplings
to be separately constrained, namely $u$ and $c$ quark FFs can be separated from one another, and
$d$, $s$ and $b$ quark FFs can be separated from one another \cite{Kretzer:2001pz}.
The {\it tagging probabilities} 
\be
\eta^h_{q_J}(x,s)=\int_{x}^{1}dx F^h_{q_J}(x,s)=
(1-x)\langle F^{h^\pm}_{q_J}\rangle (x,1,s)
\label{tagprobP}
\ee
have been measured \cite{Abbiendi:1999ry} by the OPAL collaboration for 
$q_J=u$, $d$, $s$, $c$ and $b$ individually, which in particular 
are the only data that give phenomenological constraints on the separation between the light quark flavours.
These are the only data from $e^+ e^-$ reactions which can separately constrain the $d$ and $s$ quark FFs.
However, they are rather limited in number and/or accuracy, and in particular only exist for $x>0.2$.
In addition, the experimental definition of these measurements may not coincide with the theoretical definition in 
eq.\ (\ref{tagprobP}) \cite{deFlorian:2007hc}.

Global fits to data from $e^+ e^-$ reactions have been performed in Refs.\ 
\cite{Chiappetta:1992uh,Cowan:1994dd,Binnewies:1994ju,deFlorian:1997zj,Kretzer:2000yf,Kniehl:2000fe,Bourhis:2000gs,
Bourrely:2003wi,Albino:2005me,Albino:2005mv,Hirai:2007cx,deFlorian:2007aj,deFlorian:2007hc,Albino:2008fy}
via minimization of $\chi^2$, which is typically achieved through the minimization program MINUIT \cite{minuit}.
Currently, all calculations are performed to NLO accuracy, and
competitive phenomenological extractions of $\alpha_s(M_Z)$ have simultaneously been performed in some of these fits.
Other determinations of FFs \cite{Baier:1979tp,Anselmino:1982yf,Greco:1993tt} using theoretical constraints such as
dimensional-counting rules for the large $z$ behaviour \cite{Jones:1978he,Baier:1979tp} or Monte Carlo \cite{Greco:1993tt}, 
have used such data to motivate the values of any degrees of freedom in these approaches.
Data for which the energy scale $E_s$ is less than a few GeV are excluded to avoid higher twist effects, detected hadron mass
effects, the unreliability of truncated perturbative series, and other effects beyond the standard FO approach
that may be relevant at low $E_s$.
Usually, data for which $x <0.1$, or even $x<0.05$, are excluded from fits because small $x$ logarithms in the FO calculation
may prevent this region from being described, or because of other small $x$ effects not accounted for in the calculations.
Improvements to the theoretical description of this region will be discussed in section \ref{sglres}.
Measurements of \pr\ production in $e^+ e^-$ reactions are rather inaccurate,
and therefore the much more accurate measurements of production of unidentified hadrons, 
when used together with measurements of \cpi\ and \cka\ production and/or
FFs for these two particles, would make a significant improvement to the constraints on FFs for \pr.
However, the amount of contamination by charged particles other than the light charged hadrons in unidentified hadron data is unknown
but may be significant:
As noted in Ref.\ \cite{Kniehl:2000fe}, hadron unidentified data from the ALEPH \cite{Buskulic:1994ft,Buskulic:1995aw}
and OPAL \cite{Ackerstaff:1998hz} collaborations are inconsistent 
with similar data from DELPHI \cite{Abreu:1998vq} and SLD \cite{Abe:1998zs}.
Furthermore, including such data would require uniting the fits for each hadron species into a single fit,
which is a greater computational challenge than performing these fits separately
(although the fits for different hadron species may have to be united 
if $\alpha_s(M_Z)$ is included in the list of parameters to be fitted).

Longitudinal and transverse cross section measurements for hadron production in $e^+ e^-$ reactions
have provided accurate constraints on the summed FFs $D_{q_I/\bar{q}_I}^h$.
The differences, namely the valence quark FFs $D_{\Delta_c q_I/\bar{q}_I}^h$, could be constrained by 
measurements of the asymmetry cross section $d\sigma^h_{q_J,A}/dx$, which has been performed at LEP \cite{Akers:1995wt}.
Unfortunately, no identification of the produced charged particle's species has been made in these asymmetry measurements, 
so that, as mentioned in section \ref{intro}, currently the only constraints 
on the valence quark FFs are provided by data from $pp$ reactions at RHIC
which are rather poor, and also by HERMES data for which low $E_s=Q$ effects may be important.

In global fits, the FFs are extracted at some ``initial'' or ``starting'' scale $M_f=M_0$.
The FFs $D_i$, which are usually taken to be $D_i^{h/\bar{h}}$ and the charge-sign asymmetry $D_i^{\Delta_c h/\bar{h}}$
(which in the case that $i$ is a quark are equal respectively to the summed and valence quark FFs), 
or $D_i^h$ and $D_i^{\bar{h}}$, are typically parameterized in the form
\be
D_i(z,M_0^2)=N_i z^{a_i} (1-z)^{b_i} \times f_i(z),
\label{standardFFparam}
\ee
and the parameters $N_i$, $a_i$, $b_i$, \ldots are freed in the fit.
The function $f_i(z)$, which was set to 1 in early fits, 
depends on further free parameters and is used to extend the function space available to the $D_i$.
The $(1-z)^b$ behaviour is motivated by dimensional counting rules \cite{Jones:1978he},
and is expected to set in at large enough $M_f$ due to the large $z$ behaviour of the splitting functions \cite{Albino:2000cp}.
The $z^a$ behaviour is chosen because of the behaviour of the evolution at small (but not too small) $z$,
to be discussed in section \ref{sglres}.
However, these physical arguments do not precisely follow from QCD and, furthermore,
the choice of parameterization used in a fit only needs to provide a sufficient region of function space
to the FFs for the data to be as well described as the approximation for the high energy theory allows.
Usually, $M_0$ is chosen to be below
the lowest value of $E_s$ of the data, but such that $M_0 \gg \Lambda_{\rm QCD}$ in order for the perturbative calculation of 
the DGLAP evolution to still be valid. Typically $M_0=O(1)$ GeV.
In principle, any value may be chosen. Even the choice $M_0 = O(\Lambda_{\rm QCD})$ may be justified 
since the resulting large theoretical errors in the evolution around this scale
would effectively be absorbed into the parameters on fitting,
i.e.\ these low scale effects essentially just modify the choice of parameterization,
although then the choice of parameterization in eq.\ (\ref{standardFFparam}) may not be suitable.

These choices of parameterization for each FF are usually the strongest non perturbative theoretical constraint,
if the other non perturbative constraints are exact or at least good approximations in the framework of current
experimental and theoretical information.
An example of such a reliable constraint is eq.\ (\ref{chargeconjsymm}), 
which is exact in the Standard Model (in a physical scheme like $\overline{\rm MS}$).
On the other hand, as noted for the generation of the asymmetry between
strange quark and antiquark PDFs of the proton in Ref.\ \cite{Catani:2004nc},
because $P_{\Delta_c \Sigma}\neq P_{\Delta_c {\rm NS}}$ beyond NLO (see the discussion following eq.\ (\ref{achoiceofvalenceNS})),
the condition for unfavoured FFs to obey $D_{q_I}^h=D_{\bar{q}_I}^h$ for all $M_f$ is only possible in certain cases.
For example, neglecting electroweak effects, suppose we impose the constraint 
$D_u^{\pi^+}-D_{\bar{u}}^{\pi^+}=D_{\bar{d}}^{\pi^+}-D_d^{\pi^+}$ at $M_f=M_0$,
which follows from SU(2) isospin symmetry between $u$ and $d$.
This symmetry is exact in the limit that the masses of these quarks are equal (and electroweak effects can be ignored), 
which is a reasonable assumption given that their masses are 1.5 -- 4 MeV and 4 -- 8 MeV respectively.
Then, if the charge-sign symmetry constraint $D_{q_I}^{\pi^+}=D_{\bar{q}_I}^{\pi^+}$ for $q_I=s,c,b,\ldots$ 
is also chosen to hold at $M_f=M_0$, it will hold for all $M_f$, 
as will the isospin symmetry constraint above, because $D_{\Delta_c \Sigma}$
and $D_{I,\Delta_c {\rm NS}}$, which do not mix with any other FFs on evolution,
vanish at $M_f=M_0$ and therefore all $M_f$.
In other words, the violation of SU(2) isospin is responsible 
for the asymmetry between the fragmentation to \cpi\ from an unfavoured quark and that from its antiquark.
Therefore, these unfavoured asymmetries to $\pi^+$ are expected to be larger than the unfavoured asymmetries to e.g.\ $h=K^+$.
Note that electroweak effects only predict non-zero unfavoured asymmetries for $\pi^+$ and $K^+$, but
does not which are the largest.
As for eqs.\ (\ref{momsumrule}) and (\ref{chargesumrule}), this $M_f$ independence of 
the isospin and charge-sign symmetry constraints implies that they may be regarded as ``physical'', 
but only when taken together.
Other than the constraints mentioned above,
insufficient phenomenological constraints in global fits may require imposing yet further theoretical constraints between FFs 
which have no real physical justifications and which may not hold for all $M_f$.
We will consider some examples below.

If FFs for different hadron species are related by a simple symmetry, FFs for one of these hadron species 
can be used to make predictions for, or can be constrained by, processes in which another of these hadrons is produced.
For example, SU(2) isospin symmetry implies
\be
D_i^{\pi^0}=\frac{1}{2}D_i^{\pi^\pm},\qquad
D_{u,d,s,c,b,g}^{K_S^0}=\frac{1}{2}D_{d,u,s,c,b,g}^{K^\pm},\qquad {\rm and}\qquad
D_{u,d,s,c,b,g}^{n/\overline{n}}=D_{d,u,s,c,b,g}^{p/\overline{p}}.
\label{SU2isospinres}
\ee
However, it should be noted here that some hadrons are not produced by direct partonic fragmentation,
but rather by decay from another hadron, which itself may have been produced either by partonic fragmentation or hadronic decay.
If the decay channels involved in the production of e.g.\ $\pi^0$ and \cpi\ do not respect SU(2) isospin symmetry,
the relation between their FFs above will be violated even if it is true for direct fragmentation.

A valuable consequence of the second result in eq.\ (\ref{SU2isospinres}) arises \cite{Christova:2008te}:
For any initial state, the difference between $d\sigma^{K^\pm}/2$ and $d\sigma^{K_S^0}$
depends only on the FF $(D_d-D_u)^{K^\pm}/2=(D_u-D_d)^{K_S^0}$, which is a non singlet.
Therefore, this FF is rather well constrained relative to the other FF components.
This is similar to the consequence from charge conjugation symmetry that the difference
between $d\sigma^{h^+}$ and $d\sigma^{h-}$ depends only on the charge-sign asymmetry FFs,
except for the absence of $e^+ e^-$ reaction data in that case.
The assumption $D_{s,c,b,g}^{K_S^0}=\frac{1}{2}D_{s,c,b,g}^{K^\pm}$ may be violated by those \cka\ and \nka\
arising from complicated decay channels involving other hadrons instead of from direct partonic fragmentation,
which may differ between \cka\ and \nka. 
However, SU(2) isospin suggests that these hadronic decay processes for \cka\ should be similar to those for \nka.
An indication of this is found in the AKK08 fit because the fitted masses of \cka\ and \nka\ differ from their true
masses by the same amount, as will be discussed in subsection \ref{AKK08}.
Since the non singlet splitting functions and coefficient functions for $e^+ e^-$ reactions are known to NNLO,
the non singlet $(D_u-D_d)^{K_S^0}$ can also be extracted, and through it $\alpha_s (M_Z)$, to NNLO. 
Such a procedure would be similar to the NNLO extraction of the non singlet PDF $f_u^p-f_d^p$ and $\alpha_s (M_Z)$
using data from proton and deuteron targets \cite{Blumlein:2004ip},
and would allow for a further test of perturbative stability in the timelike case.

\begin{figure}[h!]
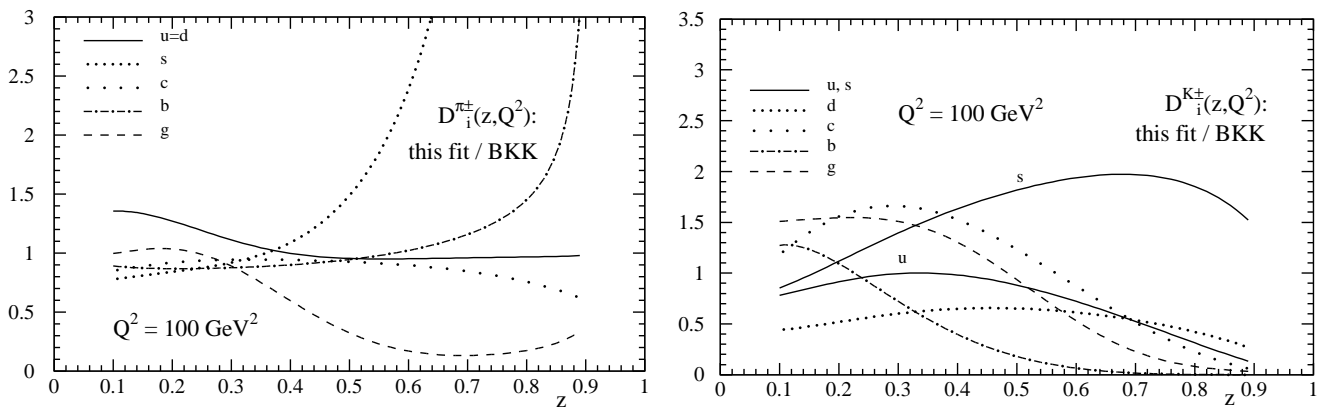

\parbox{.49\linewidth}{
\begin{center}
\includegraphics[width=8.5cm]{phd39.epsi}
\end{center}
}\hspace{0cm}
\parbox{.49\linewidth}{
\begin{center}
\includegraphics[width=8.5cm]{phd40.epsi}
\end{center}}
\caption{Ratios of Kretzer (labeled ``this fit'') FFs to BKK FFs for \cpi\ (left) 
and \cka\ (right) at $M_f^2=100$ GeV$^2$ (where $M_f$ is written
as ``$Q$''). From Ref.\ \cite{Kretzer:2000yf}. \label{phd39}}
\end{figure}

\begin{figure}[h!]
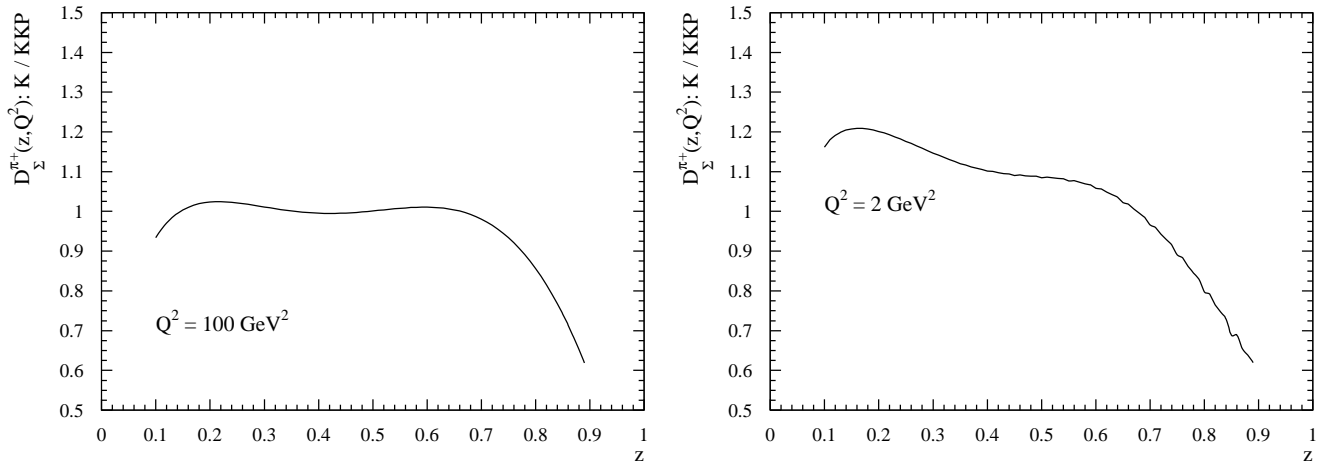

\parbox{.49\linewidth}{
\begin{center}
\includegraphics[width=8.5cm]{sing.100.epsi}
\end{center}
}\hspace{0cm}
\parbox{.49\linewidth}{
\begin{center}
\includegraphics[width=8.5cm]{sing.2.epsi}
\end{center}}
\caption{As in Fig.\ \ref{phd39}, for the ratios of the Kretzer (labeled ``K'') to the KKP values for the singlet
$D_\Sigma^{\pi^+}=D_u^{\pi^+}+D_{\bar{u}}^{\pi^+}+D_d^{\pi^+}+D_{\bar{d}}^{\pi^+}+D_s^{\pi^+}+D_{\bar{s}}^{\pi^+}$
at $M_f^2=100$ GeV$^2$ (left) and 2 GeV$^2$ (right). From Ref.\ \cite{Kretzer:2001pz}. \label{sing.x}}
\end{figure}

Aside from the rather limited measurements of tagging probabilities from OPAL \cite{Abbiendi:1999ry}, 
defined in eq.\ (\ref{tagprobP}),
measurements at LEP, taken at center-of-mass (c.m.) energies at the $Z$ pole mass, 
have not implemented tagging of individual flavours of light quarks.
Consequently, global fits to LEP data will not significantly constrain the differences between the
individual light quark flavour FFs.
If the OPAL tagging probabilities are not included in the fit, these differences will be completely unconstrained,
leading to significant discrepancies between each of the light quark FFs from different sets such as BKK and Kretzer,
as illustrated in Fig.\ \ref{phd39}.
These discrepancies are constrained solely by theoretical constraints, which differ from one set to another.
However, LEP data do constrain the sum of these FFs weighted with the
respective electroweak couplings given by eq.\ (\ref{elecweakfacttot}) (neglecting effects beyond LO).
In fact, because these electroweak couplings are approximately equal at the $Z$ pole mass according eq.\ (\ref{elecweakfacttot}) 
this sum is approximately proportional to the 3 flavour singlet quark FF \cite{Kretzer:2001pz}.
This may explain why, despite the very different theoretical constraints 
on the KKP \cite{Kniehl:2000fe} and Kretzer \cite{Kretzer:2000yf} sets, 
the values for this FF from these sets are very similar in the fit range of $x$, 
except at large $z$ because the data at large $x$ are scarce (see Fig.\ \ref{sing.x}).
Some lower energy data was also used in the extraction of these FF sets, although they are much 
lower in accuracy and number compared to the LEP data. 
Data over a range of c.m.\ energies will provide some constraints
on the difference between the $u$ quark FF and the sum of the $d$ and $s$ quark FFs.
This may explain why, in Fig.\ \ref{phd39}, the $u$ quark FFs from BKK and Kretzer are similar,
at least relative to the $s$ quark FF, and also to the $d$ quark FF for \cka.
However, because the electroweak couplings of the $d$ and $s$ quarks are equal at all energies, 
no untagged $e^+ e^-$ reaction data can constrain the difference between their FFs. 
Therefore, since the OPAL tagging probabilities were not available at the time 
that these analyses were carried out, the difference between $d$ and $s$ quark FFs 
were constrained by fixing one of these two FFs in the fit. 
For example, in the KKP fit, the $d$ quark FF for \cpi\ is fixed to the $u$ quark one, 
which according to SU(2) isospin symmetry is a very good approximation.
For \cka, the $s$ quark FF is fixed to the $u$ quark FF, which would be a valid approximation if the
$s$ and $u$ quark masses were similar.
This is clearly not the case --- the $s$ quark FF should be somewhat larger because
the $u$ ($\bar{u}$) quark must form a bound state with a heavier $\bar{s}$ ($s$) quark from the vacuum \cite{Field:1976ve}.
This strangeness suppression, measured by the {\it strangeness suppression factor} $\gamma_s$,
being the ratio of the production probability from the hadronization sea of $s$ quarks to that of $u$ and $d$ quarks,
is indeed observed in the OPAL tagging probabilities \cite{Abbiendi:1999ry}.
For \pr, the $d$ quark FF is fixed to half the $u$ quark FF merely to reflect the fact that
there are more $u$ than $d$ quarks in the proton. 
This condition is chosen to hold at $M_f=M_0$, but cannot hold for all $M_f$, and
is therefore not physical.

\begin{figure}[h!]
\includegraphics[width=8.5cm]{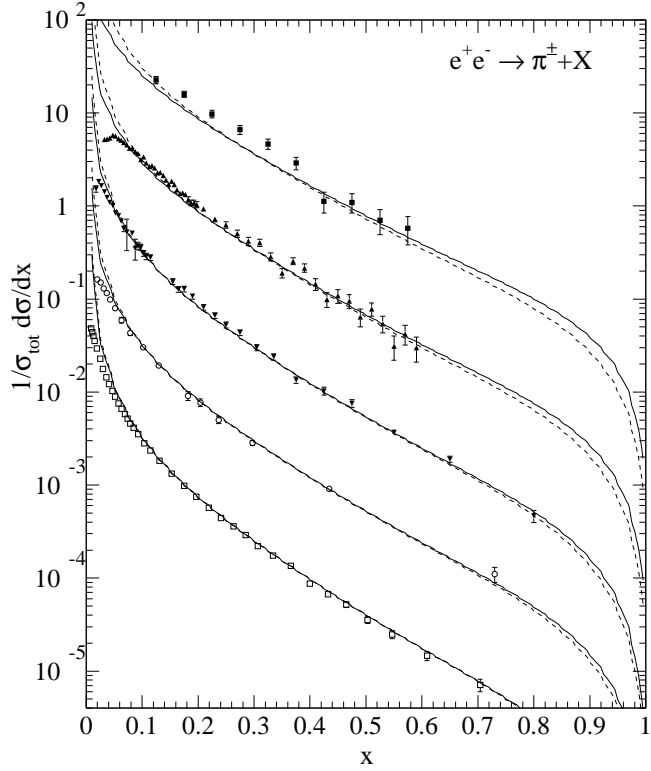}
\caption{NLO (solid) and LO (dashed) predictions using the KKP FF set for inclusive \cpi\ production 
at (from top to bottom) DASP \cite{Brandelik:1978bz}, ARGUS \cite{Albrecht:1989wd}, TPC \cite{Aihara:1988fc}, 
TASSO \cite{Braunschweig:1988hv} and SLD \cite{Abe:1998zs} at, respectively, $\sqrt{s}=5.2$, 9.98, 29, 34 and 91.2 GeV. 
Each pair of curves is rescaled relative to the nearest pair above by a factor of 1/10. From Ref.\ \cite{Kniehl:2000fe}.
\label{pionKKP}}
\end{figure}

To illustrate the reliability of the theoretical approach used in global fits (including the choice of
FF parameterization and SU(2) isospin symmetry), we show in Fig.\ \ref{pionKKP} 
the description, over a large range of c.m.\ energies and using the KKP FF set, of \cpi\ data from $e^+ e^-$ reactions, 
which provide stronger constraints on FFs for \cpi\ than other data do on other FFs. 
The description of all data except the oldest data, from DASP, is very good in the range $x>0.1$.
Data below this range were excluded in the fit, which hence the deviation there from the data is no cause for concern. 
The fact that this deviation is so large may be due to neglected theoretical effects at small $x$ such as
unresummed logarithms or mass effects of the detected hadron.

\section{Predictions from globally fitted fragmentation functions \label{predfromFFs}}

The universality of FFs between processes with different initial states as implied by the factorization theorem 
allows data from one process to be described using FFs sufficiently constrained by data from another process,
giving a test of theoretical and/or experimental results.
Essentially, such a universality test involving predictions that have been measured
would really be a test of whether all relevant physics effects
have been accounted for in the calculations used in the global fits and in the calculations for the predictions,
assuming experimental errors on FFs, to be discussed in section \ref{treaterrs}, have been propagated to these predictions.
In other words, such fits will give more of a handle on 
the importance of contributions of supposedly negligible effects such as higher twist.
In this section we discuss such programs, focusing mainly on descriptions of data from $ep$ and $pp$ reactions.

\subsection{$ep$ reactions from HERA}

\begin{figure}[h!]
\begin{center}
\parbox{.30\linewidth}{
\begin{center}
\includegraphics[width=5cm]{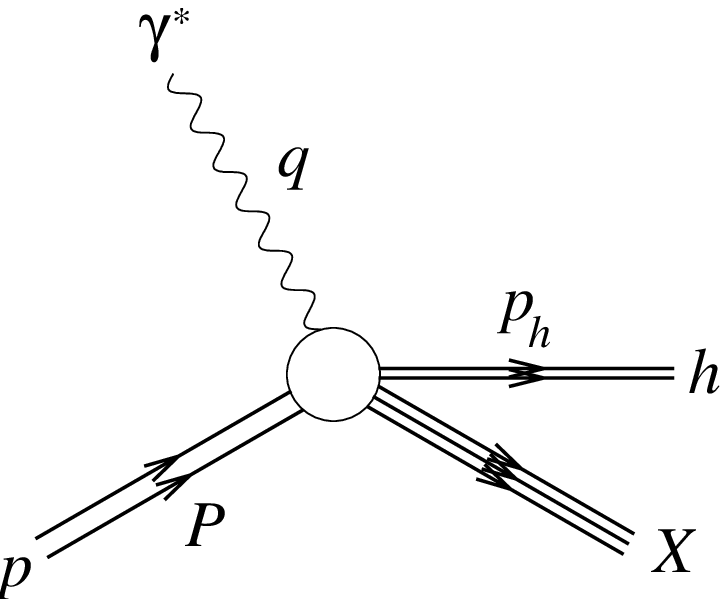}
\end{center}
}\parbox{.09\linewidth}{\huge $\longrightarrow$}
\parbox{.60\linewidth}{
\begin{center}
\includegraphics[width=10cm]{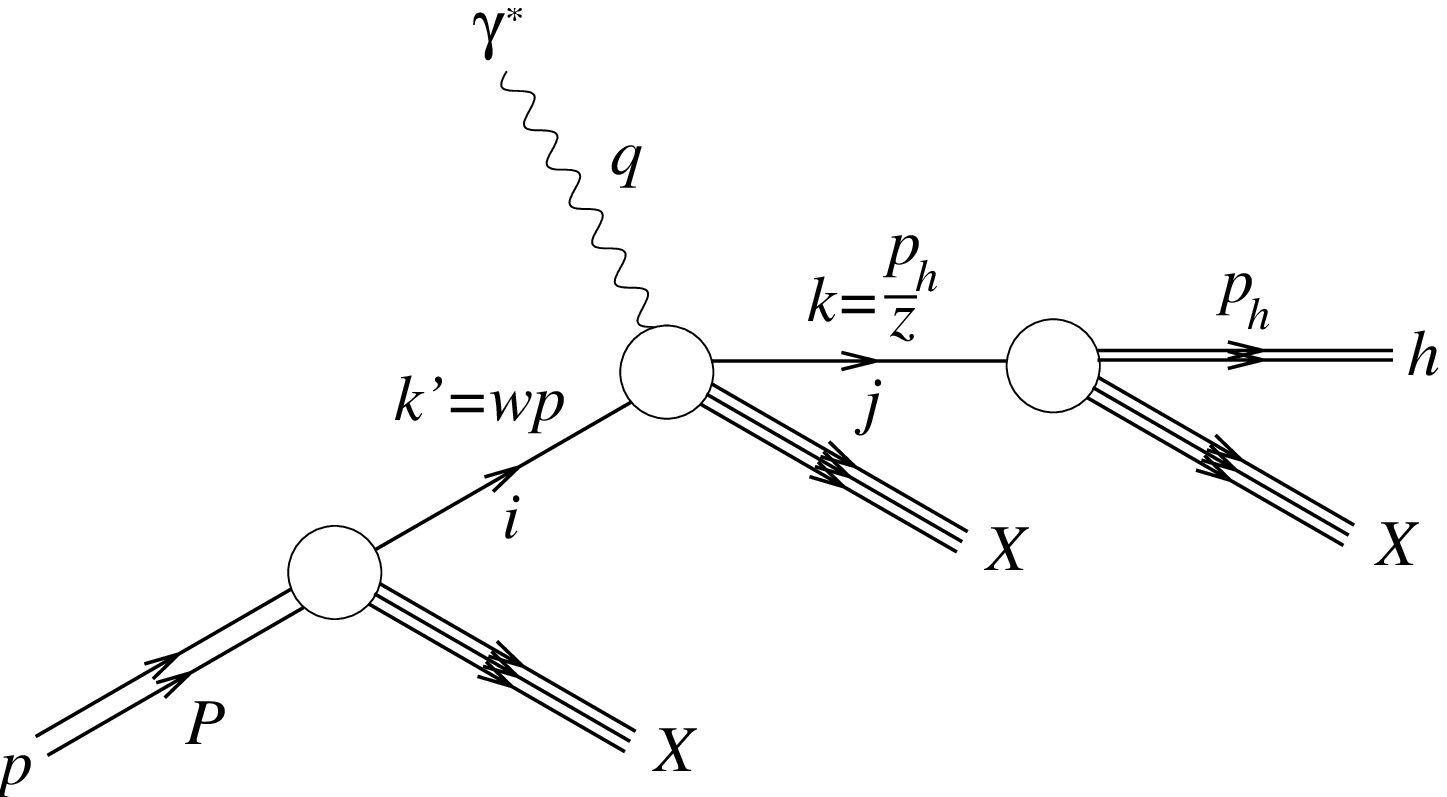}
\end{center}}
\caption{Schematic illustration of factorization in the current fragmentation region for inclusive single hadron production 
in neutral current DIS $ep \rightarrow e+h+X$, giving all relevant kinematics
including that of the initial ($k'$) and final ($k$) state real partons involved in the hard interaction.\label{kin}}
\end{center}
\end{figure}

\begin{figure}[h!]
\includegraphics[width=10cm]{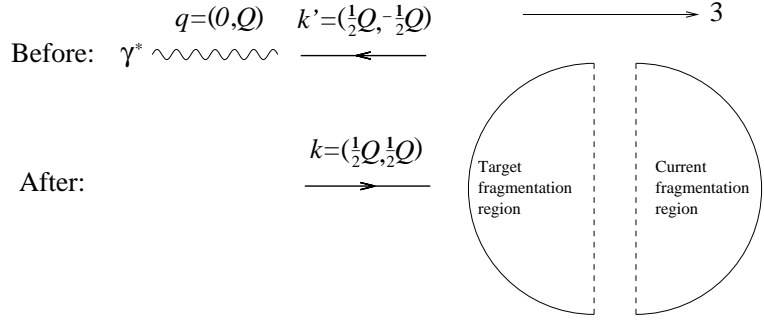}
\caption{Kinematics of the virtual photon and of the quark in the proton which it strikes, before and after this happens,
in the Breit frame.
Only the time and $3$ components of momenta are given, in that order, the $1$ and $2$ components vanishing in all cases.
At LO, all proton remnants move in the negative $3$ direction,
while the struck quark will fragment into a hadron moving in the positive $3$ direction.
At any order, most hadrons produced by fragmentation of a hard 
parton will go into the current fragmentation region, i.e.\ will have a positive $3$ component of momentum. \label{breitproc}}
\end{figure}

In this subsection we discuss inclusive single hadron production in neutral current DIS,
$ep \rightarrow e+h+X$ or, omitting the purely leptonic subprocess and assuming the contribution
from $Z$ boson exchange is negligible, $\gamma^* p \rightarrow h+X$,
whose relevant kinematics are given in Fig.\ \ref{kin} (left).
The hard scale $E_s$ of the process is provided by $Q$, where $Q^2=-q^2>0$ 
is the negative virtuality of the spacelike virtual photon $\gamma^*$.
Unlike $e^+ e^-$ reactions, $ep$ reactions are complicated by the initial state hadron, 
which at leading twist contributes partons with densities given by PDFs to the hard interaction with the virtual photon, and 
contributes hadronic remnants to the final state.
Because of the hardness of the collision, 
the hadronic final state separates into two clusters which are kinematically fairly distinguishable:
Those produced by fragmentation of a hard parton which must therefore be assigned to $h$,
and those which are proton remnants contained in $X$.
However, some hadrons cannot be unambiguously assigned to $h$ or the proton remnants,
which at the theoretical level translates into the need 
for {\it fracture functions} \cite{Trentadue:1993ka,Collins:1997sr} 
in the cross section to absorb additional potential mass singularities.
Fracture functions describe the process $p\rightarrow h+i+X$.
The detected hadron $h$ in this process is a remnant of the initial proton, and the parton $i$ is required
to connect this process with the hard interaction.
In other words, fracture functions describe the partonic structure of the initial hadron 
after it has produced the detected hadron.
In this sense, a fracture function is both an FF and a PDF and depends on two momentum fractions.
Fracture functions are non perturbative and therefore, like FFs and PDFs, 
contribute unknown degrees of freedom to the cross section.
Since on evolution they mix with both FFs and PDFs but not vice versa, one anticipates the existence of a scheme and scale
independent cross section which does not depend on them.
Indeed, fracture functions do not contribute to the cross section 
when the direction of the detected hadron's momentum
is within the {\it current fragmentation region}:
In the Breit frame, which is the frame in which the virtual photon's energy vanishes and
its spatial momentum is antiparallel with the initial proton's (see Fig.\ \ref{breitproc}), this region 
is defined by $\theta<\pi/2$, where $\theta$ is the angle 
between the spatial momentum of the detected hadron and that of the virtual photon.
Consequently, this region is essentially devoid of proton remnants, 
i.e.\ hadrons in this region are assigned to $h$ and not $X$, which results in all non perturbative
information in the calculation of the corresponding leading twist cross section being provided by the FFs and PDFs only.
In terms of the Bjorken scaling variable $x_B=Q^2/(2P\cdot q)$ and the scaled detected hadron momentum 
\be
x=\frac{2p_h \cdot q}{q^2}, 
\label{defofxinSIDIS}
\ee
the cross section after the change of integration variables $z\rightarrow x/z$ and $w\rightarrow x_B/w$ is
\be
\frac{d \sigma^h_{\rm proton}}{dx dx_B dQ^2}(x,x_B,Q^2)
=\sum_{ij} \int_x^1 \frac{dz}{z} \int_{x_B}^1 \frac{dw}{w} \frac{d \sigma^{ij}}{dz dw dQ^2}
\left(w,z,\frac{Q^2}{M_f^2},a_s(M_f^2)\right)
f_i^{\rm proton} \left(\frac{x_B}{w},M_f^2\right)
D_j^h \left(\frac{x}{z},M_f^2\right).
\label{xsinfacttheor}
\ee
Its schematic form is shown in Fig.\ \ref{kin} (right).
For simplicity we have used the same value $M_f$ for the factorization scale of the PDFs as that for the FFs.
The PDF of parton $i$ in a hadron $h$ is written $f_i^h$, and
$d \sigma^{ij}$ is the cross section for the equivalent purely partonic processes $\gamma^* i \rightarrow j+X$,
which is known to NLO \cite{Altarelli:1979kv}.
At LO it is given by
\be
\frac{d \sigma^{ij}}{dz dw dQ^2}
\left(w,z,\frac{Q^2}{\mu^2},a_s(\mu^2)\right)
=\frac{d\sigma_0}{dQ^2} (Q^2)\sum_I \delta_{ij} \delta_{iI}
e_{q_I}^2 \delta(1-w) \delta(1-z),
\ee
where $\sigma_0$ is the cross section for the elastic process
$e \mu \rightarrow e \mu$ involving one photon exchange in the $t$-channel.

The normalized cross section
\be
F^h_{\rm proton}(x,{\rm cuts})=\frac{\int_{\rm cuts}dx_B dQ^2\frac{d \sigma^h_{\rm proton}}{dx dx_B dQ^2}(x,x_B,Q^2)}
{\int_{\rm cuts}dx_B dQ^2\frac{d \sigma_{\rm proton}}{dx_B dQ^2}(x_B,Q^2)},
\label{epmeasXS}
\ee
is measured in practice, where $d \sigma_{\rm proton}$ is the DIS ($ep\rightarrow e+X$) cross section.
Some cancellation of PDF uncertainties between the numerator and denominator of eq.\ (\ref{epmeasXS}) should occur.
In any case, these uncertainties are expected to be lower than the FF ones.
The normalization of $F^h_{\rm proton}$ may be obtained 
by following similar steps to those that lead to eq.\ (\ref{normofepemXS}):
Similar to eq.\ (\ref{NhfromXS}), the number of hadrons of species $h$ produced 
with energy or momentum fraction between $x$ and $x+dx$ is $N^h(x,Q^2)dx=[dx /(d \sigma_{\rm proton}/ (dx_B dQ^2))]
d \sigma^h_{\rm proton}/(dx dx_B dQ^2)$. 
The total momentum in the $3$ direction of the hadrons in the current fragmentation region is 
$\sum_h \int_0^1 dx N^h(x,Q^2) p_h$, where $p_h$ is the $3$ component of the detected hadron's momentum.
This must equal that of the struck quark, which according to Fig.\ \ref{breitproc} is $Q/2$ in the Breit frame.
Therefore, since eq.\ (\ref{defofxinSIDIS}) implies that $x=2p_h/Q$ in this frame, 
$F^h_{\rm proton}(x,{\rm cuts})$ has the normalization
\be
\int_0^1 dx x F^h_{\rm proton}(x,{\rm cuts})=1.
\ee
From the momentum sum rule, eq.\ (\ref{momsumrule}), this gives
\be
\frac{d \sigma_{\rm proton}}{dx_B dQ^2}(x_B,Q^2)=\sum_i \int_{x_B}^1
\frac{dw}{w} \frac{d\sigma^i}{dw dQ^2}\left(w,\frac{Q^2}{\mu^2},a_s(\mu^2)\right)
f_i^{\rm proton} \left(\frac{x_B}{w},\mu^2\right),
\label{xsinfacttheordenom}
\ee
where
\be
\frac{d\sigma^i}{dw dQ^2}=\sum_j \int_0^1 dz \frac{d \sigma^{ij}}{dz dw dQ^2},
\ee
which at LO is therefore
\be
\frac{d \sigma^i}{dw dQ^2}=\frac{d\sigma_0}{dQ^2} (Q^2)\sum_I \delta_{iI}e_{q_I}^2 \delta(1-w).
\ee

\begin{figure}[h!]
\begin{center}
\includegraphics[width=12cm]{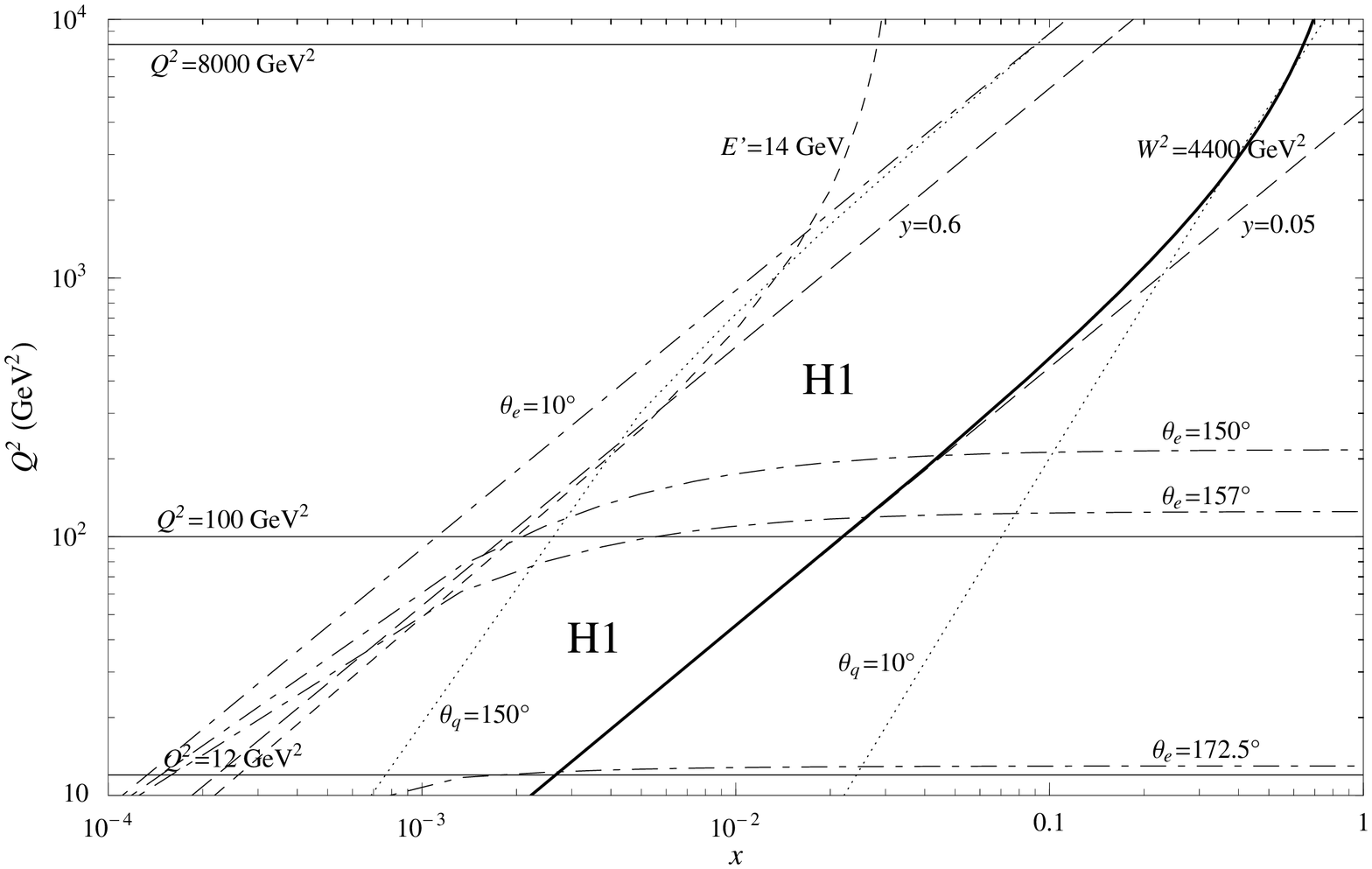}
\end{center}
\begin{center}
\includegraphics[width=12cm]{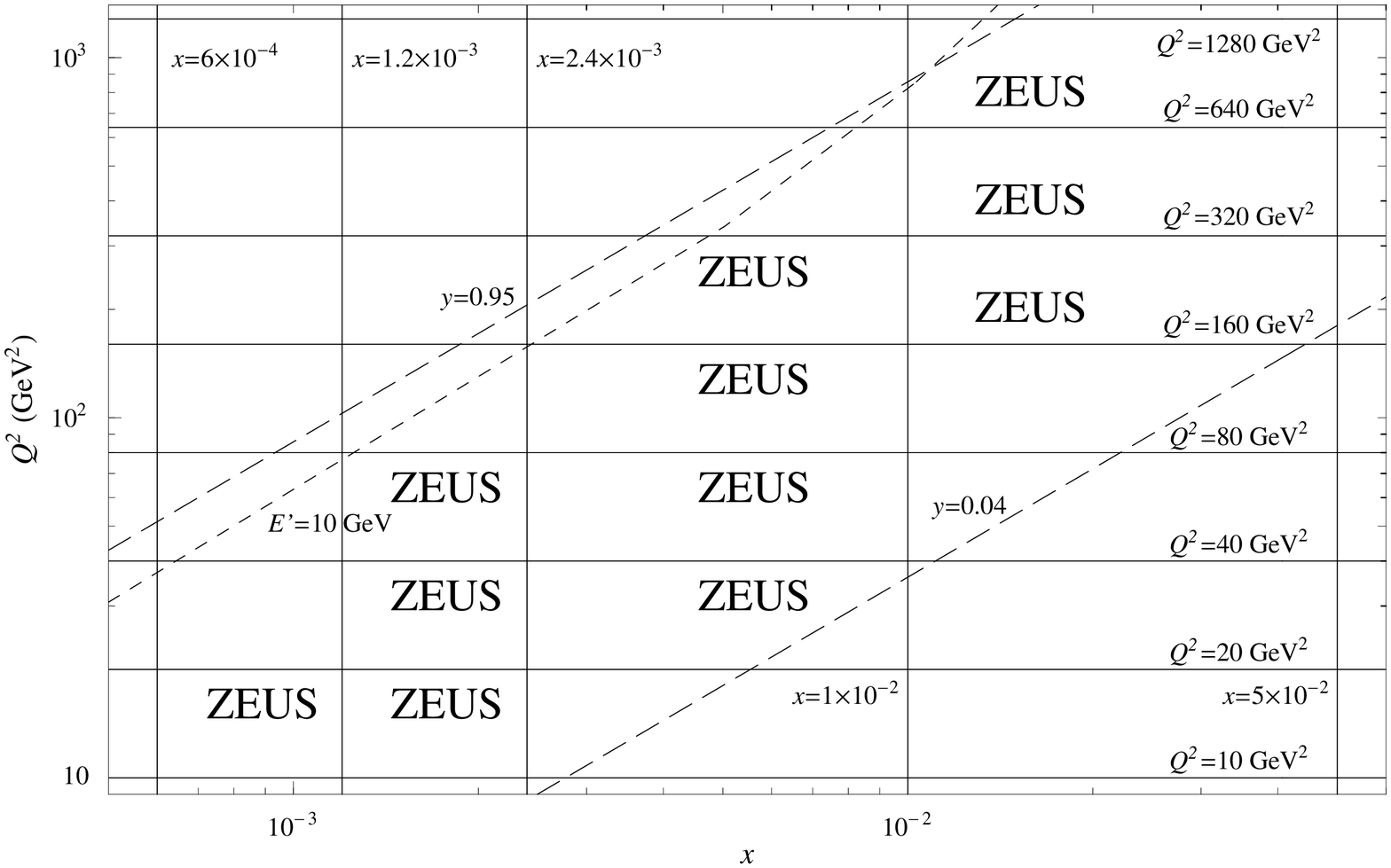}
\end{center}
\caption{Cuts in the $(x_B,Q^2)$ plane (where $x_B$ is written ``$x$'') used in the H1 analysis of Ref.\ \cite{Adloff:1997fr} (left)
with $E=27.5$ GeV and $\sqrt{s}=300.3$ GeV,
and in the ZEUS analysis of Ref.\ \cite{Derrick:1995xg} (right) with $E=26.7$ GeV and $\sqrt{s}=296$ GeV.
The low and high $Q^2$ regions used by H1 are each indicated by the label ``H1''
and the region used by ZEUS by the label ``ZEUS'', being bound in each case by the nearest cut to the label.
From Ref.\ \cite{Albino:2006wz}. \label{limits}}
\end{figure}

The region of the $(x_B,Q^2)$ plane, written ``cuts'' in eq.\ (\ref{epmeasXS}),
is usually specified by experimentalists as cuts on 
the squared c.m.\ energy of the virtual photon-proton system,
\be
W^2=(P+q)^2=Q^2 \left(\frac{1}{x_B}-1\right),
\ee
and the fraction of the energy of the initial electron/positron which is lost in the rest frame of the proton,
\be 
y_B=\frac{P\cdot q}{P\cdot k}=\frac{Q^2}{x_B s},
\ee
where $\sqrt{s}$ is the c.m.\ energy of the initial $ep$ system.
Working in the laboratory frame, a lower bound on the scattered electron's energy
\be
E'=E-Q^2\left(\frac{E}{x_B s}-\frac{1}{4E}\right),
\label{EprimeEQ2xs}
\ee
where $E$ is the energy of the initial electron/positron,
is sometimes imposed to prevent the scattered electron/positron being falsely identified 
with isolated low energy deposits in the calorimeter while
the true scattered electron/positron passes undetected down the beam pipe.
The H1 collaboration imposes additional cuts \cite{Adloff:1997fr,Kant:1995sc}
on the angle of deflection of the electron/positron and struck parton in the laboratory frame, 
respectively $\theta_e$ and $\theta_p$, to maintain good detector acceptance. 
These are given in terms of $x_B$ and $Q^2$ by
\be
\cos \theta_e
=\frac{x_B s\left(4E^2-Q^2\right)-4E^2 Q^2}{x_B s\left(4E^2+Q^2\right)-4E^2 Q^2}
\ee
and
\be
\cos \theta_p =\frac{x_B s(x_B s-Q^2)-4E^2 Q^2}{x_B s(x_B s-Q^2)+4E^2 Q^2}.
\ee
The regions in the $(x_B,Q^2)$ plane used by the H1 collaboration in Ref.\ \cite{Adloff:1997fr}
and by the ZEUS collaboration in Ref.\ \cite{Derrick:1995xg} are shown in Fig.\ \ref{limits}.
As for $F^h_{S_A}$ in $e^+ e^-$, $F^h_{\rm proton}$ is usually averaged over a finite bin width in $x$.

Conservation of energy and momentum implies that, after being struck by the virtual photon, 
the quark's spatial momentum changes in sign only, as illustrated by Fig.\ \ref{breitproc}.
Therefore, at LO, all hadrons produced in the current fragmentation region originate from fragmentation of the struck quark.
(We note in passing that, at LO, hadrons produced in the remaining region $\theta>\pi/2$, 
the {\it target fragmentation region} (see Fig.\ \ref{breitproc}), can only be proton remnants,
which theoretically means that fracture functions play a more important role than FFs (or PDFs) in this region, 
making this a perfect region for extracting fracture functions.)
Therefore, measurements in the current fragmentation region can give good constraints on the FFs.
The distribution of hadrons in this region is very similar to that in any one of the two event hemispheres
in $e^+ e^-$ reactions with $\sqrt{s}=Q$, where a hemisphere is defined to be the union of all
directions that make an angle less than $\pi/2$ with the
thrust axis of the hadron distribution of the event, 
which at LO is aligned with either the primary quark or antiquark in the case of $e^+ e^-$ reactions.
In mathematical terms, if the range in $Q^2$ is negligible, which is usually a good approximation
since $F^h_{\rm proton}$ is approximately independent of $Q$ up to $O(1/\ln Q)$ corrections,
the LO calculation
\be
F^h_{\rm proton}(x,{\rm cuts})=\frac{\sum_{J=1}^{n_f} e_{q_J}^2 (G_{q_J}(Q^2)D_{q_J}^h (x,Q^2)
+G_{\bar{q}_J}(Q^2)D_{\bar{q}_J}^h (x,Q^2))}{\sum_{J=1}^{n_f} e_{q_J}^2 (G_{q_J}(Q^2)+G_{\bar{q}_J}(Q^2))},
\label{simpformofF}
\ee
is similar to that for $e^+ e^-$ in eq.\ (\ref{epemmeasuredatLO}) with $\sqrt{s}=Q$ and tagged quarks summed over all flavours
(and with $Q_{q_J} \rightarrow e_{q_J}^2$ because the contribution to the $ep$ reaction
cross section from $Z$ boson exchange has not yet been calculated)
except for the presence of the PDF dependent factors $G_{q_J}(Q^2)=\int_{\rm cuts} dx_B\ f_{q_J}^{\rm proton}(x_B,Q^2)$.
Because of the variation among the $G_{q_J}$, 
and assuming that SU(2) isospin symmetry and charge-sign symmetry of the initial proton's sea are poor approximations,
the separation of the different quark flavour FFs, and also the valence quark FFs if the charge-sign
of the produced hadron in $ep$ reactions is identified,
can be constrained by suitable data from both $e^+ e^-$ and $ep$ reaction data, or even
by $ep$ data alone by choosing different regions in the $(x_B,Q^2)$ plane for ``cuts''.
Recall that untagged data from $e^+ e^-$ reactions can constrain neither the separation between FFs of quarks
of the same electroweak couplings,
nor, in the case of transverse and longitudinal cross sections, the valence quark FFs.

\begin{figure}[h!]
\parbox{.49\linewidth}{
\begin{center}
\includegraphics[width=8cm]{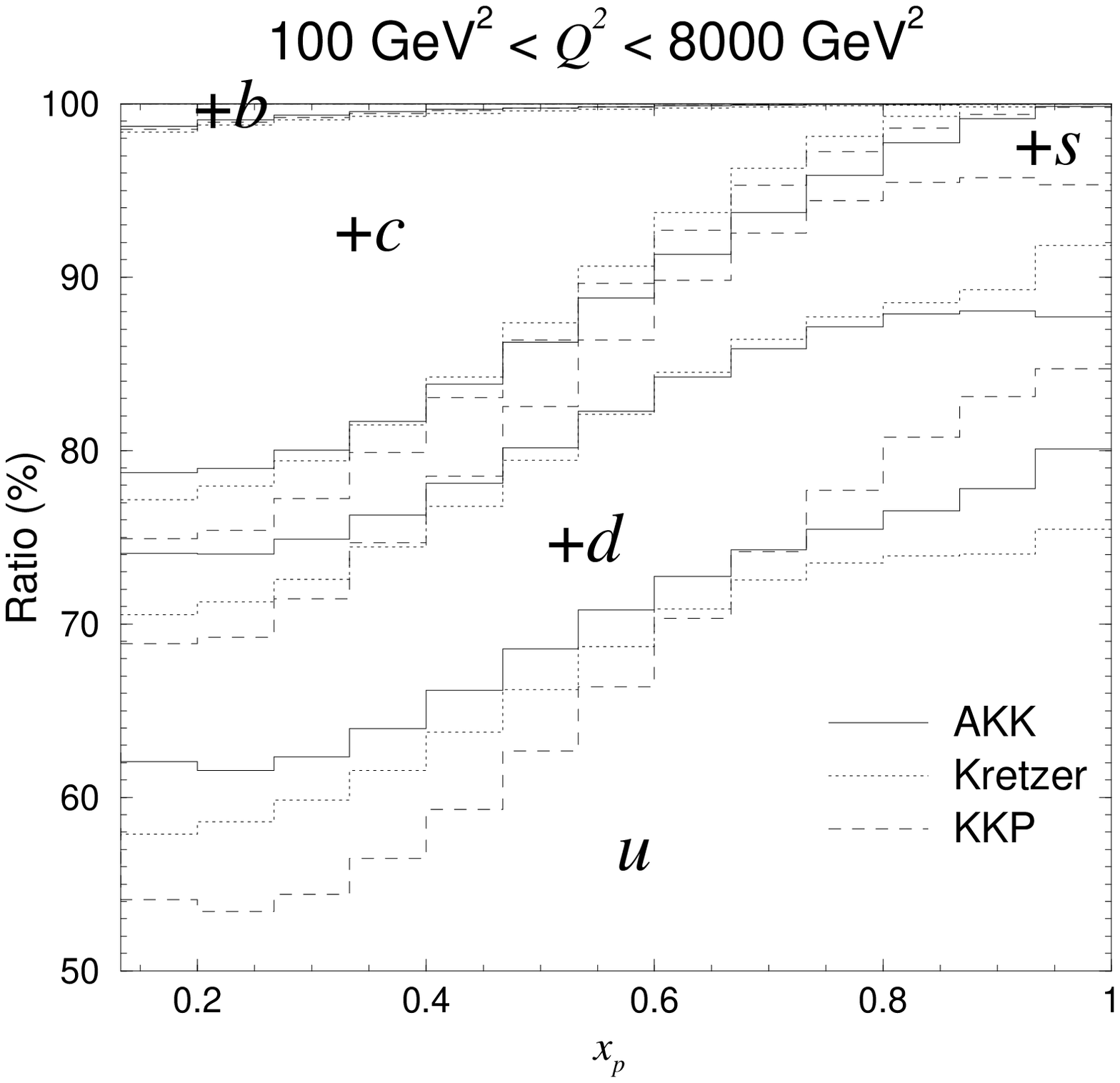}
\end{center}
}\hspace{0cm}
\parbox{.49\linewidth}{
\begin{center}
\includegraphics[width=8.2cm]{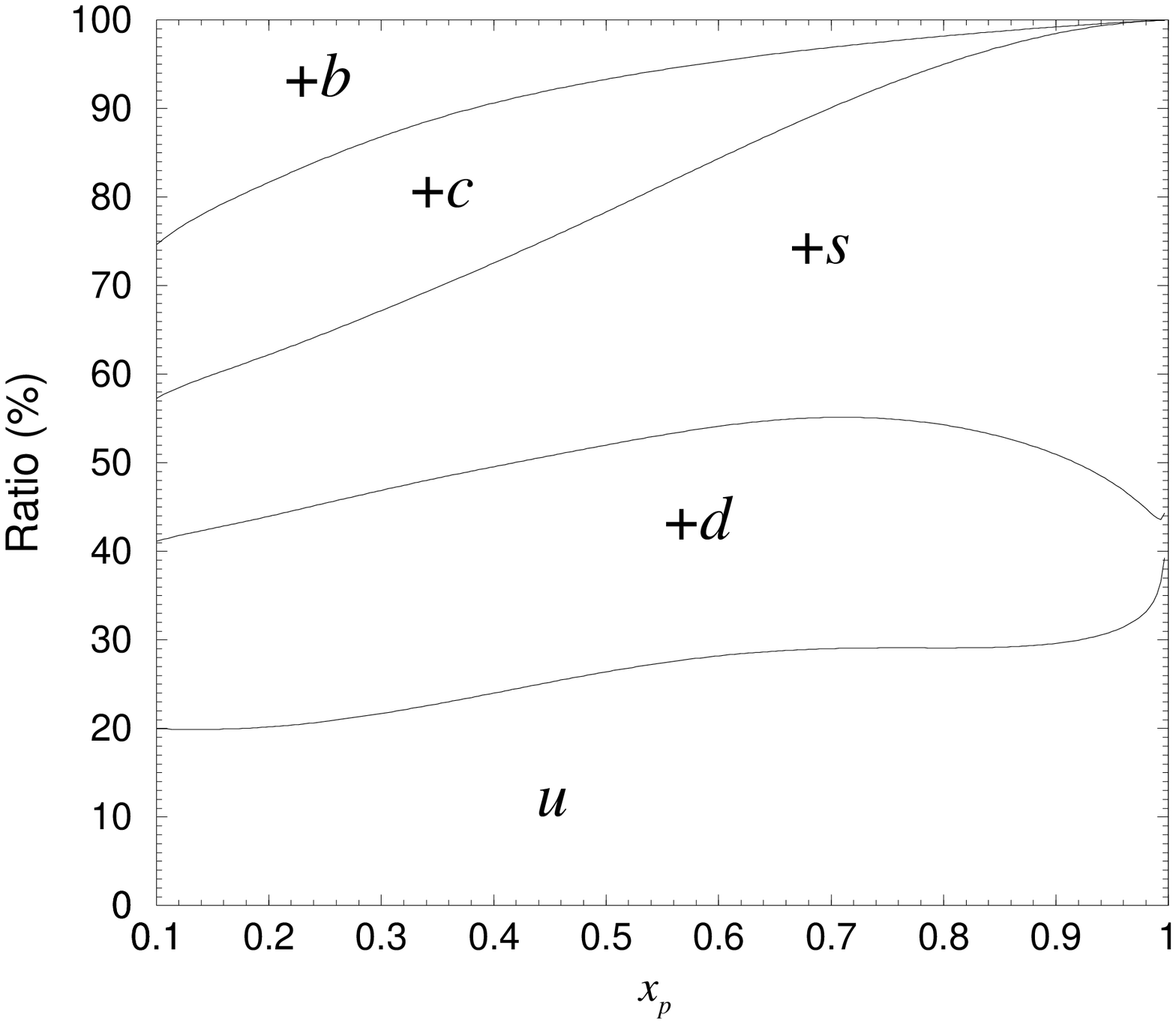}
\end{center}
}
\caption{{\bf Left}: The ratios of the quark tagged components of the cross section 
to the untagged cross section for the high $Q$ H1 data, 
using the AKK, KKP and Kretzer FF sets. The lowest 3 curves show the contribution from the $u$ quark
tagged component only, the next 3 curves above the sum of the $u$ and $d$ components, the next 3 $u$, $d$ 
and $s$ etc. 
{\bf Right}: The ratios of the quark tagged components of the $e^+ e^- \rightarrow h^\pm+X$ cross section, 
where $h^\pm$ is any light charged hadron, 
to the untagged cross section, at $\sqrt{s}=91.2$ GeV and using the AKK FF set.
Both plots, as well as the plots in Figs.\ \ref{qq_hadsep} --- \ref{ZEUSold_scale}, 
have been taken from Ref.\ \cite{Albino:2006wz}, and, in all these plots, $x$ is written ``$x_p$''.
\label{H1old_charge_xfH}}
\end{figure}

\begin{figure}[h!]
\parbox{.49\linewidth}{
\begin{center}
\includegraphics[width=8cm]{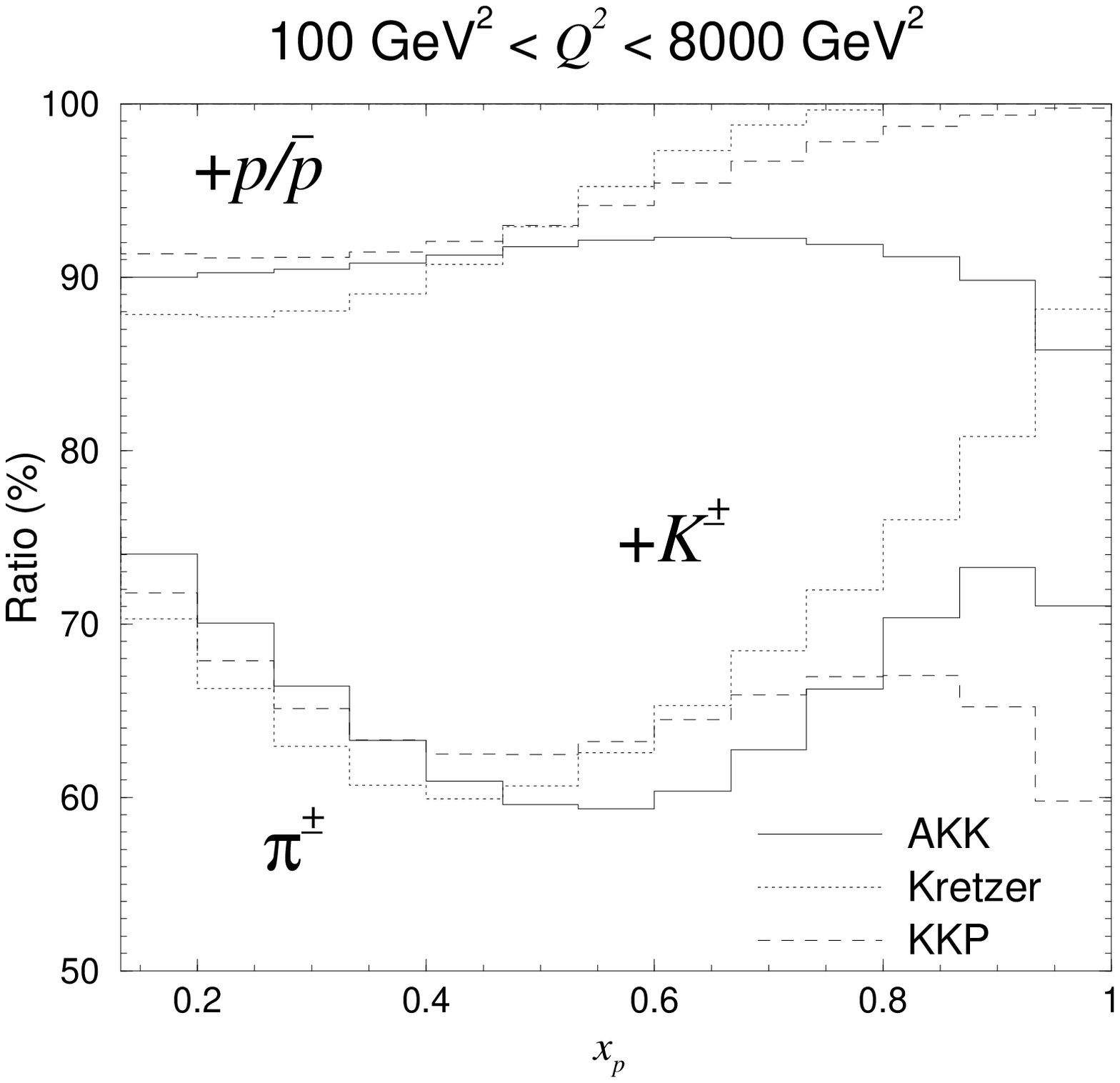}
\end{center}
}\hspace{0cm}
\parbox{.49\linewidth}{
\begin{center}
\includegraphics[width=8.2cm]{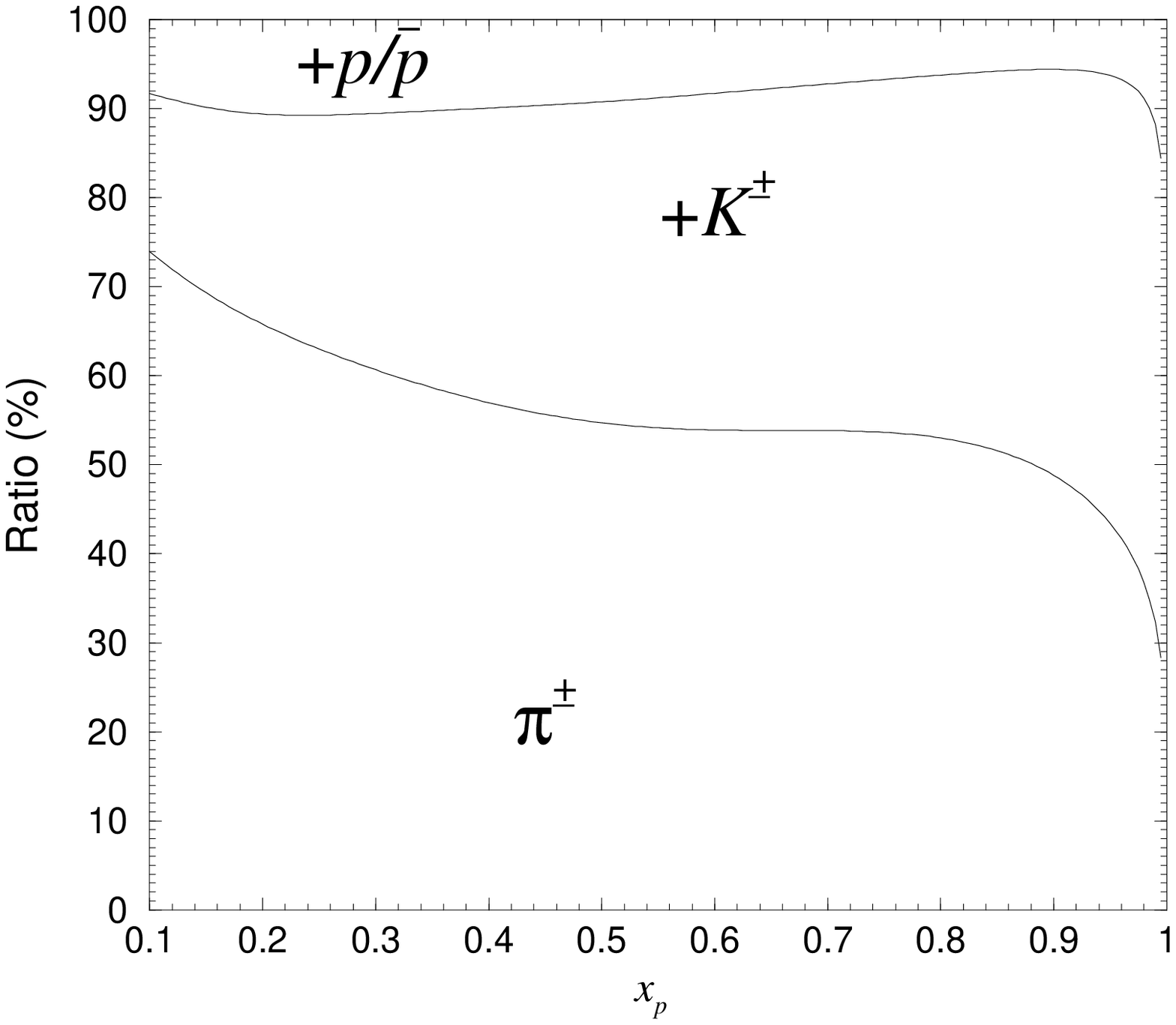}
\end{center}
}
\caption{The ratios of the individual
hadron species constituting the sample for the high $Q$ H1 data, using the AKK, Kretzer and KKP FF sets (left), 
and the ratios of the individual hadron species constituting the sample for 
the $e^+ e^- \rightarrow h^\pm+X$ cross section, where $h^\pm$ is any light charged hadron, 
to the cross section for the full sample, at $\sqrt{s}=91.2$ GeV and using the AKK FF set. \label{qq_hadsep}}
\end{figure}

The ratios of the various tagged cross sections
to the total cross section, for both $ep$ and $e^+ e^-$ reactions, are shown in Fig.\ \ref{H1old_charge_xfH} 
using the AKK \cite{Albino:2005me} FF set.
Calculations for $ep$ reactions are also performed using the KKP and Kretzer FF sets.
As for $e^+ e^-$ reactions, quark tagging for $ep$ reactions is performed by setting to zero the electroweak coupling
of all quark flavours except that of the tagged quark flavour, and is therefore ``physical'' in the sense
of being scheme and scale independent. 
However, quark tagging in $ep$ reactions may not be possible at least in the near future.
In any case, these plots highlight the relative differences in importance of the contributions
to the overall production from the fragmentations of the various quarks in both reactions.
In particular, $u$ quark fragmentation proves to be far more important in $ep$ reactions while $b$ quark fragmentation
is far more important in $e^+ e^-$ reactions, as expected from the relative magnitudes of the proton PDFs for these quarks.
According to Fig.\ \ref{qq_hadsep}, the relative yields of the various hadron species are similar.
Thus the complementary information on fragmentation between $e^+ e^-$ and $ep$ reactions
should also apply at the level of hadron identification, which is of primary interest at present.

Data from $ep$ reactions are no better than data from $e^+ e^-$ reactions at constraining the gluon FF, 
since it again only enters at NLO according to eq.\ (\ref{simpformofF}).
The most significant constraints on gluon fragmentation 
come at present from $pp(\overline{p})$ reactions, to be discussed in subsection \ref{hadhad}.

\begin{figure}[h!]
\parbox{.49\linewidth}{
\begin{center}
\includegraphics[width=7.2cm]{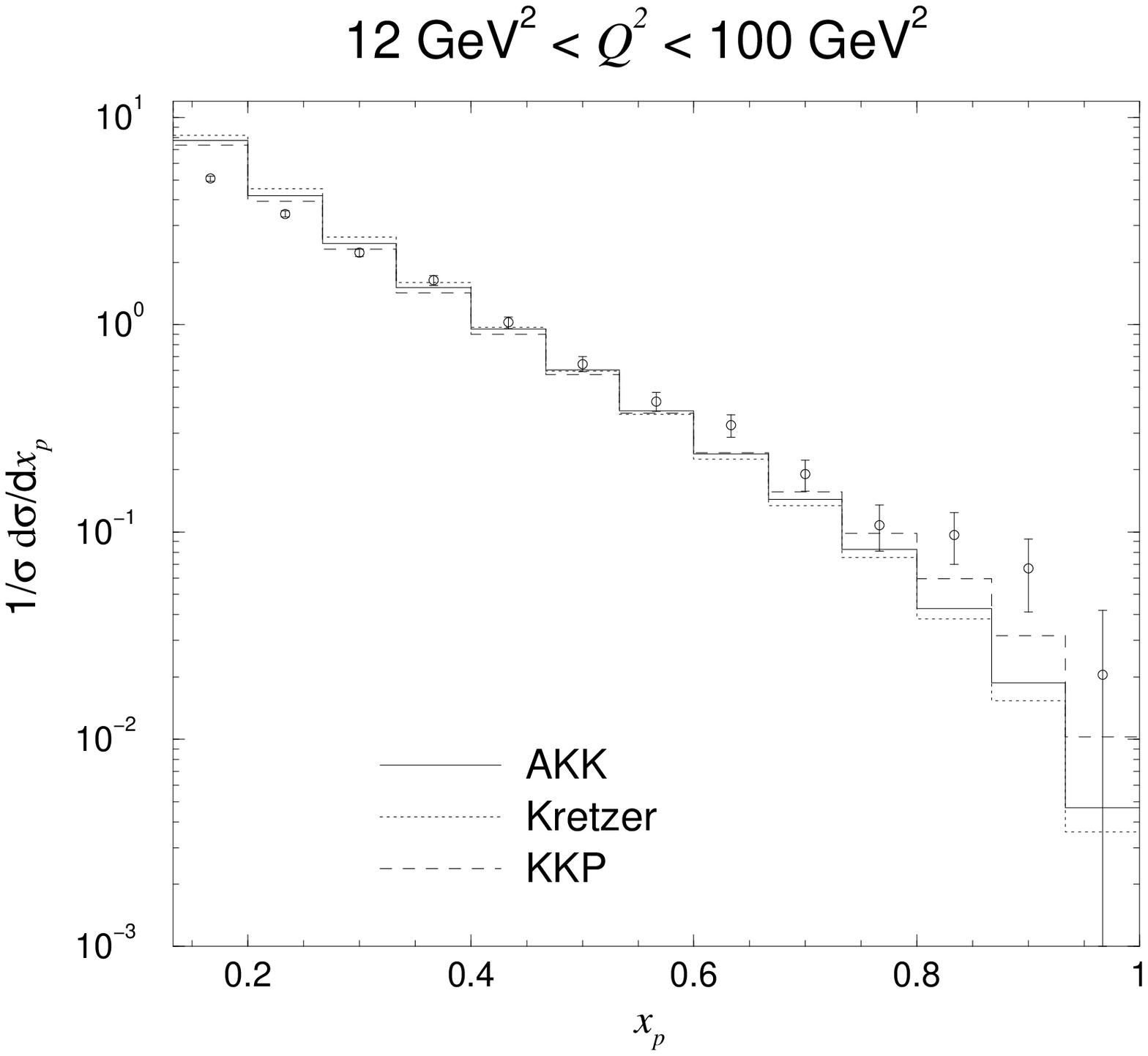}
\end{center}
}\hspace{0cm}
\parbox{.49\linewidth}{
\begin{center}
\includegraphics[width=7.2cm]{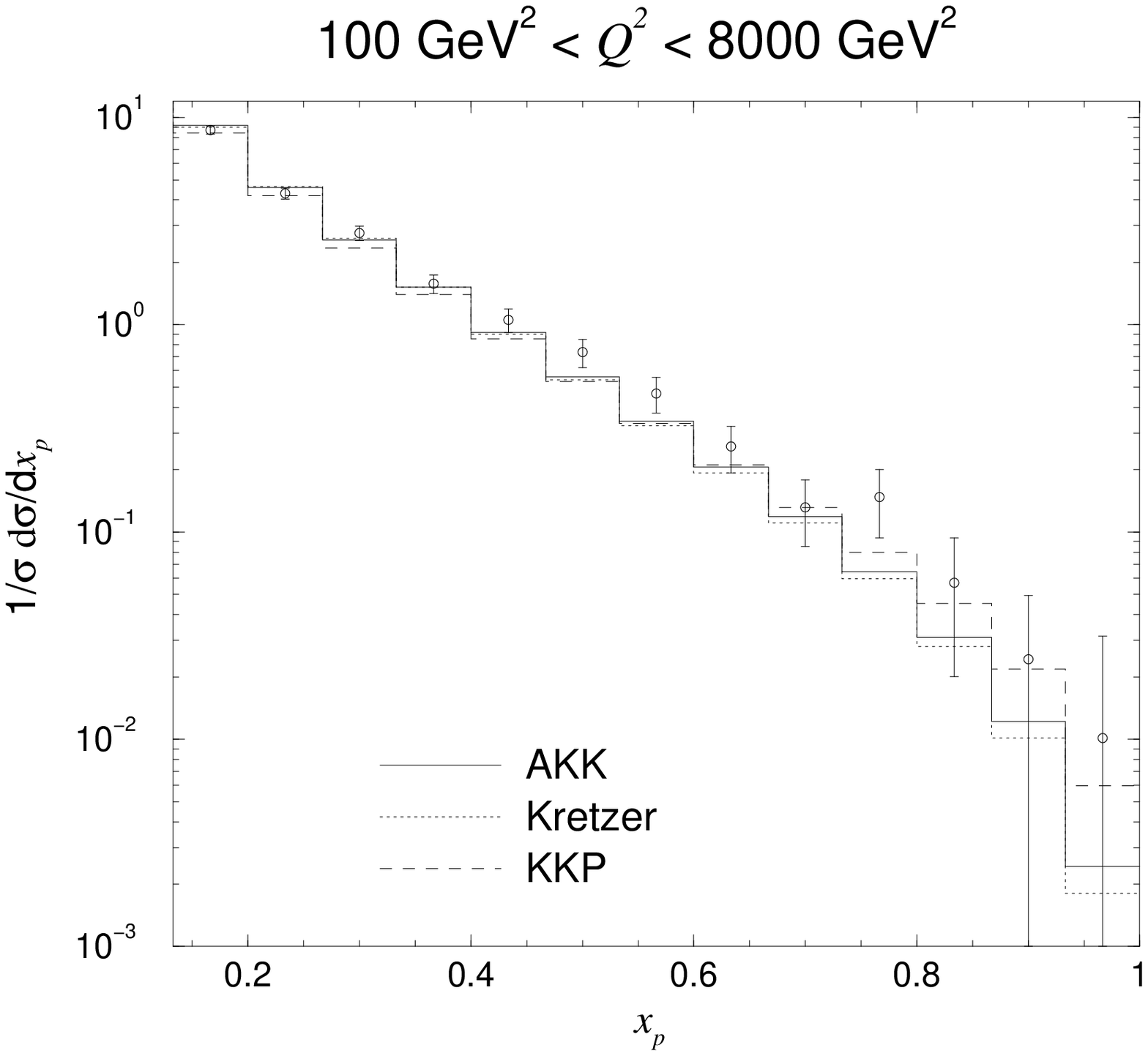}
\end{center}}
\caption{Comparisons of theoretical predictions using the AKK, Kretzer
and KKP FF sets with the $x$ distributions from H1 \cite{Adloff:1997fr}.
\label{H1old}}
\end{figure}

\begin{figure}[h!]
\parbox{.49\linewidth}{
\begin{center}
\includegraphics[width=7.2cm]{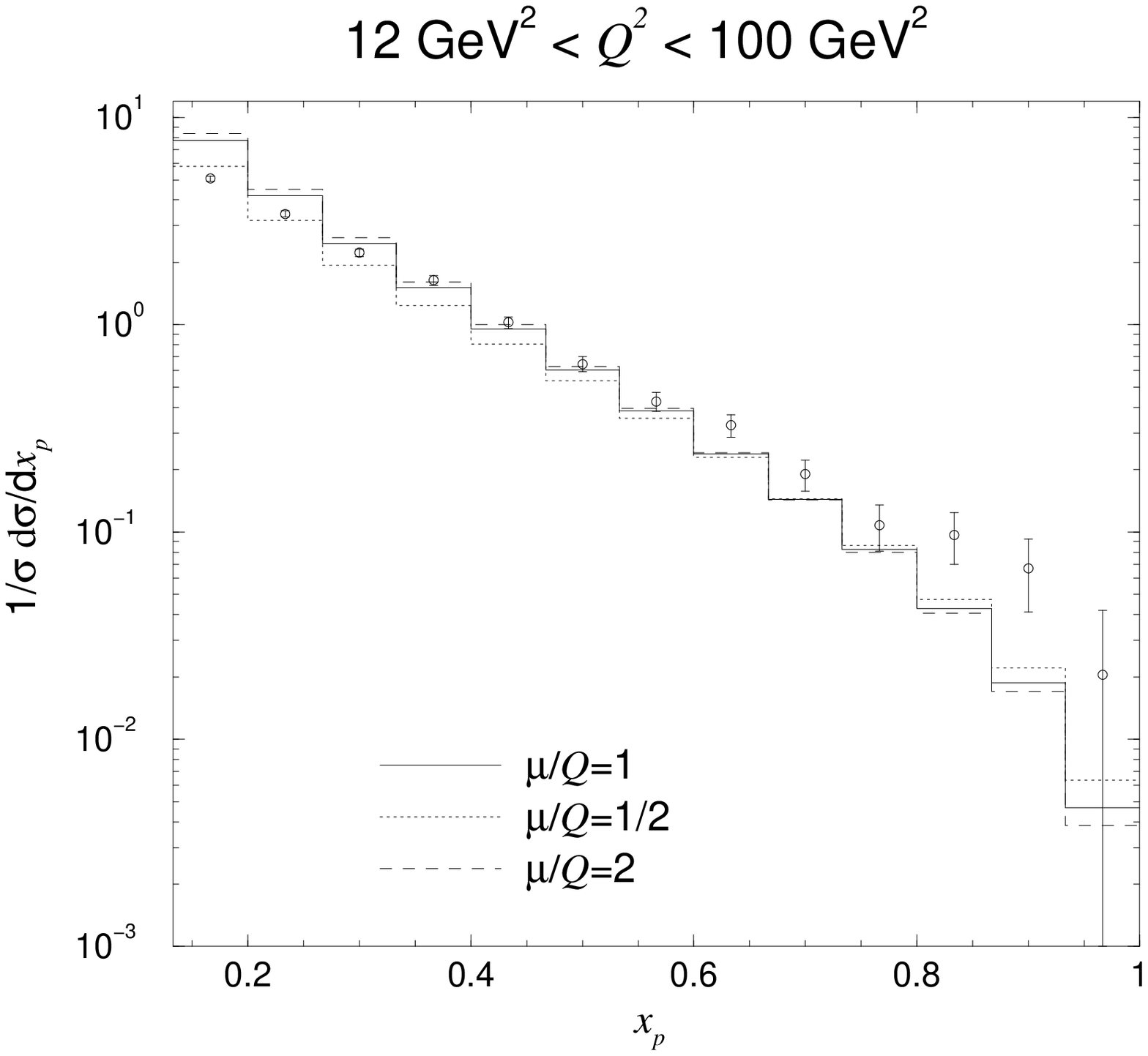}
\end{center}
}\hspace{0cm}
\parbox{.49\linewidth}{
\begin{center}
\includegraphics[width=7.2cm]{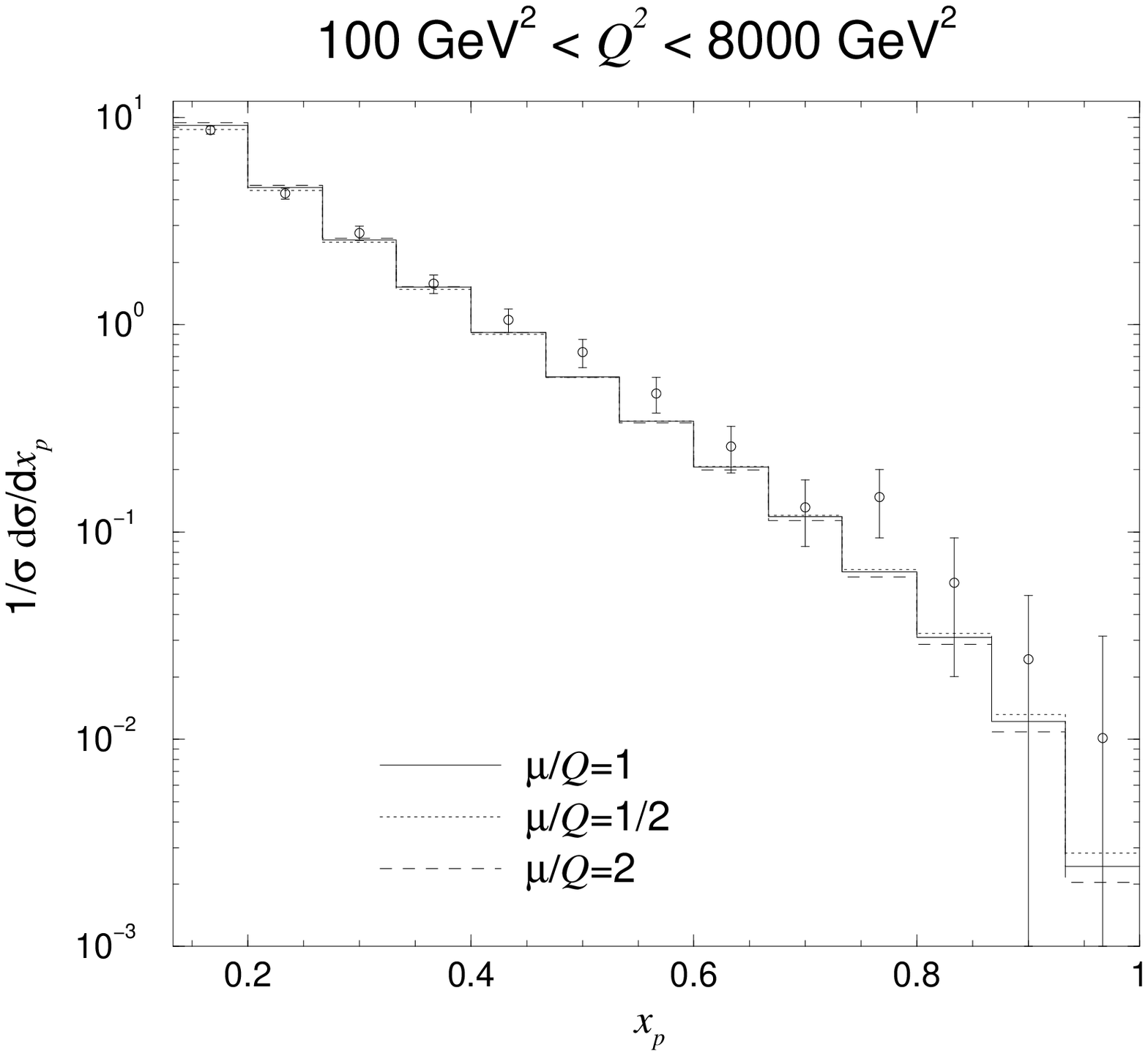}
\end{center}}
\caption{As in Fig.\ \ref{H1old}, using only the AKK FF set
and showing the modifications arising from scale variation.
\label{H1old_scale}}
\end{figure}

\begin{figure}[h!]
\begin{center}
\includegraphics[width=13cm]{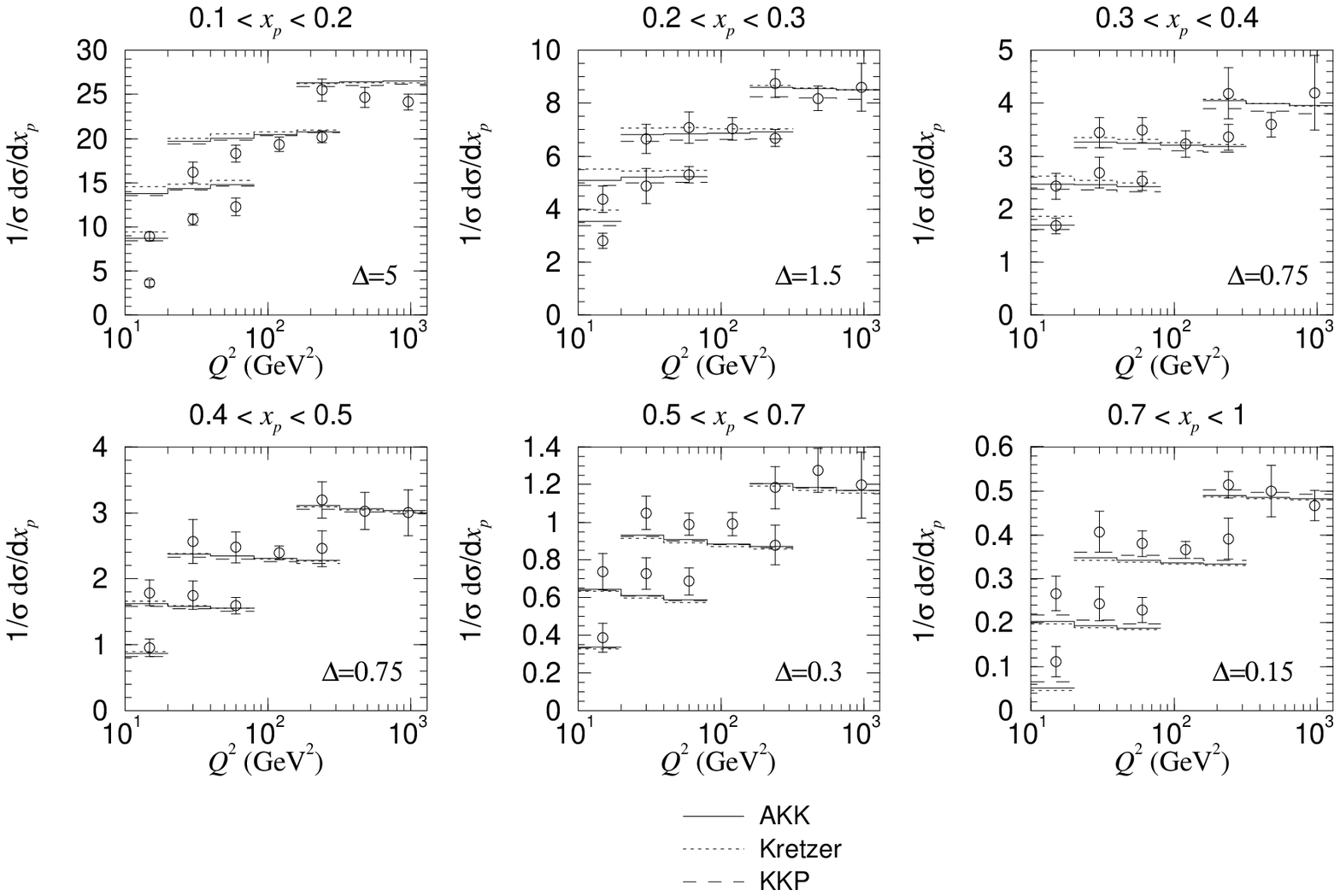}
\end{center}
\begin{center}
\includegraphics[width=13cm]{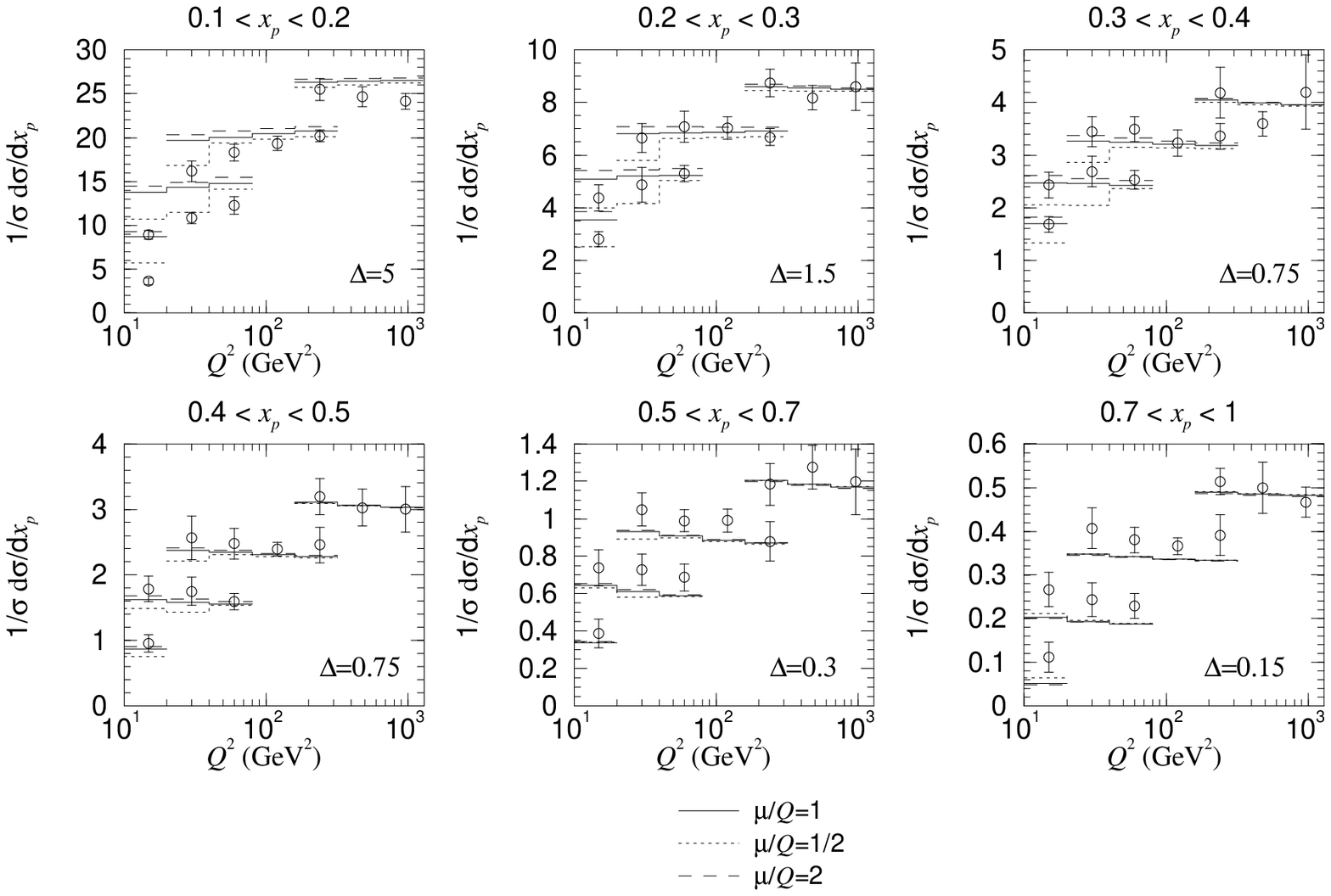}
\end{center}
\caption{Comparisons of theoretical predictions using the AKK, Kretzer
and KKP FF sets with the ZEUS data \cite{Breitweg:1997ra} (left)
and showing the modifications arising from scale variation using only the AKK FF set (right).
Each data set is measured in a specific $x$-bin and, together with its predictions, is
shifted upwards relative to the one below by the indicated value for $\Delta$.}
\label{ZEUSold_scale}
\end{figure}

Comparisons have been made in Ref.\ \cite{Albino:2006wz} with data from the H1
\cite{Adloff:1997fr} and ZEUS \cite{Derrick:1995xg,Breitweg:1997ra} experiments
using the CYCLOPS program \cite{Graudenz:1996yg}, the CTEQ6M PDF set \cite{Pumplin:2002vw} 
and the AKK, Kretzer and KKP FF sets.
The comparison with the H1 data at low and high $Q^2$ ranges is shown in Fig.\ \ref{H1old}.
The strong disagreement between the calculations from the different FF sets at large $x$ most likely arises from large experimental
errors on the FFs due to poor constraints from $e^+ e^-$ reaction data at large momentum fraction.
Otherwise, the calculations are fairly independent of the FF set used
despite very different theoretical constraints on the FFs among the different sets.
In other words, any dependence of the cross section on those FF components not well constrained by $e^+ e^-$ reaction data 
which may arise as a consequence of the differences among the $G_{q_J}$ factors in eq.\ (\ref{simpformofF})
is in fact negligible, and/or the theoretical constraints on the FFs 
in the case of the KKP and Kretzer sets and the OPAL tagging probabilities in the case of the 
AKK set are sufficiently reliable.
At high $Q^2$, the calculation for all 3 FF sets agrees well with the data.
Therefore, the disagreements at small $x$ values and, perhaps, at large $x$ values found with the lower $Q^2$ data 
may be due to neglected effects beyond the FO approach at leading twist.
For example, resummation of soft gluon emission logarithms
that become large at small and large $x$ may be necessary to improve the calculation here.
This is illustrated in Fig.\ \ref{H1old_scale} by the effect of scale variation on the calculation, being largest
at small and large $x$ and for the lower $Q^2$ range.
These observations are found to some degree in the comparison with the ZEUS data of Ref.\ \cite{Derrick:1995xg}
in Fig.\ \ref{ZEUSold_scale}, although
disagreement of the predictions with one another is largest around $x=0.3$.
The description of the data in the range $0.3<x<0.5$ is generally good,
but above this range it fails for $Q^2 <100$ GeV$^2$, which
again may be due to neglected effects at large $x$ in the calculation. 
However, the scale variation in this region is small, suggesting that the perturbative series is stable here.

\begin{figure}[h!]
\begin{center}
\includegraphics[width=11cm]{hep-ex-0706.2456fig4.epsi}
\end{center}
\begin{center}
\includegraphics[width=11cm]{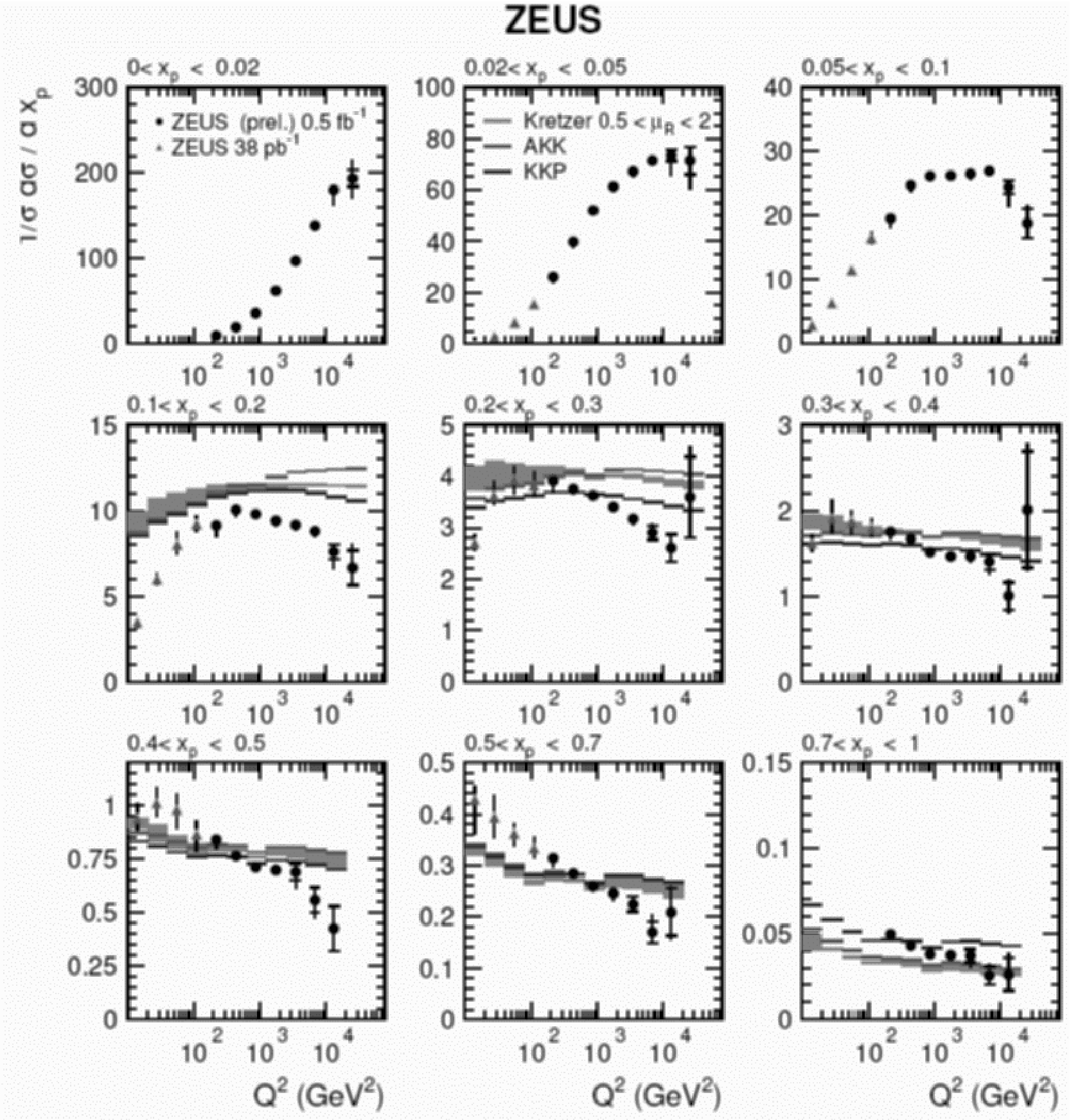}
\end{center}
\caption{Comparisons of theoretical predictions using the AKK, Kretzer
and KKP FF sets with the new data from the H1 \cite{Aaron:2007ds} (left)
and ZEUS \cite{Brzozowska:2007zz} (right) collaborations. \label{H1ZEUSnew}}
\end{figure}

Finally, we show the comparisons with the new data from the H1 \cite{Aaron:2007ds} and ZEUS \cite{Brzozowska:2007zz}
collaborations in Fig.\ \ref{H1ZEUSnew}.
The H1 collaboration reports using a sample which is a factor of 10 larger than that for the H1 data 
discussed above to extract these data, as well as a better understanding of the experimental uncertainties,
while the ZEUS collaboration reports an integrated luminosity of 0.5 fb$^{-1}$ to be compared
with 0.55 pb$^{-1}$ in the older ZEUS data of Ref.\ \cite{Derrick:1995xg} considered above.
The description of these much more accurate data is much worse compared to that of the previous HERA data discussed above,
and thus may require better constrained FFs and/or further improvements in the theory.
One possible improvement for the description of the ZEUS data at large $Q^2$ may be the inclusion of $Z$ boson effects.

In general, various low $Q^2$ effects that have been neglected in these calculations may be important,
particularly higher twist, heavy quarks and the mass of the hadron. 
The effect of the latter will be considered in subsection \ref{hadmass}.

\subsection{Hadron-hadron reactions \label{hadhad}}

Inclusive production of single hadrons in hadron-hadron reactions are important because
they can verify and improve constraints on the charge-sign and flavour separation 
of quark FFs provided by $ep$ and $e^+ e^-$ reactions. 
Perhaps most importantly, data from $pp$ reactions at RHIC should constrain gluon FFs significantly better
than data from $ep$ and $e^+ e^-$ reactions can, owing to the occurrence of the gluon FF at LO in the calculation.

The (dimensionless) quantity describing the inclusive single hadron production in the collision of 2 hadrons $h_1$ and $h_2$, 
$h_1 h_2 \rightarrow h+X$, that is measured in experiment is
\be
F^h_{h_1 h_2}(x,y,s)=s^2 E \frac{d^3 \sigma^h_{h_1 h_2}}{d{\bm p}^3}(p_T,y,s)
=s^2 \frac{1}{2\pi p_T} \frac{d^2 \sigma^h_{h_1 h_2}}{dp_T dy}(p_T,y,s).
\label{defofFhh1h2}
\ee
(exploiting azimuthal symmetry in the second equality),
where $\sqrt{s}$ is the c.m.\ energy, where $E$ and ${\bm p}$ the energy and
spatial momentum respectively of the detected hadron $h$,
where $p_T$ its transverse momentum relative to the spatial momenta of $h_1$ and $h_2$, which are antiparallel,
where the rapidity
\be
y=\frac{1}{2}\ln \frac{E+p_L}{E-p_L},
\label{defofrap}
\ee
with $p_L$ the longitudinal momentum (in the direction of the spatial momentum of $h_1$) of $h$, and
where $x$ is the scaling variable, i.e.\ in the c.m.\ frame it is given by
\be
x=\frac{2p_T}{\sqrt{s}}\cosh y=\frac{2E}{\sqrt{s}}.
\ee
We may also write
\be
x=1-V+VW,
\label{chifromVW}
\ee
where $V$ and $W$ are the variables typically used in perturbative calculations,
related to the usual Mandelstam variables $S$, $T$ and $U$ of $h$ through
\be
\begin{split}
V=&1+\frac{T}{S},\\
W=&-\frac{U}{S+T}.
\label{VWfromMandelstam}
\end{split}
\ee

As for all inclusive single hadron production processes, the cross section at leading twist factorizes according to
\be
F^h_{h_1 h_2}(x,y,s)
=\sum_i \int_x^1 dz F^i_{h_1 h_2}\left(\frac{x}{z},y,\frac{s}{M_f^2},a_s(M_f^2)\right)D_i^h\left(z,M_f^2\right),
\label{ptdistfromff}
\ee
where $F^i_{h_1 h_2}$ is the equivalent partonic production cross section.
This equation has the following physical interpretation:
In any frame related to the c.m.\ frame via a boost 
(anti-)parallel to the beam direction (for massless hadrons the precise
choice of frame is irrelevant in the study below --- later when we treat the mass effects of the 
produced hadron we will need to specify a frame), the reaction
of $h_1$ with $h_2$ results in the inclusive production of a parton $i$, which subsequently 
fragments to a hadron moving in the same direction and carrying away a fraction $z$ of the parton's momentum. 
Note that the partonic rapidity
is the same as the hadronic rapidity, since for massless hadrons $y$ can be approximated by the {\it pseudorapidity}
\be
\eta=-\ln\left(\tan\frac{\theta}{2}\right),
\ee
where $\theta$ is the angle which both the produced hadron and the 
massless fragmenting parton make with the beam in the c.m.\ frame.

The quantity $F^i_{h_1 h_2}$ in eq.\ (\ref{ptdistfromff}) depends on the non perturbative initial state hadrons $h_i$, 
$i=1,2$, through their PDFs according to
\be
\begin{split}
F^i_{h_1 h_2}\left(x,y,\frac{s}{M_f^2},a_s(M_f^2)\right)
=&\sum_{i_1 i_2}\int_x^1 dx_1 \int_{\frac{x}{x_1}}^1 dx_2 
F^i_{i_1 i_2}\left(\frac{x}{x_1 x_2},y,\frac{x_2}{x_1},
\frac{x_1 x_2 s}{M_f^2},\frac{x_1 x_2 s}{M_1^2},\frac{x_1 x_2 s}{M_2^2},a_s(M_f^2),a_s(M_1^2),a_s(M_2^2)\right)\\
&\times f^{i_1}_{h_1}(x_1,M_1^2)f^{i_2}_{h_2}(x_2,M_2^2),
\end{split}
\label{defofXSforiprodfrom2hads}
\ee
where $M_k$, of which $F^i_{h_1 h_2}$ is formally independent, is the factorization scale associated with hadron $h_k$.
This result, which also follows from the factorization theorem, can be interpreted as the inclusive production of parton $i$
from the interaction of a parton $i_k$ from one initial state hadron $h_k$, where $k=1,2$, moving parallel to it 
and carrying away a momentum fraction $x_k$, with a parton from the other.
The perturbatively calculable cross section $F^i_{i_1 i_2}$ describes the purely partonic process $i_1 i_2 \rightarrow i+X$,
and has been calculated to NLO \cite{Ellis:1985er,Ellis:1979sj}.
It depends on the c.m. energy of the partonic system $i_1 i_2$, given by the square root of $x_1 x_2 s$.
Since it is not evaluated in the c.m.\ frame of this partonic process, 
in order to make connection with the c.m.\ frame of the overall hadronic process
it must therefore depend also on the ratio $x_1/x_2$, which determines the Lorentz transformations
between the hadronic and partonic c.m.\ frames (see eq.\ (\ref{boost}) and the discussion before it).

The components of $F^i_{i_1 i_2}$ with $i=g$ are of $O(a_s^2)$, while the rest are either of this order or higher.
Therefore the gluon FF appears at LO in this cross section, in contrast to $ep$ and $e^+ e^-$ reactions, and therefore
$pp$ reactions can provide NLO constraints on the gluon.
However, large NLO corrections \cite{Ellis:1979sj} suggest that perturbative instability is a large source of error.
Indeed, the NLO cross section suffers a large scale variation, as can be seen for instance in Fig.\ \ref{res_rhic_kkp}.
This theoretical error is much greater than that coming from the propagated PDF uncertainty.

As for $ep$ reactions, the detected hadron $h$ in e.g.\ $h_1 h_2$ reactions will sometimes
be a soft remnant from one of the initial state hadrons
instead of being produced by the hard partonic
processes $i_1 i_2 \rightarrow i+X$ followed by $i\rightarrow h+X$.
As for $ep$ reactions, these production channels are accounted for by fracture functions 
for each initial state hadron \cite{Trentadue:1993ka}.
No further non perturbative input is required since it is clear from which initial hadron a remnant hadron was produced.
Mathematically, all potential mass singularities which cannot be absorbed into PDFs and FFs
can be absorbed into fracture functions.
As a result of universality, they are process independent (apart from the initial state hadron(s)), so that 
the factorized fracture functions in $h_1 h_2$ reactions, 
including the potential mass singularities that they absorb, are identical to those in $ep$ reactions 
when the same factorization scheme is used.
As for $ep$ reactions, one anticipates the existence of a scheme and scale
independent cross section for $h_1 h_2\rightarrow h+X$ which does not depend on fracture functions,
since the evolution of FFs and PDFs does not depend on them,
i.e.\ a ``current fragmentation region'' analogous to that in $ep$ reactions which is free of initial hadron remnants.
Certainly, contributions from target fragmentation to the cross section $F^h_{h_1 h_2}$ that we have been considering 
will decrease with increasing $p_T$ and decreasing $y$ \cite{Blokzijl:1975ku}, 
because the detected hadron gets further away from the beam.
Removal of these target fragmentation effects altogether may be possible 
by placing kinematic restrictions on processes with alternative final states
for inclusive single hadron production, such as the semi-inclusive Drell-Yan process $h_1 h_2\rightarrow \gamma^* +h+X$
\cite{Ceccopieri:2008fq}.

\begin{figure}[h!]
\begin{center}
\includegraphics[width=8cm]{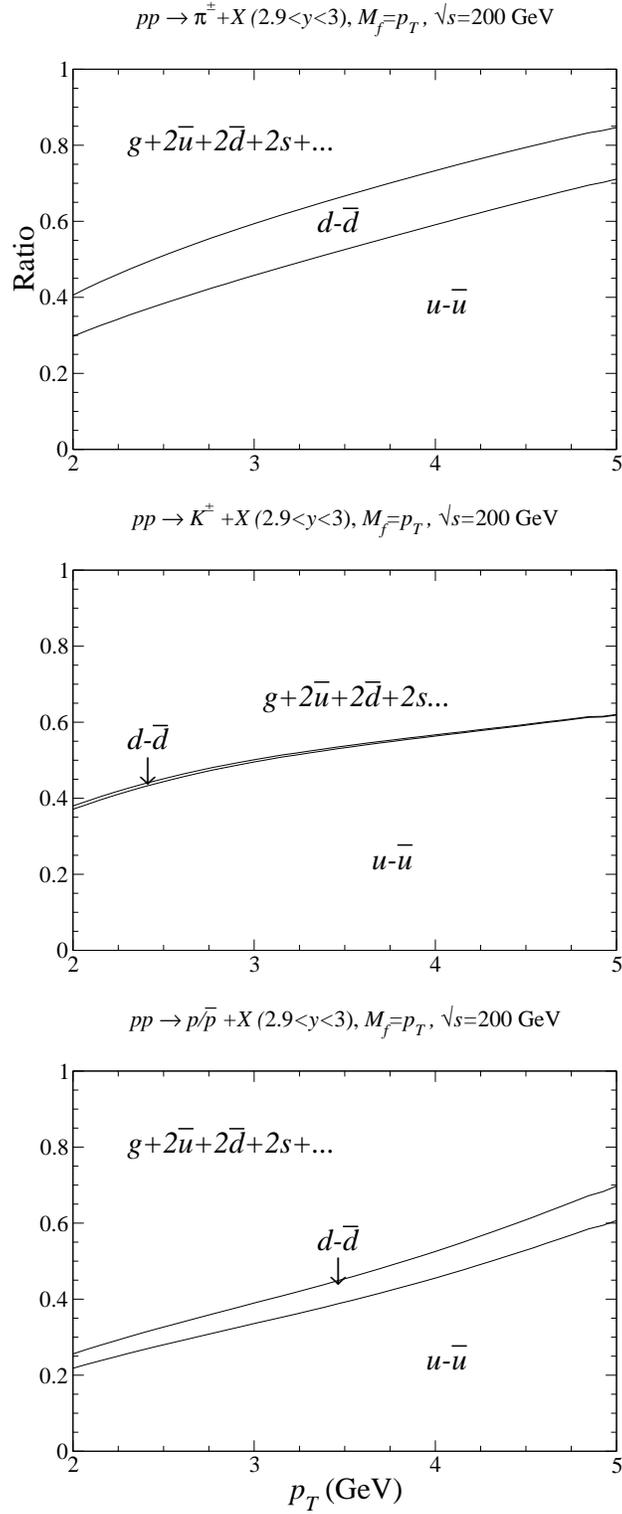}
\caption{Contributions to the production from fragmentation of the initial
state protons' valence quarks and sea partons for BRAHMS kinematics with $2.9<y<3$, using the AKK08 FF sets. 
The notation ``$u-\bar{u}$'' refers to the contribution from $u$ valence fragmentation
$\widehat{F}_u D_u^{h^\pm}-\widehat{F}_{\bar{u}}D_{\bar{u}}^{h^\pm}=\widehat{F}^{u_v} D_u^{h^\pm}$
(see eq.\ (\ref{FfromFhatandDwithindicesdecomp})), etc.
The ratios are stacked, i.e.\ for a given $p_T$ value, the distance on the $y$-axis
from zero to the first curve gives the $u-\bar{u}$ contribution, from the first to the second curve
the $d-\bar{d}$ contribution, and from the second curve to $1$ the sea parton contribution.
From Ref.\ \cite{Albino:2008fy}.
\label{ppValenceRatioBRAHMS}}
\end{center}
\end{figure}

As discussed in the beginning of subsection \ref{symmetries},
charge-sign unidentified cross sections $F^{h^\pm}=F^{h^+}+F^{h^-}$ (omitting the $h_1 h_2$ subscript 
appearing in e.g.\ eq.\ (\ref{defofFhh1h2}) for brevity) depend only on charge-sign unidentified FFs, and 
charge-sign asymmetry cross sections $F^{\Delta_c h^\pm}=F^{h^+}-F^{h^-}$ depend only on charge-sign asymmetry FFs.
In both types of observable, the contributions to the production from the fragmentations of the various
partons in the proton can be studied in a ``physical'' way, imposing tests on FFs through expectations
for these contributions:
Now restoring the $h_1 h_2$ subscript and omitting the $h^\pm$ superscript, 
we decompose the charge-sign unidentified cross section according to 
\be
F_{pp}=(F_{p\overline{p}}-F_{\overline{p}\overline{p}})|_{u_v}+(F_{p\overline{p}}-F_{\overline{p}\overline{p}})|_{d_v}
+(F_{pp}+F_{\overline{p}\overline{p}}-F_{p\overline{p}}).
\label{physdecompinvalenceandsea}
\ee
Although $F_{\overline{p}\overline{p}}=F_{pp}$,
we have not made this replacement here in order to emphasize that
the final states differ, by the interchange $h^+ \leftrightarrow h^-$.
The presence of e.g.\ ``$|_{u_v}$'' means that the valence $d$ quark PDF is set to zero, the result in Mellin space being
that the first (second) term is proportional to the square of the $u$ ($d$) valence quark PDF.
Note that these asymmetry generating terms
neither depend on the protons' sea partons nor receive contributions from interactions between valence $u$ and $d$ quarks.
The absence of the latter interactions follows
if the cross sections for the processes $ud\rightarrow h^\pm +X$ and $u\bar{d}\rightarrow h^\pm +X$ are identical,
which holds at NLO.
Because the first and second term are each scheme and scale independent, they give a physical quantification
of the contribution to the overall production 
from the fragmentations of the initial proton's $u$ and $d$ valence quarks respectively.
These quarks are the source of the charge-sign asymmetry, to be discussed in more detail
around eq.\ (\ref{FfromFhatandDwithindicesforval}) below.
Since there are more valence $u$ than $d$ quarks
in the initial protons, the first term is expected to dominate over the second for \cpi\ production
(see the first plot in Fig.\ \ref{ppValenceRatioBRAHMS}),
since the $u$ and $d$ quark fragmentations are equal, and even more so for
\pr\ production (the third plot in Fig.\ \ref{ppValenceRatioBRAHMS}), because then $u$ is larger than $d$ quark fragmentation .
For \cka\ (the second plot in Fig.\ \ref{ppValenceRatioBRAHMS}), $d$ quark fragmentation is unfavoured, and therefore 
the contribution from the protons' valence $d$ quarks is expected to be much smaller than from their valence $u$ quarks. 
The third term in eq.\ (\ref{physdecompinvalenceandsea}) corresponds to the charge-sign symmetric contribution,
becoming equal to $F_{pp}$ when the protons' valence quarks vanish.
It can be regarded as a physical quantification of the contribution to the production
from the fragmentation of the collective sea of the initial protons.
This term does not contribute to the charge-sign asymmetry cross section (to be discussed just now) because it is
charge conjugation invariant. Therefore, the larger the third term is relative to the first two,
the smaller the charge-sign asymmetry cross section is expected to be relative to the charge-sign unidentified cross section
of eq.\ (\ref{physdecompinvalenceandsea}).
For the production of hadrons which have non-zero strangeness, or which are superpositions of hadrons with non-zero strangeness,
which in our set are \cka\ and \lam\ in the first case and \nka\ in the second,
the third term is expected to dominate: Here, the favoured 
$s$ quark and $u$ and/or $d$ quark fragmentation from the abundant sea occurs, while in the first two terms only
the fragmentation from the protons' valence $u$ and/or $d$ quark contributes which necessarily 
involves the production of a heavier $s$ quark. 
This behaviour is seen in the third plot of Fig.\ \ref{ppValenceRatioBRAHMS}.
For \cpi\ and \pr\ production, for which $u$ and $d$ quark fragmentations are favoured,
it is not clear whether the first two terms are more important due to the FFs there being an order
of magnitude larger than the rest, or the third which accounts for fragmentation from the 
protons' abundant partonic sea.
In practice, it turns out that the fragmentation from the initial protons' sea partons 
(the third term) always dominates, even if the charge-sign asymmetry is very significant.
In fact the plots in Fig.\ \ref{ppValenceRatioBRAHMS} were produced using the unphysical decomposition
\be
F^{h^\pm}=\widehat{F}^{u_v} D_u^{h^\pm} +\widehat{F}^{d_v} D_d^{h^\pm} 
+\sum_{i=g,q_s}\widehat{F}^i D_i^{h^\pm},
\label{FfromFhatandDwithindicesdecomp}
\ee
where in both eq.\ (\ref{FfromFhatandDwithindicesdecomp}) and (\ref{FfromFhatandDwithindicesforval}) we omit
arguments, integration signs etc.\ for brevity, define $\widehat{F}^{q_v}=\widehat{F}^q-\widehat{F}^{\bar{q}}$
and use the label $q_s$ to refer to sea quarks,
since the results are qualitatively similar to those obtained with eq.\ (\ref{physdecompinvalenceandsea}),
except that the third term in eq.\ (\ref{FfromFhatandDwithindicesdecomp}) is much smaller than the 
third term in eq.\ (\ref{physdecompinvalenceandsea}) and therefore gives clearer plots.

\begin{figure}[h!]
\begin{center}
\includegraphics[width=8.5cm]{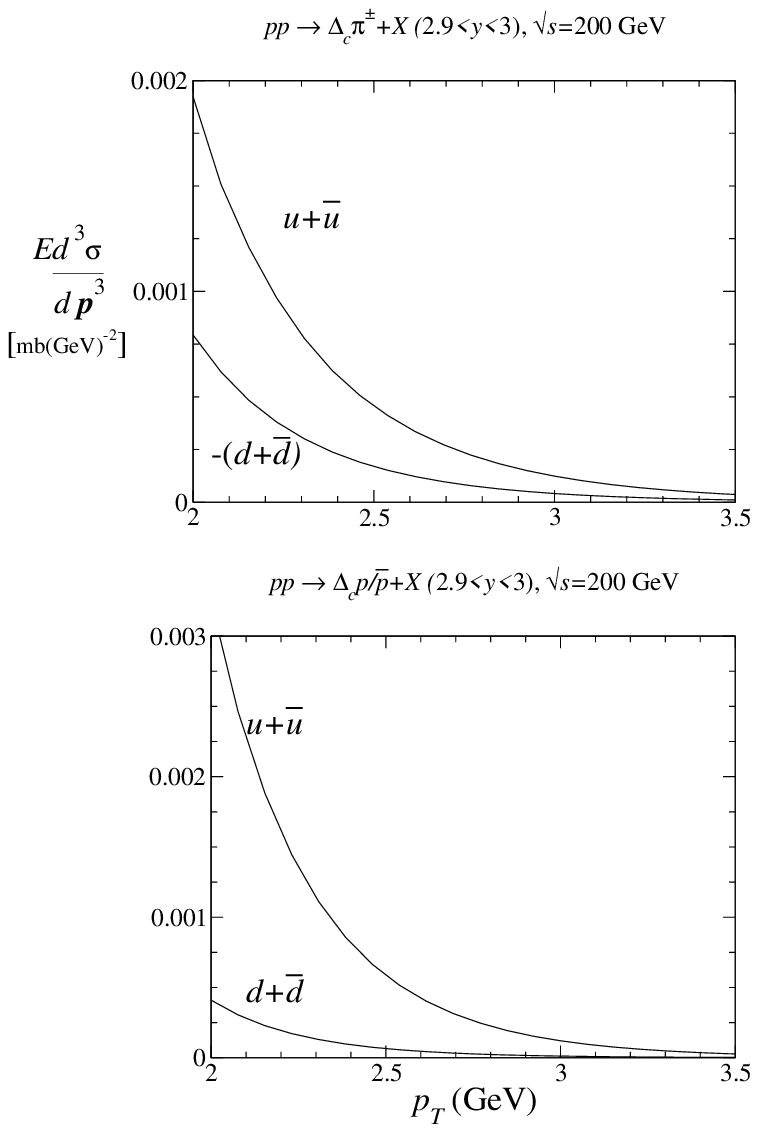}
\caption{Contributions to the charge-sign asymmetry $\Delta_c h^\pm$ from fragmentation of the initial state protons' valence quarks
for BRAHMS kinematics for which $2.9<y<3$, using the AKK08 FF sets.
The notation ``$u+\bar{u}$'' means $\widehat{F}_u D_u^{\Delta_c \pi^\pm}+\widehat{F}_{\bar{u}}D_{\bar{u}}^{\Delta_c \pi^\pm}
=\widehat{F}^{u_v} D_u^{\Delta_c h^\pm}$ (see eq.\ (\ref{FfromFhatandDwithindicesforval})), etc.
The cross section is given by the difference between the two curves.
From Ref.\ \cite{Albino:2008fy}.
\label{ppValenceRatioVal_BRAHMS}}
\end{center}
\end{figure}

The charge-sign asymmetry is determined from the FFs according to
\be
F^{\Delta_c h^\pm}=\widehat{F}^{u_v} D_u^{\Delta_c h^\pm} +\widehat{F}^{d_v} D_d^{\Delta_c h^\pm}. 
\label{FfromFhatandDwithindicesforval}
\ee
Both terms are factorization scheme and scale independent.
Note from the quark composition of \cpi\ that $D_d^{\Delta_c \pi^\pm}$ is expected to be 
negative and $D_u^{\Delta_c \pi^\pm}$ positive (this should at least be true for their first moments).
In addition, the quark composition of the proton suggest that the production of 
$u$ over $\bar{u}$ is greater than the production of $d$ over $\bar{d}$, i.e.\ $\widehat{F}^{u_v}>\widehat{F}^{d_v}>0$.
The first plot in Fig.\ \ref{ppValenceRatioVal_BRAHMS} is consistent with these observations.
This allows for the excess of $\pi^+$ over $\pi^-$,
but it should be noted that this excess is not guaranteed unless $\widehat{F}^{u_v}$ is {\it sufficiently}
greater than $\widehat{F}^{d_v}$, or the magnitude of $D_d^{\Delta_c \pi^\pm}$ is sufficiently smaller than $D_u^{\Delta_c \pi^\pm}$.
For $h^\pm =p/\overline{p}$, all 4 quantities in eq.\ (\ref{FfromFhatandDwithindicesforval}) 
are expected to be positive so that a definite excess of $p$ over $\overline{p}$ is predicted.
Because of the expectations $\widehat{F}^{u_v}>\widehat{F}^{d_v}$ and, from the quark composition of \pr,
$D_u^{\Delta_c p/\overline{p}}>D_d^{\Delta_c p/\overline{p}}$, the first term in eq.\ (\ref{FfromFhatandDwithindicesforval}) 
is expected to dominate over the second.
These expectations are observed in the second plot of Fig.\ \ref{ppValenceRatioVal_BRAHMS}.
Likewise, an excess of $K^+$ over $K^-$ is predicted, although
in this case the second term in eq.\ (\ref{FfromFhatandDwithindicesforval}) vanishes since $D_d^{\Delta_c K^\pm}=0$. 
Note that there is no dependence on $D_s^{\Delta_c K^\pm}$, 
so that this FF is not constrained by $pp(\overline{p})$ reaction measurements.

\subsection{Other processes}

Further tests of universality have been performed in Ref.\ \cite{Kniehl:2000hk} by comparing calculations
using the KKP FF set obtained by these authors with measurements of
unidentified hadron production from $p\overline{p}$ reactions
at the SPS by the UA1, UA2 collaborations and at the Tevatron by the CDF collaboration,
and from photoproduction (requiring photon PDFs) in $\gamma p$ reactions from H1 and ZEUS at HERA 
and in $\gamma \gamma$ reactions from OPAL at LEP2.
The data overall were measured over a wide range of $p_T$ and $y$.
Good agreement with all data was found provided that the theoretical error was taken to be
the variation produced in the cross section when $M_f/E_s$ is allowed to vary from 1/2 to 2.
This cross section variation is rather large, and the experimental errors are unfortunately largest at large $p_T$ where additional
non-perturbative information such as higher twist is expected to be least important. 
The photoproduction calculations also acquire large errors from the rather badly
constrained parton distribution functions (PDFs) of the photon. 
Distributions in $y$ generally have even larger theoretical errors. 
More stringent tests of universality were performed in Refs.\
\cite{Kniehl:2004hf,Daleo:2004pn,Aurenche:2003by} through analysis of
pseudorapidity and $p_T$ distributions from H1 for the
process $ep \rightarrow e+\pi^0 +X$, and in Ref.\ \cite{Kniehl:2004hf}
through analysis of $p_T$ distributions from ZEUS for
the process $ep \rightarrow e+h +X$, which did not require the use of
photon PDFs (except in the low $Q$ region \cite{Fontannaz:2004ev}). 
The disagreement found with the ZEUS data was
reduced in Ref.\ \cite{Nadolsky:2000ky} through resummation of
multiple parton radiation at low $p_T$.

\section{Improvements to the standard approach \label{impstand}}

The standard FO calculation discussed so far describes much of the data well, but is incomplete. 
The FO approximation of the high energy parts of the factorized cross section is only applicable 
to the kinematic regions for which $E_s \gg \Lambda_{\rm QCD}$ and $x$ is not too close to 0 or 1,
and the boundaries of this acceptable region are unknown.
Terms originating from the emission of soft gluons
become large for large and small $x$ and low $E_s$,
and must be summed to all orders by a procedure known as {\it resummation}.
We will discuss the well known procedure of large $x$ resummation in DGLAP evolution
and in $e^+ e^-$ reactions in subsection \ref{largexres}, and leave 
the resummation of soft gluon logarithms at small $x$ to section \ref{sglres}. 
Only the ZM-VFNS has been implemented in global fits so far,
whose error of $O(1)$ whenever $E_s=O(m_i)$ can be removed by using instead the GM-VFNS.
Even with these improvements to the perturbative approximation, the definition of the cross section we have been considering so far
itself requires various formal modifications, which usually become increasingly important with decreasing $E_s$.
The effects of the produced hadron's mass, of $O(m_h^2/E_s^2)$, will be discussed in subsection \ref{hadmass}. 
Higher twist effects, of $O(\Lambda_{\rm QCD}/E_s)$ or less, remain unknown at present.

As for eqs.\ (\ref{DGLAPmellin}) and (\ref{explevolofDmellin}), we omit parton labels from now on.
These labels may refer to individual parton species, 
or any other complete, linearly independent basis of FFs.

\subsection{Hadron mass effects \label{hadmass}}

The effect of hadron mass is one of the many important effects beyond the standard FO calculation 
in the region of low $E_s$ and small $x$.
In fact, as will be discussed in section \ref{currglobalfits},
the fitted hadron masses for \cpi, $p/\overline{p}$ and \lam\ occurring in the calculation for the 
$e^+ e^-$ reactions in global fits are approximately equal to their true values \cite{Albino:2008fy}, 
suggesting that it is the first effect to become relevant
as $E_s$ and $x$ decrease.

In this section, we derive the necessary modifications to calculations when the mass $m_h$ of the produced hadron $h$ 
is incorporated in inclusive single hadron production in $e^+ e^-$, $ep$ and hadron-hadron reactions.
The theoretical calculation of the cross section is explicitly independent of hadron mass provided that the
hadronic scaling variable $x$
is identified with the light cone scaling variables as dictated by the factorization theorem.
However, measured cross sections are usually differential in other scaling variables such as the fractions
of available energy and momentum taken away by the detected hadron.
Thus, the effects of hadron mass are a multiplicative modification to the cross section,
and a transformation of the scaling variable of the experiment to the light cone scaling variable $x$.
Otherwise the perturbative calculations are not modified, although it should be noted that the
modification factor just mentioned depends on the momenta of any initial state hadron's partons that take part in the hard interaction.
The fact that hadron mass effects are most important for low values of the scaling variable $x$ 
and of the energy scale $E_s$ of the process will be seen in the explicit results.

We begin with eq.\ (\ref{genformofinchadprodfromffs}), which is true regardless of whether the hadron mass is non zero,
provided $z$ is identified with the light cone fraction and $x$ with the light cone scaling variable.
The precise definition of $z$, and $x$ where necessary, is given for the processes of interest below.
We assume all quarks are massless, although the generalization to massive quarks is straightforward.

\noindent {\bf $e^+ e^-$ reactions} \cite{Albino:2005gd}
Using the light cone coordinates defined in the paragraph before eq.\ (\ref{masslesspartonmom}),
we work in a frame for which the virtual boson's momentum is given by
\be
q=\left(\frac{\sqrt{s}}{\sqrt{2}},\frac{\sqrt{s}}{\sqrt{2}},{\bm 0}\right).
\label{choiceofq}
\ee
and the hadron's momentum by
\be
p_h=\left(\frac{x \sqrt{s}}{\sqrt{2}},\frac{m_h^2}{\sqrt{2}x \sqrt{s}},{\bm 0}\right),
\label{phformassivehadron}
\ee
which also serves as a definition of $x$.
The momentum fraction $z$ is defined by the relation to $p_h$ of the $+$ component of each parton's momentum $k$
which contribute at leading twist: 
\be
k^+ =\frac{p_h^+}{z}.
\label{pluscomppartontohad}
\ee
Note that for massless partons, $k^-=0$. The condition ${\bm k}_T=0$ is always true.
The scaling variable $x_p$ defined before eq.\ (\ref{genformofinchadprodfromffs}) 
is given by the ratio of the spatial momentum of $p_h$, $p_h^3=(p_h^+-p_h^-)/\sqrt{2}$, to $\sqrt{s}/2$,
and so
\be
x_p=x \left(1-\frac{m_h^2}{s x^2}\right).
\ee
This can be inverted to give
\be
x=x_p \left(\frac{1}{2}+\frac{1}{2}\sqrt{1+\frac{4m_h^2}{s x_p^2}}\right).
\ee
Consequently, the experimental cross section $d\sigma^h/dx_p$, for example, can be calculated according to
\be
\frac{d\sigma^h}{dx_p}(x_p,s)=\frac{dx}{dx_p}(x_p,s)\frac{d\sigma^h}{dx}(x(x_p,s),s)=
\left(\frac{1}{2}+\frac{1}{2}\left[1+\frac{4m_h^2}{s x_p^2}\right]^{-\frac{1}{2}}\right)\frac{d\sigma^h}{dx}(x(x_p,s),s),
\ee
where, from eq.\ (\ref{genformofinchadprodfromffs}) with the transformation $z\rightarrow x/z$ made for convenience,
\be
\frac{d\sigma^h}{dx}(x,s)=\sum_i \int_x^1 \frac{dz}{z}\frac{d\sigma^i}{dz}(z,s,M_f^2)D_i^h\left(\frac{x}{z},M_f^2\right)
\label{formofepemXS}
\ee
is just the usual explicitly hadron mass independent calculation discussed in detail in subsection \ref{epemtheory},
for example eq.\ (\ref{XSfromFFs}).
The formulae above with $x_E$, defined before eq.\ (\ref{genformofinchadprodfromffs}), 
in place of $x_p$ are the same but with $m_h^2 \rightarrow -m_h^2$.

\begin{figure}[h!]
\begin{center}
\includegraphics[width=8cm]{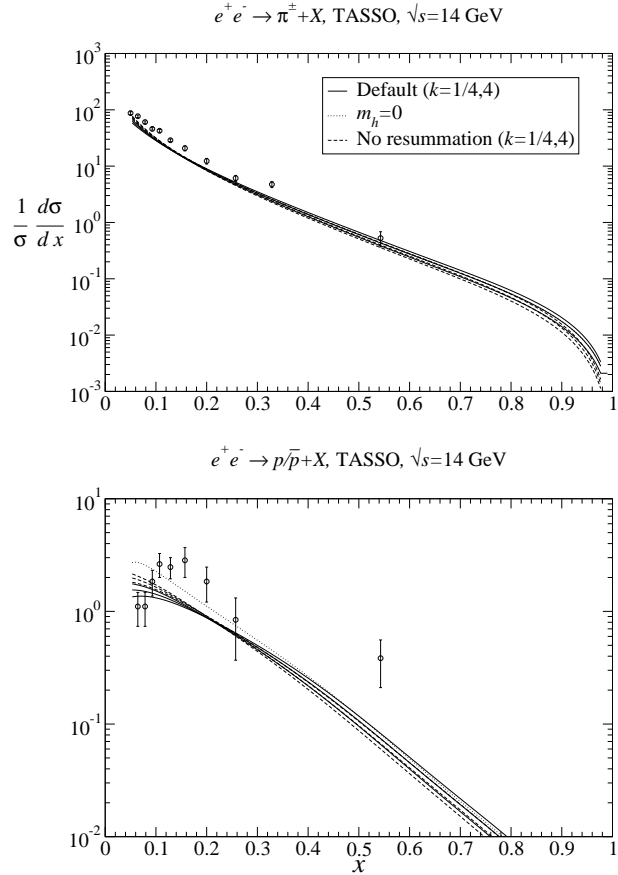}
\caption{Comparison of the calculation using the AKK08 FF sets with measurements of the invariant 
differential cross sections for inclusive production in $e^+ e^-$
reactions at $\sqrt{s}=14$ GeV from TASSO \cite{Althoff:1982dh} (labeled ``Default'').
Large $x$ resummation, to be discussed in subsection \ref{largexres}, has been implemented.
Also shown is the calculation when mass effects are
neglected (the dotted curve labeled ``$m_h=0$'') and when the ratio $k=M_f^2 /s$ is increased to 4 (lower solid curve) and decreased
to $1/4$ (upper solid curve). In the case of \cpi, the $m_h=0$ curve cannot be seen because it overlaps 
with the ``Default'' curve. From Ref.\ \cite{Albino:2008fy}.\label{TASSO14_Z-qq}}
\end{center}
\end{figure}

As Fig.\ \ref{TASSO14_Z-qq} shows, the effect of hadron mass for \cpi\ is much less than that for the heavier $p/\overline{p}$,
and hadron mass effects become more important with decreasing $x$.
In the case of the proton, the shift at small $x$ in the cross section created by hadron mass effects is significantly 
larger than the theoretical error calculated in the standard way.
It is also found that these effects become less important with increasing $\sqrt{s}$ as expected.

\noindent {\bf $ep$ reactions} \cite{Albino:2006wz}
The hadron's momentum is given by
\be
p_h=\left(\frac{m_h^2}{2x q^-},x q^-,{\bm 0}\right),
\ee
so that
\be
x_p=x \left(1-\frac{m_h^2 }{Q^2 x^2}\right),
\ee
or
\be
x=x_p \left(\frac{1}{2}+\frac{1}{2}\sqrt{1+\frac{4m_h^2}{Q^2 x_p^2}}\right).
\ee
The momentum fraction $z$ is defined by the relation $k^-=p_h^- /z$, where $k$ is the momentum of the 
fragmenting partons at leading twist. For massless partons, $k^+ =0$, and ${\bm k}_T=0$ is always true.
The cross section
\be
\frac{d\sigma^h}{dx_p dx_B dQ^2}(x_p,x_B,Q^2)
=\left(\frac{1}{2}+\frac{1}{2}\left[1+\frac{4m_h^2}{Q^2 x_p^2}\right]^{-\frac{1}{2}}\right)
\frac{d\sigma^h}{dx dx_B dQ^2}(x(x_p),x_B,Q^2),
\ee
where again $d\sigma^h/dx dx_B dQ^2$ is the usual explicitly hadron mass independent calculation,
which is given in eq.\ (\ref{xsinfacttheor}).
This modification to the normalization is equal to that of Refs.\ \cite{med} up to terms of $O((m_h^2/(x^2 Q^2))^2)$.

\begin{figure}[h!]
\hspace{-0.5cm}
\parbox{.49\linewidth}{
\includegraphics[width=8cm]{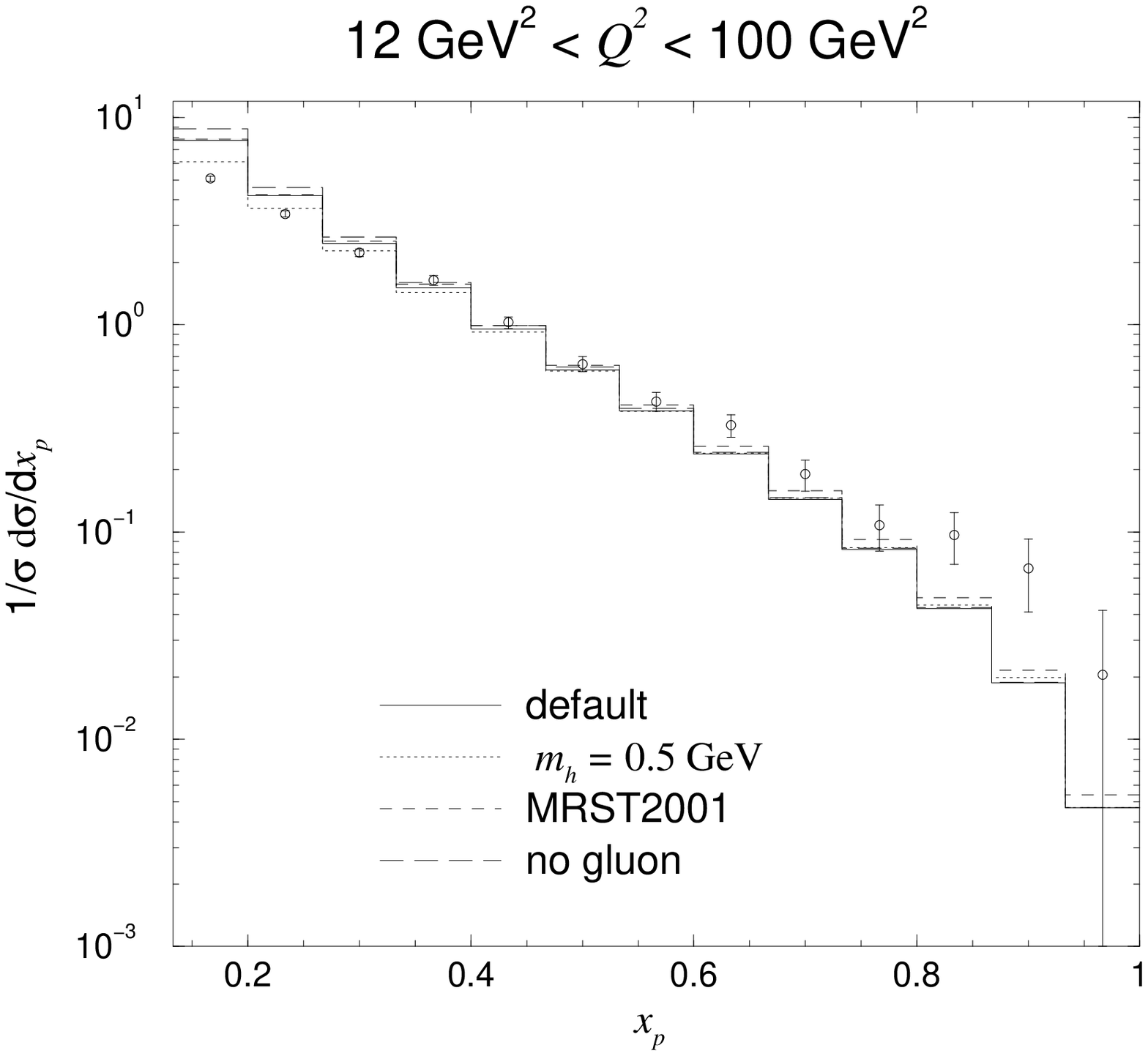}
}
\parbox{.49\linewidth}{
\includegraphics[width=8cm]{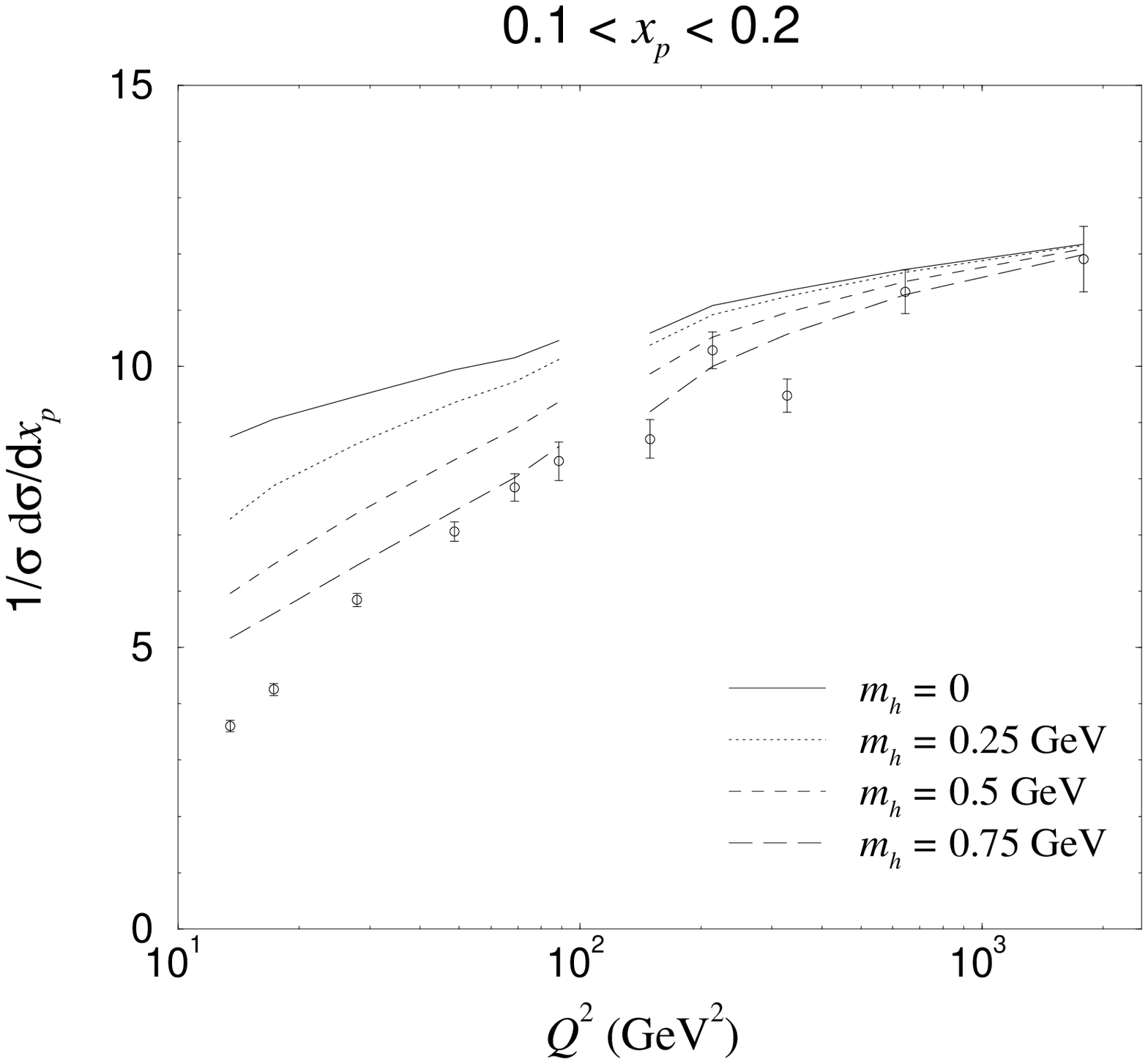}
}
\caption{{\bf Left}: 
Comparisons of theoretical predictions using the AKK FF sets with the distribution in $x$ (written ``$x_p$'' here) 
of unidentified hadrons from ZEUS \cite{Derrick:1995xg}.
The modifications to the default predictions (solid line), calculated with $m_h=0$,
arising from the replacement of the CTEQ6M PDF set of Ref.\ \cite{Pumplin:2002vw} by the
MRST2001 PDF set of Ref.\ \cite{Martin:2001es}, from the removal
of the evolved gluon, and from the incorporation of the hadron mass effect are shown.
{\bf Right}: Comparisons of theoretical predictions using the AKK FF set 
with the ZEUS data \cite{Breitweg:1997ra}, measured in the interval $0.1<x_p <0.2$
and for different values of $m_h$.
From Ref.\ \cite{Albino:2006wz}.\label{fig5}}
\end{figure}

As Fig.\ \ref{fig5} shows, hadron mass effects become important with decreasing $x$ (left) and increasing 
hadron mass (right).
Hadron mass effects are also found to become less important with increasing $Q$ (right) as expected.

Note that the effect of the initial proton's mass in HERA data is expected to approximately cancel between the 
numerator and denominator in eq.\ (\ref{epmeasXS}), and thus is not accounted for.
Furthermore, it is only important at large $x_B$, while according to Fig.\ \ref{limits} 
the HERA data are taken at relatively small $x_B$ values.

\noindent {\bf Hadron-hadron reactions} \cite{Albino:2008fy}
As for eq.\ (\ref{pluscomppartontohad}) for $e^+ e^-$ reactions above, the parton's $+$ component of momentum in the c.m.\ frame 
is given by 
\be
k^+ =\frac{p^+}{z}
\label{pluscomppartontohadforpp}
\ee
if the 3-axis is taken to be parallel with the parton's and hadron's spatial momenta, and
\be
d\sigma^h_{i_1 i_2}=\sum_i \int_x^1 dz d\sigma^i_{i_1 i_2} D^h_i(z,M_f^2).
\label{partpartXSfromfacttheo}
\ee
The measured cross section $F^h_{h_1 h_2}$ of eq.\ (\ref{defofFhh1h2}) 
is differential in the Lorentz invariant element $d^3 {\bm p}/E$, 
so we do not need to generalize the scaling variable $x$ in eq.\ (\ref{chifromVW}) to the case where $m_h \neq 0$:
Equation (\ref{ptdistfromff}) is produced by dividing the left hand side of eq.\ (\ref{partpartXSfromfacttheo}) by this element, 
which must be related to the equivalent partonic element $d^3 {\bm k}/k^0$ that we would like to appear explicitly
on the right hand side.
Assuming massless partons for simplicity, the relations $k^0=|{\bm k}|$ and $E=\sqrt{{\bm p}^2+m_h^2}$ imply firstly that
\be
\begin{split}
\frac{d {\bm p}^3}{E}=&\frac{|{\bm p}|^2}{\sqrt{|{\bm p}|^2+m_h^2}}d|{\bm p}| d\Omega,\\
\frac{d {\bm k}^3}{k^0}=&|{\bm k}|d|{\bm k}| d\Omega,
\end{split}
\ee
where $\Omega$ is the solid angle, and it has been noted that the detected hadron and fragmenting parton are spatially parallel,
and secondly that, from eq.\ (\ref{pluscomppartontohadforpp}) and using $k^+=k^0+|{\bm k}|$ and $p^+=p^0+|{\bm p}|$,
\be
2z|{\bm k}|=|{\bm p}|+\sqrt{{\bm p}^2+m_h^2}.
\ee
From these last two results we obtain our desired relation
\be
\frac{d {\bm p}^3}{E}=\frac{1}{z^2 R^2}\frac{d {\bm k}^3}{k^0},
\ee
where
\be
R=1-\frac{m_h^2}{(|{\bm p}|+\sqrt{|{\bm p}|^2 +m_h^2})^2}.
\label{eqforRinp}
\ee
Explicitly writing the PDF dependence, eq.\ (\ref{partpartXSfromfacttheo}) then becomes
\be
E\frac{d^3\sigma^h_{h_1 h_2}}{d{\bm p}^3}=\sum_{ii_1 i_2} \int dx_1 \int dx_2 
f_{i_1}^{h_1}(x_1,M_f^2) f_{i_2}^{h_2}(x_2,M_f^2) 
\int dz D_i^h(z,M_f^2) |{\bm l}| \frac{d^3\sigma^i_{i_1 i_2}}{d{\bm l}^3}\frac{1}{z^2 R^2}.
\ee
The dependence of $R$ in eq.\ (\ref{eqforRinp}) on $x_1$ and $x_2$, through $|{\bm p}|$ in
the partonic c.m.\ frame, needs to be found now.
Since $p_T$ in the c.m.\ frame is the same as that in the partonic c.m.\ frame, the definition of rapidity 
in eq.\ (\ref{defofrap}) gives
\be
|{\bm p}|=m_T \cosh y',
\ee
where 
\be
m_T=\sqrt{p_T^2+m_h^2}
\ee
and where the rapidity in the partonic c.m.\ frame
\be
y'=y+\phi.
\ee
The quantity $\phi$ appears in the boost factor $e^\phi$ which 
transforms the sum of the energy and longitudinal component of momentum between the partonic and hadronic c.m.\ frames.
To obtain the $x_1$ and $x_2$ dependence of $\phi$,
we work in the hadronic c.m.\ frame and align the 3-axis with the beams.
The $+$ ($-$) component of the parton from $h_1$ ($h_2$) is $k_1^+=x_1 P^+_1$ ($k_2^-=x_2 P^+_1$),
where $P^+_1$ is the $+$ component of momentum of $h_1$, equal to the $-$ component of momentum of $h_2$.
All other components of these partons' momenta vanish.
The quantity $k_1^+ e^\phi=k_2^- e^{-\phi}$ is respectively the $+$ and $-$ component of the parton from 
$h_1$ and the parton from $h_2$ in the partonic c.m.\ frame, so that $x_1 e^\phi =x_2 e^{-\phi}$ or, finally,
\be
\phi=\ln \sqrt{\frac{x_2}{x_1}}.
\label{boost}
\ee
Through eq.\ (\ref{boost}) and the 3 equations preceding it above, the dependence of $R$ on $x_1$ and $x_2$ is obtained.

\begin{figure}[h!]
\begin{center}
\includegraphics[width=12cm]{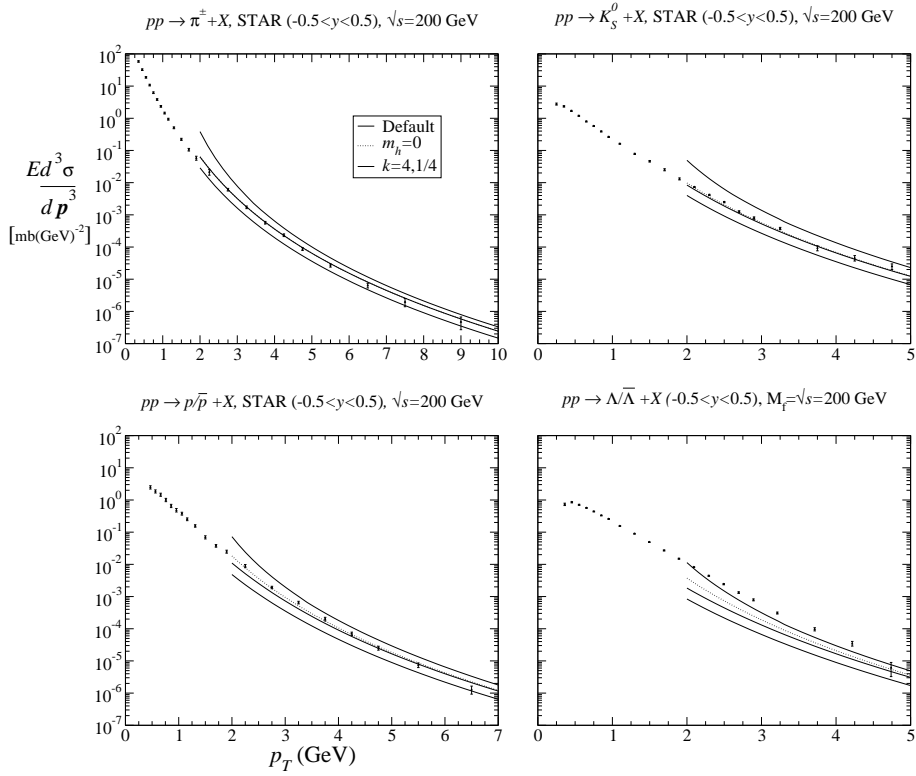}
\caption{Comparison of the calculation using the AKK08 FF sets with measurements of the invariant 
differential cross sections for inclusive production in $pp$
collisions at $\sqrt{s}=200$ GeV data from STAR \cite{Adams:2006nd,Abelev:2006cs} for which $-0.5<y<0.5$ (labeled ``Default'').
Also shown is the calculation when mass effects are
neglected (the dotted curve labeled ``$m_h=0$'') and when the ratio $k=M_f^2 /s$ is increased to 4 (lower solid curve) and decreased
to $1/4$ (upper solid curve). In the case of \cpi, the $m_h=0$ curve cannot be seen because it overlaps 
with the ``Default'' curve. From Ref.\ \cite{Albino:2008fy}.\label{ppSTAR}}
\end{center}
\end{figure}

As can be seen in Fig.\ \ref{ppSTAR}, hadron mass effects become more important as the mass of the detected hadron
increases and as $p_T$ decrease, but remain smaller than the theoretical error calculated in the standard way.
Similar results are obtained at higher rapidities, i.e.\ the size of hadron mass effects is fairly independent of rapidity.

\subsection{Large $x$ resummation \label{largexres}}

When $x$ is sufficiently large, the accuracy of the FO perturbative calculations for the equivalent hard partonic
cross section and the DGLAP evolution is worsened by unresummed
divergences occurring in the formal limit $x\rightarrow 1$.
Here, the ``large $x$'' region refers to the region where these logarithms are important, but is otherwise not more precisely
defined, and could even include rather small $x$ values.
The possible importance of large $x$ resummation effects for available experimental data was pointed out 
in Ref.\ \cite{Bourhis:2000gs}, where uncertainties due to unresummed large $x$ logarithms 
were minimized by optimization of the scale.
Resummation of these logarithms should reduce this theoretical error.
The effects of these unresummed logarithms 
should become more significant with decreasing $E_s$, where errors due to perturbative truncation are largest.
Resummation of all large $x$ logarithms that occur up to NLO in $e^+ e^-$ reactions 
is now possible using results \cite{Cacciari:2001cw} which we now present.
Resummation makes a significant improvement to current global fits \cite{Albino:2008fy}, 
which will be discussed in section \ref{currglobalfits}.

In general, in the series expansion of the hard part ${\mathcal W}(x,a_s)$ of some cross section, these
large $x$ divergences take the form $a_s^n [\ln^{n-r} (1-x)/(1-x)]_+$, where $r=0,...,n$ labels the class of divergence.
In Mellin space, these divergences therefore take the form $a_s^n \ln^{n+1-r} N$. 
They may be factored out, which results in the calculation of ${\mathcal W}$ taking the form
\be
{\mathcal W}(N,a_s)={\mathcal W}_{\rm res}(N,a_s)\left(\sum_n a_s^n {\mathcal W}_{\rm FO}^{(n)}(N)\right),
\label{genresproc}
\ee
where the FO series in parenthesis on the right hand side is free of
these divergences since they are all contained in ${\mathcal W}_{\rm res}$. 
${\mathcal W}$ at large $N$ is approximated
by ${\mathcal W}_{\rm res}$ when the divergences in ${\mathcal W}_{\rm res}$ are resummed, which involves
writing ${\mathcal W}_{\rm res}$ as an exponential and expanding the exponent in $a_s$ keeping $a_s \ln N$ fixed.
The perturbative series for the NLO quark coefficient function $C_q=C_{\rm NS}=C_{\rm S}$ for inclusive quark production
in $e^+ e^-$ reactions contains a leading (class $r=0$) 
and next-to-leading (class $r=1$) divergence, which can be replaced in the manner of eq.\ (\ref{genresproc})
by all such divergences of these two classes.
These divergences are all contained in the formula \cite{Cacciari:2001cw}
\be
\ln C_q(N,a_s(s))=\int_0^1 dz \frac{z^{N-1}-1}{1-z}\Bigg[\int_{s}^{(1-z)s} \frac{dq^2}{q^2} A(a_s(s))
+B(a_s((1-z)s))\Bigg] +O(1),
\label{genresumforCq}
\ee
where
\be
(A,B)(a_s)=\sum_{n=1}^\infty (A,B)^{(n)} a_s^n.
\ee
To determine them explicitly requires the results
\be
\begin{split}
A^{(1)}=&2 C_F,\\
A^{(2)}=&-C_F\left(C_A \left(\frac{\pi^2}{3}-\frac{67}{9}\right)+\frac{20}{9}T_R n_f \right)\ {\rm and}\\
B^{(1)}=&-\frac{3}{2}C_F.
\end{split}
\ee
Since eq.\ (\ref{genresumforCq}) is algebraically similar to the resummed quark coefficient function
of DIS \cite{Catani:1989ne}, we may obtain the divergences of classes $r=0,1$ 
directly from the $\overline{\rm MS}$ result in Ref.\ \cite{Albino:2000cp},
\be
\begin{split}
\ln C_q^{r=0,1}(N,a_s)=&\frac{A^{(1)}}{a_s \beta_0^2}\left[(1-\lambda_s)\ln (1-\lambda_s)+\lambda_s\right]
+\frac{A^{(1)}\beta_1}{2\beta_0^3}\ln^2 (1-\lambda_s)\\
&+\left(\frac{B^{(1)}}{\beta_0}-\frac{A^{(1)}\gamma_E}{\beta_0}+\frac{A^{(1)} \beta_1}
{\beta_0^3}-\frac{A^{(2)}}{\beta_0^2}\right) \ln (1-\lambda_s)
-\left(\frac{A^{(2)}}{\beta_0^2}-\frac{A^{(1)}\beta_1}{\beta_0^3}\right)\lambda_s,
\label{lnCqforr1andr0}
\end{split}
\ee
where $\lambda_s=a_s \beta_0 \ln N$. 
According to the general form of eq.\ (\ref{genresproc}),
the resulting resummed quark coefficient function that we seek is
\be
C_q=C_q^{r=0,1}\left(1+a_s(C_q^{(1)}-C_q^{r=0,1\ (1)})\right).
\label{howtoresumCq}
\ee
where 
\be
C_q^{r=0,1\ (1)}=\frac{A^{(1)}}{2}\ln^2 N+\left(A^{(1)}\gamma_E-B^{(1)}\right)\ln N,
\label{CqunreslargeN}
\ee
is the coefficient of the $O(a_s)$ term in the expansion of $C_q^{r=0,1}$ in $a_s$,
and ensures that the original NLO result is obtained when the
whole of the right hand side of eq.\ (\ref{howtoresumCq}) is expanded in $a_s$, i.e.\ when the resummation
is ``undone'', because it prevents double counting of the divergences.
Note that there are an infinite number of other schemes which are consistent with this criteria and give the
large $N$ behaviour of eq.\ (\ref{lnCqforr1andr0}), a typical feature of perturbation theory.
For example, the modification $N\rightarrow N+a$ in eqs.\ (\ref{lnCqforr1andr0}) and (\ref{CqunreslargeN}) simultaneously, 
where $a$ is a finite, possibly complex, constant, is perfectly valid provided $|a|$ is not too large.

Equation (\ref{lnCqforr1andr0}) contains a Landau pole when $\lambda_s=1$, for which $N$ is real and $\gg 1$. 
However, in the inverse Mellin transform it is not necessary for the contour 
in the complex $N$ plane to run to the right of this pole, as it should for the other poles, 
because it is unphysical, created by the ambiguity of the asymptotic series. 
In $x$ space it gives essentially a higher twist contribution and is therefore negligible.
The best convergence of the series occurs when the contour is chosen to be to the left of this pole, 
which is known as the {\it minimal prescription} \cite{Catani:1996yz}, which is also the
most efficient choice for the numerical evaluation of the inverse Mellin transform.

\begin{figure}[h!]
\includegraphics[width=8cm]{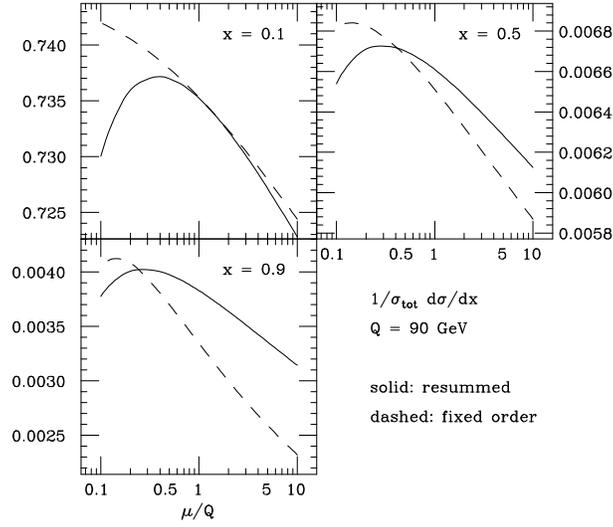}
\caption{Dependence of the unresummed (``fixed order'') and resummed cross section 
on the factorization scale $M_f$, written here as ``$\mu$''. 
The renormalization scale is fixed to the factorization scale. 
From Ref.\ \cite{Cacciari:2001cw}. \label{scale-dep}}
\end{figure}

According to Fig.\ \ref{scale-dep}, the effect of the resummation is to increase the cross section
(at the standard choice $M_f/E_s=1$), and more so with increasing $x$. 
This behaviour is also observed in Fig.\ \ref{TASSO14_Z-qq}.
This is a typical feature of soft gluon resummation in both timelike and spacelike processes.
Fig.\ \ref{scale-dep} also shows that resummation lowers the scale variation if the resummed
result is compared with the unresummed result at the same values for $M_f/E_s=\mu/Q$.
From this we can infer that resummation lowers the theoretical error,
assuming that the factorization scale in the resummed case can be given the same interpretation
as the factorization scale in the unresummed case.
However, caution is advised here because resummation is not a systematic improvement to a FO calculation in the same way that
extending a FO calculation to higher orders is.

Explicit results for the resummation of large $x$ logarithms in $ep$ reactions do not exist at present,
but, in the current fragmentation region, are likely to be similar to the results for $e^+ e^-$ reactions above
due to the physical similarities between these two processes.

\begin{figure}[h!]
\includegraphics[width=8cm]{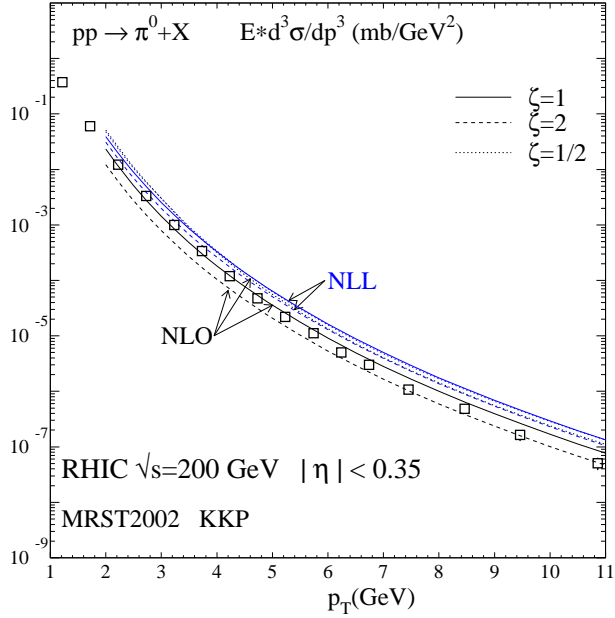}
\caption{Dependence of the unresummed (``NLO'') and resummed (``NLL'') cross section on the factorization scale $M_f$,
where $\xi=M_f/p_T$. The renormalization scale is fixed to the factorization scale.
From Ref.\ \cite{deFlorian:2005yj}. \label{res_rhic_kkp}}
\end{figure}

Large $x$ resummation is particularly valuable for $pp(\overline{p})$ reactions where the scale variation is large,
as can be seen in Fig.\ \ref{res_rhic_kkp}.
Note again the reduction in theoretical error due to the resummation.
Unfortunately, in this case formal results only exist for the case that rapidity is integrated over all values,
the results for a given finite rapidity range having to be determined approximately \cite{deFlorian:2005yj}.

\begin{figure}[h!]
\includegraphics[width=9.5cm,angle=-90]{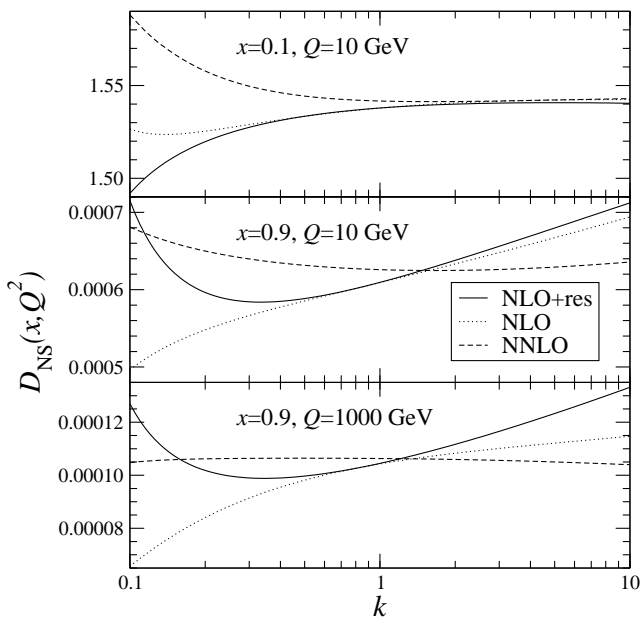}
\caption{Dependence of the unresummed (``NLO'') and resummed (``NLO+res'') non singlet FF on the renormalization scale $\mu$.
The factorization scale $M_f$ is written ``$Q$'', and $k=\mu/Q$. From Ref.\ \cite{Albino:2007ns}. \label{letter_lxr_fig}}
\end{figure}

For consistency, the large $x$ logarithms in 
the DGLAP evolution of the FFs must also be resummed.
These logarithms occur in the diagonal parts of the splitting function (including the valence and non singlet).
The FO result for $P$ is already resummed, because these divergences approach $1/(1-z)$, or $\ln N$ in Mellin space, 
at every order \cite{Korchemsky:1988si,Albino:2000cp}. 
This is because they originate from soft radiation which is classical in nature, 
and therefore can be arranged to appear at LO only by choosing a physical strong coupling \cite{Kalinowski:1980wea}.
In other words, there are no $r=0$ divergences, the divergence at LO, proportional to $a_s \ln N$, is the only $r=1$ divergence,
the divergence at NLO, proportional to $a_s^2 \ln N$, is the only $r=2$ divergence, and so on.
Up to NLLs, the matrix $U$ appearing in eq.\ (\ref{waytosolvedglap}) behaves at large $N$ like \cite{Albino:2007ns}
\be
U(N,a_s)=\sum_i \exp\left[u^{[1]}_{\alpha_i\alpha_i} a_s\ln N+O(a_s(a_s \ln N)^n)\right]M^i,
\ee
where $\alpha_+ =\Sigma$ and $\alpha_- =g$, and
\be
u^{[1]}_{\alpha_i\alpha_i}=\lim_{N\rightarrow \infty}\left(\frac{\beta_1}{\beta_0^2}P^{(0)}_{\alpha_i\alpha_i}(N)
-\frac{1}{\beta_0}P^{(1)}_{\alpha_i\alpha_i}(N)\right)/\ln N
\ee
A method for resumming these logarithms in the analytic Mellin space solution to the DGLAP equation 
of Refs.\ \cite{Furmanski:1981cw,Ellis:1993rb} discussed in subsection \ref{anmelDGLAP} is given in Ref.\ \cite{Albino:2007ns}.
As can be seen in Fig.\ \ref{letter_lxr_fig}, the effect of this resummation at $\mu/M_f=1$ is much smaller than that 
in the coefficient functions, although the modification to the theoretical error obtained by scale variation is substantial.

\section{The treatment of experimental errors \label{treaterrs}}

One of the main purposes of fitting FFs to experimental data is to make predictions for other observables 
that depend on them through the universality of FFs implied by the factorization theorem.
Assuming that the data of the fit were calculated correctly,
a disagreement between the calculation of a prediction (or predictions) and its measurement suggests
either a problem in the calculation of the prediction
or that this prediction depends on components of the FFs which were 
not well constrained by the experimental data used in the fit.
The latter scenario will be the case if the quality of the fit repeated with the inclusion of
the prediction's measurement is good, 
but the former scenario must be considered if the prediction's measurement cannot be fitted well.
More generally, constraints imposed on an as yet unmeasured prediction by a fit 
can be determined by using instead a fictitious measurement of the prediction in the fit 
and then repeating the fit multiple times for a range of values of this measurement, 
the acceptable values being those for which a good fit is obtained.
This is precisely what is achieved by the {\it Lagrange multiplier method}, to be discussed at the end of subsection \ref{errprop}.
A perhaps slightly less reliable but currently more practical alternative to constraining predictions 
is to fit only to experimental data for which all relevant physics can be reliably calculated,
determine the errors on the FFs propagated from these data, 
and then propagate those FF errors to the prediction.
This method is more practical because a fit does not need to be performed each time constraints on a prediction 
are to be determined.
In addition, if a prediction which has been measured is believed to be dependent on new physics effects, 
such as higher twist or physics beyond the Standard Model,
regions of the space of this new physics's parameters can be easily ruled out by a measurement of this prediction
by calculating the prediction's propagated error ranges for different values of these parameters
and seeing which ranges fail to overlap with the measurement within its error range.
Otherwise, if the FF errors are not known, these regions can only be ruled out by 
repeating the fit with the prediction's measurement included 
for many different values for the new physics's parameters.

In the above discussion, we have assumed that the constraints provided by the experimental data used in a fit are complete,
in the sense that all systematic effects in the experiment are available.
In practice it can happen that some important systematic effects are unknown, resulting 
in a failure to fit or to describe a measurement as discussed above even though all relevant physics effects are accounted for
in the calculation.
In this case it must be decided whether this measurement is consistent with other similar measurements from other experiments.

Most recently, a method has been developed and applied by the NNPDF collaboration 
which uses neural networks as unbiased interpolants \cite{Ball:2008by}.
This method solves a number of statistical issues in global fits.

In the rest of this section we discuss these issues in more detail.
We first discuss the meaning of $\chi^2$ and the interpretation of the parameter values at its minimum. 
The current method of presenting PDF errors such that they can be easily propagated to predictions is then discussed,
because this procedure is applicable to FFs.
Finally, we discuss the Lagrange multiplier method already applied in FF fits, since this is perhaps the most
reliable method of propagating errors to predictions.

\subsection{Relation between $\chi^2$ and probability densities}

In this subsection we define $\chi^2$ for a given set of measurements 
and explain why the location of its minimum in the space of parameters of the theory
corresponds to the most likely values of these parameters according to these measurements.
Our starting point is Bayes' theorem \cite{Amsler:2008zz,Giele:1998gw}, which states
that the probability, as a quantification of the present degree of confidence from experience, 
for the vector $f^e$ of experimental values $f^e_i$ to take
values between $f^e$ and $f^e+df^e$ and for the vector $f^t$ of ``true'' values $f^t_i$ to take
values between $f^t$ and $f^t+df^t$ is given by
\be
P(f^e,f^t)df^e df^t=P(f^t|f^e)df^t P(f^e)df^e=P(f^e|f^t)df^e P(f^t)df^t.
\label{bayestheorem}
\ee
The probability densities $P(f^t|f^e)$, $P(f^t)$, $P(f^e)$ and $P(f^e|f^t)$
appearing in eq.\ (\ref{bayestheorem}) will now be defined.
$P (f^e|f^t)df^e$ is the probability that measurements give results between $f^e$ and $f^e+df^e$ 
given ``true'' values $f^t$, and its dependence on $f^e$ is determined from experiment.
The quantity $\chi^2 (f^e,f^t)$ is defined through
\be
P (f^e|f^t) \propto e^{-\frac{1}{2}\chi^2 (f^e,f^t)}.
\label{Pfeftfromchi2}
\ee
For example, in the case that the experimental uncertainties are purely statistical and are Gaussian distributed,
\be
\chi^2=\sum_i \left(\frac{f_i^t-f_i^e}{\sigma_i}\right)^2.
\label{simpdefofchi2}
\ee
We defer to subsection \ref{syserrinchi2} the form of $\chi^2$ when systematic errors are present.
$P(f^t)df^t$ is the probability that the true results lie between $f^t$ and $f^t+df^t$,
which reflects the knowledge of these observables prior to the $f_i^e \pm \sigma_i$ measurements. 
$P(f^e)df^e$ is the probability that measurements will give results between 
$f^e$ and $f^e+df^e$ prior to the experiment taking place, and
may be determined from $P(f^e|f^t)$ and $P(f^t)$ by performing the integration in the first and second 
equalities in eq.\ (\ref{bayestheorem}) over $f^t$, using $\int P(f^t|f^e)df^t =1$ and dividing out the factor $df^e$:
\be
P(f^e)=\int P(f^e|f^t)P(f^t)df^t.
\label{Pfeelim}
\ee
Finally, $P(f^t|f^e)df^t$ quantifies the ``degree of belief'' for the true observables to take values between $f^t$ and $f^t +df^t$
as a result of the experimental results $f^e$, 
and is the quantity through which the test theoretical calculations are constrained by the experimental data by identifying
these calculations with the test true values $f^t$.
From eqs.\ (\ref{bayestheorem}) and (\ref{Pfeelim}), 
\be
P(f^t|f^e)=\frac{P(f^e|f^t)P(f^t)}{\int P(f^e|f^t)P(f^t)df^t}.
\ee
If the experimental errors are sufficiently small, $P(f^t)$ will be approximately constant in $f^t$ in the 
region of most statistical relevance, $f_i^t \pm \sigma_i$, and so approximately divides out of this expression. 
Then also using eq.\ (\ref{Pfeftfromchi2}), we have finally
\be
P(f^t|f^e) \propto e^{-\frac{1}{2}\chi^2 (f^e,f^t)}.
\label{eqforPftfromfeinchi2}
\ee
Consequently, the values of $f^t$ in which we are most confident are those for which
$\chi^2$ is a minimum, since then the likelihood $P(f^t|f^e)$ is maximal.
Since in general we are interested in experimental constraints of the vector $a$ of parameters $a_i$ of a theory, 
whose number should be much less than the number of data points or components of $f^t$ in order
to justify such tests, we should be considering probability densities in $a$ instead of in $f^t$.
Any probability distribution $p(f^t)$ in $f^t$ and the equivalent $p(a)$ in $a$
are related via $p(f^t) df^t =p(a) da$, i.e.\ 
$p(a) =p(f^t) J_{f^t}(a)$, where $J_{f^t}(a)$ is the Jacobian of $f^t$ in $a$.
Again, if the experimental errors are sufficiently small such that $J_{f^t}(a)$ is approximately constant in $a$ 
in the region of most statistical relevance, $P(a|f^e) \propto P(f^t|f^e)$ and eq.\ (\ref{eqforPftfromfeinchi2}) becomes
\be
P(a|f^e) \propto e^{-\frac{1}{2}\chi^2 (f^e,f^t(a))}.
\label{Palphafefromchi2}
\ee
Thus the minimization of $\chi^2$ with respect to $a$ gives 
the most likely values of the parameters for these experimental data.

If the theoretical model and experimental data are reliable, the expected value of the minimized $\chi^2$
is $\chi^2_{\rm min}=N_{\rm DF}\pm \sqrt{2N_{\rm DF}}$ \cite{Soper:1994km,Collins:2001es}, 
where $N_{\rm DF}$ is the number of degrees
of freedom, i.e.\ the difference between the number of data points and the number of parameters being fitted.
Often the value of the reduced $\chi^2_{\rm min}$, $\chi^2_{\rm DF}=\chi^2_{\rm min}/N_{\rm DF}$, is quoted since 
this value is always expected to be around unity, give or take $\sqrt{2/N_{\rm DF}}$.

\subsection{Incorporation of systematic errors in $\chi^2$ \label{syserrinchi2}}

An experimental systematic effect is a physical effect arising from a single experiment-dependent source 
(so this does not include physical effects which are common to all experiments measuring similar observables), 
and therefore the modification to the central value of any data point $f^e_i$ due to this systematic effect
is completely correlated with its modification to any other from the same experiment.
Thus, the $K$th source of systematic uncertainty will cause $f_i^e$ to be shifted to $f_i^e+\lambda_K \sigma_i^K$,
where $\lambda_K$, which is statistically distributed, measures the importance of the systematic effect itself 
and so is independent of the measurements,
and the $\sigma_i^K$ measures the influence of this systematic effect on each measurement.
While $\lambda_K \sigma_i^K$ is fixed, the relative normalization between $\lambda_K$ and the $\sigma_i^K$ is not.
It can be fixed by choosing the probability density in $\lambda_K$ to be proportional to $\exp[-\lambda_K^2/2]$,
assuming Gaussian systematic errors.
Therefore, the $\chi^2$ for Gaussian statistical errors in eq.\ (\ref{simpdefofchi2}) is modified to
\be
\chi^2=\sum_i \left(\frac{f_i^t-f_i^e-\sum_K \lambda_K \sigma_i^K}{\sigma_i}\right)^2+\sum_K \lambda_K^2.
\label{chi2intermsoflambdak}
\ee
For the specific theory used to calculate the data,
eq.\ (\ref{Palphafefromchi2}) allows for an estimate of the $K$th systematic effect on each data point, $\lambda_K \sigma_i^K$,
knowing only the statistically acceptable maximum size of this systematic effect on each data point, $|\sigma_i^K|$, 
and its direction relative to the other data points (the sign of $\sigma_i^K$),
through minimization of $\chi^2$ with respect to $\lambda_K$.
Doing such a minimization numerically would significantly delay termination 
of the minimization routine being used to perform the global fits.
Therefore, this minimization should be done analytically \cite{Pumplin:2002vw,Albino:2008fy}:
The most likely values of the $\lambda_K$ occur where
\be
\frac{\partial \chi^2}{\partial \lambda_K}=0.
\ee
Solving these equations for the $\lambda_K$ gives
\be
\lambda_K=\sum_i \frac{f_i^t-f_i^e}{\sigma_i^2}
\left(\sigma_i^K-\sum_{jkL}\sigma_i^L\sigma_j^L \left(C^{-1}\right)_{jk}\sigma_k^K\right),
\label{bestfitforlambda}
\ee
where the covariance matrix 
\be
C_{ij}=\sigma_i^2 \delta_{ij}+\sum_K \sigma_i^K \sigma_j^K.
\label{covmatfromsystandstaterrs}
\ee
Substituting the best fit results for the $\lambda_K$ in eq.\ (\ref{bestfitforlambda}) into eq.\ (\ref{chi2intermsoflambdak}) gives
the more familiar expression for $\chi^2$ when systematic errors are accounted for:
\be
\chi^2=\sum_{ij} (f_i^t-f_i^e)\left(C^{-1}\right)_{ij} (f_j^t-f_j^e).
\label{gendefofgausschi2}
\ee

In practice, not all information on the systematic effects in an experiment is available.
For example, often the analysis provides only the ``total'' systematic error on a data point, 
equal to all the systematic errors of the data point from all sources added in quadrature.
In this case, eq.\ (\ref{simpdefofchi2}) is typically used in global fits with the statistical error modified
to be the statistical error and this total systematic error added in quadrature.
This corresponds to setting the off-diagonal components of the covariance matrix
in eq.\ (\ref{covmatfromsystandstaterrs}) to zero, i.e.\ to neglecting correlation effects between data points.
More serious is the presence of unknown systematic effects.
Such effects have to be neglected even though they could be important, but as the number of data sets increases, 
the cancellation of these effects among different experiments will increase 
as far as the determination of the most likely value for the parameters is concerned.
Unfortunately, these unknown systematic effects have to be accounted for in the determination of parameter errors.
We will discuss common methods of doing this in subsection \ref{errprop}.

\subsection{Propagation of experimental errors to fitted fragmentation functions \label{errprop}}

For Gaussian statistical errors and no correlations between data points,
eqs.\ (\ref{eqforPftfromfeinchi2}) and (\ref{simpdefofchi2}) imply that
the standard deviation error on $f^t_i$ is $\sigma_i$ and is equal to the amount by which 
it has to increase or decrease with all other $f^t_{j\neq i}$ fixed in order for $\chi^2$ 
to increase by one from its minimum value of zero.
A more realistic procedure based on this ideal one can be applied 
to propagate experimental errors to parameters $a$ of a given model 
from the experimental data to which they have been fitted. 
For simplicity, instead of $a_i$ we work with the parameters 
\be
y_i=a_i-a_i^0,
\ee
where $a_i=a_i^0$ is the parameter space location of the minimum in $\chi^2$ for these data.
All following expressions are to lowest order in the $y_i$. 
The change in $\chi^2$ from its minimum value at the origin $y_i=0$ to any small values of the $y_i$ is approximately 
\be
\Delta \chi^2=\sum_{ij} y_i H_{ij} y_j,
\label{condony}
\ee
where the Hessian $H$ is given by
\be
H_{ij}=\frac{1}{2}\left(\frac{\partial^2 \chi^2}{\partial y_i \partial y_j}\right)\Bigg{|}_{y_k=0}.
\label{Hfrom2ndderivofchi2}
\ee
The experimental error on a given $a_i$ is equal to the magnitude of $y_i$ that is required to make
\be
\Delta \chi^2=1
\label{idealupperchi2}
\ee
when all other $y_{j\neq i}$ are fixed. 
Thus, in the absence of correlations between the $y_i$, 
so that the  off-diagonal elements of $H$ vanish, the experimental error on $y_i$ is $1/\sqrt{H_{ii}}$.
More generally, according to eq.\ (\ref{Palphafefromchi2}), $H$ is the inverse of the {\it covariance matrix for the parameters},
\be
\langle y_i y_j \rangle =(H^{-1})_{ij}.
\label{covmatonpars}
\ee
According to eq.\ (\ref{Hfrom2ndderivofchi2}), this covariance matrix 
can be determined numerically by calculating then inverting the matrix of second derivatives of $\chi^2$,
which corresponds to the {\it parabolic approximation} of eq.\ (\ref{condony}).
If this approximation is not valid, the covariance matrix 
can be determined by the criterion of eq.\ (\ref{idealupperchi2}) directly, as was done in Ref.\ \cite{Bourhis:2000gs}.

However, the criterion of eq.\ (\ref{idealupperchi2}) turns out to be far too idealistic for actual global fits.
For example, in Ref.\ \cite{Thorne:2002kk} the inconsistency between the resulting constraints on $\alpha_s(M_Z)$ 
determined from this criterion from
different data sets used in the CTEQ6 extraction of PDFs \cite{Pumplin:2002vw} is highlighted.
This is attributed in Refs.\ \cite{Thorne:2002kk} and \cite{Botje:2001fx} to various sources. 
Firstly, the published statistical information on the experimental data may not be sufficiently reliable ---
errors, including systematic ones, may be inaccurate or not known at all, or 
some errors may not be Gaussian distributed, let alone symmetric.
Thus, for example, a fit to, and extraction of parameter errors from, data from one experiment in which
the normalization error is underestimated may result in a good fit and apparently sensible results, but a similar analysis
using data from two independent experiments suffering this problem will usually not.
Secondly, the theory, including the parameterization used for the initial FFs/PDFs, 
may not be valid in all kinematic regions spanned by the data being fitted to.
The simplest way to handle the reasons for the failure of the $\Delta \chi^2 =1$ rule, whatever they may be,
is to modify the criterion of eq.\ (\ref{idealupperchi2}) to \cite{Pumplin:2000vx,Pumplin:2001ct}
\be
\Delta \chi^2 =T^2,
\label{defoftolpar}
\ee
where the {\it tolerance parameter} $T$ is a constant somewhat larger than 1.
According to eq.\ (\ref{gendefofgausschi2}), this is equivalent to multiplying all
errors, statistical and systematic, by a factor $T$.
For example, the authors of Ref.\ \cite{Hirai:2007cx} use $T^2 \sim 10$.
This is significantly less than the choice $T^2 =100$
in the CTEQ6 analysis \cite{Pumplin:2002vw}, although the number of data points in the latter analysis 
exceeds that of the former by a factor of almost 10.

An alternative approach to eq.\ (\ref{defoftolpar}) to handle unknown systematic effects is to somehow take into
account the degree of incompatibility of the data with one another. 
This is done in the neural network approach of Ref.\ \cite{Ball:2008by}.

The statistical information on parameters of a theory from experimental data can be propagated to observables as follows.
Consider a prediction $X$ which depends on the FF parameters $a_i$. We wish to determine the uncertainty
on $X$, $\sqrt{\Delta X^2}$, resulting from the uncertainties on the experimental data which constrain the $a_i$.
Since $\Delta X = \sum_i (\partial X/ \partial y_i )y_i$,
\be
\langle \Delta X^2 \rangle= \sum_{ij} \frac{\partial X}{\partial y_i}\frac{\partial X}{\partial y_j}
\langle y_i y_j \rangle.
\ee
The quantity $\langle y_i y_j \rangle = \int P(y_k) y_i y_j dy$. 
The probability distribution in $y$ is chosen to be
$P(y_k)\propto \exp[-\Delta \chi^2/(2T^2)]$, where $\Delta \chi^2$ is given by eq.\ (\ref{condony}),
which is the same as eq.\ (\ref{Palphafefromchi2}) except that
the divisor $T^2$ has been introduced to ensure that
the range of acceptable locations in parameter space are those for which $\Delta \chi^2\leq T^2$ (recall eq.\ (\ref{defoftolpar})).
This modifies the eq.\ (\ref{covmatonpars}), the covariance matrix of the parameters, to $\langle y_i y_j \rangle =T^2(H^{-1})_{ij}$. 

A practical formula for calculating the error on a prediction is
\be
\langle \Delta X^2 \rangle=\sum_{ij}\frac{X(y_k=\kappa \delta_{ki})-X(y_k=-\kappa \delta_{ki})}{2\kappa}
T^2(H^{-1})_{ij}\frac{X(y_k=\kappa \delta_{kj})-X(y_k=-\kappa \delta_{kj})}{2\kappa},
\label{defofdeltaX2}
\ee
where $\kappa$ should be chosen less than or of the order of the error on each $y_i$, but is otherwise arbitrary.
In other words, constraints on a prediction are obtained simply by calculating it
for each of the different sets of FFs for which $y_k=\pm \kappa \delta_{ki}$,
(and for which $y_k=0$ for the central value of the prediction). 
Knowledge of the fitted $a_i$ and their errors, or even the choice of parameterization, is not required.

\begin{figure}[h!]
\includegraphics[width=9.5cm]{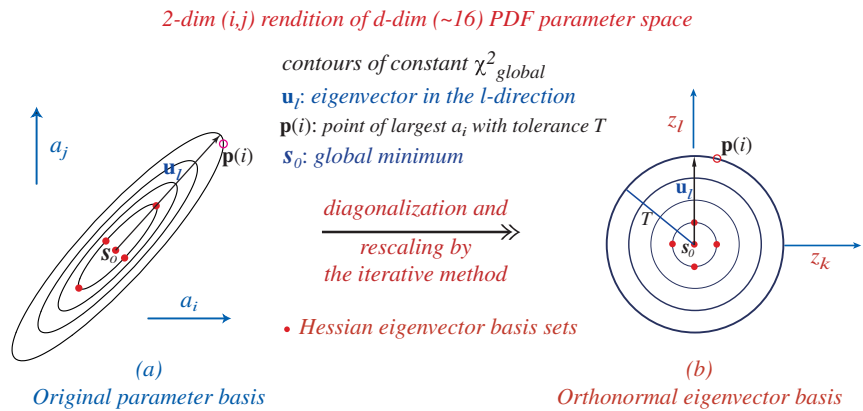}
\caption{Illustration of the Hessian diagonalization of Refs.\ \cite{Pumplin:2000vx,Pumplin:2001ct} 
for the case of 2 parameters for simplicity, showing contours of constant $\chi^2$ in parameter space.
In the new space of uncorrelated parameters (right), the contours are circles.
From Ref.\ \cite{Pumplin:2001ct}. \label{HesseMethod}}
\end{figure}

Equation (\ref{defofdeltaX2}) can be simplified by diagonalization of the Hessian \cite{Pumplin:2000vx,Pumplin:2001ct},
so that the parameters of the new basis are uncorrelated.
The normalizations of these uncorrelated parameters $z_i$ are chosen such that
\be
\Delta \chi^2=\sum_i z_i^2,
\ee
which is achieved through eq.\ (\ref{condony}) by taking
\be
z_i =\sum_j \sqrt{\epsilon_i} y_j v_{ji},
\label{zfromy}
\ee
where $v_{lk}$ is the k$th$ eigenvector of $H_{ij}$ with eigenvalue $\epsilon_k$,
\be
\sum_j H_{ij} v_{jk}=\epsilon_k v_{ik},
\ee
and the eigenvectors have been chosen orthonormal,
\be
\sum_i v_{ij} v_{ik} =\delta_{jk},
\ee
i.e.\ the matrix $v$ obeys the orthonormal condition $v^T =v^{-1}$.
This change in the parameter correlations due to the change of parameter basis is illustrated in
Fig.\ \ref{HesseMethod}.
Equation (\ref{defofdeltaX2}) is simplified to
\be
\langle \Delta X^2 \rangle=\frac{1}{2}\frac{T^2}{4t^2}\sum_i\left(X(y_k( z_l=t \delta_{li}))-X(y_k(z_l=-t \delta_{li}))\right)^2,
\label{prederr2}
\ee
where the choice $t \lesssim T$ should be made.
Thus, errors on predictions can be simply calculated from two sets $S_i^\pm$ for each $i$ of
FFs at parameter values $y_k=y_k( z_l=\pm t \delta_{li})$, as advocated in Refs.\ \cite{Soper:1994km,Pumplin:2000vx,Pumplin:2001ct},
and is the current way in which PDFs are presented.

The Lagrange multiplier method \cite{Pumplin:2000vx}, implemented in the global fit of Ref.\ \cite{deFlorian:2007hc},
is a more reliable approach to obtaining errors on predictions because it avoids the assumption 
that $\chi^2$ has a quadratic dependence on the parameters, as assumed in eq.\ (\ref{condony}).
The price is that it is computationally slower.
This method gives the range of values of a prediction $X$ such that the minimum in $\chi^2$ in which
the fitted parameters are constrained such that
$X$ is in this range is no greater than an amount $T^2$ above the unconstrained minimum $\chi^2$.
This is achieved by minimization of the function
\be
F=\chi^2+\lambda X
\label{LMM}
\ee
for a range of different values of $\lambda$, because
the corresponding $\chi^2(\lambda)$ in each case is the minimum $\chi^2$ given that
the prediction takes the value $X(\lambda)$. 
To see this, take $dF=0$ and choose the $da_i$ such that $X(\lambda)$ is fixed, i.e.\ 
such that $\sum_i (\partial X/\partial a_i)da_i=0$.
Using this same reasoning, we see that eq.\ (\ref{LMM}), which is the one used in practice, 
leads to the same results as the alternative form
\be
F=\chi^2+\left(\frac{X-X_0}{\sigma_X}\right)^2,
\label{LMM2}
\ee
with $\sigma_X$ and $X_0$ varied, and therefore the Lagrange multiplier method 
is equivalent to the method of fictitious measurements that was discussed in the introduction to this section.

Figure \ref{pion-unc} shows some results obtained with the Lagrange multiplier method, and will 
be discussed in subsection \ref{dss}.

\section{Current global fits \label{currglobalfits}}

Recent global fits \cite{Hirai:2007cx,deFlorian:2007aj,deFlorian:2007hc,Albino:2008fy}
have made use of various experimental, computational, statistical and theoretical advances.

Data from $e^+ e^-$ reactions have provided accurate constraints on charged-sign unidentified quark FFs.
$pp$ reaction data from RHIC provides better and NLO constraints on the gluon and, 
along with HERMES data from HERA, is the only data which can constrain the charge-sign asymmetry FFs.
However, due to the high accuracy and number of the $e^+ e^-$ reaction data relative to other data,
charge-sign asymmetry FFs remain much less constrained than charged-sign unidentified FFs.
Note that an FF for a particle of a given charge-sign
$D_i^{h^+(h^-)}=D_i^{h^\pm}/2+(-)D_i^{\Delta_c h^\pm}/2$ 
will carry a large error coming from the charge-sign asymmetry FF $D_i^{\Delta_c h^\pm}$.
However, this problem can be avoided.
As we saw in subsection \ref{symmetries}, the evolution of charge-sign asymmetry FFs is independent of 
the evolution of charge-sign unidentified FFs.
Furthermore, observables in general can be combined such that they either depend 
on charge-sign asymmetry FFs or charge-sign unidentified FFs, but not both, and we saw 
this for the case of $e^+ e^-$ reactions in subsection \ref{epemtheory},
and in the case of $pp$ reactions in subsection \ref{hadhad}.
Consequently, fits of charge-sign unidentified FFs and charge-sign asymmetry FFs can be performed independently of one another.

Due to the large amount of data now available for constraining FFs, it is important
that cross sections can be calculated quickly to ensure that the minimization of $\chi^2$ occurs in a reasonable time.
Calculating $d\sigma^i$ via the inverse Mellin transform is faster than via the convolution
in eq.\ (\ref{genformofinchadprodfromffs}) since
the evolved FFs in Mellin space are obtained much more quickly than the 
evolved FFs in $z$ space which are obtained via eq.\ (\ref{explevolofD}).
Although the calculations of the $d\sigma^i$ in Mellin space can be very time consuming,
especially if no analytic results exist so that the integration in eq.\ (\ref{modmeltrans}) must be performed numerically,
they are usually fixed during the fit 
and therefore they need only be performed prior to fitting,
at points in complex Mellin space which serve as the supports for the integration 
in the inverse Mellin transform \cite{deFlorian:2007aj,Vogt:2004ns}.
The same also applies for the evolution $E$ in eq.\ (\ref{explevolofDmellin}).
Thus only the initial FFs, which vary during the fit, need to be calculated each time a cross section is calculated.
This precalculation is also suitable for adaptive integrations such as that used in the function GAUSS \cite{dgauss}
in the CERNLIB package, because, for any given number of points to be used for the integration,
each and every possible positioning of these points is predictable, the number of positionings being finite in number,
and the values of the $d\sigma^i$ (multiplied by $E$) at each point 
need to be calculated only once as and when they are needed during the fitting, and then stored.
Most modern computers can manage the large amount of computer memory required for this precalculation for adaptive integrations.
Although adaptive integration routines can take longer than integration routines which use the same points each time, 
they are more reliable.
In the case that quantities such as $\alpha_s(M_z)$ and hadron mass 
are also fitted, the $d\sigma^i$ in Mellin space will not themselves be fixed during the fit. 
Some modification to the simple precalculation to handle such scenarios will usually be possible.
Finally, we note here that it is not necessary to limit the contour of integration in Mellin space to be of finite length,
because the infinite contour can be mapped to a finite range by a change of integration variable.

Fitting the $F^i_{h_1 h_2}$, with some modification to account for logarithmic singularities, 
to polynomials whose Mellin transform can be obtained analytically 
avoids the use of large amounts of computer memory, but still allows 
cross sections to be calculated sufficiently quickly for global fits to terminate in a reasonable amount of time.
The use of Chebyshev polynomials of the first kind \cite{numrec} 
proves to be the best choice for speed and accuracy \cite{Albino:2008fy}.
After performing the analytic Mellin transform of the Chebyshev expansions, 
eq.\ (\ref{ptdistfromff}) may be evaluated in Mellin space 
(there are no delta functions in $F^i_{h_1 h_2}(x,y,s/M_f^2,a_s(M_f^2))$ because these are integrated out
in the convolution with PDFs in eq.\ (\ref{defofXSforiprodfrom2hads})).
This leads to an accuracy of a few parts per mil.

As well as these new experimental results and computational methods, the latest global fits 
of Refs.\ \cite{Hirai:2007cx,deFlorian:2007aj,deFlorian:2007hc,Albino:2008fy}, discussed in 
subsections \ref{HKNS}---{AKK08}, have made use of the 
theoretical methods discussed in section \ref{impstand} 
and the statistical methods discussed in section \ref{treaterrs}.
We compare some of the results of these fits with one another in subsection \ref{compFFsets}.

\subsection{HKNS \label{HKNS}}

\begin{figure}[h!]
\hspace{-0.5cm}
\parbox{.49\linewidth}{
\includegraphics[width=7.9cm]{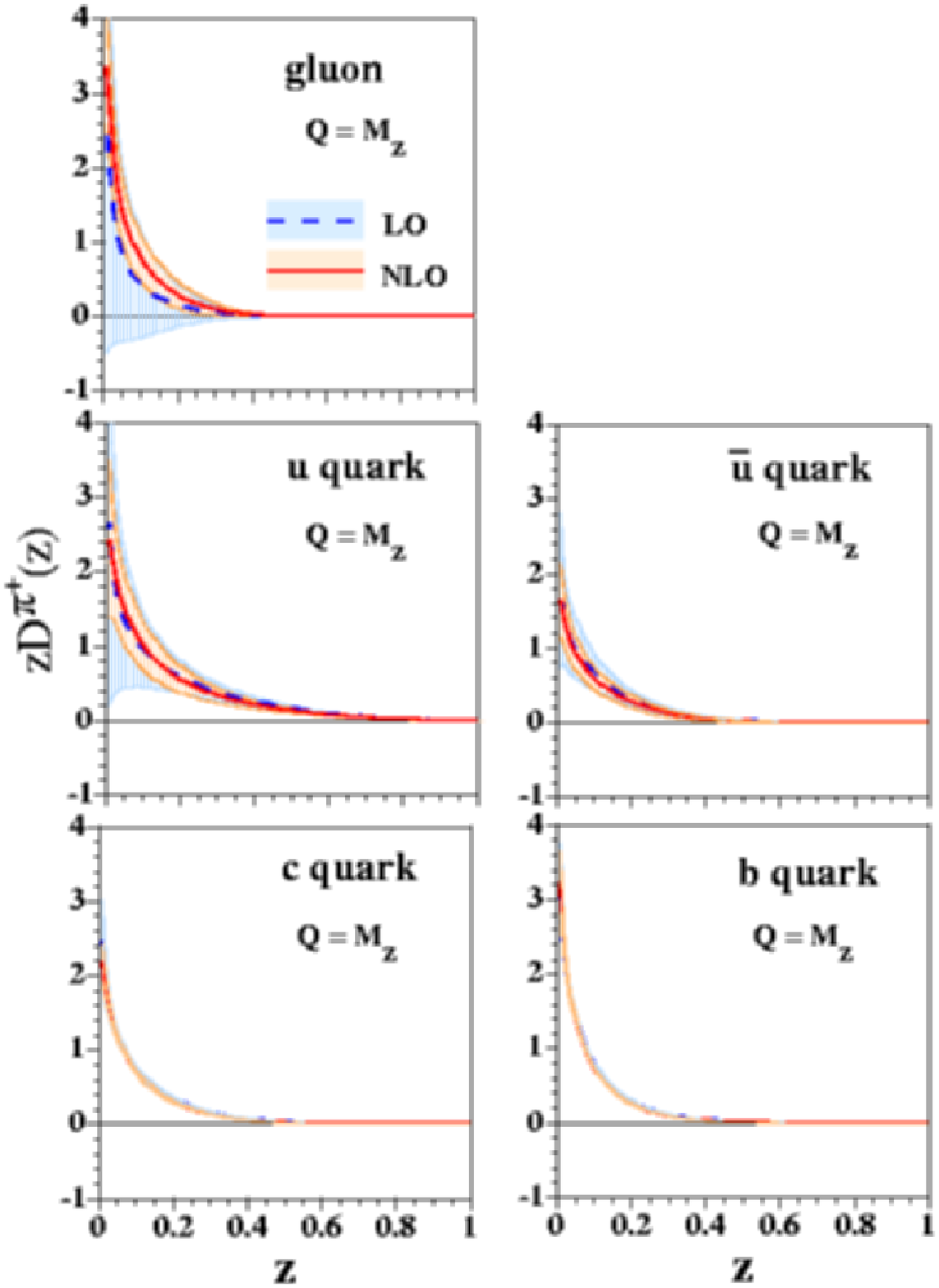}
}
\parbox{.49\linewidth}{
\includegraphics[width=7.9cm]{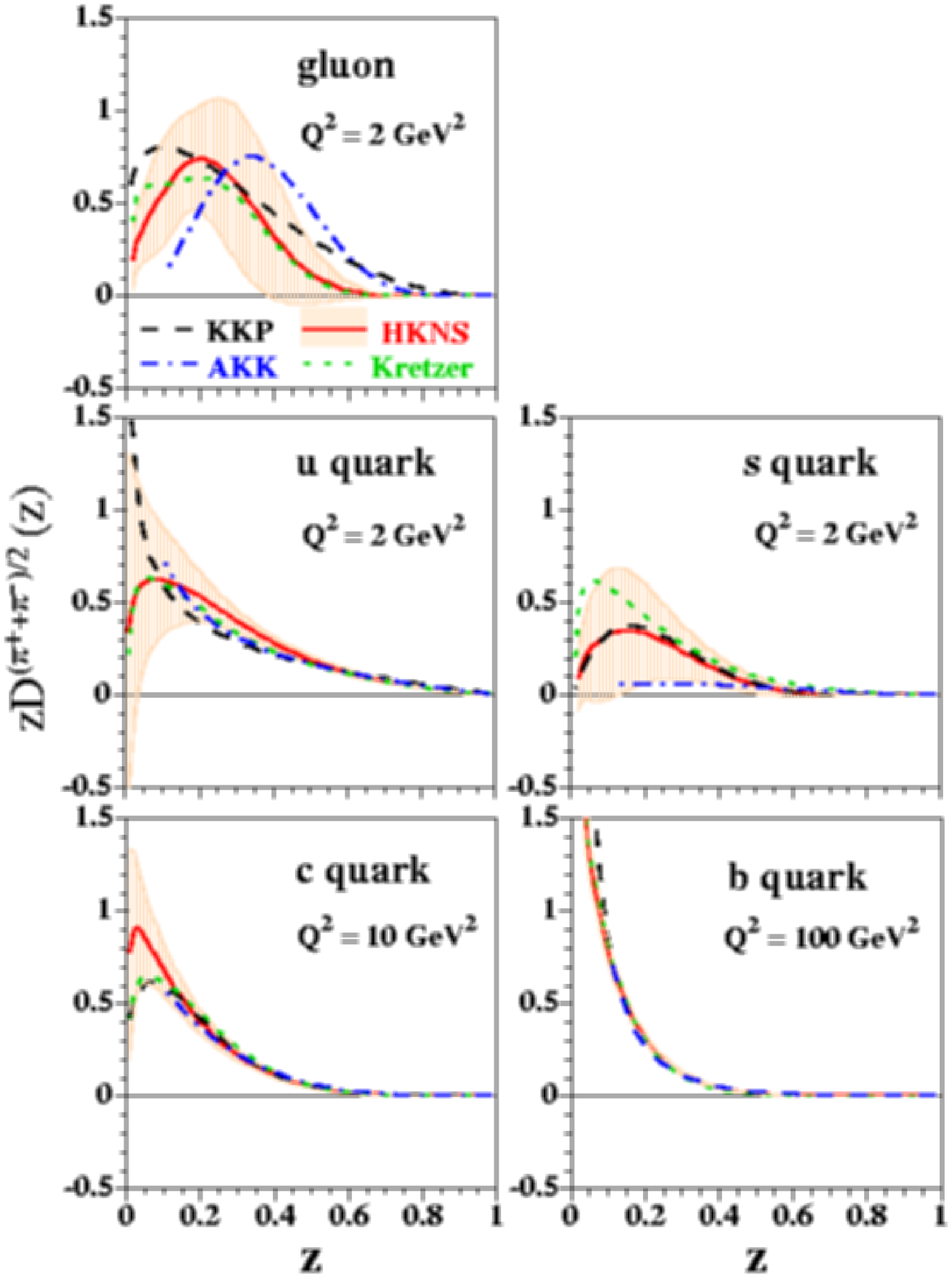}
}
\caption{Errors on the fitted HKNS FFs for \cpi\ propagated from the errors on the data. 
{\bf Left}: The results from a LO and NLO fit. {\bf Right}:
Comparison of the NLO HKNS FFs, with errors, with other FF sets. From Ref.\ \cite{Hirai:2007cx}. \label{fig_HKNS}}
\end{figure}

In Ref.\ \cite{Hirai:2007cx}, a fit of FFs for \cpi, \cka\ and \pr\ 
to all available $e^+ e^-$ reaction data, excluding particle
unidentified data because it may be contaminated with other particles beyond those just mentioned, 
and also excluding the OPAL tagging probabilities of Ref.\ \cite{Abbiendi:1999ry},
was performed and the Hessian matrix of FF parameter errors was calculated. 
We shall refer to these as the HKNS FF sets.
The results with errors for the FFs for \cpi\ are shown in Fig.\ \ref{fig_HKNS}.
The plots on the left show a reduction in experimental errors propagated to FFs on going from LO to NLO.
This is particularly significant in the case of the gluon FF, probably because at LO the gluon only contributes
through the evolution and therefore the cross section's dependence on it is less than at NLO.
Furthermore, as can be seen by comparing these plots 
with the plots on the right, FF errors are relatively higher at lower factorization scales, 
and therefore FF errors may be particularly important in $pp$ reaction data.
The plots on the right suggest that the AKK, KKP and Kretzer FF sets are generally consistent with the $e^+ e^-$ reaction data
used in the HKNS analysis.

\subsection{DSS \label{dss}}

\begin{figure}[h!]
\includegraphics[width=11cm]{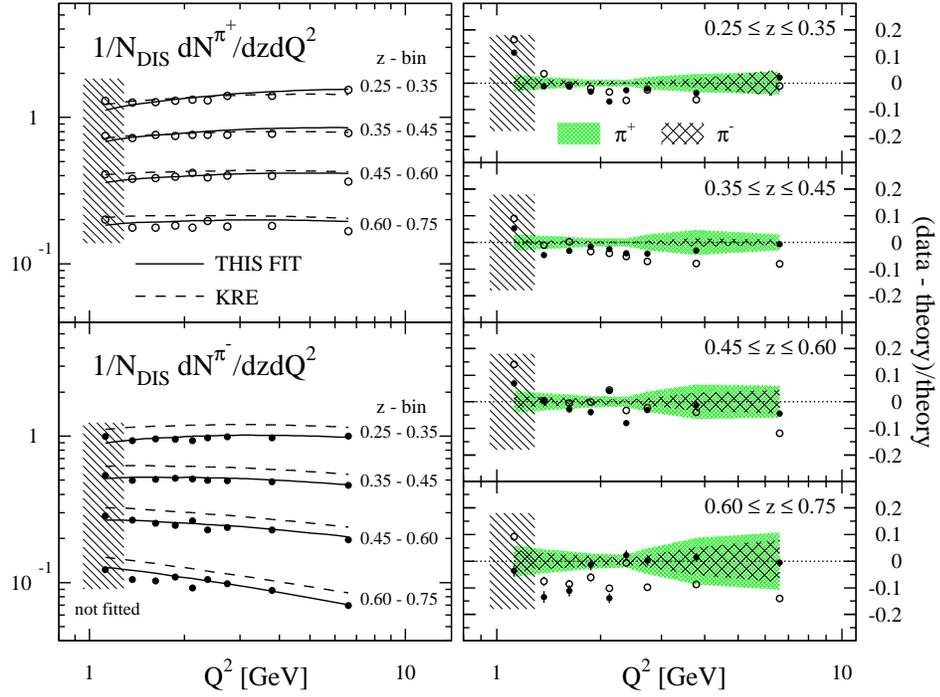}
\caption{Comparison of the calculations using the NLO DSS and Kretzer (labeled ``KRE'')
FF sets of the multiplicities for the production of positively (open circles) and negatively (full)
\cpi\ measured by HERMES. The shaded bands in the plots on the right
indicate estimates of theoretical uncertainties due to finite bin-size effects.
From Ref.\ \cite{deFlorian:2007aj}. \label{sidis-pion}}
\end{figure}

\begin{figure}[h!]
\includegraphics[width=11cm]{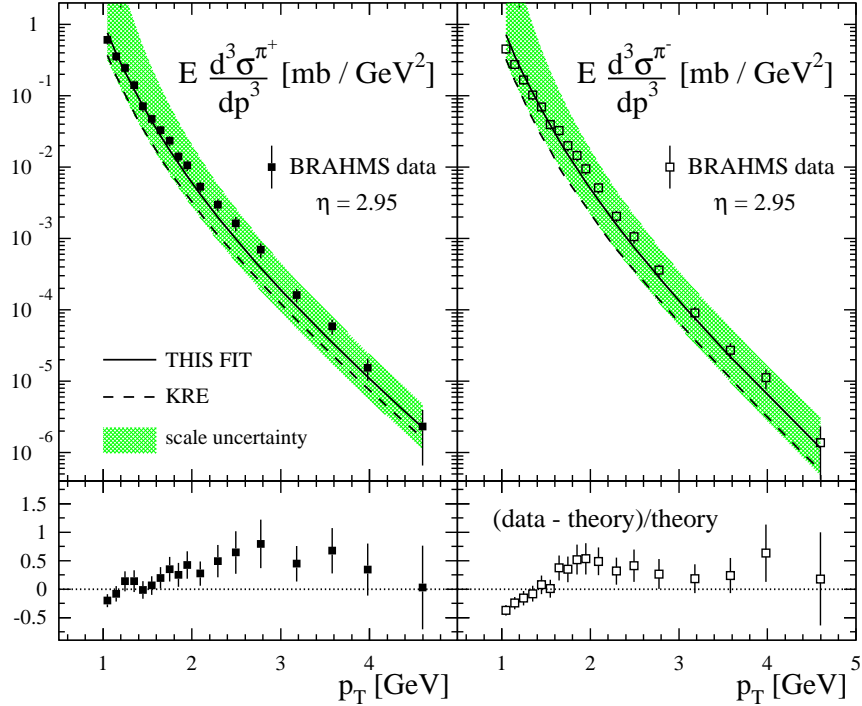}
\caption{As in Fig.\ \ref{sidis-pion}, but for the BRAHMS data. 
The shaded bands in the upper panels indicate theoretical uncertainties, and are obtained by
calculating in the range $1/2<M_f/p_T<2$, with $\mu=M_f$. From Ref.\ \cite{deFlorian:2007aj}. \label{pp-pi-brahms}}
\end{figure}

\begin{figure}[h!]
\includegraphics[width=8cm]{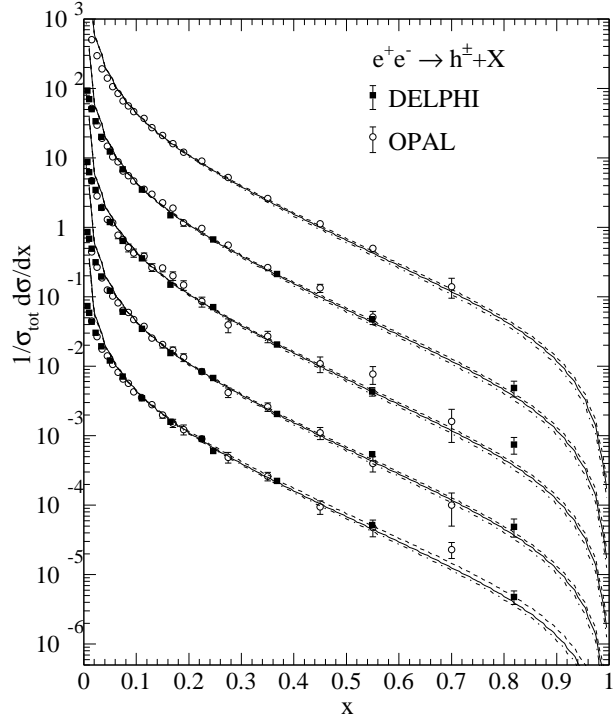}
\caption{NLO calculations using the KKP FF set 
with $M_f/\sqrt{s}=1/2$ (dashed lines), $M_f/\sqrt{s}=1$ (solid lines) 
and 2 (dot-dashed lines) of data from DELPHI \cite{Abreu:1999vs} and OPAL \cite{Alexander:1996kh,Ackerstaff:1997kk}
at energies $\sqrt{s}=133$, 161, 172, 183 and 189 GeV
(from bottom to top in this order). From Ref.\ \cite{Kniehl:2000hk}. \label{kkpdo}}
\end{figure}

\begin{figure}[h!]
\includegraphics[width=9cm]{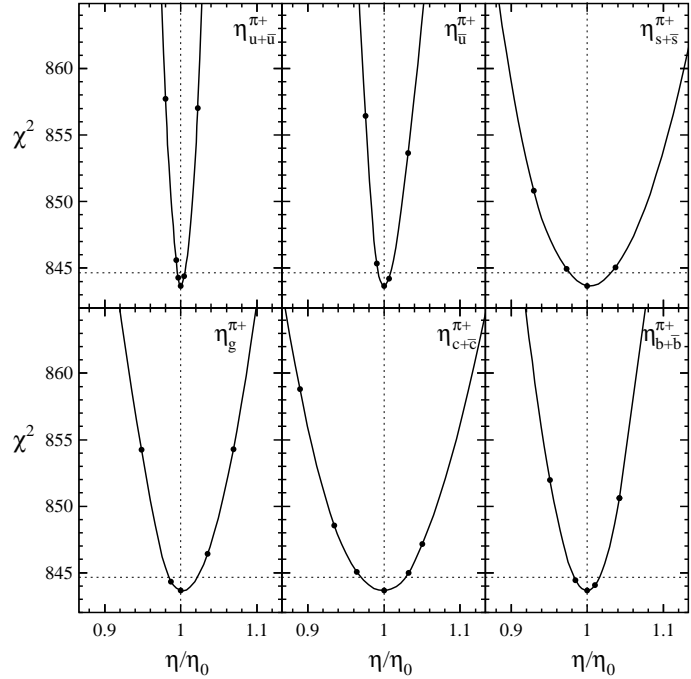}
\caption{Values of minimized $\chi^2$ subject to fixed values of $\eta_i^{\pi^+}=\int_{0.2}^1 dz D_i^{\pi^+}(z,25 {\rm GeV}^2)$,
normalized to the value of $\eta_i^{\pi^+}$, $\eta_{i0}^{\pi^+}$, obtained when $\chi^2$ is minimized without these constraints. 
From Ref.\ \cite{deFlorian:2007aj}. \label{pion-unc}}
\end{figure}

Global fits of FFs for \cpi\ and \cka\ were also performed in Ref.\ \cite{deFlorian:2007aj}, including also 
$ep$ reaction data from the HERMES collaboration \cite{Hillenbrand:2005ke} at HERA 
and $pp$ data from the BRAHMS, PHENIX and STAR collaborations at RHIC.
$pp(\overline{p})$ reaction data for \cpi\ and \cka\ 
is summarized in Tables \ref{ppPionResults} and \ref{ppKaonResults} respectively.
We shall refer to these as the DSS FF sets.
Systematic errors due to normalization uncertainties on the data were accounted for.
Although the HERMES data are measured for $Q\lesssim 2$ GeV, where low $Q$ effects may be important,
a good fit is obtained as can be seen in Fig.\ \ref{sidis-pion} for the case of \cpi.
Since the charge-sign of the particles is measured, these data also provide much needed constraints on the valence quark FFs,
or the charge-sign asymmetry FFs which are the same functions.
The difference between the calculations for the BRAHMS data 
using the DSS and Kretzer FF sets seen in Fig.\ \ref{pp-pi-brahms} indicates that $pp$ reaction data 
provide constraints not provided by $e^+ e^-$ reaction data, which were the only type of data used in the Kretzer analysis.
Note, however, that the calculation using the Kretzer FF set is within the theoretical error, which is large
compared to the theoretical error on calculations for $e^+ e^-$ reactions illustrated in Fig.\ \ref{kkpdo}.
The method of Lagrange multipliers discussed in subsection \ref{errprop} to 
determine errors on predictions was applied in this analysis using the 
FFs at $M_f=5$ GeV integrated from $z=0.2$ to 1 as example predictions
(for quark FFs, these are equal to the LO calculations of the OPAL tagging probabilities).
The result of this study in the case of $\pi^+$ is shown in Fig.\ \ref{pion-unc}.
$\eta_{u+\bar{u}}^{\pi^+}$ has the smallest error, equal to 3\%
if the tolerance parameter appearing in eq.\ (\ref{defoftolpar}) is chosen to be $T=\sqrt{15}$.
This is because all observables in the fit have a strong dependence on this FF.
The error on $\eta_{\bar{u}}^{\pi^+}$, 5\%, is not much larger. 
This quantity is expected to be well constrained by
$ep$ and $pp$ reaction data due to the large PDFs for $\bar{u}$ at low $x$ and for $u$ at high $x$.
Presumably because of the use of $pp$ reaction data, the error on $\eta_g^{\pi^+}$ relative to that on the other 
$\eta_i^{\pi^+}$ shows that the gluon FF is reasonably well constrained.
Presumably because $b$ quark tagged observables can be measured more accurately than $c$ quark tagged ones,
$\eta_{b+\bar{b}}^{\pi^+}$ has a lower error than $\eta_{c+\bar{c}}^{\pi^+}$.

A global fit of FFs for \pr\ (and, separately, for unidentified particles) was performed in Ref.\ \cite{deFlorian:2007hc}
by the same collaboration, including $pp$ data from STAR.
$pp(\overline{p})$ reaction data for \pr\ is summarized in Table \ref{ppProtonResults}.
BRAHMS data was excluded due to possible contamination of the sample by the beam due to the large rapidity.

\subsection{AKK08 \label{AKK08}}

A global fit of FFs for \cpi, \cka, \pr, \nka\ and \lam\ 
was performed in Ref.\ \cite{Albino:2008fy}, including $pp$ reaction data from RHIC and $p\overline{p}$
reaction data from the CDF collaboration at the Tevatron \cite{Acosta:2005pk}.
We shall refer to these as the AKK08 FF sets.
As illustrated in Fig.\ \ref{ppSTAR} for the case of STAR data, the description of the RHIC data was good except 
in the case of the production of \lam.
Normalization errors were accounted for as systematic errors in a covariance matrix according to the
procedure discussed in subsection \ref{syserrinchi2}.
The $\lambda_K$ at the location of the minimum value of $\chi^2$ were determined 
according to eq.\ (\ref{bestfitforlambda}), and were typically found to be in the expected range $|\lambda_K|\lesssim 2$.
As for the DSS and HKNS fits, $e^+ e^-$ reaction data for $x$ values as low as 0.05 and for $\sqrt{s}<M_Z$ was
included with the LEP data at $\sqrt{s}=M_Z$,
in contrast to the previous AKK fits \cite{Albino:2005me,Albino:2005mv} where a lower cutoff $x>0.1$ was imposed and 
only LEP data at $\sqrt{s}=M_Z$ and TPC data at $\sqrt{s}=29$ GeV was used.
Because of this additional lower $\sqrt{s}$ and smaller $x$ data, 
the effect of hadron mass discussed in subsection \ref{hadmass} was incorporated into the calculations of both 
the $pp(\overline{p})$ and the $e^+ e^-$ reaction data. 
The hadron mass appearing in the calculation of the $e^+ e^-$ reaction data was fitted, and in this way
absorbed (approximately) any other small $x$, low $\sqrt{s}$ effects such as higher twist.
The results are shown in Table \ref{hadmasses}.
\vspace{1cm}
\begin{table}[h]
\caption{Fitted particle masses used in the calculation of the hadron production from $e^+ e^-$ reactions in the AKK08 fit. 
For comparison, the true particle masses are also shown. From Ref.\ \cite{Albino:2008fy}.
\label{hadmasses}}
\begin{center}
\begin{tabular}{|c|c|c|}
\hline
Particle & Fitted mass (MeV) & True mass (MeV) \\
\hline
\input{hadmasssummary}
\end{tabular}
\end{center}
\end{table}
\vspace{1cm}
The results for \cpi, \pr\ and \lam\ suggest that hadron mass effects are the
most important small $x$, low $\sqrt{s}$ effects for the data considered.
This is also consistent with the expectation that the contributions at higher twist fall off
like $O(1/Q^2)$ \cite{Balitsky:1988fi}, in the sense that if they fell less fast, e.g.\ like $O(1/Q)$,
they would be expected to dominate over hadron mass effects, which fall like $O(1/Q^2)$.
The slight excess in the fitted masses is expected because the overall production does not
arise solely from direct partonic fragmentation but also includes contributions from decays of heavier particles.
A full error analysis is required in order to determine whether the excess seen in the fitted masses is significant.
The large undershoot in the fitted masses of \cka\ and \nka\ is very likely due to physics
effects not present in the production of \cpi, \pr\ and \lam,
the most likely effect being the complicated production mechanisms of kaons from decays of heavier hadrons.
Note that the undershoots in the fitted masses of \cka\ and \nka, $156.7$ and $154.6$ MeV respectively, are similar, 
which may be explained by similar production mechanisms, however complicated, for these two particles,
as expected from SU(2) isospin symmetry. 
As discussed in section \ref{epemXS}, this result suggests that the argument in Ref.\ \cite{Christova:2008te}
is unaffected by such production mechanisms.
Again, a full error analysis is required here.

For the reasons given at the beginning of this section,
charge-sign unidentified and charged-sign asymmetry FFs were parameterized and fitted separately.
The strongest non perturbative assumptions used in the AKK08 analysis is the parameterization
of eq.\ (\ref{standardFFparam}), taking $f_i(z)=1+c_i (1-x)^{d_i}$ for charge-sign unidentified FFs
and $f_i(z)=1$ for the less well constrained charge-sign asymmetry FFs.
No further constraints among FFs were imposed other than eq.\ (\ref{chargeconjsymm}), which is exact in QCD,
and the SU(2) isospin symmetry conditions
$D_u^{\pi^\pm/\Delta_c \pi^\pm}(z,M_f^2)=D_{\bar{d}}^{\pi^\pm/\Delta_c \pi^\pm}(z,M_f^2)$
which are also exact in QCD in the limit that the difference between $u$ and $d$ quark masses vanish.
This is in contrast to the DSS and HKNS analyses, where well justified assumptions were also imposed,
which provide additional non phenomenological constraints on FFs.
\begin{table}[h!]
\caption{The minimized $\chi^2$ values in each of the charge-sign unidentified AKK08 fits. For comparison, 
the $\chi^2$ values for the unresummed fit are shown (under ``Unres.\ fit''). From Ref.\ \cite{Albino:2008fy}.
\label{chi2summary}}
\begin{center}
\begin{tabular}{|c|c|c|}
\hline
\multirow{2}{*}{$H$} & \multicolumn{2}{c|}{\multirow{2}{*}{\vspace{0.3cm} $\chi^2$}} \\
\cline{2-3}
    & Main fit & Unres.\ fit \\
\hline \hline
\input{chi2summary}
\end{tabular}
\end{center}
\end{table}

The large $x$ resummation discussed in subsection \ref{largexres}
was incorporated into the calculations of the $e^+ e^-$ reaction data in the charge-sign unidentified fits, and makes
a significant improvement in the case of \cka, \pr\ and \lam\ as Table \ref{chi2summary} shows,
and does not worsen the fits for \cpi\ and \nka.
Further large $x$ data from e.g.\ BaBar \cite{Anulli:2004nm} would further ascertain whether 
large $x$ resummation improves the description of $e^+ e^-$ reaction data.
In this case it would be interesting to determine whether it is in fact necessary either to implement 
or to not implement large $x$ resummation.

\subsection{Comparisons of the different FF sets \label{compFFsets}}

\begin{figure}[h!]
\begin{center}
\includegraphics[width=15cm]{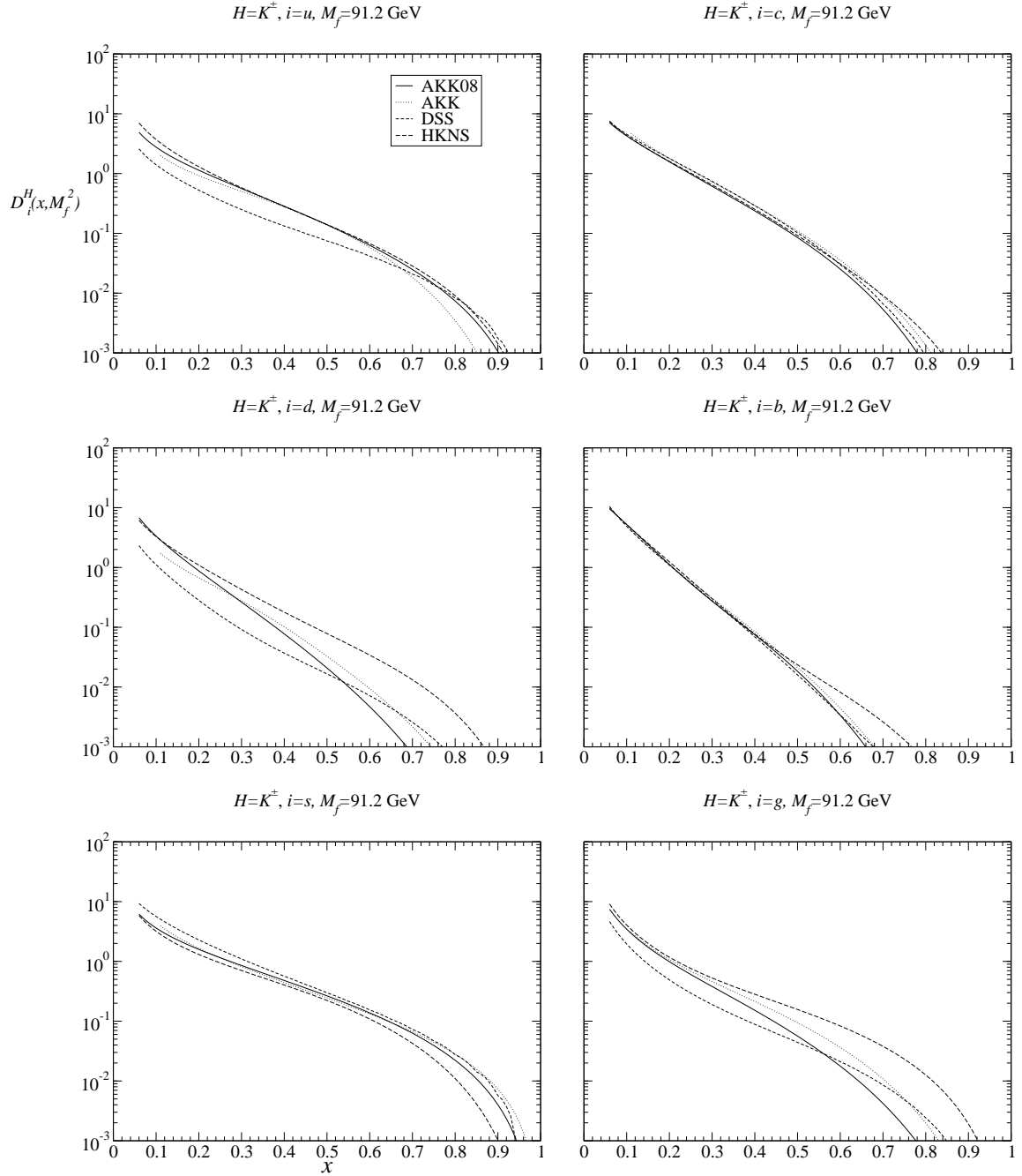}
\caption{The FFs for \cka\ at $M_f=91.2$ GeV. \label{91Kaon}}
\end{center}
\end{figure}
\begin{figure}[h!]
\begin{center}
\includegraphics[width=8.5cm]{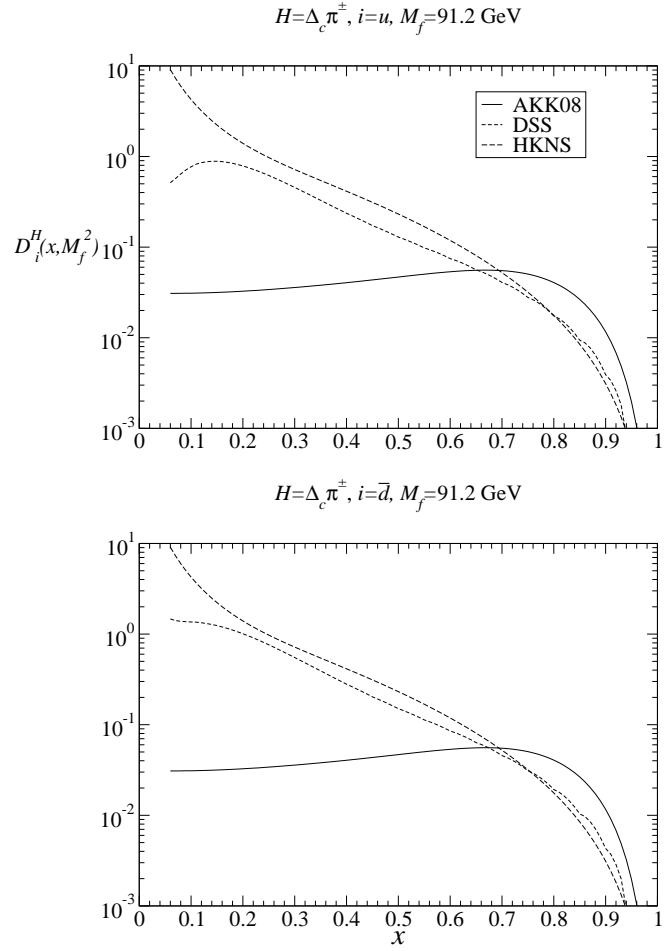}
\caption{The quark FFs for $\Delta_c \pi^\pm$ at $M_f=91.2$ GeV. \label{91PionVal}}
\end{center}
\end{figure}
\begin{figure}[h!]
\begin{center}
\includegraphics[angle=-90,width=8.7cm]{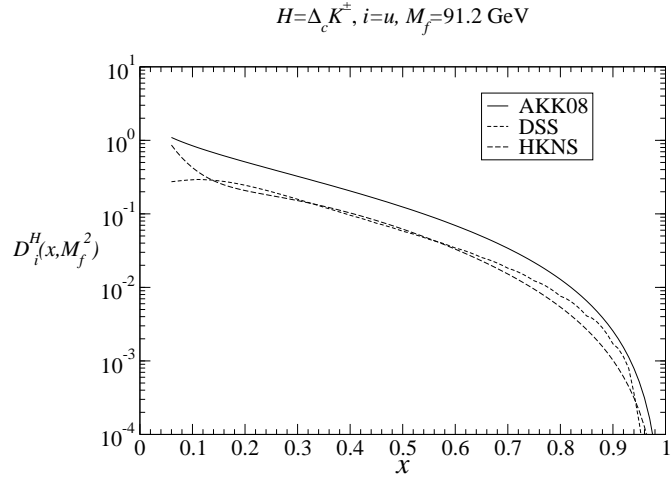}
\caption{The $u$ quark FF for $\Delta_c K^\pm$ at $M_f=91.2$ GeV. \label{up91KaonVal}}
\end{center}
\end{figure}
\begin{figure}[h!]
\begin{center}
\includegraphics[width=8.5cm]{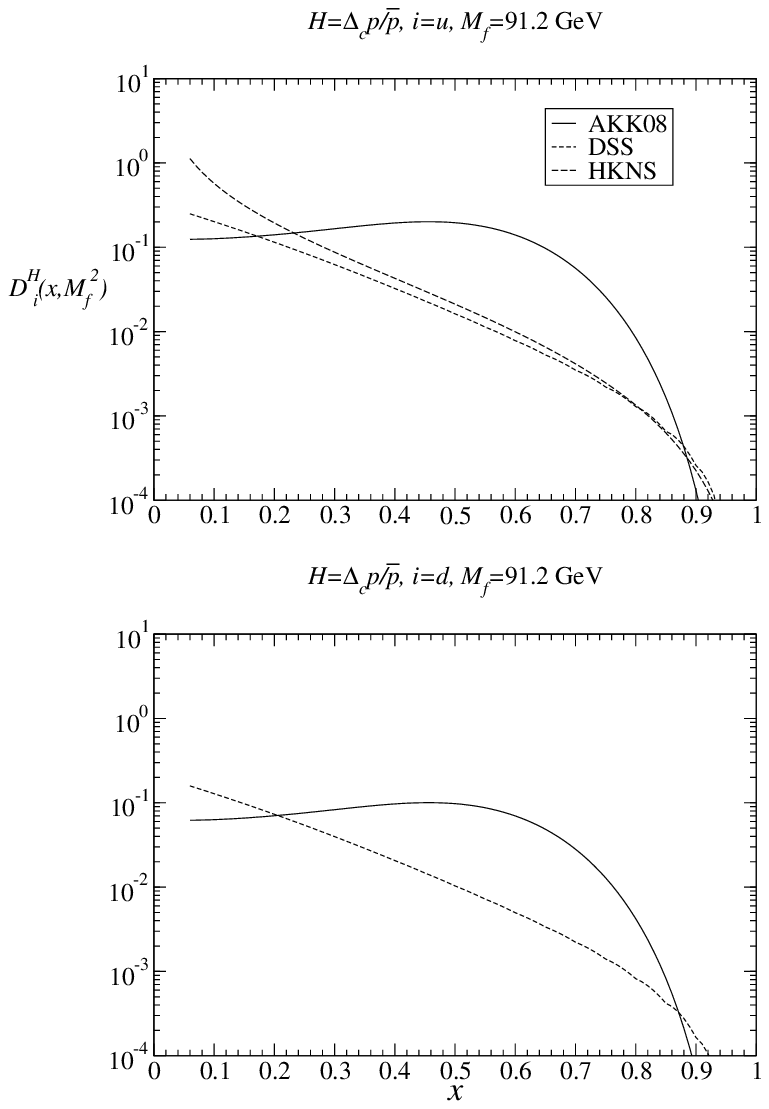}
\caption{The quark FFs for $\Delta_c p/\overline{p}$ at $M_f=91.2$ GeV. \label{91ProtonVal}}
\end{center}
\end{figure}
As an example of how the different FF sets compare, we show the FFs for \cka\ from the different sets in Fig.\ \ref{91Kaon}.
These represent an ``average'' comparison, i.e.\ better agreement is found among the different FF sets for
\cpi\ and worse for \pr\.
Similar results are obtained at intermediate and large $z$ 
for the $u$ quark (except for DSS, and for AKK at large $z$) which is favoured, and for the $c$ and $b$ quarks away from large $z$
which is well constrained by $c$ and $b$ quark tagged measurements.
For the $d$ quark, which is unfavoured, the results are all very different.
The large differences between the DSS and AKK08 gluon FFs suggests the constraints
on gluon fragmentation provided by $pp$ reaction data are not as significant as might be hoped.

In Figs.\ \ref{91PionVal}, \ref{up91KaonVal} and \ref{91ProtonVal}, we show the charge-sign asymmetry FF sets for 
\cpi, \cka\ and \pr\ respectively.
The HKNS and DSS results are generally rather similar. 
However, no phenomenological constraints on the charge-sign asymmetry FFs were applied in the HKNS analyses,
so this similarity is most likely due to the similar non perturbative assumptions used in the HKNS and DSS analyses,
indicating that most of the constraint on the charge-sign asymmetry FFs in these two analyses may come from these assumptions.
These results show that the charge-sign asymmetry FFs,
which help to extract important information on nucleon PDFs \cite{Gronau:1973gc,deFlorian:1997ie,deFlorian:2002az} 
as mentioned in the introduction, are poorly constrained,
and highlight the need for hadron species and charge-sign identified data from e.g.\ HERA and further such data from RHIC,
as well as measurements of the asymmetric cross section in $e^+ e^-$ reactions discussed in subsection \ref{epemtheory}.

\section{Improving the calculation at small $x$ \label{sglres}}

Perhaps the most serious limitation of the standard FO approach is its
inability to describe a wealth of small $x$ data from $e^+ e^-$ reactions, because 
unresummed soft gluon logarithms (SGLs) in both the evolution and the coefficient functions 
spoil the convergence of the perturbation series.
This problem even renders some well measured observables singular, such as the {\it multiplicity} of hadrons of species $h$,
being the average total number of these hadrons produced in a single event.
Using eq.\ (\ref{NhfromXS}), the multiplicity is obtained from the cross section according to
\be
\langle n^h_{q_J}(s) \rangle =\int_0^1 dx \frac{1}{\sigma(s)}\frac{d\sigma^h_{q_J}}{dx}(x,s),
\ee
This can be interpreted in a physically meaningful sense as the number of hadrons produced by fragmentation of quark $J$, 
because, at LO,
\be
\langle n^h_{q_J}(s) \rangle = \int_0^1 dx \frac{Q_{q_J}}{\sum_{I=1}^{n_f}Q_{q_I}}D_{q_J/\bar{q}_J}^h (x,s).
\label{multfromFFs}
\ee
and because to all orders it is scheme and scale independent.
The total multiplicity is of course given by $\langle n^h(s) \rangle=\sum_J \langle n^h_{q_J}(s) \rangle$.

Resummation of SGLs within this FO framework would make a major improvement to the extraction of FFs,
because it would allow small $x$ data to constrain FFs at small $z$ for the first time. 
Furthermore, the admission of these new data at small $x$ imposes further constraints on FFs at large $z$
because of the convolution over $x < z < 1$ in eq.\ (\ref{genformofinchadprodfromffs}).
As we will see, this resummation also permits a calculation of the multiplicity.

In the DGLAP evolution, the largest of these SGLs, the {\it double logarithms} (DLs), 
can be calculated to all orders and then resummed
by using the results of the {\it double logarithmic approximation} (DLA).
Since the only SGLs that appear at LO are DLs, a LO evolution of FFs that includes small $z$ is now possible.
However, at present the full DL contribution to the gluon coefficient function for $e^+ e^-$ reactions is unknown,
and is necessary for the calculation of the complete DL resummed cross section.
In a LO calculation, it may be a reasonable approximation to neglect this DL contribution because
the gluon contribution vanishes at this order.
A consistent NLO fit including small $x$ data is not possible at present
because the next largest classes of SGLs after the DLs, which in order are the
{\it single logarithms} (SLs), the sub-single logarithms (sSLs) and the sub-sub-single logarithms (ssSLs),
none of which are known in full, are all present at this order.

In addition to these shortcomings, the study of inclusive single hadron production data at small $x$
is usually studied within a different framework to the standard FO approach,
namely the framework of the {\it modified leading logarithmic approximation} (MLLA) 
\cite{Dokshitzer:1984dx,Dokshitzer:1991wu,Dokshitzer:1992iy,Khoze:1996dn},
in which both DLs and the most important contributions from the SGL-free part of evolution are accounted for,
which leads to a good description of the data over the whole $x$ range currently measured.
In addition, in practical applications the non perturbative components, the FFs, are usually fixed, up to 
an overall normalization, by the assumptions
of {\it local parton-hadron duality} (LPHD) and the {\it limiting spectrum} \cite{Azimov:1984np},
rather than being fitted to experimental data.
Perhaps the reason for this is that, until recently, 
the way in which SGLs can be incorporated into the FO evolution was not known.
A procedure for resumming SGLs in the evolution within the FO framework now exists:
As for coefficient functions, for every class of SGLs in the FO splitting functions, 
the remaining SGLs to all orders are added.
This approach reduces to the MLLA in certain approximations,
but it provides more than a mere alternative to the MLLA because it is in principle complete:
it contains the quark contribution to the evolution, 
and it can be systematically improved.
However, the simplicity of the MLLA still makes the MLLA an attractive alternative for incorporating certain types of phenomenology
such as medium effects or double hadron production at small $x$.

In this section, we discuss the present state-of-the-art in the description of 
small $x$ data from $e^+ e^-$ reactions.
We first define, in subsection \ref{smallzevol}, the SGLs appearing in FO calculations, 
and then, in subsection \ref{unifform}, derive the form taken by SGLs belonging to a specific class when they are summed
to all orders within the FO framework.
We discuss the DLA in subsection \ref{DLA}, 
which is crucial for determining the DLs to all orders in the splitting functions in the FO framework.
By direct application of the DGLAP equation,
we then generalize the DLA and the FO approximations to a single approach in subsection \ref{DLAinDGLAP}. 
This approach incorporates both approximations, and is the lowest order approximation the approach developed 
in subsection \ref{unifform}.
It can be systematically improved both in the FO part and in the SGL part.
We show in subsection \ref{MLLA} that this approach reduces to the MLLA in certain limits used in the MLLA framework.
Some discussion is then given on LPHD and the limiting spectrum which are
used to fix the remaining, non perturbative, degrees of freedom in the MLLA,
Using the approach of SGL resummation within the FO framework, 
we are then able to give a perturbative argument for the results of the LPHD
and limiting spectrum in subsection \ref{LPHDandls} by studying the moments of the evolution.

\subsection{The behaviour of the evolution at small $z$ \label{smallzevol}}

The approximation for $P$ in eq.\ (\ref{FOexpanofP}) fails at small $z$ due to the presence of terms which,
in the limit $z\rightarrow 0$, behave like $(a_s^{n}/z) \ln^{2n-m-1}z$ for $m=1,...,2n-1$. 
Such logarithms are called SGLs, and $m$ labels their class. 
The class of logarithms for which $m=1$ are the largest logarithms, known as DLs,
while the next largest, for which $m=2$, are called SLs, the next largest sSLs, the next largest ssSLs, and so on.
As $z$ decreases, these unresummed SGLs will spoil the convergence of the FO series for $P(z,a_s)$ once $\ln (1/z) \ge O(a_s^{-1/2})$.
Consequently, the evolution of the $D_i(z,M_f^2)$ will not be valid at such small values of $z$, since
the whole range $z\leq z'\leq 1$ contributes in eq.\ (\ref{DGLAP}). 
Therefore, according to eq.\ (\ref{genformofinchadprodfromffs}),
the FO approach is only a good approximation for cross sections whose $x$ value, being the minimum value of $z$,
is sufficiently large, specifically it must obey $\ln (1/x) < O(a_s^{-1/2})$.

Using the result
\be
\frac{1}{\omega^p}=\frac{1}{p!}\int_0^1 dz z^{\omega}\frac{\ln^{p-1} \frac{1}{z}}{z}
\label{meltransofsoftglogs}
\ee
for ${\rm Re} (\omega) >0$ and the integer $p\geq 1$,
we find that in Mellin space these SGLs behave like $a_s^n/\omega^{2n-m}$ where $\omega=N-1$,
i.e.\ they become singular as $\omega \rightarrow 0$.
Thus the standard FO approach in Mellin space will not be a valid approximation once $|\omega|\le O(a_s^{1/2})$.
Sometimes, those Mellin space terms for which $m=2n$, i.e.\ which behave like $a_s^n$ at small $\omega$, 
are often included in the definition of SGLs.
For example, in the MLLA, the $m=2n=2$ term in $P(z,a_s)$ at LO is referred to as an SL.
Since such terms behave like $a_s^n \delta(1-z)$ in $z$ space, i.e.\ do not affect the accuracy of the FO expansion at low $z$
and so do not need to be resummed,
we will not include such terms in our definition of SGLs, which will always be restricted to $m\leq 2n-1$.

By writing eq.\ (\ref{multfromFFs}) as
\be
\langle n^h(s) \rangle =\frac{Q_{q_J}}{\sum_{I=1}^{n_f}Q_{q_I}}D_{q_J/\bar{q}_J}^h (\omega=0,s),
\label{multfromFFsmel}
\ee
it is clear then that the multiplicity is undefined in the FO calculation because of the unresummed SGLs in the splitting functions.
(More precisely, the multiplicity at any value of $s$ cannot be determined from the multiplicity at a given value of $s$
because its $s$ dependence in the FO approach is singular if the SGLs are unresummed.)
Beyond LO, unresummed SGLs in the coefficient functions will also give singular contributions to eq.\ (\ref{multfromFFsmel}).

\subsection{A unified formalism for small and large $z$ evolution \label{unifform}}

The correct way to approximate $P(\omega,a_s)$ such that both the large and small $|\omega|$ regions 
(and hence the large and small $z$ regions) of the evolution can be described
by the DGLAP equation is to put it in the form \cite{Albino:2005gd,Albino:2005gg}
\be
P(\omega,a_s)=P^{\nSGL}(\omega,a_s) +P^{\rm SGL}(\omega,a_s),
\label{PsepinSGLsandremain}
\ee
where $P^{\nSGL}$ is equal to $P$ after all SGLs have been subtracted, 
so that it can be approximated in Mellin space for both large and small $|\omega|$ as the FO series
\be
P^{\nSGL} (\omega,a_s)=\sum_{n=1}^{\infty}a_s^n P^{\nSGL (n-1)}(\omega),
\label{expanforPFOinasomega}
\ee
truncated at some finite value for $n$,
similar to eq.\ (\ref{FOexpanofP}) (note that $P^{\nSGL (n-1)}(\omega)\propto \ln \omega$ at large $\omega$,
as discussed in subsection \ref{largexres}), and where $P^{\rm SGL}$ contains only and all the SGLs in $P$,
and which therefore can be approximated in Mellin space by the series
\be
P^{\rm SGL}(\omega,a_s)=\sum_{m=1}^{\infty}\left(\frac{a_s}{\omega}\right)^m
g_m \left(\frac{a_s}{\omega^2}\right),
\label{expanpsglinmel}
\ee
truncated at some finite value for $m$.
For a given value of $m$, $\left(\frac{a_s}{\omega}\right)^m
g_m \left(\frac{a_s}{\omega^2}\right)$ in eq.\ (\ref{expanpsglinmel}) is obtained by resummation of all class $m$ SGLs in $P$.
Equation (\ref{expanpsglinmel}) is just the general result of expanding a function of $a_s$ and $\omega$
in $a_s/\omega$ keeping $a_s/\omega^2$ fixed (although it is not completely general because the series starts at $m=1$).
Equation (\ref{expanpsglinmel}) should be a good asymptotic approximation in the region $|\omega|= O(a_s^{1/2})$,
at least for sufficiently small $a_s$.
In fact, from incomplete calculations up to the class $m=2$ \cite{Dokshitzer:1991wu}, in particular the
splitting function for the MLLA evolution of the gluon FF in eq.\ (\ref{PforDGinMLLA}) below,
$P^{\rm SGL}$ approximated as in eq.\ (\ref{expanpsglinmel}) is believed 
to be valid for the whole region $|\omega|\le O(a_s^{1/2})$ as well.
In particular, at $\omega=0$ it is expected to be a series in $\sqrt{a_s}$ with finite coefficients, beginning at $O(\sqrt{a_s})$.
Note that eq.\ (\ref{PsepinSGLsandremain}) falls to zero as $|\omega| \rightarrow \infty$,
because in this limit $P$ (and therefore $P^{\rm SGL}$) is well approximated by an expansion in $a_s$ keeping $\omega$ fixed,
wherein the SGLs vanish.
This condition is met by eq.\ (\ref{PforDGinMLLA}).

From eq.\ (\ref{meltransofsoftglogs}), the inverse Mellin transform of eq.\ (\ref{expanpsglinmel}) gives the $z$ space result
\be
P^{\rm SGL}(z,a_s)=\frac{\ln^{-1} z}{z}\sum_{m=1}^{\infty}\left(a_s \ln z\right)^m 
f_m \left( a_s \ln^2 z\right),
\label{PSGLinxspaceresummed}
\ee
which can also be obtained by summing the SGLs in $z$ space for each $m$. 
Equation (\ref{PSGLinxspaceresummed}) shows that the approximation of eq.\ (\ref{expanpsglinmel})
is valid in the evolution of $D(z,Q^2)$ in the whole region $\ln (1/z)\le O(a_s^{-1/2})$,
which includes the small, but not arbitrarily small, $z$ region $\ln (1/z)= O(a_s^{-1/2})$.
Note that the usual condition $a_s \ll 1$ must still hold in order that the approximation in eq.\ (\ref{expanforPFOinasomega})
holds everywhere in this range of $z$.
If $P^{\rm SGL}(\omega,a_s)$ approximated as in eq.\ (\ref{expanpsglinmel})
really is finite at $\omega=0$, then $P^{\rm SGL}(z,a_s)$ in eq.\ (\ref{PSGLinxspaceresummed})
must obey $z P^{\rm SGL}(z,a_s) \rightarrow 0$ as $z\rightarrow 0$.

Using the DLA, it is in fact possible to calculate the $m=1$ term in eq.\ (\ref{expanpsglinmel}) or equivalently
in eq.\ (\ref{PSGLinxspaceresummed}), i.e.\ the complete DL contribution.
We show how to do this in subsection \ref{DLA}.

\subsection{The double logarithmic approximation\label{DLA}}

The DLA \cite{Bassetto:1982ma} states that, if the evolution is written in the form
\be
\frac{d}{d \ln M_f^2}D(z,M_f^2)
=\int_z^1 \frac{dz'}{z'} \frac{2C_A}{z'}A  z^{\prime 2\frac{d}{d\ln M_f^2}}
\left[a_s(M_f^2) D\left(\frac{z}{z'},M_f^2\right)\right]
+\int_z^1 \frac{dz'}{z'}\overline{P}(z',a_s(M_f^2)) D\left(\frac{z}{z'},M_f^2\right)
\label{DGLAPandDLA1}
\ee
where $(2C_A/z)A=P^{{\rm DL}(0)}(z)$ is the LO term in the expansion in $a_s$ of the full DL contribution to $P$, $P^{\rm DL}$, 
given by the $m=1$ term in eq.\ (\ref{PSGLinxspaceresummed}), then $\overline{P}(z,a_s)$ is free of DLs.
$A$ will be given explicitly in the basis of singlet, non singlet and valence quark FFs just now,
when we will see that $A$ is a projection operator,
its normalization being chosen such that it obeys $A^2=A$, 
and this property will make resummation calculations to appear later in this section easier.
Equation (\ref{DGLAPandDLA1}) can be obtained from the generating functional technique \cite{Dokshitzer:1991wu} in the context 
of angular ordering \cite{Fadin:1983aw} of successive emissions of gluons in the process $e^+ e^- \rightarrow q+\bar{q}+Ng$,
where $Ng$ refers to $N$ gluons in the final state.
The quantity $\overline{P}$ is completely constrained by the DGLAP equation.
It can be obtained in terms of $P$ order by order in $a_s$,
by expanding the operator in eq.\ (\ref{DGLAPandDLA1}) in the form
\be
z^{\prime 2\frac{d}{d\ln M_f^2}}=\exp\left[2\ln z' \frac{d}{d\ln M_f^2}\right]
=\sum_{n=0}^{\infty}\frac{(2\ln z')^n}{n!}\left(\frac{d}{d\ln M_f^2}\right)^n
\label{expanofop}
\ee
and then repeatedly applying the evolution equations, Eqs.\ (\ref{DGLAP}) and (\ref{betafuncdef}),
to the $\left(\frac{d}{d\ln M_f^2}\right)^n \left[a_s(M_f^2) D\left(\frac{x}{y},M_f^2\right)\right]$ 
operations in eq.\ (\ref{DGLAPandDLA1}). For example, to $O(a_s^2)$ (NLO) one finds that
\be
\overline{P}(z,a_s)=P(z,a_s)-2C_A A \Bigg[\frac{a_s}{z} +2\beta(a_s)\frac{\ln z}{z}
+\int_z^1 \frac{dz'}{z'}\frac{2a_s \ln z'}{z'}P\left(\frac{z}{z'},a_s\right)\Bigg]+O(a_s^3).
\label{constraintonPbartoNLO}
\ee
In the square brackets on the RHS of eq.\ (\ref{constraintonPbartoNLO}), only the first term
contributes to the $O(a_s)$ (LO) part of $\overline{P}$, while the second and 
third term contribute to the $O(a_s^2)$ part. To this accuracy,
the third term is calculated with
$P(z,a_s)=a_s P^{(0)}(z)$ (see appendix \ref{LOsplitfunc} for the explicit functions).
From eq.\ (\ref{constraintonPbartoNLO}), we observe that $\overline{P}$ to NLO
is free of DLs, by taking the DLs in $P(z,a_s)$ for the
first term on the RHS of eq.\ (\ref{constraintonPbartoNLO}) to NLO,
\be
P^{\rm DL}(z,a_s)=2C_A\frac{A}{z}a_s- 4C_A^2 \frac{A\ln^2 z}{z}a_s^2+O(a_s^3).
\label{PDLtoNLOinx}
\ee

In eq.\ (\ref{DGLAPandDLA1}), $A$ is zero whenever $D$ is a non singlet FF or composed of valence quark FF.
In this case, eq.\ (\ref{DGLAP}) implies $\overline{P}=P$, i.e.\ the splitting functions for such FFs are free of DLs.
For $D=(D_{\Sigma},D_g)$,
\begin{eqnarray}
A=\left( \begin{array}{cc}
0 & \frac{2 C_F}{C_A} \\
0 & 1
\end{array} \right).
\end{eqnarray}
Note in general that $A^2=A$, as mentioned earlier.

For consistency, 
resummation of the DLs in the hard partonic cross sections (i.e.\ the coefficient functions) of the process under consideration
is also necessary.
In principle, this has been done for $e^+ e^-$ reactions in Ref.\ \cite{Mueller:1982cq}, the result being
that $C_{\rm NS}$ and $C_{\rm PS}$ are free of DLs while
\be
C_g(\omega,a_s)=\frac{2C_F}{C_A}\left[\frac{1}{2}\frac{\omega-\sqrt{\omega^2+16C_A a_s}}{\sqrt{\omega^2+16C_A a_s}}\right].
\ee
The expansion up to NNLO of this result,
\be
C_g(\omega,a_s)=\frac{2C_F}{C_A}\left[-4C_A \frac{a_s}{\omega^2}-48 C_A^2 \left(\frac{a_s}{\omega^2}\right)\right],
\ee
may be compared with the small $\omega$ limit of the result in Ref.\ (\cite{Blumlein:2006rr}),
\be
C_g(\omega,a_s)=\frac{2C_F}{C_A}\left[-4C_A \frac{a_s}{\omega^2}+40 C_A^2 \left(\frac{a_s}{\omega^2}\right)\right].
\ee
There is clearly agreement at NLO, but not at NNLO.
The most likely reason is the difference in the choice of scheme, or rather, since the 
DLs should be scheme independent, an $\omega$ dependent difference in the choice of normalization of $C_g$
between the two approaches.
Further studies of the factorization procedure used in Ref.\ \cite{Mueller:1982cq} are needed here.
In any case, the coefficient functions for $e^+ e^-$ reactions at LO are free of DLs and so
it may be a reasonable approximation to neglect them altogether. 
This is done in the MLLA, to be discussed in subsection \ref{MLLA}.

The DLA implies a relation among quark and gluon FFs which simplifies calculations of cross sections.
According to eq.\ (\ref{DGLAPandDLA1}) with $\overline{P}$ neglected, which is a reasonable approximation at small $z$
since it is free of DLs,
the ratio of $d D_{\Sigma}/d\ln M_f^2$ to $d D_g/d\ln M_f^2$ is $2C_F/C_A$,
while the operation of $d/d\ln M_f^2$ on the non singlet and valence quark FFs gives zero in this approximation.
Integrating these results over $\ln M_f^2$ and neglecting the constants of integration relative to the FFs gives
\be
D_{q_J}=D_{\overline{q}_{J'}} =\frac{C_F}{C_A}D_g,
\label{DLArelforDquarkandDg}
\ee
reducing the number of FFs required for the cross section to just one, $D_g$.
Although we will not use eq.\ (\ref{DLArelforDquarkandDg}) in this subsection,
we will consider an application of it in subsection \ref{DLAinDGLAP}.

Equation (\ref{DGLAPandDLA1}) is very difficult to solve numerically. 
An alternative approach \cite{Albino:2005gd,Albino:2005gg} which is appropriate for numerical work
is to use eq.\ (\ref{DGLAPandDLA1}) to obtain $P^{\rm DL}$
and then to solve eq.\ (\ref{DGLAP}) with $P$ given by eq.\ (\ref{PsepinSGLsandremain}).
$P^{\rm DL}$ is obtained by substituting eq.\ (\ref{DGLAPmellin}) into the Mellin space transform of
eq.\ (\ref{DGLAPandDLA1}):
\be
\left(\omega+2\frac{d}{d \ln M_f^2} \right) \frac{d}{d \ln M_f^2}D
=2C_A a_s A D
+\left(\omega+2\frac{d}{d \ln M_f^2}\right)\overline{P} D,
\label{DRAP}
\ee
where for brevity we have written $D=D(\omega,M_f^2)$, $a_s=a_s(M_f^2)$ and $\overline{P}=\overline{P}(\omega,a_s)$.
The result is
\be
\left(\omega+2\frac{d}{d\ln M_f^2}\right)\left(P-\overline{P}\right)+2\left(P-\overline{P}\right)P-2C_A a_s A=0,
\label{eqfordelP}
\ee
where again for brevity we have written $P=P(\omega,a_s)$.
As discussed at the end of subsection \ref{anmelDGLAP}, eq.\ (\ref{DGLAPmellin}) with $P$ unspecified is completely general,
so that at this point eq.\ (\ref{eqfordelP}) is simply an alternative way of writing eq.\ (\ref{DRAP}).
In particular, $P(\omega,a_s)$ in eq.\ (\ref{eqfordelP}) must not be seen as an expansion in $a_s$.
To obtain $P^{\rm DL}$, we split $P$ into its DL and non DL ($P^{\nDL}$) parts,
\be
P=P^{\nDL}+P^{\rm DL},
\label{splitofPintotildePPdl}
\ee
then expand eq.\ (\ref{eqfordelP}) as a series in $a_s/\omega$ keeping $a_s/\omega^2$ fixed (as in eq.\ (\ref{expanpsglinmel}),
noting that $P^{\nDL}$ and $\overline{P}$ are of $O((a_s/\omega)^2)$, 
that $\omega=(a_s/\omega)(a_s/\omega^2)^{-1}=O(a_s/\omega)$ and that 
$a_s=(a_s/\omega)^2(a_s/\omega^2)^{-1}=O((a_s/\omega)^2)$)
and finally extract the first, $O((a_s/\omega)^2)$, term to find that the constraint on $P^{\rm DL}$ is exactly
\be
2(P^{\rm DL})^2+\omega P^{\rm DL}-2C_A a_s A=0.
\label{DLAeqsimplest}
\ee
Equation (\ref{DLAeqsimplest}) gives two solutions for each component of $P$. Since
$P$ is never larger than a 2$\times$2 matrix in the basis consisting of
singlet, gluon, non-singlet and valence quark FFs, there are four solutions.
The only solution which can be expanded in $a_s$ and which reproduces the DLs in $P$ at LO and NLO, given by
\begin{eqnarray}
P^{\rm DL}(\omega,a_s)=
\left( \begin{array}{cc}
0 & a_s \frac{4 C_F}{\omega}-a_s^2 \frac{16 C_F C_A}{\omega^3} \\
0 & a_s \frac{2 C_A}{\omega}-a_s^2 \frac{8 C_A^2}{\omega^3} 
\end{array} \right)+O(a_s^3),
\label{NLODLinmelspace}
\end{eqnarray}
is
\be
P^{\rm DL}(\omega,a_s)=\frac{A}{4}\left(-\omega+\sqrt{\omega^2+16C_A a_s}\right).
\label{DLresummedinP}
\ee
Equation (\ref{DLresummedinP}) agrees with the results of Ref.\ \cite{Mueller:1982cq},
which are derived using the conventional renormalization group approach, and with the
results from the generating functional technique of Ref.\ \cite{Dokshitzer:1991wu}.

\subsection{Incorporating the DLA into the DGLAP equation \label{DLAinDGLAP}}

\begin{figure}[h!]
\includegraphics[width=8.5cm]{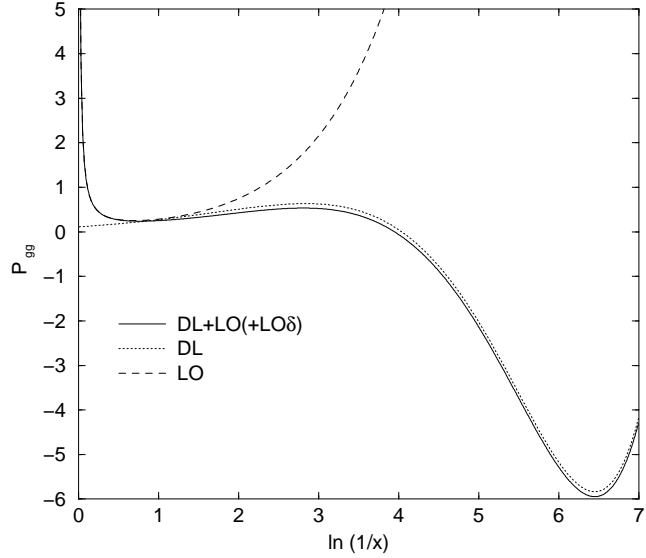}
\caption{\label{plot1} (i) $P_{gg}(z,a_s)$ calculated as in eq.\ (\ref{firstapproxforPallz})
(labeled ``DL+LO(+LO$\delta$)''),
(ii) $P_{gg}(z,a_s)$ calculated to $O(a_s)$ in the FO approach (labeled ``LO''), and 
(iii) $P_{gg}^{\rm DL}(z,a_s)$ (labeled ``DL''). $a_s=0.118/(2\pi)$ and ``$x$'' is $z$. From Ref.\ \cite{Albino:2005gd}.}
\end{figure}

To a first approximation, the evolution valid for all $\ln (1/z) \le O(a_s^{-1/2})$ takes the form
of eq.\ (\ref{DGLAP}) with
\be
P(z,a_s)=a_s P^{\nSGL (0)}(z)+P^{\rm DL}(z,a_s).
\label{firstapproxforPallz}
\ee
Here, $P^{\rm DL}$ in $z$ space (the $m=1$ term in eq.\ (\ref{PSGLinxspaceresummed})) is given by
\be
P^{\rm DL}(z,a_s)=\frac{A\sqrt{C_A a_s}}{z\ln \frac{1}{z}}
J_1\left(4\sqrt{C_A a_s}\ln \frac{1}{z}\right),
\label{allDLinzindelPclosed}
\ee
where $J_1(y)$ is the Bessel function of the first kind, which can be calculated from the result
\be
J_1(y)=\frac{1}{\pi}\int_0^{\pi}d\theta \cos (y\sin \theta -\theta).
\ee
In Fig.\ \ref{plot1}, we see that $P_{gg}(z,a_s)$ calculated as in eq.\ (\ref{firstapproxforPallz}) 
interpolates well between its $O(a_s)$ approximation
in the FO approach, $P^{(0)}_{gg}$, at large $z$ and 
$P_{gg}^{\rm DL}(z,a_s)$ at small $z$ (the small difference here
comes from $P^{\nSGL (0)}(z)$ at small $z$). DL resummation clearly makes a 
large difference to $P$ at small $z$.

Higher order calculations in the FO approach should extend the validity
of the FO calculation of the evolution to smaller $z$ values, but not by as much as the 
result in which all SGLs up to a given class are included because,
for sufficiently small $z$, the order required for the evolution in the FO approach to be accurate
will become higher than the order available.
We can use eq.\ (\ref{allDLinzindelPclosed}) to illustrate this point.
Using the expansion
\be
J_1(y)=\frac{y}{2}\sum_{r=0}^{\infty}\frac{\left(\frac{-y^2}{4}\right)^r}
{r! (r+1)!},
\ee
the series for $P^{\rm DL}(z,a_s)$ in $a_s$ reads
\be
P^{\rm DL}(z,a_s)=\frac{2C_A a_s A}{z}\sum_{r=0}^{\infty}\frac{(-1)^r}{ r! (r+1)!} (4 C_A a_s \ln^2 z)^r.
\label{allDLinzindelP}
\ee
The series in eq.\ (\ref{allDLinzindelP}) may also be obtained by
expanding eq.\ (\ref{DLresummedinP}) to infinite order in $a_s$.
For $z >0$, eq.\ (\ref{allDLinzindelP}) converges rapidly at sufficiently large $r$,
but the value of $r$ at which this convergence sets in increases with decreasing $z$.
For example, for $z=0.01$, the accuracy of the series reaches the level of a few percent only at $r= 7$.
In any case, higher order calculations in the FO approach should at least extend the validity
of the FO calculation of the evolution to lower $z$ values.

By using eq.\ (\ref{DLresummedinP}) for $P$ in the DGLAP equation, eq.\ (\ref{DGLAPmellin}), 
the dependence of the multiplicity in eq.\ (\ref{multfromFFsmel}) on $s$ is found to be 
\be
\langle n^h(s) \rangle\propto \exp\left[\frac{2\sqrt{C_A}}{\beta_0}\frac{1}{\sqrt{a_s(s)}}\right],
\ee
i.e.\ the calculation of the $s$ dependence of the multiplicity is finite once the DLs have been resummed.
Of course, the DL contribution from the gluon coefficient function has been neglected, but this is probably
negligible relative to the LO quark coefficient function.

We now discuss how eq.\ (\ref{firstapproxforPallz}) was used in the phenomenological studies in Ref.\ \cite{Albino:2005gd}. 
From eqs.\ (\ref{XSfromFFs}) (with tagged quarks $q_J$ summed over) and (\ref{coeffNLO}), the
LO cross section for $e^+ e^-$ reactions reads
\be
\frac{1}{\sigma(s)}\frac{d\sigma^h}{dx}(x,s)=\frac{1}{n_f \langle Q(s) \rangle }
\sum_{J=1}^{n_f} Q_{q_J}(s) D_{q_J/\bar{q}_J}^h(x,M_f^2),
\label{approxxs}
\ee
where $\langle Q(s) \rangle=\frac{1}{n_f} \sum_{J=1}^{n_f} Q_{q_J}(s)$ is the average electroweak coupling of quarks.
Using eq.\ (\ref{approxxs}), fits to $e^+ e^-$ reaction data have been performed in Ref.\ \cite{Albino:2005gd}.
The initial FFs are parameterized as
\be
D_i^h(z,M_0^2)=N\exp[-c\ln^2 z]z^{\alpha} (1-z)^{\beta},
\label{ParamfortestDLA}
\ee
because at large $z$ its behaviour is that in eq.\ (\ref{standardFFparam}) (with $f_i=1$),
which has so far sufficed for global fits to large $x$ data,
while at small $z$ eq.\ (\ref{ParamfortestDLA}) reduces, for $c>0$, to a Gaussian in $\ln 1/z$,
\be
D_i^h(z,M_0^2)\approx N\exp\left[-c\ln^2\frac{1}{z}-\alpha \ln \frac{1}{z}\right],
\label{GaussianformforFFs}
\ee
which is the empirical behaviour of the cross section at small $x=z$.
As will be shown at the end of subsection \ref{LPHDandls},
eq.\ (\ref{GaussianformforFFs}) is also the large $M_0$ behaviour of the FFs predicted by the DLA / MLLA,
with $\alpha <0$ to ensure the center in $\ln 1/x$ is at $-\alpha/(2c)>0$.
The value of $\alpha$ and of $c$ is chosen to be independent of the FF, as dictated by the DLA result 
in eq.\ (\ref{DLArelforDquarkandDg}).

\begin{figure}[h!]
\parbox{.49\linewidth}{
\begin{center}
\includegraphics[width=8.5cm]{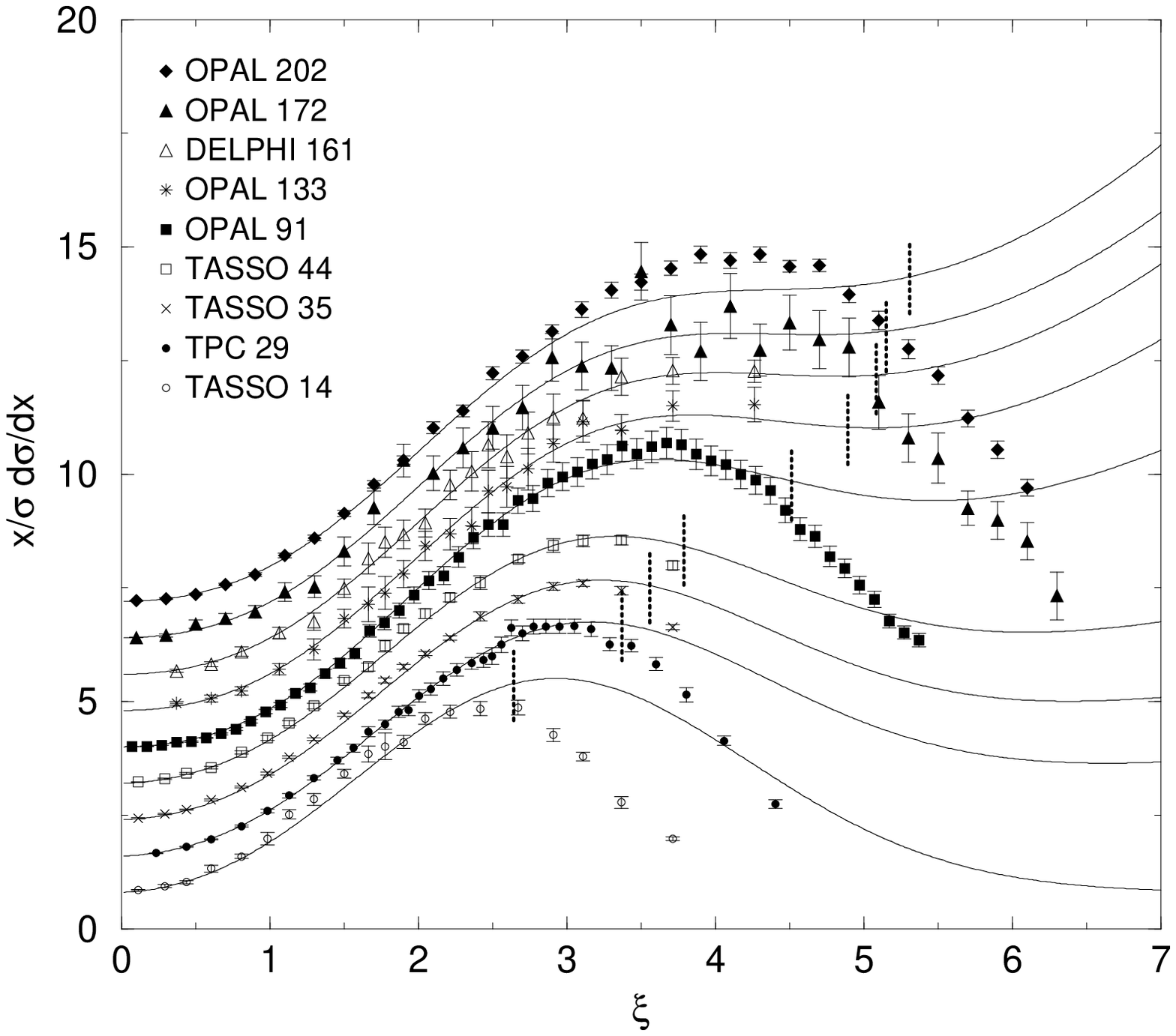}
\end{center}
}
\parbox{.49\linewidth}{
\begin{center}
\includegraphics[width=8.5cm]{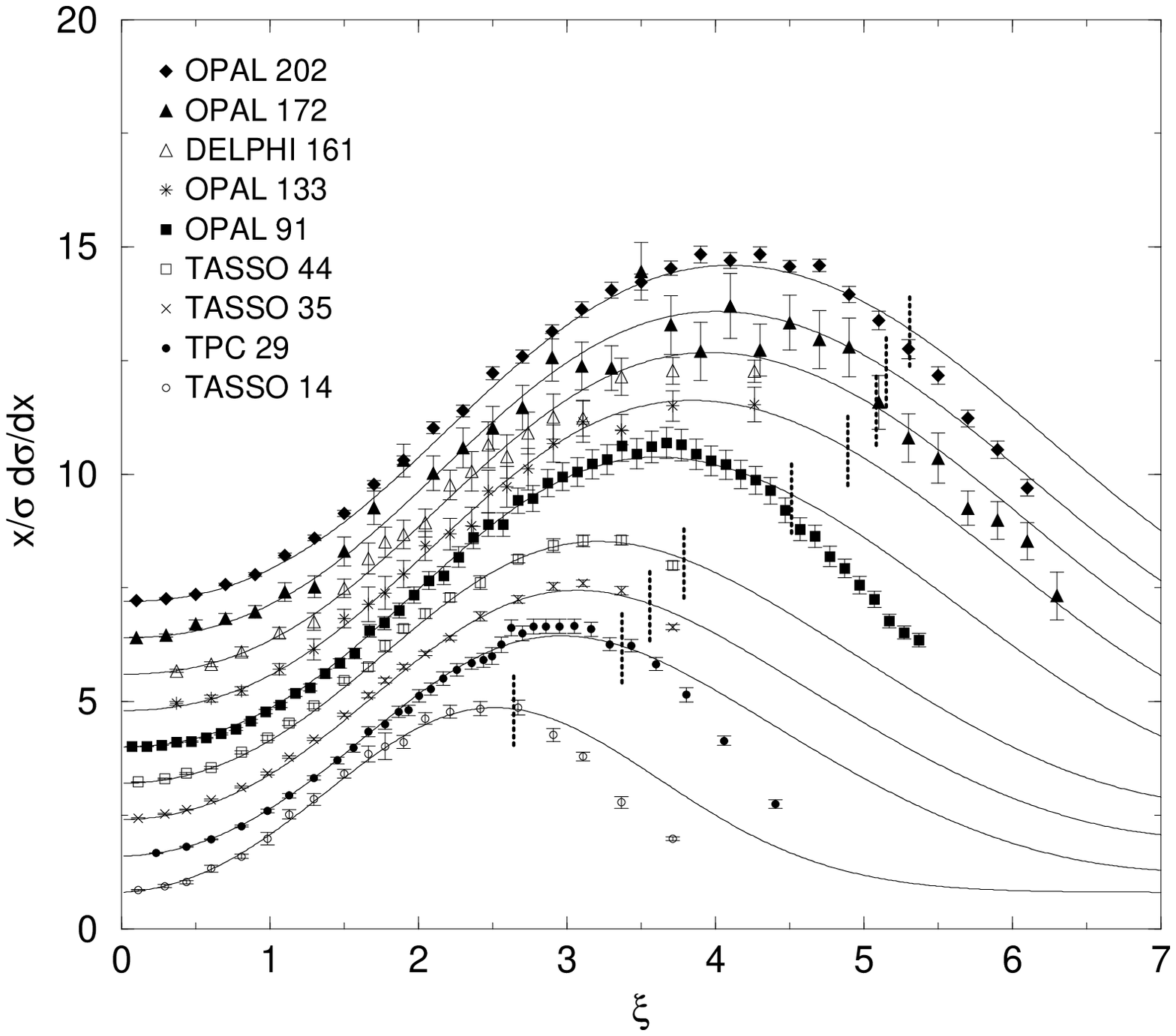}
\end{center}}
\caption{\label{figffd} Fit to data using DGLAP evolution in the FO approach to LO (left) 
and using DGLAP evolution in the FO approach to LO but with DLs resummed (right). 
The variable $\xi=\ln 1/x$.
Only some of the data sets used for the fit are shown,
together with their theoretical predictions from the results of the fit. Data to the right
of the horizontal dotted lines have not been used in the fit. Each curve is shifted up by 0.8 for clarity.}
\end{figure}

Data for unidentified hadron production at $e^+ e^-$ reactions over a large range of values of $\sqrt{s}$ was used,
being composed of the sets from TASSO at $\sqrt{s}=14$, 35, 44 GeV \cite{Braunschweig:1990yd}
and 22 GeV \cite{Althoff:1983ew}, MARK II \cite{Petersen:1987bq} and
TPC \cite{Aihara:1988su} at 29 GeV, TOPAZ at 58 GeV \cite{Itoh:1994kb}, 
ALEPH \cite{Barate:1996fi}, DELPHI \cite{Abreu:1996na}, L3 \cite{Adeva:1991it},
OPAL \cite{Akrawy:1990ha} and SLC 
\cite{Abrams:1989rz} at 91 GeV, ALEPH \cite{Buskulic:1996tt} and OPAL \cite{Alexander:1996kh} at 133 GeV,
DELPHI at 161 GeV \cite{Ackerstaff:1997kk} and
OPAL at 172, 183, 189 GeV
\cite{Abbiendi:1999sx} and 202 GeV \cite{Abbiendi:2002mj}.
These data also span a wide range in $x$, from large to small values,
as is usually the case with inclusive single hadron production measurements.
The level of accuracy of the calculations in Ref.\ \cite{Albino:2005gd} was such that
the possible contamination of these hadron unidentified data as discussed in section
\ref{epemXS} was not considered a problem.
However, such data are probably not suited to a NLO fit.
The {\it unresummed} fit, i.e.\ using $P=a_s P^{(0)}$ in eq.\ (\ref{DGLAP})
fails at the small $x$ values around and to the right of the maximum, as can be seen in Fig.\ \ref{figffd} (left).
On the other hand, the {\it resummed} fit, i.e.\ using eq.\ (\ref{firstapproxforPallz}) for the splitting
functions appearing in the evolution, of Fig.\ \ref{figffd} (right) succeeds much better at these small $x$ values,
while the large $x$ description remains intact.
The choice $M_f=\sqrt{s}$ was made, as well as the large value $M_0=14$ GeV to ensure that
the form of eq.\ (\ref{GaussianformforFFs}) predicted by the DLA at large $M_0$ was valid.
The fitted value of $\Lambda_{\rm QCD}$ in the unresummed fit
was 388 MeV, which is fairly reasonable, but the resummed fit
gave $\Lambda_{\rm QCD}=801$ MeV, which is somewhat larger than expected.
This may not be serious since $\Lambda_{\rm QCD}$ carries a multiplicative error of $O(1)$:
Multiplying $\Lambda_{\rm QCD}$ by a factor is equivalent to dividing the choices above for $M_0$ and $M_f$ by this factor,
which by factorization scale independence and perturbative convergence is allowed provided that this factor is of $O(1)$.
To further improve the small $x$ description, hadron mass effects were accounted
for in the manner discussed in subsection \ref{hadmass}.
In the case of the unresummed fit, the description of all fitted data was as good 
as the resummed fit of Fig.\ \ref{figffd} (right), while the resummed fit did not improve substantially.
However, the resummed fit gave the reasonable values $\Lambda_{\rm QCD}=399$ MeV and $m_h=252$ MeV
(assuming the hadron sample is composed mostly of \cpi),
while the unresummed fit gave the clearly unreasonable value $\Lambda_{\rm QCD}=1308$ MeV and perhaps
slightly too large value $m_h=408$ MeV.
Thus the conclusion from these analyses is that both DL resummation {\it and} treatment of hadron mass effects 
are needed in order to achieve a reasonable fitted value of $\Lambda_{\rm QCD}$ {\it and} 
in order to describe the data down to small $x$ values.

The fact that eq.\ (\ref{expanpsglinmel}) is expected to be valid for all $\omega$ including
as $\omega \rightarrow 0$ suggests that the failure to describe the very small $x$ data 
when both DLs and $O(1/s)$ effects are treated
is due to the neglect of the complete SL contribution to the evolution 
and/or the complete DL contribution in the gluon coefficient function.
There may however also be a discrepancy between the theoretical and experimental definitions of 
the cross section which become more important with decreasing $x$.

\subsection{The modified leading logarithmic approximation \label{MLLA}}

Equation (\ref{DGLAP}) with $P$ given by eq.\ (\ref{firstapproxforPallz}) reduces to the MLLA 
master equation \cite{Dokshitzer:1992iy},
differentiated with respect to the factorization scale \cite{Lupia:1997hj},
if $P^{\nSGL (0)}$ in eq.\ (\ref{firstapproxforPallz}) is replaced, in Mellin space, by its value at $\omega=0$, i.e.\ 
\be
P(\omega,a_s)=a_s P^{\nSGL (0)}(\omega=0)+P^{\rm DL}(\omega,a_s).
\label{PinMLLA}
\ee
In discussions of the MLLA, 
\begin{eqnarray}
P^{\nSGL (0)}(\omega=0)=
\left( \begin{array}{cc}
0 & -3C_F \\
\frac{2}{3}T_R n_f \ & -\frac{11}{6}C_A-\frac {2}{3}T_R n_f
\end{array} \right)
\label{singlogsatLO}
\end{eqnarray}
is usually referred to as the SL term, as discussed after eq.\ (\ref{meltransofsoftglogs}).
At small $\omega$, the accuracy of eq.\ (\ref{PinMLLA}) is similar to the accuracy of eq. (\ref{firstapproxforPallz}),
while for all $\omega$, eq.\ (\ref{PinMLLA}) is still a better approximation for $P$ than $P=P^{\rm DL}$ is.
By coincidence \cite{Dokshitzer:1991wu}, eq.\ (\ref{PinMLLA}) leads to a similar result at large $z$ as 
eq. (\ref{firstapproxforPallz}) or (\ref{FOexpanofP}) do.
In any case, the approach of eq.\ (\ref{firstapproxforPallz}), or more generally 
eq.\ (\ref{PsepinSGLsandremain}), incorporates more information than the MLLA, in particular the large $\omega$
region, and is therefore more accurate. 

In many applications of the MLLA, eq.\ (\ref{DLArelforDquarkandDg}) is used.
Together with the gluon component of eq.\ (\ref{DRAP})
with $D=(D_{\Sigma},D_g)$, it implies that the gluon FF evolves according to
\be
D_g(\omega,M_f^2)=E_{gg}^\prime(\omega,a_s(M_f^2),a_s(M_0^2))D_g(\omega,M_0^2)
\label{evolofDginMLLA}
\ee
where, writing
\be
E_{gg}^\prime(\omega,a_s(M_f^2),a_s(M_0^2))=\exp \left[\int_{M=M_0}^{M=M_f}d\ln M_f^2 P_{gg}^\prime(\omega,a_s(M_f^2))\right],
\ee
the splitting function $P_{gg}^\prime$ is given by \cite{Dokshitzer:1991wu}
\be
\begin{split}
P_{gg}^\prime(\omega,a_s)=&\frac{1}{4}\left(-\omega +\sqrt{\omega^2+16C_A a_s}\right)\\
&+a_s\left[\left(\frac{11C_A}{6}-\frac{2n_f T_R}{3}\right)\frac{4C_A a_s}{\omega^2+16C_A a_s}
-\left(\frac{11C_A}{6}-\frac{2n_f T_R}{3C_A^2}\right)\left(1+\frac{1}{\sqrt{\omega^2+16C_A a_s}}\right)\right].
\label{PforDGinMLLA}
\end{split}
\ee
As for $P^{\rm DL}$, $P_{gg}^\prime(\omega,a_s)$ is finite as $\omega \rightarrow 0$.
Furthermore, eq.\ (\ref{DLArelforDquarkandDg}) reduces eq.\ (\ref{approxxs}) to
\be
\frac{1}{\sigma(s)}\frac{d\sigma^h}{dx}(x,s)=\frac{2C_F}{C_A}D_g(x,s).
\label{XSinMLLA}
\ee

We have derived these MLLA results using the results of the DLA and the FO approximation,
but they were first derived as a correction to the derivation of the DLA using the generating functional technique
mentioned in subsection \ref{DLA}.

It is often convenient and simpler to study the {\it moments} of the cross section.
The $n$th moment of a function $f(z)$ is given by
\be
K_n=\left(-\frac{d}{d\omega}\right)^n\ln f(\omega)\bigg{|}_{\omega=0}.
\label{defofmoments}
\ee
According to eq.\ (\ref{evolofDginMLLA}), the $M_f^2$ dependence of the $n$th moment of $f(z)=D(z,M_f^2)$
is given by
\be
K_n(M_f^2)=\Delta K_n(a_s(M_f^2),a_s(M_0^2)) +K_n(M_0^2),
\label{delkapinindefint}
\ee
where $\Delta K_n(a_s(M_f^2),a_s(M_0^2))$ is the $n$th moment of $E_{gg}^\prime(z,a_s(M_f^2),a_s(M_0^2))$,
given by
\be
\Delta K_n(a_s(M_f^2),a_s(M_0^2))=\int_{M=M_0}^{M=M_f}d\ln M_f^2 \left(-\frac{d}{d\omega}\right)^n P_{gg}^\prime(\omega,a_s(M_f^2)).
\ee
which from eq.\ (\ref{PforDGinMLLA}) gives, for $n\geq 1$,
\be
\Delta K_n(a_s(M_f^2),a_s(M_0^2))= a_s^{-\frac{n+1}{2}}(M_f^2)
\left(C_n^{(0)}+C_n^{(1)}a_s^{\frac{1}{2}}(M_f^2)+O\left(a_s\right)\right)-\big{\{}a_s(M_f^2) \leftrightarrow a_s(M_0^2)\big{\}}.
\label{delkapatsmallalphas}
\ee
The explicit results for the $C_n^{(0,1)}$ can be calculated from eq.\ (\ref{PforDGinMLLA}), and are presented in
Ref.\ \cite{Fong:1989qy} for the first few values of $n$.
For $n\geq 3$ and odd, $C_n^{(0)}=0$. 
Equation (\ref{delkapatsmallalphas}) also applies for $n=0$, but with the presence of a term proportional to $\ln a_s$.

\subsection{Local parton-hadron duality and the limiting spectrum \label{LPHDandls}}

Perturbative QCD is incomplete in that it cannot describe the physics of hadrons entirely.
An intuitive solution to this problem is provided by 
the LPHD, which states
that the distribution of partons in
inclusive processes with a sufficiently low energy scale is similar to the distribution of
hadrons, up to the number of particles actually produced (the multiplicity).
This implies that the hadronic FF is proportional to the partonic FF at a factorization scale of $O(\Lambda_{\rm QCD})$.
In other words, assuming that eq.\ (\ref{DLArelforDquarkandDg}) is valid at such low scales,
we must ensure that our initial gluon FF obeys
\be
D_g(z,M_0^2)=N \delta(1-z)
\label{FFsfromLPHD}
\ee
if we choose $M_0=O(\Lambda_{\rm QCD})$.
Ordinarily, evolution at such low scales will not be possible in perturbation theory, 
because the convergence of the perturbation series is spoilt, and can even be singular.
However, by using hypergeometric functions, it is possible to write the MLLA evolution of $D_g$, and therefore
the $s$ dependence of the cross section in eq.\ (\ref{XSinMLLA}), 
such that the calculation is well defined at $M_f=\Lambda_{\rm QCD}$.
This calculation is known as the {\it limiting spectrum}.
In this subsection we show how the limiting spectrum and the result of the LPHD in eq.\ (\ref{FFsfromLPHD}) 
arise as accidental consequences of truncated perturbation theory.

According to eq.\ (\ref{delkapatsmallalphas}) and the result $a_s(M^2)\rightarrow 0$ as $M \rightarrow \infty$, 
for sufficiently large $M_f$ we may neglect $K_n(M_0^2)$ relative to $\Delta K_n(a_s(M_f^2),a_s(M_0^2))$,
except for $n=0$ because the cross section is very sensitive to $N=\exp[K_0(M_0^2)]$.
From eq.\ (\ref{defofmoments}) with $f(z)=D(z,M_0^2)$ and $K_n(M_0^2)=0$ for $n\geq 1$, 
we see that $D(z,M_0^2)$ takes the form in eq.\ (\ref{FFsfromLPHD}).
In other words, at high energies, it {\it appears as if} the initial FF is a delta function, even though 
it could be a very different function.

The second, $M_0$ dependent, part of $K_n$ eq.\ (\ref{delkapatsmallalphas}) may also be neglected relative
to the first, $M_f$ dependent, part for sufficiently large $M_f$,
because in this limit $a_s(M_f^2)\ll a_s(M_0^2)$. 
Coincidentally, because 
$a_s(M^2)\rightarrow \infty$ as $M\rightarrow \Lambda_{\rm QCD}$,
the neglect of the second, $M_0$ dependent, part of $K_n$ eq.\ (\ref{delkapatsmallalphas})
is equivalent to choosing $M_0=\Lambda_{\rm QCD}$,
provided that the series in $a_s^{1/2}$ is terminated at $O(a_s^{n/2})$, because the next term will
be finite and the terms following that will be singular.
In other words, at high energies, it {\it appears as if} the choice $M_0=\Lambda_{\rm QCD}$ is justified,
even though it is in fact not.

We conclude that the LPHD and limiting spectrum in the context of hadron production
can only be verified using low energy data, because their consequences at high energy 
are also implied by perturbation theory alone.

With the two approximations above, eq.\ (\ref{delkapinindefint}) finally becomes
\be
K_n(M_f^2)\approx \Delta K_n(a_s(M_f^2),\infty)\ {\rm for}\ n\geq 1,
\label{delkapinindefint2}
\ee
which are the moments of the limiting spectrum.
Equation (\ref{delkapinindefint2}) implies an interesting large $M_f$ behaviour of the gluon FF, and hence,
from eq.\ (\ref{DLArelforDquarkandDg}), of all quark FFs.
The inverse Mellin transform of the gluon FF,
\be
zD_g(z,M_f^2)=\frac{1}{2\pi i}\int_C d\omega \exp[\omega \xi]D_g(\omega,M_f^2),
\label{invLT}
\ee
where $\xi =\ln (1/z)$, may, by making the replacement $y=i\omega \sigma$, where $\sigma^2=K_2(M_f^2)$,
be written
\be
zD_g(z,M_f^2)=\frac{{\mathcal N}}{\sigma\sqrt{2\pi }}\exp\left[-\frac{\delta^2}{2}\right] R(\delta,\{\kappa_n\}),
\label{FuptoksinvLT}
\ee
where ${\mathcal N}=\exp[K_0(M_f^2)]$ (so that ${\mathcal N}(M_0^2)=N$ in eq.\ (\ref{FFsfromLPHD})),
where $\delta=(\xi -\overline{\xi})/\sigma$ with $\overline{\xi}=K_1(M_f^2)$ the average value of $\xi$, 
where the real quantity $R$ is given by
\be
R(\delta,\{\kappa_n\}) =\frac{e^{\delta^2/2}}{\sqrt{2\pi}}
\int_{-\infty}^{\infty}dy\exp\left[\sum_{n=3}^{\infty}\kappa_n\frac{(-iy)^n}{n!}\right]
\exp\left[iy\delta-\frac{y^2}{2}\right],
\label{defofR}
\ee
and where
\be
\kappa_n=\frac{K_n(M_f^2)}{K_2^{\frac{n}{2}}(M_f^2)}=a_s^{\frac{n-2}{4}}(M_f^2)\left[1
+O\left(a_s^{\frac{1}{2}}(M_f^2)\right)\right].
\label{pertserfortilk}
\ee
Equation (\ref{delkapatsmallalphas}) has been used for the second equality in eq.\ (\ref{pertserfortilk}).
Because of that property, we can expand the $\kappa_n$-dependent exponential in eq.\ (\ref{defofR})
in powers of the $\kappa_n$ up to the required accuracy and perform the integral for each term. 
Then, writing $R$ as an exponential of the form
\be
R=\exp\left[\sum_{i=0}^\infty A_i \delta^i\right],
\label{generaldg}
\ee
we find that the series in the exponent will terminate at a finite value of $i$ when it is expanded
in $a_s$ to some finite order, even if $\delta=O(1)$.
In particular, if $\kappa_n=0$ for $n\geq 3$, which is always a reasonable approximation because of eq.\ (\ref{pertserfortilk}),
then all $A_i=0$ so that $R=1$.
Therefore, from eq.\ (\ref{FuptoksinvLT}), the gluon FF at large $M_f$ and therefore the cross section 
at large $\sqrt{s}$ will be a (distorted) Gaussian over the range from large to small $x$.
This justifies the choice for the small $z$ behaviour of eq.\ (\ref{GaussianformforFFs}) 
used for the FFs in the fits of Ref.\ \cite{Albino:2005gd}.

\section{Outlook \label{outlook}}

\subsection{Possible experimental results for the future}

\begin{figure}[h!]
\includegraphics[width=8.5cm]{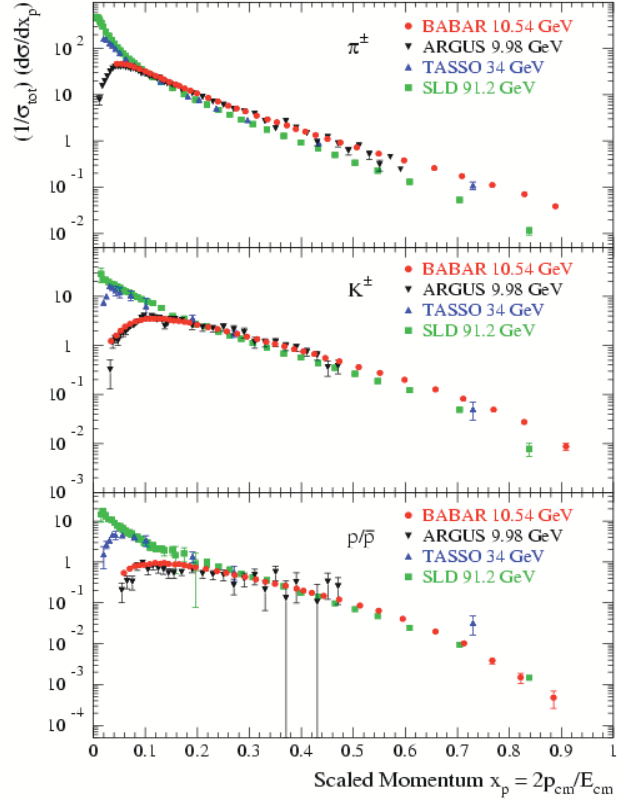}
\caption{Preliminary data from $e^+ e^-$ reactions at BaBar for \cpi, \cka\ and \pr.
Also shown are some published data for comparison. From Ref.\ \cite{Arleo:2008dn}.
\label{babar}}
\end{figure}

Very accurate measurements of light charged and neutral hadron production at HERA are now possible.
As noted in Refs.\ \cite{Albino:2006wz,Jung:2008tq},
such data in which the hadron species is identified, which does not yet exist,
would make a significant improvement to the current knowledge of FFs.
Furthermore, together with all the similar data from $e^+ e^-$ reactions already used in global fits,
the flavour separation of the FFs for each hadron species could be significantly improved,
and the validity of the quark tagging often performed in measurements of $e^+ e^-$ reactions can be studied.
By identification of the charge-sign of the produced hadrons as well, 
perhaps the greatest improvement to the current constraints on FFs that HERA could provide
are on the valence quark FFs, which at present are only constrained by similar but relatively 
much less accurate data from RHIC.
Further accurate measurements of light charged hadron production in 
$e^+ e^-$ reaction data from BaBar at $\sqrt{s}=10.54$ GeV \cite{Anulli:2004nm},
shown in its preliminary version in Fig.\ (\ref{babar}),
is expected to be published in the next year or so, 
which will provide much needed constraints on FFs at large $z$
that are not provided by any current data sets.
Because these data are taken at a different $\sqrt{s}$ to that of the currently most accurate data, which is at $\sqrt{s}=91.2$ GeV,
such data will also significantly improve the constraints on $\alpha_s(M_Z)$.
In addition, new more accurate $pp$ reaction data is being taken at RHIC which includes larger $p_T$ values than before,
and hence will also improve the large $z$ constraints.

In the further future, accurate measurements in $e^+ e^-$ reactions may be taken at CLEO, Belle and LEP.
Valence quark FFs could also be constrained by measurements of the asymmetric cross section 
in $e^+ e^-$ reactions, but to a much lesser degree than by measurements of charge-sign identified data from $ep$ reactions.
Further $ep$ reaction data could be taken at the proposed eRHIC and LHeC colliders.
Finally, measurements of $pp$ reactions at the LHC would likely be of sufficiently high accuracy
to significantly improve the FF constraints. 

\subsection{Future theoretical input to global fits}

A number of improvements which global fits may benefit from are possible, 
and we list those which are most likely to be available first:

\newcounter{qcounter}
\begin{list}{\arabic{qcounter}.}
{\usecounter{qcounter}
\setlength{\rightmargin}{\leftmargin}}
\item The inclusion of heavy quarks and heavy quark masses in the partonic cross sections.
\item The application of large $x$ resummation to $pp$ reaction calculations for a finite rapidity range, and also for $ep$
reactions both with and without cuts in $p_T$.
\item The calculation of the NNLO off-diagonal splitting functions, as well as the NNLO partonic cross sections
in $ep$ and $pp$ reactions.
\item The calculation of the full DL contribution to the gluon coefficient function in $e^+ e^-$ reactions data, 
and possibly further SGLs here and in the splitting functions, 
with the incorporation of these improvements in global fits to $e^+ e^-$ 
reaction data including more of the data at small $x$ that have previously been excluded.
The resummation of the same classes of SGLs in other types of processes may also be beneficial,
if there exist measurements of those processes whose calculations would improve due to the resummation.
\end{list}

Of course, a simultaneous treatment of all the above improvements in a single global fit would be an additional goal,
but one which is further into the future.

Global fits of FFs can be further improved by fitting other quantities not determinable in perturbation theory, 
such as hadron mass, higher twist contributions, intrinsic FFs, fracture functions 
and the modifications associated with hadronic decay channels,
whose results in the case of the latter four quantities would contribute significantly to the current knowledge of QCD physics.

\section{Summary \label{summary}}

We have given a comprehensive account of the most important methods and results in the phenomenological
extraction of FFs from experimental data.
Many FF components for many particles are now very well constrained.
FFs have been useful in the study of other phenomena and in the extraction of other physical quantities,
of which some examples are given in section \ref{intro}, and the number and intensity of such applications
is expected to increase as the constraints on FFs improve.
The best constraints on FFs at present come from $e^+ e^-$ reaction data
However, these data do not constrain the valence quark FFs (or charge-sign asymmetry FFs), 
and to a large degree the gluon FF, which can be at least weakly constrained by current data from RHIC.
Theoretically, calculations of inclusive single hadron production in all reactions that have been performed experimentally
can be calculated to NLO, and the effects of large $x$ resummation in $e^+ e^-$ reactions
and heavy quarks effects in all reactions can be included at this level of accuracy.
The deviation of the fitted detected hadron mass from its true value gives a quantification of the importance
of the production of this particle from decays of heavier hadrons instead of from direct partonic fragmentation.
In global fits, systematic errors are now accounted for through a covariance matrix, 
and various sound approaches for propagating experimental errors from measurements to predictions have been applied.

However, there is alot of room for improvement in this program, 
perhaps the most important being the eventual incorporation in global fits 
of future accurate measurements of hadron production at HERA in which both the species and charge-sign of the
hadrons is measured, in order to significantly improve the constraints on the differences between
FFs of different quark flavours (the non singlet quark FFs) and of different quark charge-signs (the valence quark FFs) 
respectively.
Such FF components are currently constrained by theoretically sound but experimentally untested non perturbative assumptions.
A significant improvement in the extraction of $\alpha_s(M_Z)$ would also result from such fits relative to previous fits.
Because of the theoretical similarity between charge-sign unidentified data from $e^+ e^-$ and $ep$ reactions,
further tests on FF universality would also result from the inclusion of hadron species identified data from HERA
in global fits together with data from $e^+ e^-$ reactions.

Note that FF universality tests are provided by the inclusion of $pp(\overline{p})$ reaction data, but, because of their
much lower accuracy, to a much lesser degree
than by the inclusion of $ep$ reaction data.
At present, $pp(\overline{p})$ reaction data are crucial for the extraction of the currently very badly constrained gluon 
and valence quark FFs,
and therefore future accurate measurements from RHIC and the LHC will be most welcome.
We note that the constraints on these FFs could also be significantly improved by measurements
of, respectively, the longitudinal and asymmetric cross sections (with hadron species identification, of course)
in $e^+ e^-$ reactions.
\begin{acknowledgments}
I thank B.\ A.\ Kniehl and G.\ Kramer for a thorough reading of the manuscript,
and J.\ C.\ Collins for valuable discussions and criticisms concerning appendix \ref{outlinederivfactTheo}.
\end{acknowledgments}

\appendix

\renewcommand{\thesection}{\Alph{section}}
\renewcommand{\thesubsection}{\Alph{section}.\arabic{subsection}}
\setcounter{section}{0}

\section{Derivation of the QCD factorization theorem \label{outlinederivfactTheo}}

In this appendix, we discuss the factorization theorem applied to inclusive single hadron production.
We will restrict our discussion to the process $e^+ e^- \rightarrow \gamma^*\rightarrow h+X$ of Fig.\ \ref{epem},
although it can be applied to other inclusive single hadron production processes.
This cross section can be decomposed as
\be
d\sigma=L^{\mu \nu} W_{\mu \nu},
\ee
where $L^{\mu \nu}$ is the well known tensor describing the process $e^+ e^- \rightarrow \gamma^*$
(we work to LO in QED), while the tensor
\be
W_{\mu \nu}(q,p_h)=\frac{1}{2\pi}\int d^4 x e^{iq\cdot x} \langle 0 | J_\mu(x) a_h^{\dagger}(p_h) | X\rangle
\langle X| a_h (p_h) J_\nu(0)|0\rangle,
\ee
where $a_h^{(\dagger)}(p_h)$ is the annihilation (creation) operator 
for a hadron $h$ with momentum $p_h$ (we do not assume a massless hadron in this appendix),
describes the hadronic process $\gamma^*\rightarrow h+X$ and is the quantity in which we are interested.
$W$ may be partially calculated using perturbation theory by
decomposing it into the form (omitting the spacetime indices $\mu \nu$ for brevity from now on)
\be
W=w D +r,
\label{twistexpan}
\ee
where the remainder $r$ is {\it power suppressed}, i.e.\ 
\be
r=O\left(\left(\frac{\Lambda_{\rm QCD}}{\sqrt{s}}\right)^p\right)
\label{psupofr}
\ee
where $w$ is a vector of the equivalent, factorized, full partonic processes $w_i$ 
(where the parton species label $i$ includes (combinations of) the spin state(s) and charge(s) 
as discussed in subsection \ref{symmetries})
for the processes $\gamma^*\rightarrow i+X$, and where $D$ is a vector of factorized FFs $D_i$. 
Equation (\ref{twistexpan}) is then the same as eq.\ (\ref{genformofinchadprodfromffs}).
The ``product'' $wD$ involves sums and integrations over all species of the ``detected'' final state parton of $w$,
which is a real particle moving spatially parallel to the detected hadron.
Note that the momentum integration is precisely the convolution of eq.\ (\ref{genformofinchadprodfromffs}),
i.e.\ over the momentum fraction $z$, and not over all four components of momentum.
In the language of the {\it operator product expansion} (OPE)  \cite{Moch:1999eb}, 
eq.\ (\ref{twistexpan}) corresponds to a {\it twist expansion},
the term $wD$ being referred to as the {\it leading twist} component. 
The variable $p\geq 1$ in eq.\ (\ref{psupofr}), 
so at large enough energy the {\it higher twist} terms are smaller than the radiative corrections of $O(1/\ln \sqrt{s})$.

In the remainder of this appendix, we briefly but comprehensively outline the formal derivation 
of the factorization theorem of Ref.\ \cite{Collins:1998rz},
in which full details and more references may be found.
In subsection \ref{twistexpand}, we outline the steps given in Ref.\ \cite{Collins:1998rz} for separating the non leading twist part, 
whose order of magnitude can be reliably estimated,
from the cross section such that it is suitable for factorization. 
In subsection \ref{CollinsOR}, we discuss the factorization approach of Ref.\ \cite{Collins:1998rz}, which 
is essentially a generalization of former approaches.
Then, in subsection \ref{CFP}, we attempt to connect it with the
older factorization approach of Refs.\ \cite{Ellis:1978ty,Curci:1980uw,Furmanski:1981cw}.
Because the manner in which a parton should be treated in factorization 
depends on whether its mass is much greater than the hard scale, different schemes
are appropriate for different energies and 
therefore, in subsection \ref{mathcond}, we discuss matching conditions between quantities
in these different schemes.
However, we note there that the correct treatment of quarks with mass much greater than the hard scale
has not been specified in the literature so far, and we indicate how this may be remedied.
Finally, in subsection \ref{openissues}, we summarize the open issues remaining in the factorization theorem.

\subsection{Twist expansion \label{twistexpand}}

\begin{figure}[h!]
\includegraphics[width=15cm]{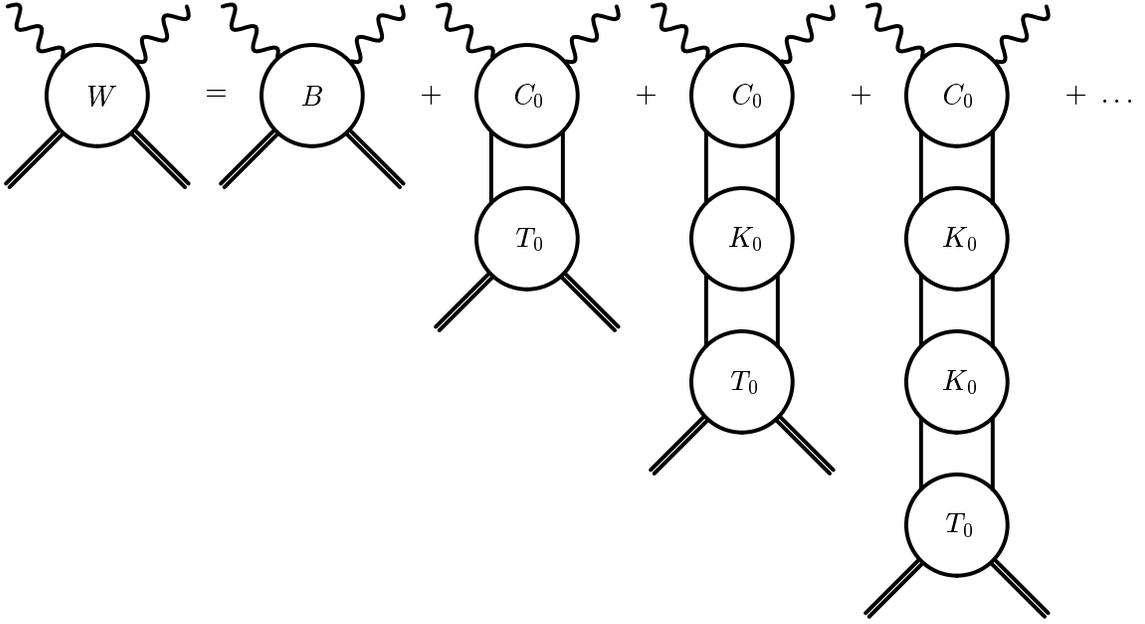}
\caption{2PI $t$-channel expansion of the cross section for $e^+ e^- \rightarrow \gamma^* \rightarrow h+X$ 
(with the electron and positron legs removed for simplicity).
The left hand side is the modulus squared of the amplitude shown in Fig.\ \ref{epem},
i.e.\ the external wavy legs represent the virtual photon $\gamma^*$ and
the external double lines the produced hadron $h$.
On the right hand side, the lines between 2PI graphs are virtual partons in loops 
(i.e.\ summed and integrated over all quantum numbers).
$C_0$ and $K_0$ represent all 2PI $t$-channel contributions to the processes 
$\gamma^* \rightarrow j+X$ and $i \rightarrow j+X$ respectively.
$B$ and $T_0$ represent all 2PI $t$-channel contributions to $\gamma^* \rightarrow h+X$ and $j\rightarrow h+X$, where in this
case ``2PI'' must be defined by non perturbative external field methods.
Strictly speaking, all graphs have a cut down the middle signifying the final state and on which all particles must be real.
Alternatively, we may forget the cut if these graphs may be regarded as forward amplitudes, 
whose imaginary component also gives the cross section as dictated by the optical theorem.
\label{2PIexpan}}
\end{figure}

\begin{figure}[h!]
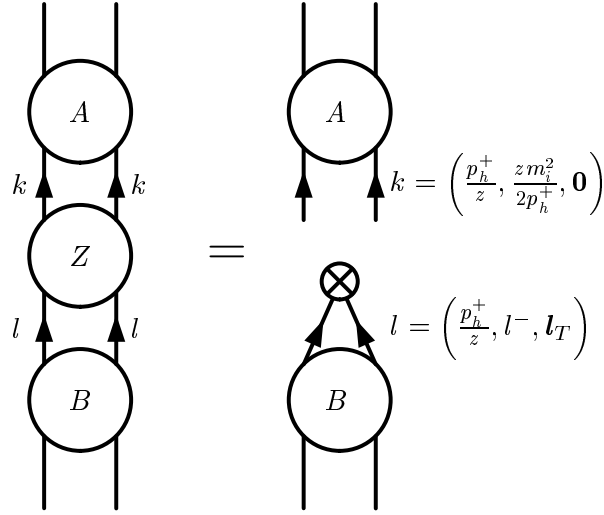

\parbox{.18\linewidth}{
\begin{center}
\includegraphics[width=1.8cm]{insertZ1.epsi}
\end{center}
}\parbox{.04\linewidth}{\huge $=$}
\parbox{.27\linewidth}{
\begin{center}
\includegraphics[width=4.2cm]{insertZ2.epsi}
\end{center}}
\caption{The action of $Z$ on the connecting partons of mass $m_i$ between any graphs $A$ and $B$. The momentum integration is 
over the variables $z$, $l^-$ and ${\bm l}_T$. \label{insertZ}}
\end{figure}

For now, let us assume that all partons have masses less than or of the order of the hard scale $\sqrt{s}$.
We will return to the case that there are also quarks with masses much greater than $\sqrt{s}$ in subsection \ref{NPQ}.
In Ref.\ \cite{Collins:1998rz}, the form of eq.\ (\ref{twistexpan}) is derived for DIS,
$e h\rightarrow e+X$ (in which case $D$ is a vector of PDFs), by starting from an expansion in graphs which are 
two-particle irreducible (2PI), i.e.\ cannot be disconnected by cutting through two internal lines, in the $t$-channel,
and then using the result that all graphs that contribute at leading twist are two-particle reducible \cite{Libby:1978qf}.
There are many subtleties involved in getting to this result, 
and the reader is referred to subsection IV A of Ref.\ \cite{Collins:1998rz} for details and references.
By repeating these arguments but with the negative virtuality of the exchanged boson replaced with positive virtuality 
and the incoming hadron replaced with an outgoing hadron, one obtains the factorized form 
of the cross section $e^+ e^- \rightarrow \gamma^*\rightarrow h+X$.
The expansion is shown in Fig.\ \ref{2PIexpan} which, using the shorthand $1+K_0 +K_0^2 +\ldots =1/(1-K_0)$, may be represented as
\be
W=C_0 \frac{1}{1-K_0} T_0 +B.
\label{2PIexpanofFalg}
\ee
As it stands, eq.\ (\ref{2PIexpanofFalg}) is no use for practical calculations. 
Firstly, $T_0$ and $B$ contain hadron legs for which no perturbative or other analytic representation exists.
Secondly, perturbation theory cannot be applied even to calculate the equivalent purely partonic cross section $C_0/(1-K_0)$:
In general, each parton species contributes {\it potential mass singularities},
which are logarithms of each parton mass that becomes singular as that mass approaches zero. 
The potential mass singularities of any quark, then, will not be singular if all quarks are taken to be massive,
but they will be large if this quark's mass $m_i$ is much less than $\sqrt{s}$, i.e.\ in the limit that this
mass be neglected, at the expense of a relative error of $O(m_i^2/s)$ on the cross section as dictated by the 
{\it decoupling theorem} \cite{Appelquist:1974tg}.
Because the magnitude of potential mass singularities increases with the order in $a_s$,
for sufficiently large $s$ they will cause the perturbative calculation of any part of $W$ to diverge.
Thirdly, the products in eq.\ (\ref{2PIexpanofFalg}) are rather complicated, 
containing sums and integrations over all virtual partons 
connecting 2PI graphs, which we will call {\it connecting partons}.
To tackle the second and third problem, we begin with the observation that, 
in a physical gauge such as the light-cone gauge, which we use from now on, the 2PI graphs are free of
potential mass singularities \cite{Ellis:1978ty},
which therefore arise from those connecting partons which are in the vicinity of being real, i.e.\ on shell and with physical spin,
and moving spatially parallel to the detected hadron.
These 2PI graphs are also free of UV singularities after renormalization of the strong coupling and the quark masses.
Thus we introduce a projection operator $Z=Z^2$ which projects onto
the connecting partons in a manner which includes all those partons with momenta responsible for the potential mass singularities.
The operator $(1-Z)$ annihilates the potential mass singularity between any two 2PI graphs.
Let us now give one explicit definition of such an operator.
We use light cone coordinates $V=(V^+=(V^0+V^3)/\sqrt{2},V^-=(V^0-V^3)/\sqrt{2},{\bm V}_T=(V_1,V_2))$,
and work in a frame for which the detected hadron's momentum is given by
eq.\ (\ref{phformassivehadron}) and the virtual boson's momentum by eq.\ (\ref{choiceofq}).
Considering only massless partons for simple illustration,
mass singularities arise from connecting partons with physical spins and physical momenta obeying
\be
k=\left(\frac{p_h^+}{z},0,{\bm 0}\right).
\label{masslesspartonmom}
\ee
Then
\be
Z_{\alpha \alpha';\beta,\beta'}(k,l)=\frac{1}{4}\gamma_{\alpha \alpha'}^- \gamma_{\beta \beta'}^+
(2\pi)^4 \delta (k^+-l^+)\delta(k^-)\delta^{(2)}({\bm k}_T)
\label{defofZformasslesspartons}
\ee
and something similar for gluons. 
Note that eq.\ (\ref{defofZformasslesspartons}) is consistent with the projection operator property $Z=Z^2$.
The extension of eq.\ (\ref{defofZformasslesspartons}) to massive quarks (we assume all quarks' and
the detected hadron's masses are non negligible) is quite straightforward, in particular the ``$-$'' component
of eq.\ (\ref{masslesspartonmom}) becomes $zm_i^2/(2p_h^+)$ and the mass singularities become potential mass singularities.
Like $K_0$, $Z$ can be regarded as a graph with two partons at the top, each with momentum $k$,
and two with momentum $l$ at the bottom.
As illustrated in Fig.\ \ref{insertZ},
the insertion of $Z$ between two graphs reduces the sums and integrations in the product between them
to an integration over $z$ of the pair of connecting partons at the bottom of the top graph, 
which have momenta given by eq.\ (\ref{masslesspartonmom}), i.e.\ they are on-shell,
and an integration over the momenta $l^-$ and ${\bm l}_T$ of the pair of connecting partons at the top of the bottom graph,
whose $+$ component of momentum is the same as that in eq.\ (\ref{masslesspartonmom}), i.e.\ $l^+=p_h^+/z$.
This is the case, for example, in the product between $wD$ in eq.\ (\ref{twistexpan}) ---
in practice it is usually not necessary to show explicitly 
the $l^-$ and ${\bm l}_T$ integrations over the initial state partons in $D$.

The $Z$ operator is used to put eq.\ (\ref{2PIexpanofFalg}) into the form 
of eq.\ (\ref{twistexpan}) in subsection V B of Ref.\ \cite{Collins:1998rz},
where the remainder $r$ is shown to be power suppressed (i.e.\ to obey eq.\ (\ref{psupofr})).
A general graph, such as $W$, containing a hard and a soft scale such as $\sqrt{s}$ and $\Lambda_{\rm QCD}$ respectively
will in general contain a leading region, i.e.\ a part which is not power suppressed,
coming from graphs which can be expressed as 
a graph coupled to the hard scale, the {\it hard graph} joined to a graph 
coupled to the soft scale, the {\it soft graph}, by 2 partons 
of mass much less than the hard scale \cite{Libby:1978qf}.
Being 2PI, $B$ does not contain such graphs and therefore does not contain a leading region, i.e.\ it is power suppressed.
This will not be the case in a general gauge, because additional gluon exchanges between the hard and soft graphs are allowed.
However, it is assumed, without proof, that a suitable light-cone gauge exists such that such graphs are power suppressed.
This is true for the $A^+=0$ gauge that we have been using, at least in the case that these gluons' momenta are almost
collinear with the detected hadron's.
In the leading region, lines in the hard graph will have much larger virtualities than lines in the soft graph,
so lines in a 2PI graph must have similar virtualities.
Therefore, because each $(1-Z)$ insertion between two graphs gives a suppression of 
order (highest virtuality in graph below / lowest virtuality in graph above)$^p$ \cite{Collins:1998rz},
successive insertions of $(1-Z)$ in between all 2PI graphs in eq.\ (\ref{2PIexpanofFalg}) will give successive factors 
of this suppression leading an overall power suppression.
Consequently, 
\be
r=C_0 (1-Z) \frac{1}{1-K_0 (1-Z)} T_0 +B
\label{defofr}
\ee
obeys eq.\ (\ref{psupofr}), so $r$ may be taken to represent all higher twist terms in $W$.
All potential mass singularities in $r$ are annihilated by the $(1-Z)$ insertions.
All UV singularities in $r$ cancel: Acting on partons from the left with $Z$ introduces a UV singularity,
which is then canceled by the $1-Z$ factors in essentially the same way as these factors remove the potential mass singularities.

The leading twist component of $W$ is therefore $W-r$, which we will now factorize 
in order to express it in its familiar form in eq.\ (\ref{twistexpan}).

\subsection{The modern approach to factorization \label{CollinsOR}}

The approach to factorization of Ref.\ \cite{Collins:1998rz} that we consider in this subsection
is a generalization of previous approaches such as the OPE.
Most importantly, it implicitly uses the cut vertices \cite{Mueller:1978xu} and nonlocal operators \cite{Balitsky:1988fi}
that were necessary in modifying the Wilson expansion in local operators on the light cone,
which can be applied to e.g.\ the time ordered product of currents appearing in $ep \rightarrow e+X$,
to the time ordered product of currents appearing in $e^+ e^- \rightarrow h+X$,
which is subtle due its nonlocal operator structure.
The result that $r=O((\Lambda_{\rm QCD}/Q)^2)$, i.e.\ that $p=2$ in eq.\ (\ref{twistexpan}), 
found in Ref.\ \cite{Balitsky:1988fi} using an expansion in nonlocal operators,
should be obtainable using the approach of Ref.\ \cite{Collins:1998rz}.
Using the $Z$ operator, eq.\ (\ref{2PIexpanofFalg}) may be rewritten in the form 
\be
W-r=w_B D_B.
\label{OPEformoffact}
\ee
In this expression, the vector
\be
w_B =C_0 \frac{1}{1-(1-Z)K_0} Z
\ee
contains the equivalent ``UV bare'' (containing UV divergences) partonic processes. 
It is clearly free of potential mass singularities.
Note that the ``detected'' parton is real due to the presence of $Z$ on the right hand side.
The UV bare FFs
\be
D_B=Z \frac{1}{1-K_0} T_0
\label{defofbareFFs}
\ee
appearing in eq.\ (\ref{OPEformoffact}) describe the fragmentation of partons.
Clearly, $D_B$ contains only and all processes contributing to the process $i\rightarrow h+X$ 
($i$ is in fact a virtual parton, but its $+$ component of momentum is fixed to that of the real parton lines at the 
top of $Z$, while its remaining components are integrated out), 
and may therefore be written as a matrix element of unrenormalized operators. 
In a general gauge, it is given by \cite{Collins:1981uk,Collins:1981uw,Collins:1992kk}
\be
\begin{split}
D_{B q_I}^h(z)=&\frac{z^{d-3}}{4\pi}\int dx^- e^{-iP^+ x^-/z}\frac{1}{3}{\rm tr}_{\rm color} \frac{1}{3}{\rm tr}_{\rm spin}
\bigg{\{}\gamma^+\langle 0| \psi_{q_I}^{(0)}(0,0,{\bm 0})\overline{\mathcal T}\exp\left[ig^{(0)}
\int_0^\infty dy^- A_a^{(0)+}(0,y^-,{\bm 0})t_a^T\right]\\
&\times a_h^\dagger (P^+,0,{\bm 0})a_h(P^+,0,{\bm 0}) {\mathcal T}
\exp\left[-ig^{(0)}\int_{x^-}^\infty dy^- A_a^{(0)+}(0,y^-,{\bm 0})t_a^T\right]
\overline{\psi}_{q_I}^{(0)}(0,x^-,{\bm 0})|0\rangle \bigg{\}},
\label{matelemforquarkFFs}
\end{split}
\ee
for a quark of flavour $q_I$,
where $\psi_{q_I}$ is the spinor field for quarks of flavour $I$, 
$g$ is the strong coupling, $t_a=\lambda_a/2$ where $\lambda_a$ are the Gell-Mann matrices, 
and $A_a^\mu$ is the field for gluons of colour charge $a$.
The operator ${\mathcal T}$ ($\overline{\mathcal T}$) orders the field operators $A_a^+(0,y^-,{\bm 0})$ in the products 
in each term of the expansion of the exponential in $g$ such that an operator is always to the left (right) of any other operator
with a lower (higher) value of $y^-$. 
The superscript ``$(0)$'' on e.g.\ $\psi_{q_I}^{(0)}$ implies that the object is unrenormalized.
The expression for the gluon FF is similar, 
obtained essentially by replacing $\psi_{q_I}^{(0)}$ everywhere with the gluon field strength tensor.
The Wilson lines are not present in eq.\ (\ref{matelemforquarkFFs}) in the $A^+=0$ gauge that we have been using, 
i.e.\ the exponentials are equal to unity. Their direct derivation involves going beyond the 2PI expansion in a non-trivial way.
Since the UV divergences of the bare operators are subtracted by ordinary multiplicative renormalization,
the renormalized FFs take the form
\be
D=G D_B,
\label{ffactfromfbareUV}
\ee
where $G=GZ$ renormalizes UV divergences in the operator in $D_B$, in the standard way of operator 
renormalization (see e.g. \cite{Peskin:1995ev}), 
and therefore introduces a dependence of $D$ on some operator renormalization scale $M_f$.
$G$ is non singular as parton masses vanish, as will be explicitly shown later 
(in the sentence containing eq.\ (\ref{defofGintermsofP})).
Dimensional regularization is necessary to implement the CWZ schemes, in which case $G$ will be singular
as the number of dimensions approaches 4.
As usual in operator renormalization, the dependence of $G$ on $M_f$ is governed by
the renormalization group equation
\be
\frac{d}{d\ln M_f^2}G =P G,
\ee
where the components of the matrix $P$ are the relevant perturbatively calculable anomalous dimensions for the operators in $D_B$,
and are known as {\it splitting functions}. 
Therefore from eq.\ (\ref{ffactfromfbareUV}),
\be
\frac{d}{d\ln M_f^2}D =P D,
\label{DGLAPderived}
\ee
which is the DGLAP equation.
In studies using the OPE, eq.\ (\ref{DGLAPderived}) appears in Mellin space.
where the Mellin space variable $N$ is an integer 
equal to the spin of the given operator in the OPE, which is equivalent to eq.\ (\ref{DGLAPderived}) in $z$ space that 
we have just derived by applying the inverse Mellin transform of eq.\ (\ref{invmelltrans}).

We finally arrive at eq.\ (\ref{twistexpan}) from eq.\ (\ref{OPEformoffact}) by making the identity
\be
w =w_B G^{-1},
\label{UVdefofw}
\ee
where $G^{-1}$ is defined such that $G G^{-1}=Z$.
As for $w_B$, the factorized partonic processes $w$ are also free of potential mass singularities.
In practice, $w$ can be calculated as follows:
We may write 
\be
w_0=w_B K_B,
\ee
where 
\be
w_0=C_0 \frac{1}{1-K_0} Z
\label{expldefofF0}
\ee
is precisely the equivalent partonic process in which the ``detected'' parton is real,
which is clearly free of UV divergences, but contains potential mass singularities,
and
\be
K_B=Z\frac{1}{1-K_0}Z
\ee
are the bare FFs for partons to fragment to partons.
The renormalized equivalent of these FFs is
\be
\Gamma=G K_B, 
\label{def1ofGamma}
\ee
which is non singular as the regulator of UV singularities is removed, but is singular as any parton masses vanish.
Note that $K_B$ and $\Gamma$ are written in Ref.\ \cite{Collins:1998rz} as $A_{Bp}$ and $A_{Rp}$ respectively.
Then, instead of calculating $w$ via eq.\ (\ref{UVdefofw}), $w$ may be explicitly calculated from $w_0$ according to
\be
w=w_0 \Gamma^{-1},
\label{factFhatfrombareFhat}
\ee
where $\Gamma^{-1}$ is defined such that $\Gamma \Gamma^{-1}=\Gamma^{-1} \Gamma =Z$.
Equation (\ref{factFhatfrombareFhat}), which is equivalent to eq.\ (65) of Ref.\ \cite{Collins:1998rz}
(after expanding in $a_s$),
makes it clear that the purpose of $\Gamma$ is to remove from $w_0$ the potential mass singularities. 
In other words, each potential mass singularities in $w_0$ is canceled by {\it counterterms} in $\Gamma$.
$\Gamma$ may be chosen, through the choice of $G$ and $Z$ in eq.\ (\ref{def1ofGamma}), to depend
on parton masses only through mass logarithms, which will be the case for the CWZ schemes.
The precise form of this $\Gamma$ (eq.\ (\ref{pertexpandofGamma})) will be derived later.

To summarize, $w$ and the $M_f$ dependence of $D$ are perturbatively calculable.
The components of $w$ are the standard coefficient functions in the literature, up to electroweak couplings etc.
Like the $D_B$, the factorized FFs $D$ in eq.\ (\ref{ffactfromfbare}) are clearly universal, 
since all details of the initial state are contained in $C_0$.

\subsection{Connection with the older approach to factorization \label{CFP}}

Finally, we make some connections between the above approach 
and the earlier approach of Refs.\ \cite{Ellis:1978ty,Curci:1980uw,Furmanski:1981cw}.
The latter approach was formulated for the case of massless quarks only, 
so we will need to make some modifications to account for massive quarks as well.
In addition, our definitions here of bare FFs will need to differ from the earlier approach in order
to obtain reliable results.
We begin with an alternative to eq.\ (\ref{OPEformoffact}) 
which corresponds to the starting point of calculations in the literature (i.e.\ eq.\ (\ref{factFhatfrombareFhat})):
\be
W-r=w_0 D_0,
\ee
where the ``bare'' FF in this case is
\be
D_0 =Z\frac{1}{1-K_0 (1-Z)} T_0,
\label{altbareff}
\ee
which is free of UV singularities because the UV singularity introduced by acting from the left with $Z$
is canceled by $(1-Z)$ insertions occurring to the right, similar to the cancellation of UV singularities
in $r$ discussed in the paragraph following its definition in eq.\ (\ref{defofr}).
We define a projection operator ${\mathcal P}={\mathcal P}^2$
which projects in the same way as $Z$ on parton lines above (so that $Z{\mathcal P}={\mathcal P}$),
and which projects in a similar way as $Z$ on parton lines below except that, in contrast to $Z$, it becomes 
sufficiently suppressed for increasing momenta of these lines such that it does not introduce UV singularities.
The scale at which this suppression sets in is called the {\it factorization scale}, 
which will be written $M_f$ since it will turn out to be identical to the operator renormalization scale defined above.
Thus, for example, we may choose ${\mathcal P}$ (in the case of massless quarks for simple illustration) to be 
given by eq.\ (\ref{defofZformasslesspartons})
multiplied by a function $f(k^-/M_f,{\bm k}_T/M_f)$ such that $f(0,0)=1$ and $f(\infty,\infty)=0$.
A more explicit definition of ${\mathcal P}$ (or $f$) is not required since an implicit definition will be given later.
Note that, like $Z$, ${\mathcal P}$ is flavour diagonal. 
As a projection operator, ${\mathcal P}^2={\mathcal P}$, and insertion of $1-{\mathcal P}$ between two 2PI graphs
annihilates the potential mass singularity due to the connecting partons.

The momentum dependence of ${\mathcal P}$ can lead to problems with gauge invariance and with rapidity divergences 
\cite{Collins:2003fm}. 
Such problems were not considered in Refs.\ \cite{Ellis:1978ty,Curci:1980uw,Furmanski:1981cw}, which was limited to
the case of massless quarks. 
We will assume that some suitable choice for the remaining degrees of freedom in
${\mathcal P}$ exists such that these problems do not arise in the case of massive quarks.
Further work is needed here to prove that such a ${\mathcal P}$ exists.

The dependence of $\Gamma$, which was introduced in eq.\ (\ref{def1ofGamma}), on ${\mathcal P}$ is
\be
\Gamma=\left(1+{\mathcal P}K_0 \frac{1}{1-K_0}\right)Z,
\label{firstdefofGamma}
\ee
which can be interpreted as the FFs for hard partons fragmenting to on-shell soft partons.
Using eq.\ (\ref{def1ofGamma}), we may write 
\be
G=\left(1+({\mathcal P}-Z)K_0 \frac{1}{1-(1-Z)K_0}\right)Z,
\label{defofGintermsofP}
\ee
i.e.\ $G$ is free of potential mass singularities because, in the $(Z-{\mathcal P})$ operator,
those potential mass singularities projected out by $Z$ are identical to those projected out by ${\mathcal P}$.
This behaviour was noted just after eq.\ (\ref{ffactfromfbareUV}).
According to eqs.\ (\ref{twistexpan}) and (\ref{factFhatfrombareFhat}), the factorized FFs defined 
in eq.\ (\ref{ffactfromfbareUV}) may also be written
\be
D=\Gamma D_0.
\label{ffactfromfbare}
\ee
Using eq.\ (\ref{factFhatfrombareFhat}), 
the hard partonic cross section in terms of ${\mathcal P}$ reads
\be
w=C_0 \frac{1}{1-(1-{\mathcal P})K_0}Z.
\label{factFhatfrombareFhat2}
\ee
which is clearly free of both UV singularities and potential mass singularities.
$\Gamma=Z\Gamma =\Gamma Z$ is a function only of the ratio 
of the $+$ component of light cone momenta of its initial and final partons,
as well as of parton masses and indices.
By giving this function explicitly, eq.\ (\ref{firstdefofGamma}) defines ${\mathcal P}$ implicitly.
The simplest possible choice for $\Gamma$, 
which is the case for the CWZ schemes, is that for which the coefficients in the perturbative series for $\Gamma$
are themselves finite series in $\ln M_f^2/m_i^2$, where $i$ runs over all partons.
Although the gluon mass must be zero for renormalizability,
connecting gluons can be given a small mass after
renormalization has been performed within the 2PI graphs.
Although the perturbative approximation for $\Gamma$ fails in the numerical sense, its $M_f$ dependence does not, because
\be
\frac{d}{d\ln M_f^2}\Gamma =P \Gamma,
\label{partonversDGLAP}
\ee
where the splitting functions
\be
P=\left(\frac{d}{d\ln M_f^2}{\mathcal P}\right)K_0\frac{1}{1-(1-{\mathcal P})K_0}Z
\label{defofsplitfuncfromdiags}
\ee
are clearly free of UV and potential mass singularities.
This is obviously true since the perturbatively calculable $w$ must obey a similar equation 
to eq.\ (\ref{partonversDGLAP}) in order for eq.\ (\ref{twistexpan}), in particular $wD$, to be independent of $M_f$.
Thus, when $\Gamma$ is chosen to depend on parton masses 
only through finite powers of potential mass singularities,
as discussed in the paragraph preceding eq.\ (\ref{partonversDGLAP}), $P$ will be independent of all parton masses.
The DGLAP equation, eq.\ (\ref{DGLAPderived}), may be obtained from
eqs.\ (\ref{ffactfromfbare}) and (\ref{partonversDGLAP}).

\subsection{Treatment of non partonic quarks \label{NPQ}}

Up to this point, all results are only valid when no parton mass is much greater than $\sqrt{s}$.
Because the potential mass singularities of quarks that do not fall into this category are power suppressed by $O(s/m_i^2)$,
where $m_i$ is the mass of any such quark, they must not be subtracted by a large counterterm, 
i.e.\ they must be treated differently. 
Thus partons must be distinguished according to how they are treated: 
the partons that are treated as discussed in the last 3 subsections are called
{\it active} partons, while the rest, including in particular those for which $m_i \gg \sqrt{s}$, 
are called {\it non partonic} quarks.
Note that a quark for which $m_i=O(\sqrt{s})$ can be treated as either active or non partonic,
since its potential mass singularities and their corresponding counterterms are not large.
We define the 2PI graphs to be 2PI in non partonic quarks as well.
The choice of $Z$ on non partonic quarks must be such that $r$ remains both power suppressed and free of UV singularities,
i.e.\ $Z$ on non partonic quarks must be such that the arguments around eq.\ (\ref{defofr}) still hold.
Choosing $Z=0$ when acting on non partonic quark lines as is done in Ref.\ \cite{Collins:1998rz} leads to 
UV singularities in $r$, and therefore is not a valid possibility.
One possibility which is valid is to choose $Z(k,l)$ when the magnitude of $l$ is large 
to behave in the same way on non partonic quark lines as on active quark lines,
i.e.\ to choose eq.\ (\ref{defofZformasslesspartons}) to hold for non partonic quarks as well,
thus guaranteeing the cancellation of UV divergences in $r$, 
while the $Z(k,l)$ on non partonic quark lines vanishes as $l$ becomes small to both guaranteeing the power suppression of $r$
and ensuring no large counterterms are introduced for potential mass singularities due to non partonic quark, which are
suppressed or not large.
In other words, $Z$ on non partonic quark lines may be chosen in a similar way to how ${\mathcal P}$ acting on active quark lines
was in the paragraph following eq.\ (\ref{altbareff}), except that now $f(k^-/M_f,{\bm k}_T/M_f)$
is replaced with another function, $g(k^-/\mu_f,{\bm k}_T/\mu_f)$ say, where $\mu_f$ is another arbitrary scale
at which the suppression of $Z$ acting on non partonic quark lines sets in,
which obeys $g(0,0)=0$ and $g(\infty,\infty)=1$.
In this case, $Z$ would no longer be a projection operator, 
and $r$ is now scheme and scale dependent, i.e.\ it depends on the number of flavours and on $\mu_f$ respectively.
Neither of these two facts presents any problems.
However, as noted in the paragraph preceding the paragraph containing eq.\ (\ref{firstdefofGamma}),
introducing a momentum dependence into a projection operator can lead to problems with gauge invariance and with rapidity divergences,
so further work is needed to prove that a $Z$ exists for which these problems do not arise.

For now, for simplicity, to avoid possible problems with gauge invariance and with rapidity divergences, and
to ensure that $r$ is free of UV singularities, we will choose $Z$ on non partonic quark lines to be the same
as $Z$ on active quark lines 
(i.e.\ the massive quark equivalent of eq.\ (\ref{defofZformasslesspartons})).
However, in this case $r$ is no longer power suppressed.
Then, in the presence of non partonic quarks, all results derived so far still hold, except that
the sum over partons in all products such as that in eq.\ (\ref{twistexpan}) includes non partonic quarks.
In certain ``physical'' schemes, defined in subsection \ref{sumrules} to be schemes such as the CWZ schemes in which
no subtraction is made on quantities which are free of UV singularities, 
$G$ depends on non partonic quark masses only, and is singular as such masses vanish (recall from
subsection \ref{CollinsOR} that $G$ is non singular as active parton masses vanish), 
while $\Gamma$ is independent of non partonic quark masses.
When acting on non partonic quark lines, ${\mathcal P}$ must be chosen 
such that it does not introduce large counterterms to cancel power suppressed potential mass singularities,
e.g.\ ${\mathcal P}$ can be chosen to vanish on non partonic quark lines.

The other type of potential mass singularity that a non partonic quark contributes, namely logarithms of 
its mass $m_{q_I}$ that become singular as $m_{q_I}\gg \sqrt{s}$ 
are absorbed into the strong coupling constant in e.g.\ a CWZ scheme.
Note that, for active partons, for which $m_i\ll \sqrt{s}$, such logarithms are power suppressed by $O(m_i^2/s)$.

In calculations of light hadron production, it may or may not 
be possible to neglect the contribution of {\it intrinsic fragmentation},
namely the non partonic components in the product in eq.\ (\ref{twistexpan}).
In the case of PDFs, according to the decoupling theorem
those graphs that contain a non partonic quark $i$ will be suppressed by a power of $\Lambda_{\rm QCD}/m_i$ 
\cite{Witten:1975bh},
which is therefore the relative error on the cross section due to the neglect of the intrinsic PDF of parton $i$.
If such a suppression occurs also for FFs then, as for PDFs,
$D_{n_f+1}$ can be neglected, in which case eq.\ (\ref{FFmatch}) implies that
$D^\prime_{n_f+1}$ is determined entirely from the $D_i$ for $i=0,\ldots,n_f$.
This component of $D^\prime_{n_f+1}$ describes the {\it extrinsic fragmentation} of parton $n_f+1$.
However, the results of Ref.\ \cite{Witten:1975bh} only apply to the case of local operators,
such as those relevant to DIS, but not necessarily to FFs for which non local operators must be used, as was mentioned earlier.
If the intrinsic fragmentation of a non partonic quark is deemed important (e.g.\ charm fragmentation in a 3 flavour scheme), 
calculations would need to be made in which $Z$ acting on non partonic lines takes the required form.
In the meantime, a quark whose intrinsic fragmentation is large must be treated as an active parton, with 
an intrinsic FF that can be fitted to experimental data as is the case for the light partons,
and thus $\sqrt{s}$ cannot be too small relative to its mass.

\subsection{Matching conditions \label{mathcond}}

Next we consider the relation, or {\it matching conditions}, between $D$, and $w$,
in schemes that differ by the number of active partons.
Taking all quantities above to be defined for $n_f$ active quark flavours and using primes to denote quantities 
defined with $n_f+1$ active quark flavours, 
eq.\ (\ref{ffactfromfbareUV}) implies that the matching conditions for FFs and coefficient functions is, respectively,
\be
D'=A D,
\label{FFmatch}
\ee
and
\be
w'=w A^{-1},
\label{matchFFs}
\ee
where $A=G'G^{-1}=\Gamma' \Gamma^{-1}$
is a function only of the ratio of the $+$ component of light cone momenta of its initial and final partons,
as well as of parton masses and indices. 
Diagrammatically it is given by
\be
A=Z\left[1+({\mathcal P}'-{\mathcal P})\frac{1}{1-K_0 (1-{\mathcal P})}K_0\right]Z.
\ee
Note that $A$ remains non singular as the regulator of UV singularities is removed, in contrast to $G$,
and, as for $G$, contains no potential mass singularities due to active parton masses.
$G$ can be chosen such that $A$ depends on non partonic quark masses only, and only through mass logarithms.
This is the case in the CWZ schemes.
Recall that some modification of our incomplete
results in this subsection are needed to ensure that $r$ is properly power suppressed.
Again, similar to $P$ above, 
if $\Gamma$ is chosen to have the simplest possible form as discussed in the paragraph preceding eq.\ (\ref{partonversDGLAP}),
the only parton mass that $A$ can depend on is $m_{n_f+1}$. 
Since the series for $K_0$ starts at $O(a_s)$, since ${\mathcal P}'-{\mathcal P}$ projects out a potential mass singularity from the
$n_f+1$th quark, and since $n$ factors of $K_0$ contain $n-1$ pairs of connecting partons and therefore
a product of $n-1$ potential mass singularities, $A$ in Mellin space and as a matrix in parton species
takes the form
\be
A(N,M_f^2)={\bm 1}+\sum_{n=1}^\infty a_s^n(M_f^2) \sum_{m=0}^n A^{(n)}_m(N) \ln^m \frac{M_f^2}{m_{n_f+1}^2}.
\label{Aexpan}
\ee
From this we obtain the simplest possible form for $\Gamma$:
\be
\Gamma=\prod_{i=0}^{n_f}\left({\bm 1}+\sum_{n=1}^\infty a_s^n(M_f^2) \sum_{m=0}^n A^{(n)}_m(N) \ln^m \frac{M_f^2}{m_i^2}\right).
\label{pertexpandofGamma}
\ee
In practice, the $A^{(n)}_m$ are chosen such that $w$ in eq.\ (\ref{factFhatfrombareFhat}) 
is finite in the limit that active parton masses vanish 
(although these limits do not of course need to be taken in the actual cross section calculations).
The CWZ scheme is obtained by choosing the $A^{(n)}_m$ such that $w$ in eq.\ (\ref{factFhatfrombareFhat}) 
reduces to the $\overline{\rm MS}$ scheme for a theory with $n_f$ massless flavours only in the limits that 
active parton masses vanish and non partonic masses approach infinity
(which again do not need to be taken in actual applications).

\subsection{Open issues \label{openissues}}

We have given a brief summary of the current status of factorization which was developed in Ref.\ \cite{Collins:1998rz}.
This approach both generalizes earlier approaches and solves a number of issues therein.
However, some issues remain. 
In particular, the modification that we have proposed here to the behaviour of $Z$ on non partonic quark lines 
given in Ref.\ \cite{Collins:1998rz},
as well as our proposed relations between the results of Ref.\ \cite{Collins:1998rz} and those of the earlier approach of 
Refs.\ \cite{Ellis:1978ty,Curci:1980uw,Furmanski:1981cw} in terms of ${\mathcal P}$,
need more explicit study in order to ensure that there are no problems with gauge invariance or rapidity divergences,
which are a general consequence of momentum dependent projection operators.
Furthermore, while it has been proved that graphs in which there is an exchange between the hard and soft graphs of gluons
collinear to the detected hadron are power suppressed, a proof is lacking for gluons which are not collinear.

\section{Leading order splitting functions \label{LOsplitfunc}}

The LO coefficients of the splitting functions are given by
\be
\begin{split}
P^{(0)}_{\Sigma \Sigma}(z)=&\ C_F\left(-1-z+2\left[\frac{1}{1-z}\right]_+ + \frac{3}{2}\delta(1-z)\right),\\
P^{(0)}_{\Sigma g}(z)=&\ 2C_F\frac{1+(1-z)^2}{z},\\
P^{(0)}_{g\Sigma}(z)=&\ T_R n_f(z^2+(1-z)^2),\\
P^{(0)}_{gg}(z)=&\ 2C_A \left(\frac{1}{z}-2+z-z^2+\left[\frac{1}{1-z}\right]_+\right)
+\left(\frac{11}{6}C_A-\frac{2}{3}T_R n_f\right)\delta(1-z),
\end{split}
\label{A1}
\ee
where $T_R=1/2$ and, for the color gauge group SU(3), $C_A=3$ and $C_F=4/3$.
The $[f(z)]_+$ operation occurs often in perturbative calculations, and can be most usefully defined by its behaviour
in a convolution (in which it will always appear):
\be
\int_x^1 \frac{dz}{z}\left[f(z)\right]_+ D\left(\frac{x}{z}\right)
=\int_x^1 \frac{dz}{z} f(z)\left[D\left(\frac{x}{z}\right)-zD(x)\right]-D(x) \int_0^x dz f(z) 
\label{defofplus}
\ee
for any functions $f(z)$ and $D(z)$. 
Supposing that $f(z)$ may contain a singularity as $z\rightarrow 1$ but nowhere else,
eq.\ (\ref{defofplus}) is non singular provided $f(z)$ is less singular than $1/(1-z)^2$,
because the quantity in square brackets in the first integral on the right hand side falls to zero as fast as $(1-z)$ in this limit.
On the other hand, without the $[]_+$ operation, eq.\ (\ref{defofplus}) is only non singular provided 
$f(z)$ is less singular than $1/(1-z)$.
By choosing $D(z)=z^{-N}$, removing the common factor $x^{-N}$ 
and then replacing the lower limit of the integration on the left hand side with $0$, 
we find the Mellin transform of $\left[f(z)\right]_+$ to be
\be
\left(\left[f(z)\right]_+\right) (N)=\int^1_0 dz \left(z^{N-1}-1\right)f(z).
\ee

Transforming eq.\ (\ref{A1}) to Mellin space gives
\be
\begin{split}
P^{(0)}_{\Sigma \Sigma}(N)=&\ C_F\left[\frac{3}{2}+\frac{1}{N(N+1)}-2S_1(N)\right], \\
P^{(0)}_{\Sigma g}(N)=&\ 2C_F\frac{N^2+N+2}{(N-1)N(N+1)}, \\
P^{(0)}_{g \Sigma}(N)=&\ T_R n_f\frac{N^2+N+2}{N(N+1)(N+2)}, \\
P^{(0)}_{gg}(N)=&\ 2C_A\bigg[\frac{11}{12}+\frac{1}{N(N-1)}+\frac{1}{(N+1)(N+2)}-S_1(N)\bigg]-\frac{2}{3}T_R n_f,
\label{P0NLOmel}
\end{split}
\ee
where, for integer $N$, the harmonic sum
\be
S_1(N)=\sum_{k=1}^N \frac{1}{k}.
\label{S1defintN}
\ee
Equation (\ref{S1defintN}) can be analytically continued to complex $N \neq -1,-2,\ldots$ \cite{Albino:2005me} 
by making the replacement
$\sum_{k=1}^N \rightarrow \sum_{k=1}^\infty -\sum_{k=N+1}^\infty$ and then making the replacement $k \rightarrow k-N$ in 
the second sum. The result is
\be
S_1(N)=\sum_{k=1}^\infty \frac{N}{k(k+N)}.
\ee
This sum converges, but rather slowly. 
For numerical work, $S_1(N)$ should be calculated using the result \cite{Albino:2005me}
\be
S_1(N)=S_1(N+r)-\sum_{k=1}^{r}\frac{1}{k+N}
\label{SnNandSnNprrel}
\ee
that follows from eq.\ (\ref{S1defintN}), where $r$ should be chosen such that $|N+r|$ is large,
and then calculating $S_1(N+r)$ as an expansion in $1/(N+r)$ \cite{Abramowitz:1968}.

\section{Mellin space \label{mellinspace}}

Any succession of convolutions, which may be written as
\be
f(z)=\int_z^1 \frac{d z_1}{z_1} 
\int_{z_1}^1 \frac{d z_2}{z_2} \ldots \int_{z_{n-2}}^1 \frac{d z_{n-1}}{z_{n-1}}
f_1\left(\frac{z}{z_1}\right) f_2\left(\frac{z_1}{z_2}\right) \ldots f_{n-1}\left(\frac{z_n}{z_{n-1}}\right) f_n(z_{n-1}),
\label{multconvol}
\ee
is converted by the {\it Mellin transform}, defined by the integral transformation
\be
f(N)=\int^1_0 \frac{dz}{z} z^N f(z),
\label{defofmeltrans}
\ee
into an analytically more manageable succession of products
\be
f(N)=f_1(N)f_2(N)\ldots f_n(N),
\ee
which is most easily proved by applying the Mellin transform to an alternative form of eq.\ (\ref{multconvol}),
\be
f(z)=\int_0^1 dz_1 \int_0^1 dz_2 \ldots \int_0^1 dz_n \delta(z-z_1 z_2 \ldots z_n) f_1 (z_1) f_2 (z_2) \ldots f_n (z_n).
\ee
The Mellin transform is invertible via the {\it inverse Mellin transform}
\be
f(z)=\frac{1}{2\pi i}\int_C dN z^{-N} f(N),
\label{invmelltrans}
\ee
where $C$ is a contour in complex $N$ space which starts from a point at ${\rm Im}(N)=-\infty$, ends at
a point at ${\rm Im}(N)=\infty$, and passes to the right of all poles in $f(N)$.
According to Cauchy's theorem, the contour $C$ may be deformed provided that it does not pass through any poles in the process.
The numerical evaluation of the integration in eq.\ (\ref{invmelltrans}) converges fastest when
a contour for which ${\rm Re}(-N)\rightarrow \infty$ is used, because then $z^{-N}$ falls exponentially to zero along it.

Often in a convolution of two functions,
\be
g(z)=\int_z^1 \frac{dz'}{z'}g_1(z')g_2\left(\frac{z}{z'}\right)=\frac{1}{2\pi i}\int_C dN z^{-N} g_1(N) g_2(N),
\label{convofh1h2}
\ee
the analytic Mellin transform of one of these functions, $g_1$ say, may not be calculable.
(Note that eq.\ (\ref{convofh1h2}) includes the cases of multiple convolutions in eq.\ (\ref{multconvol}), 
since $g_1$ can be equated with a subset of the convolutions in eq.\ (\ref{multconvol}) and $g_2$ with the rest.)
An example is the $F^i_{h_1 h_2}$ appearing in eq.\ (\ref{defofXSforiprodfrom2hads}).
In this case the Mellin transform of $g_1=F^i_{h_1 h_2}$ may be performed numerically, but only after dealing with a subtlety:
Equation (\ref{defofmeltrans}) with $f\rightarrow g_1$ implies that the second equality in eq.\ (\ref{convofh1h2}) has
a divergent contribution proportional to $\int_C dN (z/z')^{-N} g_2(N)$ whenever $0< z'<z$,
if the contour $C$ is chosen as discussed at the end of the previous paragraph.
However, inspection of the first equality in eq.\ (\ref{convofh1h2}) reveals that $f(z)$ is independent of 
$g_1(z')$ in this region. Thus $g_1(N)$ in the second equality in eq.\ (\ref{convofh1h2}) must be replaced with
the modified Mellin transform
\be
g_1(N;z)=\int_z^1 \frac{dz'}{z'}z^{\prime N}g_1(z').
\label{modmeltrans}
\ee
It can sometimes happen that the analytic Mellin transform of the other function, $g_2$, also cannot be obtained.
Unfortunately, eq.\ (\ref{modmeltrans}) cannot be used to also calculate the Mellin transform of $g_2$, 
because then the inverse Mellin transform does not converge:
As well as eq.\ (\ref{modmeltrans}), assume a second result which is the same as 
eq.\ (\ref{modmeltrans}) but with $g_1 \rightarrow g_2$ and $z' \rightarrow z''$. 
Then eq.\ (\ref{convofh1h2}) has a divergent contribution proportional to 
$\int_C dN (z/(z'z''))^{-N}$ from the region $z'z''<z$, which exists even though $z'> z$ and $z'' >z$.
One solution is instead to
approximate $g_2$ in $z$ space as e.g.\ an expansion in Chebyshev polynomials, and then analytically Mellin transform it.

\pagebreak
\section{Summary of inclusive single hadron production measurements \label{sumofexp}}

Only \cpi, \cka, \pr, \nka\ and \lam\ data 
which can be reasonably reliably calculated, and which are therefore suitable for global fits, 
are listed here.

\subsection{$e^+ e^-$ reactions}

\begin{table}[h!]
\parbox{8cm}{
\caption{Summary of the measurements for inclusive single \cpi\ production in $e^+ e^-$ reactions.
The column labeled ``\# data'' gives the number of data for which $x \geq 0.05$.
The last column refers to the normalization uncertainty on the data. \label{PionResults}}
\begin{center}
\begin{tabular}{|c|c|c|c|c|}
\hline 
\multirow{2}{*}{Collaboration} & \multirow{2}{*}{Tagging}
& \multirow{2}{*}{\vspace{0.3cm} $\sqrt{s}$} & \multirow{2}{*}{\vspace{0.3cm} \#} & \multirow{2}{*}{\vspace{0.3cm} Norm.\ } \\
& & (GeV) & data & (\%) \\
\hline \hline
\input{PionTable}
\hline
\end{tabular}
\end{center}}
\parbox{8cm}{
\caption{As in Table \ref{PionResults}, but for \cka. \label{KaonResults}}
\begin{center}
\begin{tabular}{|c|c|c|c|c|}
\hline 
\multirow{2}{*}{Collaboration} & \multirow{2}{*}{Tagging}
& \multirow{2}{*}{\vspace{0.3cm} $\sqrt{s}$} & \multirow{2}{*}{\vspace{0.3cm} \#} & \multirow{2}{*}{\vspace{0.3cm} Norm.\ } \\
& & (GeV) & data & (\%) \\
\hline \hline
\input{KaonTable}
\hline
\end{tabular}
\end{center}}
\end{table}

\pagebreak
\begin{table}[h!]
\parbox{8cm}{
\caption{As in Table \ref{PionResults}, but for \pr. \label{ProtonResults}}
\begin{center}
\begin{tabular}{|c|c|c|c|c|}
\hline 
\multirow{2}{*}{Collaboration} & \multirow{2}{*}{Tagging}
& \multirow{2}{*}{\vspace{0.3cm} $\sqrt{s}$} & \multirow{2}{*}{\vspace{0.3cm} \#} & \multirow{2}{*}{\vspace{0.3cm} Norm.\ } \\
& & (GeV) & data & (\%) \\
\hline \hline
\input{ProtonTable}
\hline
\end{tabular}
\end{center}}
\parbox{8cm}{
\caption{As in Table \ref{PionResults}, but for \nka. \label{K0SResults}}
\begin{center}
\begin{tabular}{|c|c|c|c|c|}
\hline 
\multirow{2}{*}{Collaboration} & \multirow{2}{*}{Tagging}
& \multirow{2}{*}{\vspace{0.3cm} $\sqrt{s}$} & \multirow{2}{*}{\vspace{0.3cm} \#} & \multirow{2}{*}{\vspace{0.3cm} Norm.\ } \\
& & (GeV) & data & (\%) \\
\hline \hline
\input{K0STable}
\hline
\end{tabular}
\end{center}}
\end{table}

\pagebreak
\begin{table}[h!]
\caption{As in Table \ref{PionResults}, but for $\Lambda/\bar{\Lambda}$. \label{LambdaResults}}
\begin{center}
\begin{tabular}{|c|c|c|c|c|}
\hline 
\multirow{2}{*}{Collaboration} & \multirow{2}{*}{Tagging}
& \multirow{2}{*}{\vspace{0.3cm} $\sqrt{s}$} & \multirow{2}{*}{\vspace{0.3cm} \#} & \multirow{2}{*}{\vspace{0.3cm} Norm.\ } \\
& & (GeV) & data & (\%) \\
\hline \hline
\input{LambdaTable}
\hline
\end{tabular}
\end{center}
\end{table}

\pagebreak
\subsection{$pp(\overline{p})$ reactions}

\begin{table}[h!]
\parbox{8.6cm}{
\caption{As in Table \ref{PionResults}, but for $pp(\overline{p})$ reactions. 
In the case of the BRAHMS data, the values in brackets are the normalization errors below 3 GeV. \label{ppPionResults}}
\begin{center}
\begin{tabular}{|c|c|c|c|c|}
\hline 
\multirow{2}{*}{Collaboration} & \multirow{2}{*}{Rapidity}
& \multirow{2}{*}{\vspace{0.3cm} $\sqrt{s}$} & \multirow{2}{*}{\vspace{0.3cm} \#} & \multirow{2}{*}{\vspace{0.3cm} Norm.\ } \\
& & (GeV) & data & (\%) \\
\hline \hline
\input{ppPionTable}
\hline
\end{tabular}
\end{center}}
\parbox{8.6cm}{
\caption{As in Table \ref{ppPionResults}, but for \cka. \label{ppKaonResults}}
\begin{center}
\begin{tabular}{|c|c|c|c|c|}
\hline 
\multirow{2}{*}{Collaboration} & \multirow{2}{*}{Rapidity}
& \multirow{2}{*}{\vspace{0.3cm} $\sqrt{s}$} & \multirow{2}{*}{\vspace{0.3cm} \#} & \multirow{2}{*}{\vspace{0.3cm} Norm.\ } \\
& & (GeV) & data & (\%) \\
\hline \hline
\input{ppKaonTable}
\hline
\end{tabular}
\end{center}}
\end{table}

\begin{table}[h!]
\caption{As in Table \ref{ppPionResults}, but for $p(\overline{p})$. \label{ppProtonResults}}
\begin{center}
\begin{tabular}{|c|c|c|c|c|}
\hline 
\multirow{2}{*}{Collaboration} & \multirow{2}{*}{Rapidity}
& \multirow{2}{*}{\vspace{0.3cm} $\sqrt{s}$} & \multirow{2}{*}{\vspace{0.3cm} \#} & \multirow{2}{*}{\vspace{0.3cm} Norm.\ } \\
& & (GeV) & data & (\%) \\
\hline \hline
\input{ppProtonTable}
\hline
\end{tabular}
\end{center}
\end{table}

\begin{table}[h!]
\parbox{8.5cm}{
\caption{As in Table \ref{ppPionResults}, but for \nka. \label{ppK0SResults}}
\begin{center}
\begin{tabular}{|c|c|c|c|c|}
\hline 
\multirow{2}{*}{Collaboration} & \multirow{2}{*}{Rapidity}
& \multirow{2}{*}{\vspace{0.3cm} $\sqrt{s}$} & \multirow{2}{*}{\vspace{0.3cm} \#} & \multirow{2}{*}{\vspace{0.3cm} Norm.\ } \\
& & (GeV) & data & (\%) \\
\hline \hline
\input{ppK0STable}
\hline
\end{tabular}
\end{center}}
\parbox{8.5cm}{
\caption{As in Table \ref{ppPionResults}, but for \lam. \label{ppLambdaResults}}
\begin{center}
\begin{tabular}{|c|c|c|c|c|}
\hline 
\multirow{2}{*}{Collaboration} & \multirow{2}{*}{Rapidity}
& \multirow{2}{*}{\vspace{0.3cm} $\sqrt{s}$} & \multirow{2}{*}{\vspace{0.3cm} \#} & \multirow{2}{*}{\vspace{0.3cm} Norm.\ } \\
& & (GeV) & data & (\%) \\
\hline \hline
\input{ppLambdaTable}
\hline
\end{tabular}
\end{center}}
\end{table}

\end{document}

%% file: hadmasssummary
$\pi^\pm$ &  154.6 &  139.6 \\
\hline
$K^\pm$ &  337.0 &  493.7 \\
\hline
$p/\bar{p}$ &  948.8 &  938.3 \\
\hline
$K^0_S$ &  343.0 &  497.6 \\
\hline
$\Lambda/\overline{\Lambda}$ & 1127.0 & 1115.7 \\
\hline

%% file: chi2summary
$\pi^\pm$ &  518.7 &  519.0 \\
\hline
$K^\pm$ &  416.6 &  439.4 \\
\hline
$p/\bar{p}$ &  525.2 &  538.0 \\
\hline
$K^0_S$ &  317.2 &  318.7 \\
\hline
$\Lambda/\overline{\Lambda}$ &  273.1 &  325.7 \\
\hline

%% file: PionTable
TASSO \cite{Brandelik:1980iy}    & untagged &   12  & 5  & 20  \\
\hline
TASSO \cite{Althoff:1982dh}      & untagged &   14  & 10 & 8.5 \\
\hline
TASSO \cite{Althoff:1982dh}      & untagged &   22  & 1  & 6.3 \\
\hline
HRS \cite{Derrick:1985wd}        & untagged &   29  & 6  &    \\
\hline
TPC \cite{Aihara:1986mv}         & l tagged &   29  & 9  &    \\
\hline
TPC \cite{Aihara:1986mv}         & c tagged &   29  & 9  &    \\
\hline
TPC \cite{Aihara:1986mv}         & b tagged &   29  & 9  &    \\
\hline
TPC \cite{Aihara:1988fc}         & untagged &   29  & 27 &    \\
\hline
TASSO \cite{Brandelik:1980iy}    & untagged &   30  & 4  & 20  \\
\hline
TASSO \cite{Braunschweig:1988hv} & untagged &   34  & 10 & 6   \\
\hline
TASSO \cite{Braunschweig:1988hv} & untagged &   44  & 7  & 6   \\
\hline
TOPAZ \cite{Itoh:1994kb}         & untagged &   58  & 8  &    \\
\hline
ALEPH \cite{Buskulic:1994ft}     & untagged & 91.2  & 22 & 3   \\
\hline
DELPHI \cite{Abreu:1998vq}       & l tagged & 91.2  & 17 &    \\
\hline
DELPHI \cite{Abreu:1998vq}       & b tagged & 91.2  & 17 &    \\
\hline
DELPHI \cite{Abreu:1998vq}       & untagged & 91.2  & 17 &    \\
\hline
OPAL \cite{Abbiendi:1999ry}      & u tagged & 91.2  & 5  &    \\
\hline
OPAL \cite{Abbiendi:1999ry}      & d tagged & 91.2  & 5  &    \\
\hline
OPAL \cite{Abbiendi:1999ry}      & s tagged & 91.2  & 5  &    \\
\hline
OPAL \cite{Abbiendi:1999ry}      & c tagged & 91.2  & 5  &    \\
\hline
OPAL \cite{Abbiendi:1999ry}      & b tagged & 91.2  & 5  &    \\
\hline
OPAL \cite{Akers:1994ez}         & untagged & 91.2  & 20 &    \\
\hline
SLD \cite{Abe:2003iy}            & l tagged & 91.2  & 28 &    \\
\hline
SLD \cite{Abe:2003iy}            & c tagged & 91.2  & 28 &    \\
\hline
SLD \cite{Abe:2003iy}            & b tagged & 91.2  & 28 &    \\
\hline
SLD \cite{Abe:2003iy}            & untagged & 91.2  & 28 &    \\
\hline
DELPHI \cite{Abreu:2000gw}       & untagged & 189   & 3  &    \\
\hline \hline
Total      &      &     & 338 & \\

%% file: KaonTable
TASSO \cite{Brandelik:1980iy}    & untagged &   12  & 3  & 20  \\
\hline
TASSO \cite{Althoff:1982dh}      & untagged &   14  & 9 & 8.5 \\
\hline
TASSO \cite{Althoff:1982dh}      & untagged &   22  & 7  & 6.3 \\
\hline
HRS \cite{Derrick:1985wd}        & untagged &   29  & 7  &    \\
\hline
MARKII \cite{Schellman:1984yz}   & untagged &   29  & 2  & 12  \\
\hline
TPC \cite{Aihara:1988fc}         & untagged &   29  & 26  &    \\
\hline
TASSO \cite{Brandelik:1980iy}    & untagged &   30  & 2  & 20  \\
\hline
TASSO \cite{Braunschweig:1988hv} & untagged &   34  & 5 & 6   \\
\hline
TOPAZ \cite{Itoh:1994kb}         & untagged &   58  & 5  &    \\
\hline
ALEPH \cite{Buskulic:1994ft,Barate:1996fi}     & untagged & 91.2  & 18 & 3   \\
\hline
DELPHI \cite{Abreu:1998vq}       & l tagged & 91.2  & 17 &    \\
\hline
DELPHI \cite{Abreu:1998vq}       & b tagged & 91.2  & 17 &    \\
\hline
DELPHI \cite{Abreu:1998vq}       & untagged & 91.2  & 17 &    \\
\hline
OPAL \cite{Abbiendi:1999ry}      & u tagged & 91.2  & 5  &    \\
\hline
OPAL \cite{Abbiendi:1999ry}      & d tagged & 91.2  & 5  &    \\
\hline
OPAL \cite{Abbiendi:1999ry}      & s tagged & 91.2  & 5  &    \\
\hline
OPAL \cite{Abbiendi:1999ry}      & c tagged & 91.2  & 5  &    \\
\hline
OPAL \cite{Abbiendi:1999ry}      & b tagged & 91.2  & 5  &    \\
\hline
OPAL \cite{Akers:1994ez}         & untagged & 91.2  & 10 &    \\
\hline
SLD \cite{Abe:2003iy}            & l tagged & 91.2  & 28 &    \\
\hline
SLD \cite{Abe:2003iy}            & c tagged & 91.2  & 28 &    \\
\hline
SLD \cite{Abe:2003iy}            & b tagged & 91.2  & 28 &    \\
\hline
SLD \cite{Abe:2003iy}            & untagged & 91.2  & 28 &    \\
\hline
DELPHI \cite{Abreu:2000gw}       & untagged & 189   & 3  &    \\
\hline \hline
Total      &      &     & 286 &    \\

%% file: ProtonTable
TASSO \cite{Brandelik:1980iy}    & untagged &   12  & 3  & 20  \\
\hline
TASSO \cite{Althoff:1982dh}      & untagged &   14  & 9 & 8.5 \\
\hline
TASSO \cite{Althoff:1982dh}      & untagged &   22  & 9  & 6.3 \\
\hline
HRS \cite{Derrick:1985wd}        & untagged &   29  & 7  &    \\
\hline
TPC \cite{Aihara:1988fc}         & untagged &   29  & 20  &    \\
\hline
TASSO \cite{Brandelik:1980iy}    & untagged &   30  & 3  & 20  \\
\hline
JADE \cite{Bartel:1981sw}    & untagged &   34  & 2  & 14  \\
\hline
TASSO \cite{Braunschweig:1988hv} & untagged &   34  & 7 & 6   \\
\hline
TOPAZ \cite{Itoh:1994kb} & untagged &   58  & 5 &    \\
\hline
ALEPH \cite{Buskulic:1994ft,Barate:1996fi}     & untagged & 91.2  & 18 & 3   \\
\hline
DELPHI \cite{Abreu:1998vq}       & l tagged & 91.2  & 17 &    \\
\hline
DELPHI \cite{Abreu:1998vq}       & b tagged & 91.2  & 17 &    \\
\hline
DELPHI \cite{Abreu:1998vq}       & untagged & 91.2  & 17 &    \\
\hline
OPAL \cite{Abbiendi:1999ry}      & u tagged & 91.2  & 5  &    \\
\hline
OPAL \cite{Abbiendi:1999ry}      & d tagged & 91.2  & 5  &    \\
\hline
OPAL \cite{Abbiendi:1999ry}      & s tagged & 91.2  & 5  &    \\
\hline
OPAL \cite{Abbiendi:1999ry}      & c tagged & 91.2  & 5  &    \\
\hline
OPAL \cite{Abbiendi:1999ry}      & b tagged & 91.2  & 5  &    \\
\hline
OPAL \cite{Akers:1994ez}         & untagged & 91.2  & 10 &    \\
\hline
SLD \cite{Abe:2003iy}            & l tagged & 91.2  & 29 &    \\
\hline
SLD \cite{Abe:2003iy}            & c tagged & 91.2  & 29 &    \\
\hline
SLD \cite{Abe:2003iy}            & b tagged & 91.2  & 29 &    \\
\hline
SLD \cite{Abe:2003iy}            & untagged & 91.2  & 29 &    \\
\hline
DELPHI \cite{Abreu:2000gw}       & untagged & 189   & 3  &    \\
\hline \hline
Total      &      &     & 289 &  \\

%% file: K0STable
TASSO \cite{Althoff:1984iz}    & untagged &   14  & 8  & 15  \\
\hline
TASSO \cite{Braunschweig:1989wg}      & untagged &   14.8  & 8 &  \\
\hline
TASSO \cite{Braunschweig:1989wg}      & untagged &   21.5  & 5 &  \\
\hline
TASSO \cite{Althoff:1984iz}      & untagged &   22  & 5  & 15 \\
\hline
HRS \cite{Derrick:1985wd}        & untagged &   29  & 12  &    \\
\hline
MARK II \cite{Schellman:1984yz}      & untagged &   29  & 17  & 12 \\
\hline
TPC \cite{Aihara:1984mk}         & untagged &   29  & 7  &    \\
\hline
TASSO \cite{Brandelik:1981ta} & untagged &   33.3  & 7 & 15   \\
\hline
TASSO \cite{Althoff:1984iz} & untagged &   34  & 13 & 15   \\
\hline
TASSO \cite{Braunschweig:1989wg} & untagged &   34.5  & 13 &    \\
\hline
CELLO \cite{Behrend:1989ae} & untagged &   35  & 9 &    \\
\hline
TASSO \cite{Braunschweig:1989wg} & untagged &   35  & 13 &    \\
\hline
TASSO \cite{Braunschweig:1989wg} & untagged &   42.6  & 13 &    \\
\hline
TOPAZ \cite{Itoh:1994kb}         & untagged &   58  & 4  &    \\
\hline
ALEPH \cite{Barate:1996fi}     & untagged & 91.2  & 16 & 2   \\
\hline
DELPHI \cite{Abreu:1994rg}       & untagged & 91.2  & 13 &    \\
\hline
OPAL \cite{Abbiendi:1999ry}      & u tagged & 91.2  & 5  &    \\
\hline
OPAL \cite{Abbiendi:1999ry}      & d tagged & 91.2  & 5  &    \\
\hline
OPAL \cite{Abbiendi:1999ry}      & s tagged & 91.2  & 5  &    \\
\hline
OPAL \cite{Abbiendi:1999ry}      & c tagged & 91.2  & 5  &    \\
\hline
OPAL \cite{Abbiendi:1999ry}      & b tagged & 91.2  & 5  &    \\
\hline
OPAL \cite{Abbiendi:2000cv}         & untagged & 91.2  & 16 &  6  \\
\hline
SLD \cite{Abe:1998zs}            & l tagged & 91.2  & 9 &    \\
\hline
SLD \cite{Abe:1998zs}            & c tagged & 91.2  & 9 &    \\
\hline
SLD \cite{Abe:1998zs}            & b tagged & 91.2  & 9 &    \\
\hline
SLD \cite{Abe:1998zs}            & untagged & 91.2  & 9 &    \\
\hline
DELPHI \cite{Abreu:2000gw}       & untagged & 183   & 2  &    \\
\hline
DELPHI \cite{Abreu:2000gw}       & untagged & 189   & 3  &    \\
\hline \hline
Total      &      &     & 252 & \\

%% file: LambdaTable
TASSO \cite{Althoff:1984iz}      & untagged &   14  & 3 & 20 \\
\hline
TASSO \cite{Althoff:1984iz}      & untagged &   22  & 4  & 20 \\
\hline
HRS \cite{Baringer:1986jd}  & untagged &   29  & 12  &    \\
\hline
MARK II \cite{delaVaissiere:1984xg}  & untagged &   29  & 15  &    \\
\hline
TASSO \cite{Brandelik:1981ta}    & untagged &   33.3  & 6  & 15  \\
\hline
TASSO \cite{Althoff:1984iz} & untagged &   34  & 6 & 20   \\
\hline
TASSO \cite{Braunschweig:1988wh} & untagged &   34.8  & 9 & 9   \\
\hline
CELLO \cite{Behrend:1989ae}         & untagged &   35  & 7  &    \\
\hline
TASSO \cite{Braunschweig:1988hv} & untagged &   42.1  & 4  & 9   \\
\hline
ALEPH \cite{Barate:1996fi}     & untagged & 91.2  & 16 & 4   \\
\hline
DELPHI \cite{Abreu:1993mm}       & untagged & 91.2  & 7 &    \\
\hline
OPAL \cite{Abbiendi:1999ry}      & u tagged & 91.2  & 5  &    \\
\hline
OPAL \cite{Abbiendi:1999ry}      & d tagged & 91.2  & 5  &    \\
\hline
OPAL \cite{Abbiendi:1999ry}      & s tagged & 91.2  & 5  &    \\
\hline
OPAL \cite{Abbiendi:1999ry}      & c tagged & 91.2  & 5  &    \\
\hline
OPAL \cite{Abbiendi:1999ry}      & b tagged & 91.2  & 5  &    \\
\hline
OPAL \cite{Alexander:1996qj}         & untagged & 91.2  & 12 &    \\
\hline
SLD \cite{Abe:1998zs}            & l tagged & 91.2  & 4 &    \\
\hline
SLD \cite{Abe:1998zs}            & c tagged & 91.2  & 4 &    \\
\hline
SLD \cite{Abe:1998zs}            & b tagged & 91.2  & 4 &    \\
\hline
SLD \cite{Abe:1998zs}            & untagged & 91.2  & 9 &    \\
\hline
DELPHI \cite{Abreu:2000gw}       & untagged & 183   & 3  &    \\
\hline
DELPHI \cite{Abreu:2000gw}       & untagged & 189   & 3  &    \\
\hline \hline
Total      &      &     & 145 &  \\

%% file: ppPionTable
\multirow{2}{*}{\vspace{-0.6cm} BRAHMS \cite{Arsene:2007jd}}  &
\multirow{2}{*}{\vspace{0.3cm} $y\in [2.9,3]$} & \multirow{2}{*}{200}
& 8  & \multirow{2}{*}{\vspace{0.3cm} 11,7,8(13),} \\
\cline{2-2} \cline{4-4} & $y\in [3.25,3.35]$ &
& 7  & \multirow{2}{*}{\vspace{0.3cm} 2,1(3)} \\
\hline
PHENIX \cite{Adler:2003pb} ($\pi^0$) & $|\eta|<0.35$ & 200  & 13  & 9.7 \\
\hline
STAR \cite{Adams:2006uz} ($\pi^0$) & $\eta=3.3$ & 200   & 4  & 16  \\
\hline
STAR \cite{Adams:2006uz} ($\pi^0$) & $\eta=3.8$ & 200   & 2  & 16  \\
\hline
STAR \cite{Adams:2006nd} & $|y|<0.5$ & 200 & 10 & 11.7  \\
\hline \hline
Total      &      &     & 44 &     \\

%% file: ppKaonTable
\multirow{2}{*}{\vspace{-0.6cm} BRAHMS \cite{Arsene:2007jd}}  &
\multirow{2}{*}{\vspace{0.3cm} $y\in [2.9,3]$} & \multirow{2}{*}{200}
& 8  & \multirow{2}{*}{\vspace{0.3cm} 11,7,8(13),} \\
\cline{2-2} \cline{4-4} & $y\in [3.25,3.35]$ &
& 6  & \multirow{2}{*}{\vspace{0.3cm} 2,1(3)} \\
\hline
CDF \cite{Acosta:2005pk} ($K_S^0$) & $|\eta|<1$ & 630 & 37 & 10 \\
\hline \hline
STAR \cite{Adams:2006nd} ($K_S^0$) & $|y|<0.5$ & 200 & 9 & 11.7 \\
\hline \hline
Total      &      &     & 60 &      \\

%% file: ppProtonTable
\multirow{2}{*}{\vspace{-0.6cm} BRAHMS \cite{Arsene:2007jd}}  &
\multirow{2}{*}{\vspace{0.3cm} $y\in [2.9,3]$} & \multirow{2}{*}{200}
& 7  & \multirow{2}{*}{\vspace{0.3cm} 11,7,8(13),}  \\
\cline{2-2} \cline{4-4} & $y\in [3.25,3.35]$ &
& 5  & \multirow{2}{*}{\vspace{0.3cm} 2,1(3)}  \\
\hline
STAR \cite{Adams:2006nd} & $|y|<0.5$ & 200 & 8 & 11.7 \\
\hline \hline
Total      &      &     & 20 &  \\

%% file: ppK0STable
\multirow{2}{*}{\vspace{-0.4cm} BRAHMS \cite{Arsene:2007jd}} ($K^\pm$) &
\multirow{2}{*}{\vspace{0.3cm} $y\in [2.9,3]$} & \multirow{2}{*}{200}
& 8  & \multirow{2}{*}{\vspace{0.3cm} 11,7,8(13),}  \\
\cline{2-2} \cline{4-4} & $y\in [3.25,3.35]$ &
& 6  & \multirow{2}{*}{\vspace{0.3cm} 2,1(3)}   \\
\hline
CDF \cite{Acosta:2005pk} & $|\eta|<1$ & 630 & 48 & \\
\hline
STAR \cite{Adams:2006nd} & $|y|<0.5$ & 200   & 9  & 11.7  \\
\hline \hline
Total      &      &     & 71 &    \\

%% file: ppLambdaTable
CDF \cite{Acosta:2005pk} & $|\eta|<1$ & 630 & 34 & 10  \\
\hline
STAR \cite{Abelev:2006cs} & $|y|<0.5$ & 200 & 9 & 11.7  \\
\hline \hline
Total      &      &     & 43 &  \\